\documentclass{article}
\usepackage{graphicx,color}
\usepackage{deleq}
\usepackage[latin1]{inputenc}
\usepackage{amsmath}

\newcommand{\bk}{{\mathbf k}}
\newcommand{\bR}{{\mathbf R}}
\newcommand{\br}{{\mathbf r}}
\newcommand{\phan}{{$^{\phantom {P^{P^P}}}_{\phantom {P_{P_P}}}$}}

\newcommand{\eq}[1]{(\ref{#1})}
\newcommand{\fig}[1]{figure \ref{#1}}
\newcommand{\tab}[1]{table \ref{#1}}
\newcommand{\chap}[1]{section \ref{#1}}
\newcommand{\sect}[1]{section \ref{#1}}

\newcommand{\bit}{\begin{itemize}}
\newcommand{\eit}{\end{itemize}}

\newcommand{\etal}{{\em et al.}}

\begin{document}

\begin{center}
{\itshape \large Advances in Physics:}\\[0.3cm]
{\Large \bf Gutzwiller-RVB Theory of}\\[0.1cm]
{\Large \bf High Temperature Superconductivity:}\\[0.15cm]
{\Large \bf Results from Renormalised Mean Field Theory}\\[0.1cm]
{\Large \bf and Variational Monte Carlo Calculations}\\[0.4cm]

{\large B. Edegger$^*$, V.N. Muthukumar$^\dagger$, and C. Gros$^*$}\\[0.1cm]
{\footnotesize $^*$ Institute for Theoretical
Physics, Universität Frankfurt, D-60438 Frankfurt, Germany\\[-0.1cm]
$^\dagger$ Department of Physics,
Princeton University, Princeton, NJ 08544, USA}
\end{center}


\begin{abstract}
\noindent We review the Resonating Valence Bond (RVB) theory of
high temperature superconductivity using Gutzwiller projected wave
functions that incorporate strong correlations. After a general
overview of the phenomenon of high temperature superconductivity,
we discuss Anderson's RVB picture and its implementation by
renormalised mean field theory (RMFT) and variational Monte Carlo
(VMC) techniques. We review RMFT and VMC results with an emphasis
on recent developments in extending VMC and RMFT techniques to
excited states. We compare results obtained from these methods
with angle resolved photoemission spectroscopy (ARPES) and
scanning tunnelling microscopy (STM). We conclude by summarising
recent successes of this approach and discuss open problems that
need to be solved for a consistent and complete description of
high temperature superconductivity using Gutzwiller projected wave
functions.
\end{abstract}

\tableofcontents

\section{Introduction}
\label{chap_intro}

This paper reviews developments in the use of Gutzwiller projected
wave functions and the Resonating Valence Bond (RVB) theory in the
context of high temperature superconductivity. We attempt to
review comprehensively, both the general framework of the
Gutzwiller-RVB theory and to summarise several recent results in
this field. Though many of these results were indeed motivated by
the phenomenon of high temperature superconductivity and the rich
phase diagram of these compounds, it is not our intention to
review high temperature superconductivity \textit{per se}.
Nonetheless, it is well nigh impossible, if not meaningless, to
attempt to write a review of this nature without discussing
certain key experimental results. Our choice in this matter is
dictated by the fact that most techniques used in the study of
Gutzwiller projected wave functions address the calculation of
single particle spectral features. Consequently, after discussing
some basic facts and a historical perspective of the
Gutzwiller-RVB concept, we present an overview of experimental
results from angle resolved photoemission spectroscopy (ARPES) and
scanning tunnelling microscopy (STM) within this introductory
section. We also discuss briefly, a few alternative theories based
on repulsive electronic models, to illustrate the complexity of
the subject.


\subsection{High temperature superconductivity}

\begin{figure}
\centering
\includegraphics[width=0.75\textwidth]{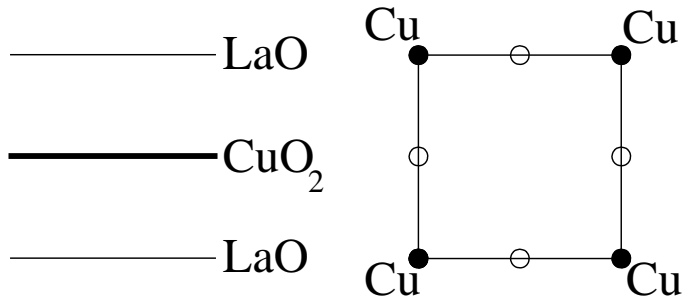}
\caption{Crystal structure of La$_2$CuO$_4$. Left panel shows the
layer structure along the c-axis, the right panel the structure of
the CuO$_2$ plane. From \cite{Norman03}.} \label{structure}
\end{figure}

Twenty years ago Bednorz and Müller \cite{Bednorz86} discovered
high temperature superconductivity in Sr-doped La$_2$CuO$_4$.
Subsequently high temperature superconductivity was reported in
many other Cuprates. These compounds have a layered structure made up
of one or more copper-oxygen planes (see \fig{structure}). It was
soon realised that many of the HTSC have an insulating
antiferromagnetic parent compound that becomes superconducting when
doped with holes or electrons. This is fundamentally
different from, say, superconductivity in alkaline metals and clearly
calls for a novel mechanism.

These unusual observations stimulated an enormous amount of
experimental as well as theoretical works on HTSC, which brought about
numerous new insights into these fascinating compounds. The $d$-wave
nature of the superconducting pairs \cite{Tsuei00} as well as the
generic temperature-doping phase diagram (\fig{PhaseDiagram}) are
now well established. On the theoretical front,
several approaches successfully describe at least some features of
the HTSC. In addition, new sophisticated numerical techniques provide
us with a better understanding of strong correlation effects
that are clearly present in the HTSC. Progress
in the field of high temperature superconductivity has also
influenced many other fields in condensed matter physics greatly.
Research on HTSC has a very fruitful history, and continues
to broaden our knowledge of strongly correlated
electron systems.

Given the numerous theories advanced to explain the phenomenon
of high temperature superconductivity \cite{Scalapino07}, it
is important to examine carefully the strengths and weaknesses
of any given theoretical approach and its relevance to
experimental observations. In this review, we
examine the resonating valence bond (RVB) scenario which
proposes a simple, yet nontrivial wave function to
describe the ground state of Mott Hubbard superconductors, {\it i.e.}
superconductors
that are obtained by doping a Mott Hubbard insulator.
We discuss various theoretical calculations based on the so called Gutzwiller-RVB
wave function both in the context of our work
\cite{Edegger05a,Fukushima05,Edegger06a,Edegger06b,Gros06,Edegger06c}
and other recent developments.

The Gutzwiller-RVB theory provides a direct description of strongly correlated
superconductors. An advantage of this approach is that the theory can be studied
by a variety of approximate analytical techniques as well as numerical methods. We will
discuss later how the theory yields many results that are in broad agreement with various
key experimental facts. However, to
%
obtain a more complete description of HTSC, the Gutzwiller-RVB calculations
need to be be extended to be able to describe finite temperature and dynamic effects.
This review should provide an adequate starting point for further
extensions of this method as well as phenomenological calculations of
various physical quantities that are relevant to the phenomenon of
high temperature superconductivity.

\subsection{A historical perspective}

The notion of Resonating Valence Bonds was introduced by
Pauling \cite{Pauling38,Pauling48} in the context
of the Heitler-London approximation for
certain types of non-classical molecular structures.
Anderson and Fazekas \cite{Anderson73,Fazekas74}
then generalised this concept
to the case of frustrated magnetism of localised
spin-1/2 moments. The RVB theory came to a first
full bloom with the discovery of high-temperature
superconductivity when
Anderson \cite{Anderson87} suggested that
an RVB state naturally leads to
incipient superconductivity from preformed
singlet pairs in the parent insulating state.

A detailed account of the progress made after
Anderson's seminal RVB proposal will be
presented in this review in subsequent sections.
At this point we will make a few comments regarding
the general lines of development of the theory.

The core of the RVB concept is variational in
nature; the RVB state may be regarded
as an unstable fixed point leading to
various instabilities, such as antiferromagnetic order,
superconductivity, \textit{etc.}, very much like the Fermi-liquid
state. However, in contrast to Fermi-liquid theory, there
is no simple Hamiltonian known for which
the RVB states discussed in this review are exact solutions.
For this reason, the theory developed historically
along several complementary lines. The first
one is the quantification of the variational
approach \textit{via} the variational Monte Carlo method (VMC).
This approach was initially hampered by the problem
of implementing the numerical evaluation of a
general RVB wave function algorithmically \cite{GrosJoynt87}.
But when this problem was solved \cite{Gros88}, the method evolved
quickly into a standard numerical technique.

Very early on it was realized \cite{Kotliar88}, that
essential aspects of the RVB concept could be formulated
within a slave-boson approach, which led to
the development of gauge theories for strongly
correlated electronic systems in general, and
high temperature superconductivity in particular.
This line of thought has been reviewed comprehensively by
Lee, Nagaosa, and Wen
\cite{Lee06}.

The superconducting state is an ordered state and this
statement applies also to the case of the high temperature
superconductors. Where there is an order parameter, there
is a mean field and it was felt early on that a suitable
mean-field theory should be possible when formulated in
the correct Hilbert space, using the appropriate
order parameters. This line of thought led to the development
of the renormalised mean-field theory (RMFT) \cite{Zhang88b}.
This theory will play a prominent role in this review, as it
allows for qualitative analytical predictions and, in some
cases, also for quantitative evaluations of experimentally
accessible response functions.

There is a certain historical oddity concerning the development of
the RVB concept and of the theory. After an initial flurry, there
was relatively little activity in the 1990's and the
Gutzwiller-RVB approach returned into the center of scientific
interest only in the last decade with the evaluation of several
new response functions \cite{Paramekanti01}, allowing for a
detailed comparison with the (then) newly available experimental
results. In retrospect, is not quite clear why this particular
approach lay idle for nearly a decade. It is tempting to speculate
that perhaps the concept was too successful initially, predicting
$d$-wave superconductivity in the cuprates at a time when
available experimental results favoured an $s$-wave.

\subsection{Experiments}
\label{experiments}

\begin{figure}
\centering
\includegraphics[width=0.75\textwidth]{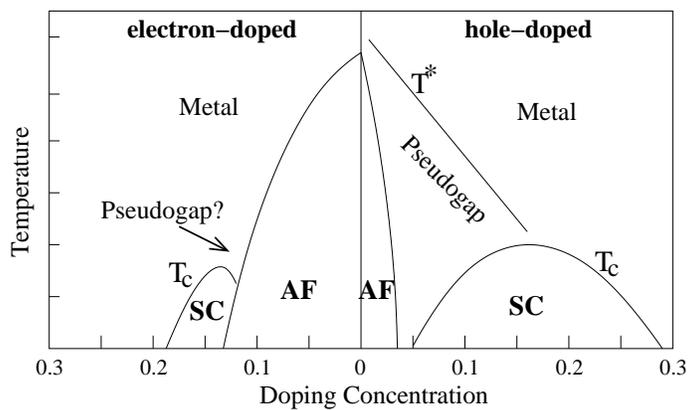}
\caption{Generic phase diagram for the high temperature
superconductors (antiferromagnetic region AF, superconducting
phase SC). The temperature below which superconductivity (a
pseudogap) is observed is denoted by $T_c$ ($T^*$). $T^*$
is possibly a crossover temperature, though some experiments
(compare figure \ref{BCS_ratio}) indicate a relation to
a mean-field like second order transition.
               }
\label{PhaseDiagram}
\end{figure}

The discovery of high temperature superconductivity stimulated the
development of several new experimental techniques. Here,
we shall mention some key experimental facts concerning
the HTSC and refer the reader to more detailed
summaries of experimental results, available in the literature
\cite{Norman03,Tsuei00,Lee06,Norman05,Damascelli03,Campuzano04}.

An early and significant result was the realization that
HTSC are doped Mott insulators, as shown in the
generic
temperature-doping phase diagram (see \fig{PhaseDiagram}). The figure
shows the antiferromagnetic phase in the undoped
(half-filled\footnote{The copper ion is in a $d^9$ configuration,
with a single hole in the $d$-shell per unit cell.
As shown by Zhang and Rice \cite{Zhang88a} this situation
corresponds to a half-filled band in an effective single-band
model.}) compound with a Neel temperature of about $T_N \approx
300 K$. Upon doping, antiferromagnetism is suppressed and
superconductivity emerges. The behaviour of $T_c$ with doping
exhibits a characteristic ``dome''. While
electron- and hole-doped HTSC share
many common features, they do exhibit some significant
differences, {\it e.g.} the antiferromagnetic region persists to
much higher doping levels for electron-doped Cuprates.

\begin{figure}
\centering
\includegraphics*[width=0.45\textwidth]{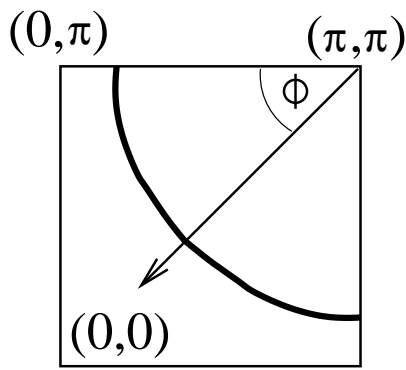}
\caption{A schematic picture of the 2D Fermi surface (thick black
line) of HTSC in the first quadrant of the first Brillouin zone.
The lattice constant $a$ is set to unity. The $\phi$ defines the
Fermi surface angle.} \label{Bzone}
\end{figure}

We will restrict our attention to the hole-doped compounds,
partly because they are better characterised and more extensively
investigated, and also because the hole-doped HTSC exhibit
a so-called pseudogap phase (with a partially gapped excitation
spectrum) above the superconducting dome. The onset
temperature of the pseudogap decreases linearly with doping
and disappears in the overdoped\footnote{The
superconducting phase is often divided into an optimal doped
(doping level with highest $T_c$), an overdoped (doping level
higher than optimal doped), and an underdoped (doping level lower
than optimal doped) regime.} regime.
The origin of the pseudogap
is one of the most controversial topics in the high-$T_c$ debate.
The relationship between the pseudogap and other important features
such as the presence of a Nernst phase \cite{Ong04,Ong03}, charge
inhomogeneities \cite{Kivelson03}, the neutron scattering
resonance \cite{Tranquada05}, marginal Fermi liquid behaviour
\cite{Varma89}, or disorder \cite{Dagotto05}. For a detailed
discussion of the pseudogap problem, we refer to a recent
article by Norman, Pines, and Kallin \cite{Norman05}.

We now discuss some
results from angle resolved photoemission spectroscopy (ARPES) and
scanning tunnelling microscopy (STM), since they are
immediately relevant to the
theoretical considerations and results presented in the
later sections. These two techniques have seen significant advances
in recent years and provided us new insights on the nature of
the pseudogap, superconducting gap and quasiparticles in the superconducting state.
As we will show
in the following sections, many features reported by these
experiments can be well understood within the framework of the
Gutzwiller-RVB theory.

\subsubsection{Angle resolved photoemission
spectroscopy (ARPES)}

\begin{figure}
\centering
\includegraphics*[width=0.45\textwidth]{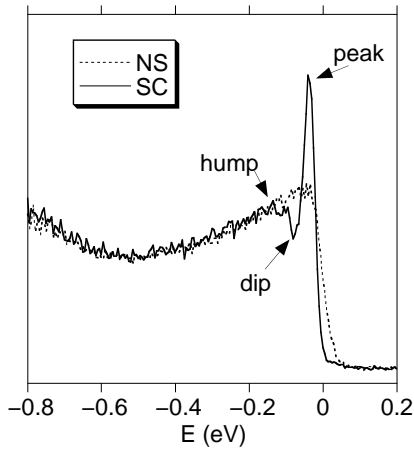}
\caption{Energy distribution curve (EDC) at fixed momentum
$\bk=(\pi,0)$ for an overdoped (87K)
Bi$_2$Sr$_2$CaCu$_2$O$_{8+\delta}$ (Bi2212) sample in the normal
state (NS) and superconducting state (SC). From \cite{Norman03}.}
\label{typEDC}
\end{figure}

By measuring the energy and momentum of photo-electrons, ARPES
provides information about the single particle spectral function,
$A(\bk,\omega)$. The latter quantity is related to the electron Green's function
by $A(\bk,\omega)=-\frac 1 \pi {\rm Im}\, G(\bk,\omega)$ \cite{Doniach82}.
In this subsection, we summarise some key results from ARPES that any
theory of HTSC has to address. The reader is referred to
the extensive ARPES reviews by
Damascelli, \etal \ \cite{Damascelli03} and Campuzano, \etal \
\cite{Campuzano04} for a discussion on experimental detail.

In \fig{Bzone} we illustrate a schematic picture of the
two-dimensional (2D) Fermi surface (FS) of HTSC in the first
quadrant of the first Brillouin zone. It can be obtained by ARPES
scans along different angles $\phi$. The FS for each $\phi$ is
then determined in general (but not in the underdoped region
\cite{Gros06}) by looking at the minimum energy of the photoelectron
along this direction in momentum space. A typical energy
distribution curve (EDC), {\it i.e.} photoemission intensity as
a function of energy at fixed momentum, from an ARPES experiment is
shown in \fig{typEDC}. The figure shows the photoemission
intensity at the $(\pi,0)$-point of a photoelectron
in the superconducting state ($T \ll T_c$) and
in the normal state ($T_c>T$). In the superconducting
state, one sees the characteristic peak-dip-hump structure; the
peak can be associated with a coherent quasiparticle. Above $T_c$,
coherence is lost and the sharp peak disappears.

\begin{figure}
\centering
\includegraphics[width=0.5\textwidth]{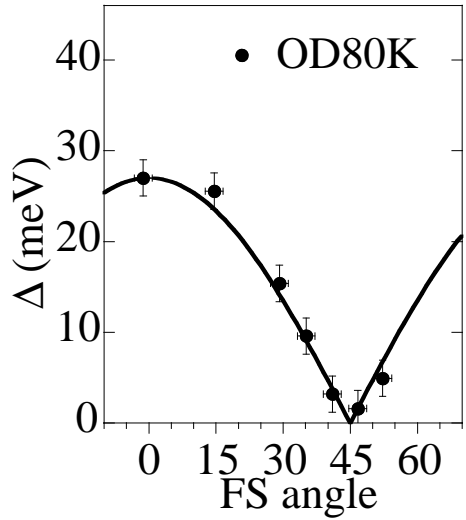}
\caption{Momentum dependence of the spectral gap $\Delta$ at the
FS in the superconducting state of an overdoped Bi2212 sample from
ARPES. The black line is a fit to the data. For a definition of
the FS angle $\phi$ see \fig{Bzone}. From \cite{Mesot99}. }
\label{pairing}
\end{figure}

\begin{figure}
\centering
\includegraphics[width=0.7\textwidth]{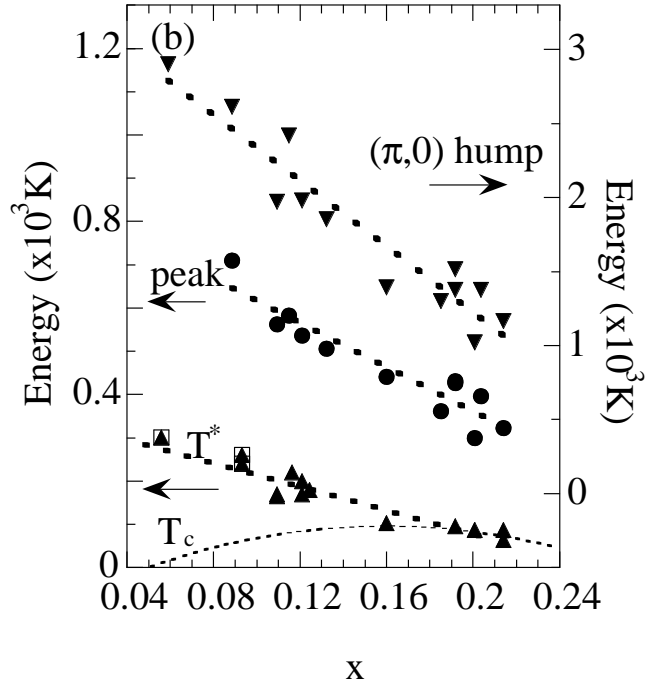}
\caption{Doping dependence of $T^*$ (the onset of the pseudogap,
compare with \fig{PhaseDiagram}), and of the peak and hump binding
energies in the superconducting state (see \fig{typEDC}). The
empirical relation between $T_c$ and doping $x$ is given by
$T_c/T^{\rm max}_c = 1 - 82.6(x - 0.16)^2$ with $T^{\rm max}_c 95$
K. Data for Bi2212, from \cite{Campuzano99}.} \label{dopingARPES}
\end{figure}

In the early years following the discovery of high temperature
superconductivity, it was unclear if the pairing symmetry were
isotropic ($s$-wave like), as in conventional phonon-mediated
superconductors, or anisotropic. Later experiments have
consistently confirmed an anisotropic gap with $d$-wave symmetry
\cite{Tsuei00}. The angular dependence of the gap function is
nicely seen in ARPES measurements on HTSC (\fig{pairing}), which
accurately determine the superconducting gap $|\Delta_\bk|$ at the
FS. As illustrated in \fig{pairing} for a
Bi$_2$Sr$_2$CaCu$_2$O$_{8+\delta}$ (Bi2212) sample, the gap
vanishes for $\phi=45^\circ$. This direction is often referred to
as the `nodal direction', the point at the FS is then called the
`nodal point' or `Fermi point'. In contrary, the gap becomes
maximal for $\phi=0^\circ,\, 90^\circ$, {\it i.e.} at the
`anti-nodal point'.

\begin{figure}
\centering
\includegraphics[width=0.75\textwidth]{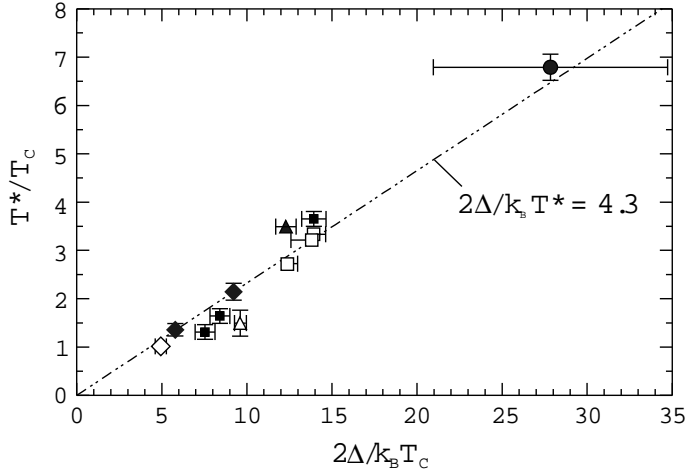}
\caption{$T^*/T_c$ versus $2 \Delta/(k_B T_c)$ for various
Cuprates compared to the mean field relation, $2 \Delta/(k_B
T^*)=4.3$, valid for $d$-wave superconductivity \cite{Won94},
where $T^*$ replaces $T_c$. From \cite{Kugler01}.}
\label{BCS_ratio}
\end{figure}

Another feature well established by ARPES is the doping dependence
of the superconducting gap and the opening of the pseudogap at a
temperature $T^* > T_c$. Unlike conventional superconductors, HTSC
exhibit a strong deviation from the BCS-ratio\footnote{The weak
coupling BCS-ratio for $s$-wave superconductors, $2\Delta/(k_B
T_c) \approx 3.5$.} of $2\Delta/(k_B T_c) \approx 4.3$ for
superconductors with a $d$-wave gap function. In HTSC, this ratio
is strongly doping dependent and becomes quite large for
underdoped samples, where the transition temperature $T_c$
decreases, while the magnitude of the superconducting gap
increases. As illustrated in \fig{dopingARPES} for a Bi2212
sample, the binding energy of the peak at $(\pi,0)$, {\it i.e.}
the superconducting gap\footnote{When speaking about (the
magnitude of) the superconducting gap $\Delta$ in a $d$-wave state
without specifying the momentum $\bk$, we mean the size of the gap
$|\Delta_\bk|$ at $\bk=(\pi,0)$.}, increases linearly (with
doping) while approaching the half filled limit. Interestingly,
the opening of the pseudogap at temperature $T^*$ seems to be
related to the magnitude of the gap. The modified ratio $2
\Delta/(k_B T^*)$ is a constant for HTSC at all doping levels and
the constant is in agreement with the BCS ratio, $4.3$ (see
\fig{BCS_ratio}), with $T_c$ substituted by $T^*$. This
experimental result is as a remarkable confirmation of early
predictions from Gutzwiller-RVB theory, as we will discuss in
further detail in latter sections. Figure \ref{dopingARPES} also
reveals that the hump feature (see EDC in \fig{typEDC}) scales
with the binding energy of the peak at $(\pi,0)$.

An additional doping dependent feature extracted from ARPES data
is the spectral weight of the coherent quasiparticle (QP) peak.
Feng, \etal \ \cite{Feng00} defined a superconducting peak ration
(SPR) by comparing the area under the coherent peak with that of
the total spectral weight. Figure \ref{SPR_Science} depicts EDCs
at several doping levels together with the computed SPR as a
function of doping. The QP spectral weight strongly decreases with
decreasing doping and finally vanishes \cite{Feng00,Ding01}.
Such a behaviour is well understood by invoking
the projected nature of the superconducting
state as we will discuss in the following sections.

\begin{figure}
\centering
\caption{
 a) Doping dependence of the superconducting state spectra in Bi2212
 at $(\pi,0)$ taken at $T \ll  T_c$. The doping level is decreasing form the top curve downwards.
 Samples are denoted by OD (overdoped), OP (optimal doped),
 and UD (underdoped), respectively,
 together with their $T_c$ in Kelvin, {\it e.g.} OD75 denotes an overdoped sample with $T_c=75 K$.
 b) The doping dependence of superconducting peak ratio (spectral weight of coherent peak with respect
 to the total spectral weight)
 is plotted over a typical Bi2212 phase diagram for the spectra
 in a). AF, antiferromagnetic regime; SC, superconducting regime.
 From \cite{Feng00}.}
\label{SPR_Science}
\end{figure}

Since ARPES is both a momentum and energy resolved probe, it allows
for the measurement of the dispersion of the coherent peak. Here,
we concentrate on the nodal point, where the excitations are
gapless even in the superconducting state, owing to the $d$-wave
symmetry of the gap. The dispersion around the nodal point is well
approximated by Dirac cones, whose shape is characterised by two
velocity, $v_F$ and $v_\Delta$. The Fermi velocity $v_F$ is
determined by the slope of the dispersion along the nodal
direction at the nodal point, whereas the gap velocity $v_\Delta$
is defined by the slope of the `dispersion' perpendicular to the
nodal direction at the nodal point. Since all other $\bk$-points
are gapped, the shape of the Dirac-like dispersion around the
nodal point is of particular importance for the description of any
effect depending on low-lying excitations.

\begin{figure}
\centering
\caption{Electron dynamics in the La$_{2-x}$Sr$_x$CuO$_4$ (LSCO)
system. a) Dispersion energy, E, as a function of momentum, $\bk$,
of LSCO samples with various dopings measured along the nodal
direction. The arrow indicates the position of the kink that
separates the dispersion into high-energy and low-energy parts
with different slopes. E$_F$ and $\bk_F$, are Fermi energy and
Fermi momentum, respectively. b) Scattering rate as measured by
MDC width  of the LSCO (x=0.063).
From \cite{Zhou03}.} \label{vF}
\end{figure}

Figure \ref{vF}(a) illustrates the slope of the
dispersion along the nodal direction for La$_{2-x}$Sr$_x$CuO$_4$
(LSCO) samples at various dopings. The ARPES data reveals a
significant splitting in high-energy and low-energy parts,
whereas the low-energy part corresponds to the Fermi velocity
$v_F$. Within ARPES data [see \fig{vF}(a)] the Fermi velocity
$v_F$ is only weakly doping-dependent. ARPES can also determine
the gap velocity $v_\Delta$ by looking at the spectral gap
along the Fermi surface as done in \fig{pairing}.
Together with the $v_F$, the $v_\Delta$ determines the shape
of the Dirac cones, which, according to ARPES, is quite
anisotropic ($v_F/v_\Delta \approx 20$ around optimal doping)
\cite{Mesot99}. This result is confirmed by thermal conductivity
measurements \cite{Chiao00}, that yield similar asymmetries as
in ARPES. Another generic feature of HTSC is a kink seen in
the ARPES nodal dispersion as shown in \fig{vF}(a). This kink also effects
the scattering rate of the coherent quasiparticles as measured by
the momentum distribution curves (MDC) width, see \fig{vF}(b)
and \cite{Damascelli03,Campuzano04}.

\begin{figure}
\centering
\includegraphics*[width=0.8\textwidth]{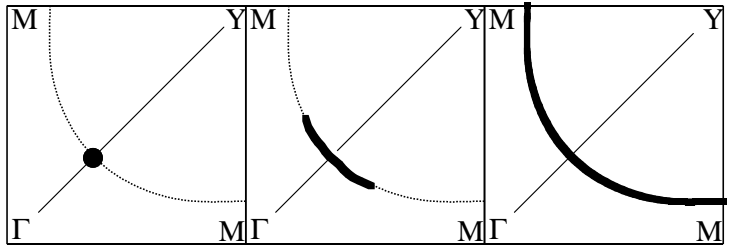}
\caption{Schematic illustration of the temperature evolution of
the Fermi surface in underdoped Cuprates as observed by ARPES. The
$d$-wave node below $T_c$ (left panel) becomes a gapless arc above
$T_c$ (middle panel) which expands with increasing $T$ to form the
full Fermi surface at $T^*$ (right panel). From
\cite{Norman98}.} \label{FermiArc}
\end{figure}

An interesting feature seen in ARPES is the shrinking of the Fermi
surface when the pseudogap opens at $T^*$. With decreasing
temperature, more and more states around the antinodal region
become gapped and the Fermi surface becomes continuously
smaller. Instead of a full Fermi surface, the pseudogapped state
exhibits Fermi arcs \cite{Loeser96,Marshall96,Ding96,Norman98,Kanigel06},
that finally collapse to single nodal Fermi points at $T=T_c$
(see \fig{FermiArc}). For a detailed discussion on this and
related ARPES observations, we refer the reader to the
ARPES reviews in the literature\cite{Damascelli03,Campuzano04}.

\subsubsection{Scanning tunnelling microscopy (STM)}

\begin{figure}
\centering
\includegraphics*[width=0.75\textwidth]{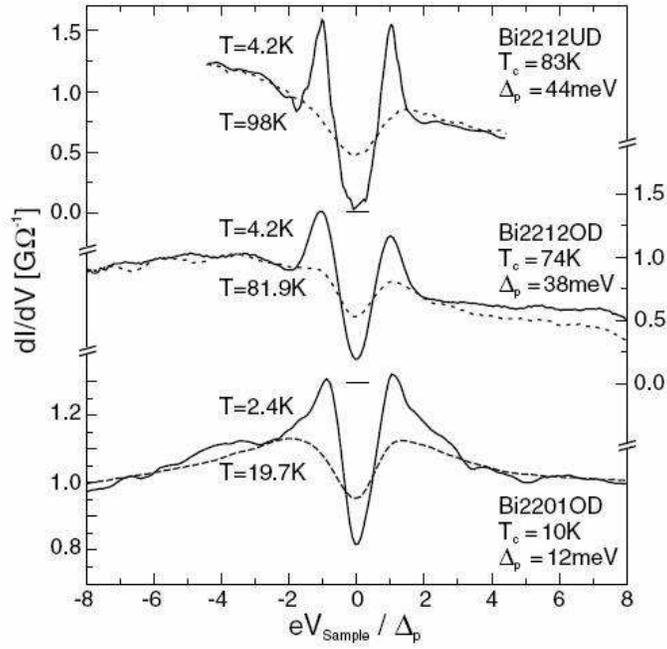}
\caption{STM data for underdoped (UD) and overdoped (OD) Bi2212,
and overdoped Bi2201; comparison between the pseudogap (dashed
line, $T > T_c$) and the gap in the superconducting state (solid
line, $T<T_c$). The underdoped data exhibit a significant
asymmetry between positive and negative bias voltages. For
an analysis of the temperature-dependent pseudogap see
figure \ref{BCS_ratio}.
From \cite{Kugler01}. } \label{STM_asymmetry}
\end{figure}

In contrast to ARPES, STM is a momentum integrated probe. However,
its ability to measure the local density of occupied as well as
unoccupied states with a high energy resolution gives very
valuable insights into HTSC. An example for a STM study of
Bismuth-based HTSC is shown in \fig{STM_asymmetry}. The data in
the superconducting state reveals a density of states, which is
characteristic of a $d$-wave gap, {\it i.e.} there is no full gap
in contrast to $s$-wave superconductivity. In the pseudogap state
(above $T_c$) the density of states is still suppressed around
$\omega=0$ (zero voltage), however, the characteristic peaks
disappear. Another interesting feature seen in \fig{STM_asymmetry}
is the striking asymmetry between positive and negative voltages,
which becomes more pronounced for the underdoped sample. An
explanation for this generic property of HTSC will be discussed in
detail in the following sections.

A key advantage of STM is the possibility to obtain spatial
information. For example, STM experiments allow for the investigation
of local electronic structure around impurities
\cite{Hudson99,Yazdani99,Pan03}, and around vortex cores
\cite{Maggio95,Renner98,Pan00} in the superconducting state. Two
other interesting features recently reported by STM are a
checkerboard like charge density wave
\cite{Vershinin04,Hanaguri04} and the existence of spatial
variations in the superconducting gaps \cite{McElroy05}. The
origin of these observations is currently being debated intensely.

\subsection{Theories}
\label{theories}

It is beyond the scope of this article to provide an overview of
various theories of high temperature
superconductivity that have been put forward in the literature.
Due to the enormous
complexity of the experimentally observed features, it is not easy
to agree on the key ingredients necessary for setting up a
comprehensive theory. Further,
the decision to trust new experimental results is
often difficult, since the sample quality, experimental resolution,
and the way the data is extracted are often not completely clear.
Not surprisingly perhaps, these circumstances have allowed for diverse
theoretical approaches, motivated
respectively by distinct aspects of the HTSC. In the following,
we summarise a few theoretical approaches where the proximity
of a superconducting phase to a Mott insulator and / or antiferromagnetism
plays an important role.

\subsubsection{Electronic models}

To find an appropriate microscopic reference model is the first
step in formulating any theory. Such a model should be simple
enough on the one hand to be treated adequately, but should
also be complex enough to explain the relevant properties.
In the case of the HTSC, it is widely accepted that strong
correlations in the two-dimensional (2D) layers play an essential
role. The copper-oxygen layers are appropriately described by a
three-band Hubbard model, which includes the Cu
$d_{x²-y²}$-orbital and the two O $p$-orbitals
\cite{Emery87,Varma87}. Its simplified version is an one-band
Hubbard model\footnote{Henceforth we refer to the
one-band Hubbard model by the phrase ``Hubbard model''.},
where each site corresponds to a copper orbital
with repulsive on-site interaction between electrons
\cite{Zhang88a}. The derivation of this model Hamiltonian can
be found in the reviews of Lee, \etal \ \cite{Lee06} and
Dagotto \cite{Dagotto94}.

\subsubsection{Resonating valence bond picture}

Soon after the discovery of high $T_c$ superconductivity, Anderson
suggested the concept of a resonating valence bond (RVB) state
\cite{Anderson87} as relevant for the HTSC. In this picture, the
half-filled Hubbard model is a Mott insulator with
one electron per site. The charged states, doublons and holons,
form bound charge-neutral excitations in the Mott insulating state
and leads to the vanishing of electrical conductivity. Equivalently
one can talk of virtual hopping causing a superexchange
interaction $J$ between the electrons at the copper sites.
Therefore, the half-filled systems can be viewed as Heisenberg
antiferromagnet with a coupling constant $J$.

Anderson proposed
that upon doping quantum fluctuations melt the antiferromagnetic
Neel lattice and yield a spin liquid ground state (denoted the RVB
state) in which the magnetic singlet pairs of the insulator become
the charged superconducting pairs. We will show in the following
sections that the RVB picture provides a natural explanation
for several key features of the HTSC such as the
$d$-wave pairing symmetry, the shape of the superconducting
dome, the existence of a pseudogap phase, the strong deviations
from the BCS-ratio, and the singular $\bk$-dependence of the
one-particle self-energy when approaching half-filling.

\subsubsection{Spin fluctuation models}

While the RVB idea approaches the problem from the strong coupling
limit, {\it i.e.} large on-site electron repulsion $U$, spin
fluctuation models\footnote{For more details we refer to the
review articles by Moriya and Ueda \cite{Moriya00}, Yanase, \etal
\ \cite{Yanase03}, and Chubukov, \etal \ \cite{Chubukov02}.} start
from the weak coupling (small $U$) limit. The technique extends
the Hartree Fock (HF) random phase approximation and leads to a
pairing state with $d$-wave symmetry. Within this picture,
superconductivity is mediated by the exchange of antiferromagnetic
spin fluctuations.

Weak-coupling approaches such as spin fluctuation models essentially
remain within the context of Landau theory of Fermi liquids for
which the quasiparticle renormalisation is $Z=m/m^*$, when the
self-energy is not strongly $\bk$-dependent. Here, $m^* \sim
v_F^{-1}$ and $m$ is the bare band mass. The Fermi
liquid relation $Z \sim  v_F$, is however difficult to reconcile
with experimental results for the HTSC, as $Z \to 0$ and
$v_F \to {\rm const}$ for doping $x \to 0$, as we
will discuss in more detail in \sect{QP_halffilled}.

\subsubsection{Inhomogeneity-induced pairing}

Within this class of theories, the proximity of high
temperature superconductivity to a Mott insulator plays an
important role.
It is postulated that the superconducting pairing is
closely connected to a spontaneous tendency of the
doped Mott insulator to phase-separate
into hole rich and hole poor regions at low doping. The
repulsive interaction could then lead to a form of local
superconductivity on certain mesoscale structures, ``stripes''.
Calculations show that the strength of the pairing tendency
decreases as the size of the structures increases. The
viewpoint of the theory is as follows: Below a critical
temperature, the fluctuating mesoscale structures condense into a
global phase-ordered superconducting state. Such a condensation
is facilitated if the system were more homogeneous, however,
more homogeneity leads to larger mesoscale structures, and thus weaker pairing.
Therefore, the optimal $T_c$ is obtained at an optimal
inhomogeneity, where mesoscale structures are large enough to
facilitate phase coherence, but also small enough to induce
enough pairing. Within the phase-separation scenario
spontaneous inhomogeneities tend to increase even
in clean systems when approaching half-filling. In this
framework, the
pseudogap in the underdoped regime can be understand as a phase,
which is too granular to obtain phase coherence, but has strong
local pairing surviving above $T_c$. These ideas are
reviewed in detail by Kivelson and collaborators
\cite{Kivelson03, Kivelson05,Carlson02}.

\subsubsection{$SO(5)$ - theory}

Motivated by the vicinity of antiferromagnetism and
superconductivity in the phase diagram of the HTSC, the
$SO(5)$-theory \cite{Demler04} attempts to unify these collective
states of matter by a symmetry principle. In the $SO(5)$ picture,
the $5$ stands for the five order parameters used to set up the
theory; three degrees of freedom for antiferromagnetic state
($N_x$, $N_y$, $N_z$) and two degrees of freedom for the
superconducting state (real and imaginary parts of the
superconducting order parameter). The theory aims to
describe the phase diagram of HTSC with a single low-energy
effective model. A so-called projected $SO(5)$-theory has been proposed
to incorporate strong correlation effects. Several studies
have also examined the microscopic basis for the $SO(5)$ theory
(see review by Demler, \etal\ \cite{Demler04}).

\subsubsection{Cluster methods}

Though numerical methods such as Lanczos (exact diagonalisation)
and quantum Monte Carlo have been very popular \cite{Dagotto94},
they are limited by the (small) cluster size. All
statements concerning the thermodynamic limit become imprecise due to
significant finite size effects. The ``quantum cluster'' method which aims
to mitigate finite size effects in numerical methods, has been used
by several groups to study strongly correlated electronic systems.
These methods treat
correlations within a single finite size cluster explicitly.
Correlations at longer length scales are treated either perturbatively
or within a mean field approximation \cite{Maier05}.
In recent years, this method has been used in several studies to extract the
ground state properties of the Hubbard model. They reproduce several
features of the Cuprate phase diagram and report $d$-wave pairing
in the Hubbard model. However, even these sophisticated numerical
methods are not accurate enough to determine the ground state
of the Hubbard model unambiguously.

\begin{figure}
\centering
\includegraphics*[width=0.75\textwidth]{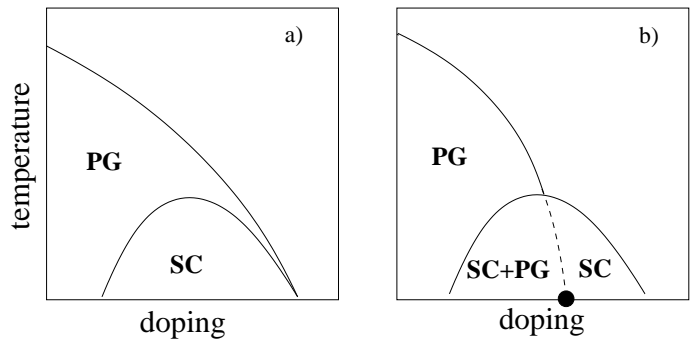}
\caption{Two proposed theoretical phase diagrams for the Cuprates.
a) RVB picture; b) Competing order scenario: the pseudogap (PG)
ends in a quantum critical point (black dot); pseudogap and
superconducting state (SC) can coexist (SC+PG).}
\label{two_diagrams}
\end{figure}

\subsubsection{Competing order}

In most of the theories outlined above, the pseudogap phase
is characterised by the existence of preformed pairs. Hence, there are two relevant
temperature scales in the underdoped regime. Pairs form at a (higher) temperature
$T^*$, and the onset of phase coherence at $T_c$ leads to
superconductivity. However, there are other
theories that take the opposing point of view; \emph{viz.}, the pseudogap and
superconductivity are two phases that compete with each other.
In these scenarios, the pseudogap is characterised by another order
parameter, {\it e.g.} given by an orbital current state
\cite{Varma06} or a $d$-density wave \cite{Chakravarty02}.
Thus, the pseudogap suppresses
superconductivity in the underdoped regime, and can also partially
survive in the superconducting state. These approaches
predict that the pseudogap line ends in a quantum critical point
inside the superconducting dome. These two scenarios are
contrasted in \fig{two_diagrams}.

\subsubsection{BCS-BEC crossover}

In this picture, the pseudogap is explained by a crossover
from BCS to Bose Einstein condensation (BEC) \cite{Randeria98,Chen05}. While
in the BCS limit the fermionic electrons condensate to a
superconducting pair state, the BEC limit describes the
condensation of already existing pairs. In the crossover regime,
one expects a behaviour very similar to that observed in the pseudogap
of HTSC; formation of pairs with a corresponding excitation gap
occurs at a temperature $T^*$, and the pairs condense at
a lower temperature $T_c<T^*$.
It is interesting to note that the physics behind
this idea can be described by a generalisation of the BCS
ground state wave function, $|\Psi_0\rangle$ \cite{Chen05}. It is
however unclear how to incorporate the antiferromagnetic
Mott-Hubbard insulating state close to half-filling within a
BCS-BEC crossover scenario.

\section{Resonating valence bond (RVB) theories}
\label{RVB}

The resonating valence bond (RVB) state describes a liquid of spin
singlets, and was proposed originally as a variational ground state
of the spin $S=1/2$ Heisenberg model (which describes the low
energy physics of the Hubbard model at half filling).
Anderson originally proposed that the magnetic singlets of the RVB
liquid become mobile when the system is doped and form charged superconducting pairs.
As we will discuss in this section, this idea
has led to a consistent theoretical framework to describe superconductivity
in the proximity of a Mott transition. In this section,
we will discuss possible realisations of RVB
superconductors along with the predictions of the theory.
We also give a outlook on the implementations of the RVB
picture by Gutzwiller projected wave functions, slave boson
mean field theory and the bosonic RVB approach.

\subsection{The RVB state - basic ideas}

\begin{figure}
\centering
\includegraphics*[width=0.85\textwidth]{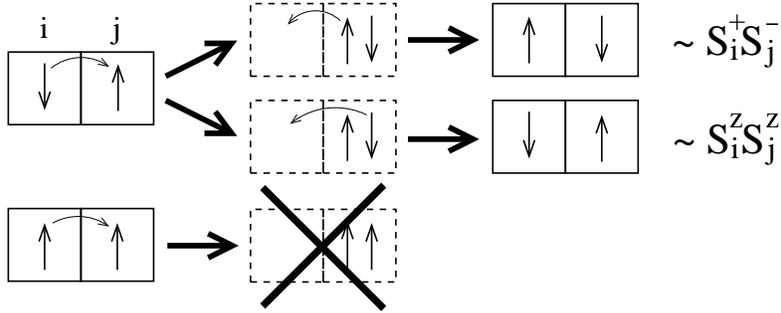}
\caption{Hopping processes with a virtual doubly occupied site
corresponding to the $S^z_i S^z_j$ and  $S^+_i S^-_j$ term of the
Heisenberg Hamiltonian, respectively; in the case of parallel
spins virtual hopping is not possible.} \label{hopping}
\end{figure}

Within the resonating valence bond (RVB) picture, strong electron
correlations are essential for superconductivity in the Cuprates.
The Hubbard model is viewed as an appropriate microscopic basis
and the corresponding many-body Hamiltonian is given by,
\begin{equation}
H=\,-\sum_{\langle ij\rangle,\sigma} t_{(ij)} \left(
c_{i\sigma}^\dagger c_{j\sigma} + c_{j\sigma}^\dagger c_{i\sigma}
\right) \, +\,U\,\sum_i\,n_{i\uparrow} n_{i\downarrow}\ , \label{HubbardH}
\end{equation}
where $c^\dagger_i$ creates and $c_i$ annihilates an electron on
site $i$. The hopping integrals, $t_{(ij)}$, connect sites $i$ and
$j$. We shall restrict our attention to nearest neighbour hopping
$t$ for the moment and will also discuss the influence of
additional hopping terms subsequently. The operator $n_{i\sigma}\equiv
c_{i\sigma}^\dagger c_{i\sigma} $ denotes the local density of
spin $\sigma=\downarrow,\uparrow$ on site $i$. We consider an
on-site repulsion $U\gg t$, {\it i.e.} we work in the strong
coupling limit, which is a reasonable assumption for the HTSC.

\subsubsection{RVB states in half-filled Mott-Hubbard insulators}

\begin{figure}
\centering
\includegraphics*[width=0.85\textwidth]{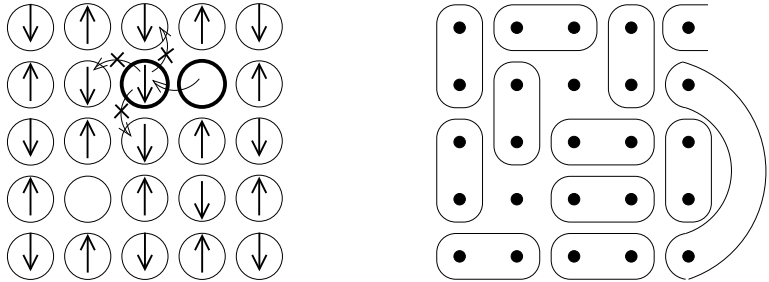}
\caption{Left: Antiferromagnetic Neel lattice with some holes. The
motion of a hole (consider bold circles) frustrates the
antiferromagnetic order of the lattice. Right: Snapshot of the RVB
state. A configuration of singlet pairs with some holes is shown.
The RVB liquid is a linear superposition of such configurations.}
\label{AF_RVB}
\end{figure}

Let us first consider the half-filled case. Since $U$ is
much larger than $t$ the mean site occupancy is close to charge
neutrality, namely one. It costs energy $U$ for an electron to hop to a
neighbouring site. This potential
energy is much higher than the energy the electron can gain by
the kinetic process. Thus, the motion of electrons is frozen and
the half-filled lattice becomes a Mott-Hubbard insulator. However,
there are virtual hopping processes, where an electron hops to its
neighbouring site, builds a virtual doubly occupied site, and hops
back to the empty site. Such virtual hoppings lower the energy by
an amount of the order $J=4t^2/U$. Pauli exclusion principle
allows double occupancy only for electrons with opposite spin (see
\fig{hopping}). Thus, virtual hopping favours anti-parallel spins
of neighbouring electrons and we obtain an effective
antiferromagnetic Heisenberg Hamiltonian,
\begin{equation}
H=\,J\, \sum_{\langle ij\rangle} \mathbf{S_i}\cdot \mathbf{S_j}~,\qquad  J>0~,
\label{HeisenbergH}
\end{equation}
with an antiferromagnetic exchange constant $J=4t^2/U$, the
spin-operator $\mathbf{S_i}$ on site $i$, and $\langle ij\rangle$
denoting a sum over nearest neighbour sites. At the level of mean field
theory, {\it i.e.} treating the spins semiclassically, the 2D Heisenberg
model on a square lattice has an antiferromagnetic Neel ground
state with long range order and broken symmetry (left panel of
\fig{AF_RVB}). This molecular-field prediction is experimentally
(by neutron scattering studies \cite{Endoh88}) as well as
theoretically (by a quantum nonlinear $\sigma$ model
\cite{Chakravarty89}) well established.

\begin{figure}
\centering
\includegraphics*[width=0.8\textwidth]{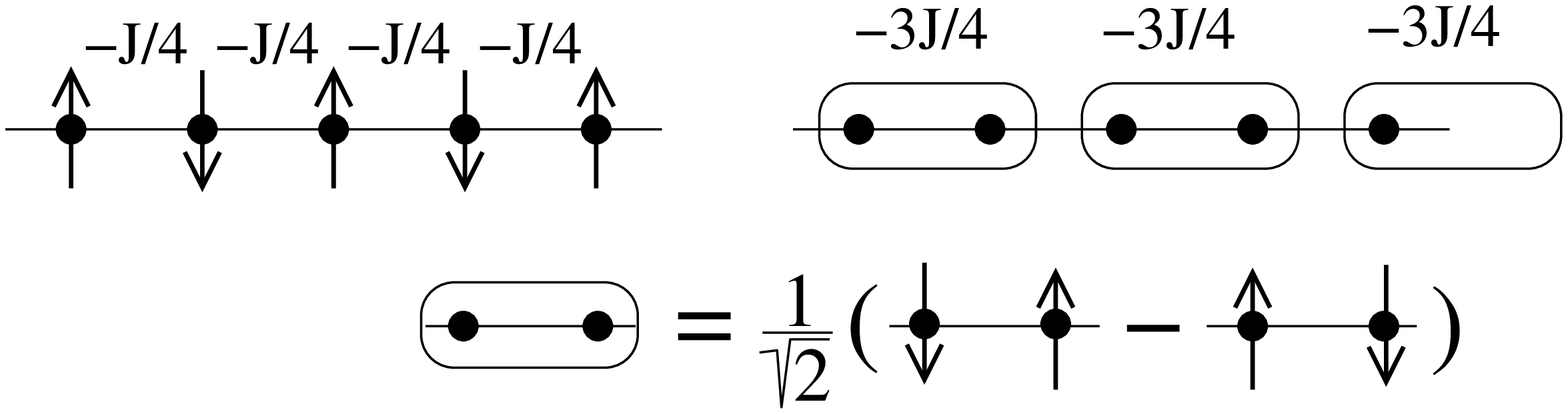}
\caption{Neel state (left) and singlet state (right) for 1D
antiferromagnetic spin $S=1/2$ chain.} \label{1Dchain}
\end{figure}

Anderson \cite{Anderson87} suggested that a resonating valence
bond (RVB) liquid\footnote{Long before the discovery of HTSC
Anderson and Fazekas \cite{Anderson73,Fazekas74} proposed the RVB
liquid as a possible ground state for the
Heisenberg model on a 2D triangular lattice.} is very
close in energy to the Neel state for undoped Cuprates. Instead of
a Neel state with broken symmetry, a fluid of singlet pairs is
proposed as the ground state; {\it i.e.}, the ground state is described by a
phase coherent superposition of all possible spin
singlet configurations (see right panel of \fig{AF_RVB}). For
spin $S=1/2$, quantum fluctuations favour such singlets than
classical spins with Neel order. To
see this, consider a one-dimensional (1D)
chain (see \fig{1Dchain}). In this case, a Neel state with $S_z=±
1/2$ gives an energy of $-J/4$ per site. On the other hand, the
ground state of two antiferromagnetic coupled spins $S=1/2$ is a
spin singlet with $-S\,(S+1)\,J=-3/4\,J$. It follows that a chain
of singlets (see \fig{1Dchain}) has an energy of $-3/8\,J$ per
site, much better than the Neel-ordered state. This simple
variational argument shows that a singlet state is superior in 1D.
Similar considerations for the 2D Heisenberg model give the
energies $-1/2\,J$  per site for the Neel lattice, the singlet
state remains at $-3/8\,J$ per site. Following this reasoning we
find that singlets become much worse than the Neel state in higher
dimensions.

Liang, Doucot, and Anderson \cite{Liang88} showed that the singlet
`valence bonds' regain some of the lost antiferromagnetic exchange
energy by resonating among many different singlet configurations
and become therefore competitive with the Neel state in 2D. The
resonating singlets are very similar to benzene rings
with its fluctuating C-C links between a single and a double bond;
an analogy that motivated the term `RVB'.

\subsubsection{RVB spin liquid at finite doping}

Though an antiferromagnetically long range ordered state is realised in the undoped insulator,
the order melts with only a few percent of doped holes. To understand this,
consider the example shown in  \fig{AF_RVB} (left). The figure shows that moving holes cause
frustration in the antiferromagnetic but not the RVB state,
\fig{AF_RVB} (right). A single hole moving in the background of a Neel state
was studied extensively by several authors\footnote{The
single hole problem together with the corresponding literature is
discussed in \cite{Lee06} in more detail.}, and analytical
calculations showed that the coherent hole motion is strongly
renormalised by the interactions with the spin excitations
\cite{Kane89,Gros89b}. When more holes are injected into the
system, the interaction of the holes with the spin background
completely destroys the antiferromagnetic Neel state and an RVB
liquid (or spin liquid) state becomes superior in energy. Then the
singlet pairs of the RVB liquid are charged and may condense to a
superconducting ground state.

\subsection{Realisations and instabilities of the RVB state}

\begin{figure}
  \centering
  \includegraphics*[width=\textwidth]{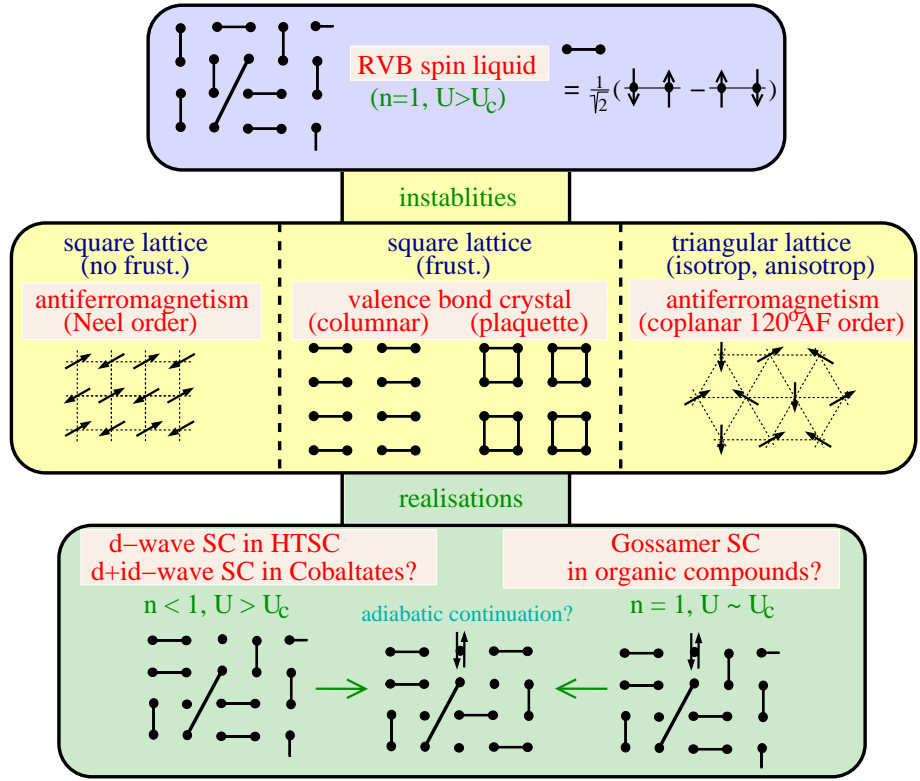}
  \caption{\label{fig:RVBoverview}%
Schematic picture of instabilities and realisations of the RVB
spin liquid state, {\it viz.} of the RVB state as an unstable
fixpoint. The top panel shows an RVB spin liquid at half-filling
in the Mott-Hubbard insulating limit ($U>U_c$). The middle panel
illustrates  instabilities of the RVB liquid state in a square
lattice, a frustrated square lattice, and a triangular lattice in
the half-filled limit. The lower panel shows realisations of the
RVB liquid, which are realised at finite doping or close to the
Mott-Hubbard transition ($U \sim U_c$).}
\end{figure}

Whether there exist two dimensional models with an RVB
ground state is still an open question. We may however regard the
RVB state as an unstable fixpoint \cite{Anderson02b} prone to
various instabilities. The situation is then analogous to that of
the Fermi liquid, which becomes generically unstable in the
low-temperature limit either towards superconductivity or
various magnetic orderings. For instance, Lee and Feng
studied numerically how a paramagnetic RVB state can be modified
to become a long range (antiferromagnetically) ordered state by
introducing an additional variational parameter \cite{lee_88}. In this
view of antiferromagnetism, the ``pseudo Fermi surface'' of the
insulating RVB state undergoes a nesting instability to yield
long range antiferromagnetic order \cite{hsu_90, ho_01}.
In \fig{fig:RVBoverview} we present an
illustration of the concept of the RVB state as an unstable
fixed point. In the following, we discuss this point further.

Besides the square lattice with nearest neighbour hopping,
the RVB spin liquid was proposed as a ground state on
a square lattice with further neighbour hopping as well as in a
triangular lattice. Experiments \cite{shimizu_03} indicate that such a
spin liquid state may be realized in the organic compound
$\kappa$-(BEDT-TTF)$_2$Cu$_2$(CN)$_3$ which is an insulator in
the proximity of a Mott transition. Trial spin liquid
wave functions using Gutzwiller projected RVB states have
been proposed in this context by Motrunich \cite{motrunich_05}.
A $U(1)$ gauge theory of
the Hubbard model has also been invoked to study this system
\cite{sslee_05}.
Although the simple Neel ordered state is
destroyed due to frustration in these cases, the RVB spin liquid
(at $n=1$) does not become the ($T=0$) ground state, which is
either a valence bond crystal state
\cite{Zhitomirsky96,Capriotti00,Takano03,Capriotti01,Mambrini06}
or a coplanar $120^o$ antiferromagnetic ordered state
\cite{Weber06a}, respectively. In addition, instabilities against
inhomogeneous states like stripes \cite{Kivelson03,
Kivelson05,Carlson02} are conceivable, and are not explicitly included
in \fig{fig:RVBoverview}. A recent ARPES study  on
La$_{2-x}$Ba$_x$CuO$_4$ \cite{Valla06}, which exhibits static
charge order and suppressed superconductivity around doping
$x=1/8$, supports the idea that the superconducting RVB state can
be continuously connected and unstable against a charge ordered
state.

Nevertheless an RVB state can be realised if a finite number of
hole is induced into the system, {\it viz.} when the bosonic spin
state realised at half-filling turns into a free fermionic state
by the introduction of charge carriers. The hopping processes then
destroy above instabilities towards magnetic or valence bond
crystal ordering and a superconducting RVB state can be
stabilised. A schematic picture of this scenario is presented in
\fig{fig:RVBoverview}.

In the case of HTSC, holes are created by changing the doping
concentration. A similar mechanism was proposed for
superconductivity in the triangular lattice based Cobaltates
\cite{Baskaran03,Kumar03}. Within RMFT calculations such a
triangular model would result a $d+id$-wave pairing state
\cite{Ogata03a}. On the other hand, an RVB superconducting state
at half-filling just below the Mott transition \cite{Imada98}
was recently suggested for organic superconductors
\cite{Powell05,Gan05b,Liu05}. Here, the necessary holes could
result from a finite number of conducting doubly occupied sites as
illustrated in \fig{fig:RVBoverview}.

To summarise, an RVB superconductor could emerge by two different
mechanisms starting from a Mott insulating system ($n=1$ and
$U>U_c$); either upon doping ($n \neq 1$), or from
self doping a half-filled
system close to the the Mott-Hubbard transition ($U
\sim U_c$). In this review, we focus our attention on the former
possibility, {\it
i.e.} the occurrence of an RVB superconductor in a doped
Mott-Hubbard insulator.

\subsection{Predictions of the RVB hypothesis for HTSC}

In this subsection we discuss some predictions from RVB
theory, which agree well with experimental observations. As we
will show in the following sections, the arguments we present here
are substantiated by more detailed microscopic calculations.

\begin{figure}
\centering
\includegraphics*[width=0.63\textwidth]{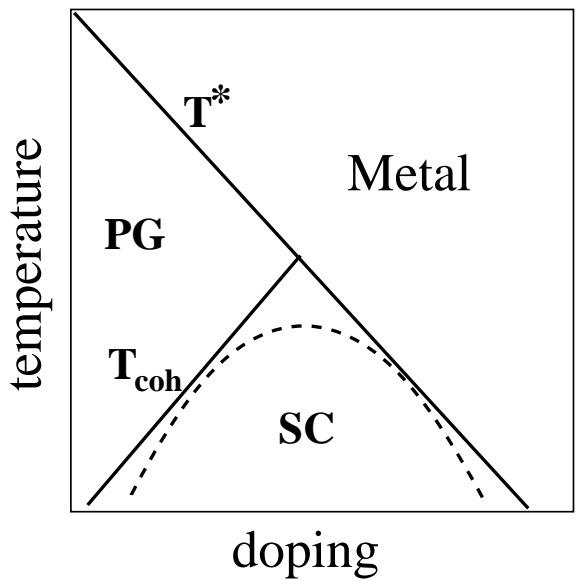}
\caption{RVB phase diagram with singlet pairing temperature $T^*$
and phase coherence temperature $T_{\rm coh}$ (superconducting
state SC, pseudogap PG).} \label{RVBphase}
\end{figure}

Within the RVB picture, a possible explanation for the
temperature-doping phase diagram is obtained by considering two
temperature scales (\fig{RVBphase}). The singlets of the RVB
liquid form at temperature $T^*$, a temperature scale which
decreases away from half-filling \cite{Baskaran87} owing to
the presence of doped and mobile holes.
Holes, on the other hand, allow for particle number fluctuations,
which are fully suppressed at half-filling, and thus enhance the
stability of the superconducting state against thermal
fluctuations. This results in a second temperature, $T_{\rm coh}$,
which increases with doping and below which the superconducting
carriers become phase coherent. The superconducting transition
temperature $T_c$ is therefore determined by the minimum of $T^*$
and $T_{\rm coh}$ as shown in \fig{RVBphase} \cite{Baskaran87}.

It is evident from the above picture that a pseudogap forms for
$T_{\rm coh}<T<T^*$, {\it i.e.} for underdoped samples. In this
state, although phase coherence is lost, the RVB singlet pairs
still exist. Therefore, we have to break a pair to remove an
electron from the copper-oxygen layers within the pseudogap
regime. The resulting excitation gap manifests itself, {\it e.g.}
in the $c$-axis conductivity or in ARPES measurements.

These schematic explanations are confirmed to a certain extent by
analytical as well as numerical calculations (at zero temperature).
RMFT and VMC methods show an increase of the
superconducting gap, but a vanishing superconducting order
parameter, when approaching half-filling. This behaviour is
in complete agreement with the $T\to 0$ observations in experiments. It also
explains the strong deviation from the BCS-ratio in the underdoped
regime of the HTSC, if the superconducting order parameter is related to $T_c$.
On the other hand, the doping dependence of the onset
temperature of the pseudogap $T^*$ can be related
to the magnitude of the gap at $T=0$
(in agreement with experiments, see \fig{BCS_ratio}).

Perhaps the most remarkable prediction of the RVB theory was the $d$-wave nature of the superconducting state.
A $d$-wave superconducting state was
predicted by RVB based studies as early as in 1988
\cite{Zhang88b,Kotliar88,Suzumura88,Gros88,Yokoyama88}, long
before the pairing symmetry was experimentally established. These
early calculations also correctly described the vanishing of
superconductivity above about $30 \%$ doping.

Implementing the RVB idea by projected wave functions, one finds
a natural explanation of
the suppression of the Drude weight and of the superfluid density
in the underdoped regime as well as the particle-hole asymmetry in
the density of single particle states. Further successes of the RVB theory
are calculations that predict a weakly doping dependent nodal Fermi velocity, but a
quasiparticle weight which is strongly doping dependent- the quasiparticle weight
decreases with doping $x$ in agreement with
ARPES experiments. These effects can be understood by a decrease
in the density of freely moving carriers at low doping, which results in a
dispersion mainly determined by virtual hopping processes
(proportional to the superexchange $J$). In the half-filled limit,
this behaviour results in a divergence of the $\bk$-dependence of
the electron's self-energy, $\lim_{\omega \to 0} {\partial
\Sigma(\omega,\bk=\bk_F)}/{\partial \omega} \sim 1/x \to \infty~,$
which transcends the nature of orthodox Fermi liquids. These will be
discussed in more detail in sections \ref{EX_RMFT} and
\ref{EX_VMC}.

In addition to the above key features of HTSC, RVB theory has
also been successfully applied to several other phenomena such as charge density patterns
\cite{Anderson04b,Huang05,Poilblanc05,Li06},
the interplay between superconductivity and antiferromagnetism
\cite{Himeda99,Chen90,Giamarchi91,Ivanov04,Shih04a,Ogata03b},
impurity problems \cite{Tsuchiura99,Tsuchiura00,Tsuchiura01},
and vortex cores \cite{Himeda97,Tsuchiura03}.

In conclusion, analytical and numerical results provide significant
support to the RVB concept. However, most RVB studies are
restricted to zero temperature\footnote{A possible ansatz for
finite temperatures was recently proposed by Anderson
\cite{Anderson05}. He suggests a spin-charge locking mechanism
within the Gutzwiller-RVB theory to describe the pseudogap phase
in the underdoped Cuprates as a vortex liquid state.}, making
the finite temperature picture detailed above, somewhat speculative.
Extending the calculations to finite temperature is an important and
open problem in the theory of RVB superconductivity.
A related issue is the destruction of superconductivity in the
underdoped samples where we expect phase fluctuations
to play an increasingly important role at low temperatures \cite{Emery95,Lee00}
since particle number fluctuations are frozen in the proximity of the
Mott insulator. It is presently an
unsettled question to which extent this picture is equivalent to
alternative formulations, such as an increase of inhomogeneities
(like in the `inhomogeneity-induced pairing' picture
\cite{Kivelson03, Kivelson05,Carlson02}) or a destruction of the
superfluid density due to nodal quasiparticle excitations (see
\sect{QPcurrent}), that were also proposed to describe
the transition from the superconducting state to the pseudogap
state in the underdoped regime. Further work is necessary to
clarify this point.

\subsection{Transformation from the Hubbard to the $t$-$J$ model}
\label{transtJ}

The RVB scenario is based on the existence of a strong
antiferromagnetic superexchange, $J$. The superexchange process via
virtual hopping processes results in an effective Heisenberg
Hamiltonian as discussed earlier (see \fig{hopping}). We now present
a more formal and systematic derivation of a low energy theory
starting from the Hubbard Hamiltonian in the strong coupling limit
($U \gg t$). The basic idea is to make the theory ``block diagonal''; \emph{i.e.},
subdivide the Hamiltonian matrix elements into processes that preserve
the local number (diagonal processes) and those that do not (off-diagonal)
by a unitary transform. Since we will be interested in
the strong coupling limit, off-diagonal
processes will be removed as such (high energy) configurations are
not allowed in the Hilbert space of the effective (low energy) theory.

The unitary transformation, $e^{-iS}$ to lowest order in
$t/U$ \cite{Gros87,Paramekanti04} can be obtained as follows.
First we assume that $S$ is of
the order ${\mathcal{O}(t/U)}$ and expand the
transformed Hamiltonian,
\begin{deqarr}
H^{(eff)}& =& e^{iS}He^{-iS} = e^{iS} (\hat T+ \hat U) e^{-iS}  \\
&=& \hat T + \hat U\,+\, i\,[S,\ \hat T+ \hat U] \,+\, \frac{i^2} 2 \,[S,[S,\hat T+ \hat U]]\,+\,\dots \\
&=& \hat U + \underbrace{\hat T + i [S,\hat U]}_{\mathcal{O}(t)} \,+
\, \underbrace{i\,[S,\hat T] \, + \, \frac{i^2} 2 \, [S,[S, \hat U]]}_{\mathcal{O}(t^2/U)} +\underbrace{\dots }_{
\mathcal{O}(t^3/U^2)} \, . \label{trafo}
\end{deqarr}
Here, we split the Hubbard Hamiltonian $H$ into the kinetic energy
part $\hat T$, the first term of \eq{HubbardH}, and the potential
energy part $\hat U$, the second term of \eq{HubbardH} (includes
the parameter $U$). In \eq{trafo} we have ordered the terms in powers
of $t/U$. For a block diagonal Hamiltonian
$H^{(eff)}$ to order ${\mathcal{O}(t/U)}$, the term, $\hat T +
i\,[S,\hat U]$, in \eq{trafo} may not contain any (real) hopping processes
changing the total number of doubly occupied sites.
An appropriate choice for $S$ is given by,
\begin{equation}
S=-i\,\sum_{\langle ij \rangle,\sigma} \frac{t_{(i,j)}}U
\left(a_{i,\sigma}^\dagger d_{j,\sigma} + a_{j,\sigma}^\dagger
d_{i,\sigma} \ - \ {\rm h.c.}\,\right)\label{S1} \ ,
\end{equation}
since,
\begin{equation}
\hat T + i\,[S,\hat U] = -\sum_{\langle ij\rangle,\sigma} t_{(ij)} \left(
a_{i\sigma}^\dagger a_{j\sigma} + d_{i\sigma}^\dagger d_{j\sigma}+ \ {\rm h.c.} \,
 \right) \ , \label{part_t}
\end{equation}
does not involve hopping process changing the number of double
occupancies. Here, we used the operators
$a^\dagger_{i,\sigma}\equiv (1-n_{i,-\sigma})c^\dagger_{i,\sigma}$
and $d^\dagger_{i,\sigma}\equiv n_{i,-\sigma}
c^\dagger_{i,\sigma}$. Equation \eq{part_t} is block diagonal and verifies the
choice of $S$ in \eq{S1}.

The full form of $H^{(eff)}$ is now obtained by evaluating all
${\mathcal{O}(t^2/U)}$-terms in \eq{trafo} with $S$ from \eq{S1}.
By restricting ourselves to the subspace of no double occupancies
(the low energy subspace or the lower Hubbard band), we find the $t$-$J$ Hamiltonian,
\begin{equation}
H_{t-J} \equiv P_G \, H^{(eff)} \, P_G = P_G\, (\, T + H_J + H_3 \,) \,P_G \ , \label{tJH}
\end{equation}
where,
\begin{equation}
P_G=\sum_i (1-n_{i\uparrow}\,n_{i\downarrow}) \ ,
\end{equation}
is the Gutzwiller projection operator that projects out all doubly
occupied sites. The terms of the Hamiltonian are given by,
\begin{eqnarray}
T&=&\,-\, \sum_{\langle i,j \rangle,\sigma} {t_{(i,j)}} \left(
c_{i,\sigma}^\dagger c_{j,\sigma} + c_{j,\sigma}^\dagger
c_{i,\sigma} \right)~, \label{kin_tJ}\\
H_J &=& \, \sum_{\langle i,j \rangle} J_{(i,j)} \left( {\bf S}_i
\, {\bf S}_j -
\frac 1 4 n_i n_j \right) ~, \label{Jterm} \\
H_3 &=&- \sum_{i,\tau_1 \neq \tau_2, \sigma} \frac
{J_{(i+\tau_1,i,i+\tau_2)}} 4  c_{i+\tau_1,\sigma}^\dagger
c_{i,{-\sigma}}^\dagger c_{i,{-\sigma}} c_{i+\tau_2,\sigma} \nonumber \\
& & + \sum_{i,\tau_1 \neq \tau_2, \sigma} \frac
{J_{(i+\tau_1,i,i+\tau_2)}} 4 c_{i+\tau_1,{-\sigma}}^\dagger
c_{i,\sigma}^\dagger c_{i,{-\sigma}} c_{i+\tau_2,\sigma}
\label{H2_tJ} \ ,
\end{eqnarray}
where $J_{(i,j)}=4t_{(i,j)}^2/U$ and $J_{(i,j,l)}=4
t_{(i,j)}t_{(j,l)}/U$. $\langle i,j \rangle$ are pairs of neighbour
sites and $i+\tau_{(1,2)}$ denotes a neighbour site of $i$.
Equation \eq{tJH}, together with \eq{kin_tJ}-\eq{H2_tJ}, gives the
full form of $t$-$J$ Hamiltonian. However, the so-called
correlated hopping or three-site term $H_3$ is often ignored
since its expectation value is proportional both to
$t^2/U$ and the doping level $x$. Further, the density-density
contribution $n_i n_j$ is sometimes neglected within the
superexchange term $H_J$, as it is a constant at half-filling.
Note that \eq{kin_tJ} is equivalent to \eq{part_t} due to the
projection operators $P_G$ occurring in the definition \eq{tJH} of
the $t$-$J$ Hamiltonian.

The unitary transformation illustrates the relationship between
superexchange and the physics of the (strong coupling) Hubbard model.
We see that as a result of the unitary transform, the low energy
model is given by the
$t$-$J$ Hamiltonian \eq{tJH} which does not allow for double occupancies.
At half filling, each site is singly occupied and
the hopping of electrons is frozen since real hopping now leads
to states in the upper Hubbard band. As a result, the
kinetic energy term in the Hamiltonian vanishes,
and the $t$-$J$ Hamiltonian reduces to an
antiferromagnetic Heisenberg model \eq{HeisenbergH}.

The original Hamiltonian relevant for the
cuprates contains three bands per unit cell,
one copper band and two oxygen-derived bands.
One band only crosses the Fermi surface with
a single effective degree of freedom per
unit cell, the Zhang-Rice singlet \cite{Zhang88a},
corresponding to an empty site in $t-J$
terminology. Using this venue, the hopping
matrix elements and the
superexchange parameters relevant for
the $t-J$ model could be derived directly.
The Hubbard-$U$ entering the relations
derived above then takes the role of
an effective modelling parameter.

\subsection{Implementations of the RVB concept}

The $t$-$J$ Hamiltonian
\eq{tJH} is more suitable than the Hubbard model for studying
RVB superconductivity, because it
includes the superexchange term explicitly, and it is
this term which is responsible for the formation of singlets.
However, for exact
numerical methods, the $t$-$J$ Hamiltonian provides only a minor
simplification over the Hubbard Hamiltonian, and one must turn to
approximate schemes for any calculations on sufficiently large
clusters. In the following, we start with the $t$-$J$ Hamiltonian as an
appropriate microscopic model for HTSC, and briefly discuss three
schemes that allow for systematic calculations of
the RVB state.

\subsubsection{Gutzwiller projected wave functions}
\label{projectedWave}

Anderson \cite{Anderson87} proposed projected BCS wave functions
as possible RVB trial states for the $t$-$J$ model. These states
provide a suggestive way to describe an RVB liquid in an elegant and compact
form\footnote{For a real space representation of equation
\eq{Psi_RVB} we refer to \sect{realspaceRVB}.},
\begin{equation}
|\Psi_{\rm RVB}\rangle= P_N\,P_G\,|{\rm BCS} \rangle
\label{Psi_RVB} \ ,
\end{equation}
with the BCS wave function
\begin{equation}
 |{\rm BCS} \rangle=  \prod_{\bk}\left(
u_\bk+v_\bk c_{\bk\uparrow}^\dagger c_{-\bk\downarrow}^\dagger
\right)|0\rangle~, \label{Psi_BCS}
\end{equation}
which constitutes a singlet pairing state. Here, the operator
$P_G$ (Gutzwiller projection operator) projects out double
occupancies and the $P_N$ fixes the particle number to $N$;
$u_\bk$ and $v_\bk$ are the variational parameters with the
constraint, $u_\bk^2+v_\bk^2\equiv1$. The form of
$|\Psi_{\rm RVB}\rangle$ provides a unified description
of the Mott insulating phase and the doped conductor. It
immediately suggests the presence of singlet correlations in the undoped
correlations and relates them to a superconducting state away from
half filling.

Projected wave functions were originally proposed by Gutzwiller in
1963 to study the effect of correlations presumed to induce
ferromagnetism in transition metal compounds \cite{Gutzwiller63}.
In subsequent years, these wave functions were applied to study
the Mott-Hubbard metal insulator transition \cite{Brinkman70} and
for a description of liquid $^3$He as an almost localised Fermi
liquid \cite{Gros87,Vollhardt84,Seiler86}, \emph{etc.}
However, these early studies considered only a projected Fermi sea,
\begin{equation}
P_G |\Psi_{{\rm FS}}\rangle\ =\ P_G\, \prod_{\bk<\bk_F}
 c_{\bk\uparrow}^\dagger c_{\bk\downarrow}^\dagger |0\rangle~,
\label{psi0_FS}
\end{equation}
in the Hubbard model, whereas Anderson \cite{Anderson87} suggested
a projected BCS paired wave function for the $t$-$J$ model.

To calculate the variational energy of a projected state $|\Psi
\rangle \equiv P_G |\Psi_0 \rangle$, expectation values of the
form
\begin{equation}
\frac{\langle \Psi_0|\,P_G\,\hat O \,P_G\,|\Psi_0\rangle} {\langle
\Psi_0|P_GP_G|\Psi_0\rangle}\  \label{matele}
\end{equation}
must be considered, where $\hat O$ is the appropriate operator. Here,
$|\Psi_0 \rangle$ can be any wave function with no restriction
in the number of double occupancies, {\it viz.} it lives in the
so-called `pre-projected' space. The choice of $|\Psi_0$ In our case we concentrate on
$|\Psi_0 \rangle=|{\rm BCS} \rangle$. In \sect{other_trial} we
will review a few other types of trial wave functions
used to study correlated electron systems. The exact evaluation of
\eq{matele} is quite sophisticated and requires variational Monte
Carlo (VMC) techniques that will be discussed in \chap{VMC}.
However, approximate analytical calculations can be done by a
renormalisation scheme based on the Gutzwiller approximation (GA).
The GA will will be outlined in the sections \ref{GutzApp} and
\ref{RMFT}. Within this approximation, the effects of projection
on the state $|\Psi_0 \rangle$ are approximated by a classical
statistical weight factor multiplying the expectation value with
the unprojected wave function \cite{Vollhardt84}, {\it i.e.}
\begin{equation}
\frac{\langle \Psi_0|\,P_G\, \hat O \,P_G\,|\Psi_0\rangle}{\langle
\Psi_0|P_G P_G|\Psi_0\rangle}\ \approx \ g_O\, \frac{ \langle \Psi_0|
\hat{O} |\Psi_0\rangle}{\langle \Psi_0|\Psi_0\rangle}~ \ .
\label{renorm}
\end{equation}
The so-called Gutzwiller renormalisation factor $g_O$ only depends
on the local densities and is derived by Hilbert space counting
arguments \cite{Zhang88b,Vollhardt84,Ogawa75} or by considering
the limit of infinite dimensions ($d=\infty$)
\cite{Metzner88,Gebhard90,Buenemann05a,Buenemann05b}. The GA shows
good agreement with VMC results (see \cite{Zhang88b}) and is
discussed detailed in section \ref{GutzApp}.

Gutzwiller projected wave functions thus have the advantage that they
can be studied both analytically (using the GA and extensions thereof)
and numerically (using VMC techniques and exact diagonalisation). Since
these wave functions provide a simple way to study correlations such as
pairing correlations, magnetic correlations \emph{etc.} in the presence
of a large Hubbard repulsive interaction, they have been used
extensively in the literature. As we will show
in the following sections, the Gutzwiller-RVB theory of superconductivity
explains several key features of the HTSC . More generally, we believe this approach
is sufficiently broad that it could be used to study a wide range of
physical phenomena in the proximity of a Mott transition.

\subsubsection{Slave boson mean field theory (SBMFT) and RVB gauge theories}
Another representation of the $t$-$J$ Hamiltonian, equation
\eq{tJH}, is obtained by removing the projection operators $P_G$,
and replacing the creation and annihilation operators by
\begin{deqarr}
c^\dagger_{i,\sigma}\ &\to& \ \tilde c^\dagger_{i,\sigma} c^\dagger_{i,\sigma} \,(1-n_{i,-\sigma}) \ ,\ {\rm and}\  \\
c_{i,\sigma}\ &\to& \ \tilde c_{i,\sigma} = c_{i,\sigma} \,(1-n_{i,-\sigma})  \ ,
\end{deqarr}
with $\sigma=\uparrow,\downarrow$ and $-\sigma$ denoting the
opposite spin of $\sigma$. In this form the restriction to no
double occupation is fulfilled by the projected operators $\tilde
c^\dagger_{i,\sigma}$ and $\tilde c_{i,\sigma}$. Thus, only empty
and single occupied sites are possible, which can be expressed by
the local inequality
\begin{equation}
\sum_{\sigma} \ \langle \tilde c^\dagger_{i,\sigma} \tilde
c_{i,\sigma}\rangle \leq 1.
\end{equation}
However, the new operators do not satisfy the fermion
commutation relations, which makes an analytical treatment
difficult. The slave-boson  method
\cite{Barnes76,Coleman84,Kotliar86} handles this problem by
decomposing $\tilde c^\dagger_{i,\sigma}$ into a fermion operator
$f_{i,\sigma}^\dagger$ and a boson operator $b_{i}$ via
\begin{equation}
\tilde c^\dagger_{i\sigma} = f_{i,\sigma}^\dagger b_{i} \ .
\end{equation}
The physical meaning of $f_{i,\sigma}^\dagger$ ($f_{i,\sigma}$) is
to create (annihilate) a single occupied site with spin $\sigma$,
those of $b_{i}$ ($b_{i}^\dagger$) to annihilate (create) an empty
site. Since every site can either by single occupied by an
$\uparrow$-electron, single occupied by a $\downarrow$-electron,
or empty the new operators must fulfill the condition
\begin{equation}
f_{i\uparrow}^\dagger f_{i\uparrow} + f_{i\downarrow}^\dagger
f_{i\downarrow} + b^\dagger_{i} b_{i} = 1 \ \ . \label{constSB}
\end{equation}
When writing the Hamiltonian in terms of the slave fermion and
boson operators the constraint \eq{constSB} is implemented by
a Lagrangian multiplier $\lambda_i$. In the slave-boson
representation, the $t$-$J$ model is thus written as,
\begin{eqnarray}
H_{t-J}\,=\,&-& \sum_{\langle i,j \rangle,\sigma} {t_{(i,j)}}
\left( f_{i,\sigma}^\dagger b_i b_j^\dagger f_{j,\sigma} +
f_{j,\sigma}^\dagger  b_j b_i^\dagger f_{i,\sigma}\right)
 \label{HamiltonianSB} \\
&-& \sum_{\langle i,j \rangle} J_{(i,j)} \left(
f^\dagger_{i\uparrow} f^\dagger_{j\downarrow} -
f^\dagger_{i\downarrow} f^\dagger_{j\uparrow} \right) \left(
f_{i\downarrow} f_{j\uparrow} - f_{i\uparrow} f_{j\downarrow}
\right)
 \nonumber \\
&-& \mu_0\sum_{i,\sigma}\, f_{i,\sigma}^\dagger f_{i,\sigma} +
\sum_i \lambda_i \,( f_{i\uparrow}^\dagger f_{i\uparrow} +
f_{i\downarrow}^\dagger f_{i\downarrow} + b^\dagger_{i} b_{i} - 1)
 \ , \nonumber
\end{eqnarray}
where the Heisenberg exchange term,
$$
{\bf S}_i \, {\bf S}_j - \frac 1 4 n_i n_j = - \left(
f^\dagger_{i\uparrow} f^\dagger_{j\downarrow} -
f^\dagger_{i\downarrow} f^\dagger_{j\uparrow} \right) \left(
f_{i\downarrow} f_{j\uparrow} - f_{i\uparrow}
f_{j\downarrow}\right)~,
$$
is a function of fermion operators only,
since superexchange does not lead to charge fluctuations.
\cite{Baskaran87}. Furthermore,
a chemical potential term, $-\mu_0\sum_{i,\sigma}\,
f_{i,\sigma}^\dagger f_{i,\sigma}$, is included within
the grand canonical ensemble.

The advantage of this representation is that the operators
($f_{i\sigma}$, $b_i$) obey standard algebra and can thus be
treated using field theoretical methods. The partition function
$Z$ of \eq{HamiltonianSB} can be written as a functional integral
over coherent Bose and Fermi fields, allowing to calculate
observables in the original Hilbert space. The Fermi fields can be
integrated out using standard Grassmann variables. Then carrying
out a saddle-point approximation for the Bose fields reproduces
the mean field level. The incorporation of Gaussian fluctuations
around the saddle point approximation provides a possibility for
systematic extensions of the SBMFT. One way to implement the constraint
of single occupancy is to formulate the problem as a gauge theory.

The development of RVB correlations and a superconducting phase in a lattice model
as a gauge theory was first studied by Baskaran and Anderson \cite{Baskaran87b}
These authors noted that the Heisenberg Hamiltonian has a local $U(1)$ gauge symmetry, which arises
precisely because of the constraint of single occupancy. One may then develop an effective
action which obeys this local symmetry and use it to calculate various averages. Since the free
energy exhibits the underlying gauge symmetry, it is possible to go beyond mean field theory
when calculating averages of physical quantities. Doping turns the local gauge
symmetry into a (weaker) global $U(1)$ symmetry which can be broken spontaneously, leading
to superconductivity. Subsequently, Wen and Lee introduced an $SU(2)$ gauge theory which leads
to RVB correlations and superconductivity in a doped Mott insulator \cite{Wen96}. These
approaches are reviewed in a recent work by Lee, \etal \ \cite{Lee06}. It should be noted
that the Gutzwiller approximation, the SBMFT (which is the mean field solution about
which gauge theories are constructed) are similar in the sense that both model the
doped Mott insulator. In particular, real kinetic energy is frozen as one approaches
half filling, and enhanced RVB correlations. In general, the
results from SBMFT are quite similar to those from RMFT, {\it
e.g.} the early prediction of $d$-wave superconductivity in the
$t$-$J$ model rests on very similar gap equations in both schemes.
The SBMFT result showing $d$-wave pairing by Kotliar and Liu
\cite{Kotliar88} and by Suzumura, \etal \ \cite{Suzumura88} nearly
simultaneously appeared with the respective RMFT study by Zhang,
\etal \ \cite{Zhang88b}. These studies followed an earlier work of
Baskaran \etal \ \cite{Baskaran87}, who initially developed a
slave boson theory for the $t$-$J$ model. For a more detailed
review on SBMFT we refer to \cite{Lee06}. The SBMFT and Gutzwiller
approaches differ in the way the
local constraint is treated and consequently, there are quantitative discrepancies between
these approaches. Some of these will be highlighted in subsequent sections of this review.


\subsubsection{The bosonic RVB theory}

As the name indicates, this approach is based on a bosonic description
of the $t$-$J$ model. The advantage of this method is that it accounts
well for the antiferromagnetic correlations of the Heisenberg model at half filling
as well as of the hole doped $t$-$J$ model. At half filling, the ground state
of the bosonic RVB theory (b-RVB) is related to the RVB wave function of
Liang \textit{et al.} \cite{Liang88} which is the best variational wave
function available for the Heisenberg model. The basic premise of the b-RVB
theory is that hole doping an insulator with AF correlations (not necessarily
long ranged) lead to a singular effect called the ``phase string'' effect
\cite{weng_99}. A hole moving slowly in a closed path acquires a nontrivial
Berry's phase. As this effect is singular at the length scales of a lattice
constant, its topological effect can be lost in conventional mean field
theories. So, the theory proposes to take this effect into account explicitly
before invoking mean field like approximations. The electron operator is
expressed in terms of bosonic spinon and holon operators, and a topological
vortex operator as,
$$
c_{i\sigma} = h^\dagger_i b_{i\sigma} e^{i\hat{\Theta}_{i\sigma}}~.
$$
The phase operator $\hat{\Theta}_{i\sigma}$ is the most important ingredient
of the theory and reflects the topological effect of adding a hole to an AF
background. The effective theory is described by holons and spinons coupled
to each other by link fields.

Away from half filling, the ground state of the b-RVB theory
is described by a holon condensate and an RVB paired state of spinons.
The superconducting order parameter is characterised by phase vortices
that describe spinon excitations and the superconducting transition occurs
as a binding/unbinding transition of such vortices \cite{vnm_02a}. The
theory leads naturally to a vortex state above $T_c$ of such spinon
vortices \cite{vnm_02b}. Bare spinon and holon states are
confined in the superconducting state and nodal (fermionic) quasiparticles are obtained
as composite objects \cite{zhou_03}.

The b-RVB theory realises transparently, the original idea of Anderson of holes moving
in a prepaired RVB state. As mentioned above, the theory leads to definite and verifiable
consequences such as a vortex state of spinons above $T_c$ and spinon excitations trapped
in vortex cores. However, the exact relationship between the b-RVB ground state and the
simple Gutzwiller projected BCS wave function has not been clarified yet \cite{Ivanov06}.

\subsection{Variational approaches to correlated electron systems}
\label{other_trial}

In this subsection, we briefly discuss how projected states,
\begin{equation}
 |\Psi \rangle = P_G |\Psi_0 \rangle \ , \label{PW}
\end{equation}
can be extended to study a wide variety of
strongly correlated systems.
variational basis. Apart from the HTSC, these wave functions
have been used in the description of Mott
insulators \cite{Capello05}, superconductivity in organic
compounds \cite{Liu05,Watanabe06}, Luttinger liquid
behaviour in the $t$-$J$ model \cite{Hellberg91,Valenti92}.

\subsubsection{Order parameters}

A simple extension of the trial state \eq{PW} is to allow
for additional order parameters in the mean field wave function
$|\Psi_0 \rangle$. In \sect{projectedWave}, we restricted
ourselves to a superconducting BCS wave function $|\Psi_0
\rangle=|{\rm BCS} \rangle$. However, antiferromagnetic
\cite{lee_88,Chen90,Himeda99,Giamarchi91,Ivanov04}, $\pi$-flux
\cite{lee_90,Ivanov04,Ivanov03}, or charge ordered
\cite{Anderson04b,Huang05,Poilblanc05,Li06} mean field wave
functions can also be used for $|\Psi_0 \rangle$. In addition,
a combination of different kind of orders is possible. As an
example, consider the trial wave function,
\begin{equation}
 |\Psi_0\rangle=  \prod_{\bk}\left(
u_\bk+v_\bk \, b_{\bk\uparrow}^\dagger \, b_{-\bk\downarrow}^\dagger
\right)|0\rangle~ \ ,
\label{Psi_BCSAF}
\end{equation}
with
\begin{equation}
b_{\bk\sigma}=\alpha_\bk  c_{\bk \sigma} +  \sigma  \beta_\bk
c_{\bk+{\mathbf Q}\sigma} \ .
\end{equation}
Equation \eq{Psi_BCSAF} includes finite superconducting as well as
antiferromagnetic order \cite{Chen90}. Here, $b_\bk$ is the
Hartree-Fock spin-wave destruction operator with ${\mathbf
Q}=(\pi,\pi)$ as required for a commensurate antiferromagnet. The
parameters, $\alpha_\bk$ and $\beta_\bk$ are related to the
antiferromagnetic order parameter $\Delta_{\rm AF}$ by usual mean
field relations; similarly, the superconducting order parameter
determines the values of $v_\bk$ and $u_\bk$. In the sections
\ref{RMFT} and \ref{VMC}, we will discuss applications of above
wave function for the HTSC.

We note that $|\Psi_0 \rangle$ is applicable to all
lattice geometries. It has been used, for instance,
to study superconductivity in
triangular lattice based Cobaltates
\cite{Weber06a,Baskaran03,Kumar03,Ogata03a} and organic compounds
\cite{Powell05,Gan05b,Liu05,Watanabe06}. Recent calculations show
that projected states also provide a competitive energy on more
exotic models such as a spin-1/2 Heisenberg model on a Kagome
lattice \cite{Ran07}.

\subsubsection{Jastrow correlators}

The incorporation of Jastrow correlator  ${\cal J}$
\cite{Jastrow55} provides an additional powerful way to extend
the class of (projected) trial wave functions. In \eq{PW}, the original Gutzwiller projector
$P_G$ can be viewed as the simplest form of a Jastrow correlator,
\begin{equation}
P_G = {\cal J}_g = g^{\sum_i n_{i,\uparrow} n_{i,\downarrow}} = \prod_i
\left ( 1 - ( 1- g)n_{i,\uparrow} n_{i,\downarrow} \right) \ .
\label{GutzJastrow}
\end{equation}
So far we only considered $P_G$ in the fully projected limit,
which corresponds to $g \to 0$ in $J_g$. However, when using
\eq{GutzJastrow} in the Hubbard model, $g$ becomes a variational
parameter that determines the number of doubly occupied sites.

The variational freedom of the trial wave function can be
increased by including further Jastrow correlators,
\begin{equation}
 |\Psi \rangle \,=\,
{\cal J}_{s} \, {\cal J}_{hd} \,{\cal J}_d\, P_G \,|\Psi_0 \rangle\,
= \,{\cal J}_{s}\, {\cal J}_{hd} \,{\cal J}_d\, {\cal J}_g \,|\Psi_0 \rangle \ . \label{PWex}
\end{equation}
Popular choices of Jastrow correlators are the density-density correlator ${\cal J}_d$,
\begin{equation}\label{densityjastrow}
{\cal J}_d = \exp \left ( -\sum_{(i,j)} v_{ij} (1-n_i)(1-n_j)
\right ),
\end{equation}
the holon-doublon correlator ${\cal J}_{hd}$,
\begin{equation}\label{holondoubonjastrow}
{\cal J}_{hd} = \exp \left (-\sum_{(i,j)} w_{ij} ( h_i d_j + d_i
h_j) \right ),
\end{equation}
with $h_i=(1-n_{i\uparrow})(1-n_{i\downarrow})$ and
$d_i=n_{i\uparrow} n_{i\downarrow}$, and the spin-spin correlator
${\cal J}_s$,
\begin{equation}\label{spinjastrow}
{\cal J}_s = \exp \left ( - \sum_{(i,j)} u_{ij} S^z_i S^z_j \right )
\ .
\end{equation}
The corresponding variational parameter are given by
$v_{ij}$, $w_{ij}$, and $u_{ij}$, respectively.

Since the generalised trial wave function \eq{PWex}
includes a very high number of
variational parameters, one invariably chooses
a small set depending on the problem at hand.
In the case of the
$t$-$J$ model the situation is slightly simplified,
because double occupancies are forbidden and thus $g \to 0 $ and $w_{ij}=0$.

We now discuss the properties of the density-density correlator in
\eq{densityjastrow} and assume $u_{ij}=w_{ij}=0$ for a moment. A
positive $v_{ij}$ implies density-density repulsion, a negative
$v_{ij}$ means attraction and may lead to phase separation.
Several studies indicate the importance of long range
density-density Jastrow correlators for improving the variational
energy. Hellberg and Mele \cite{Hellberg91} showed that the
one-dimensional $t$-$J$ model can be accurately described when
$v_{i,j} \sim \log|i-j|$, {\it i.e.} when the Jastrow correlator
is scale invariant. The incorporation of long-ranged
density-density correlations induces Luttinger liquid like
behaviour in the $t$-$J$ model \cite{Hellberg91,Valenti92}. In the
one-dimensional Hubbard model an appropriate choice of the
density-density correlator in momentum space allows to distinguish
between metallic and insulating behaviour \cite{Capello05}. In the
two-dimensional $t$-$J$ model, ${\cal J}_d$ is often used to
improves the variational energy of a projected superconducting
state \cite{Sorella02a,Yunoki05a} as we will discuss in
\sect{VMC_groundstate}.

The holon-doublon Jastrow correlator ${\cal J}_{hd}$ is important
for studying the repulsive Hubbard model on a variational basis. A
negative $w_{i,j}<0$ implies attraction of empty and doubly
occupied sites which ultimately may lead to a Mott-Hubbard insulating state
(the Mott transition) \cite{Liu05,Watanabe06}. In two dimensions, a negative nearest
neighbour $w_{i,j} \sim - \delta_{\langle i j \rangle}$,
substantially decreases the variational energy
\cite{Liu05,Watanabe06}, since these states occur as intermediate
states during the superexchange process (compare \fig{hopping}).
Combining these effect with a superconducting wave function
$|\Psi_0 \rangle=|{\rm BCS} \rangle$ then explains key aspects of
superconductivity in organic compounds near the Mott-Hubbard
transition \cite{Liu05,Watanabe06}. The $w_{ij}$ seems to be less
important for one dimension, and is likely to be a consequence of the very
good spin-spin correlation energy of the Gutzwiller wave function
in one dimension \cite{Gros87,Gros89a}.

The spin-spin Jastrow correlator ${\cal J}_{s}$ is not as often
used as the density-density and the holon-doublon Jastrow
correlators (${\cal J}_{d}$ and ${\cal J}_{hd}$). However, recent
studies show that the inclusion of ${\cal J}_{s}$ is important
when considering charge fluctuations within the two-dimensional
$t$-$J$ model \cite{Lugas06}. An appropriate spin-spin Jastrow
correlator ${\cal J}_{s}$ can also create antiferromagnetic order
in a non-magnetic wave function, an example for the ability
of Jastrow correlators to induce a new long-range
order not manifest in the unprojected wave function.

\section{Gutzwiller approximation}
\label{GutzApp}

The Gutzwiller approximation (GA) is a straightforward method to
handle Gutzwiller projected wave functions, that incorporate
strong electron correlations by prohibiting doubly occupied sites.
Within the GA, the effects of projection are absorbed by statistical
weight factors (Gutz­willer renormalisation factors), which then
allow for an analytical treatment of strongly correlated
Gutzwiller wave functions.

In this section, we present the derivation of the Gutzwiller
factors by Hilbert space counting argument as well as
considering the limit of infinite dimensions. Further, we discuss
the importance of fugacity factors in the GA when comparing
analytical results with
variational Monte Carlo (VMC) calculations in
the canonical and grand canonical scheme, respectively.  As we
will show in the last part of this section, the GA can also be
extended to the case of partially projected wave functions, where
the projection operator does not act on a single ``reservoir'' site
in the system.

\subsection[Basic principles of the Gutzwiller approximation]
{Basic principles of the Gutzwiller approximation\footnote{To
avoid confusion, in this section we denote density operators with a `hat', and write, {\it e.g.} $\hat
n_{i\sigma}=c^\dagger_{i\sigma} c_{i\sigma} $.}}
\label{GutzPrinc}

The Gutzwiller approximation (or Gutzwiller renormalisation
scheme) constitutes the basis of the RMFT and is a successful
method to treat Hilbert space restrictions due to strong electron
correlations. It was applied originally \cite{Gutzwiller63,Brinkman70}
to calculate the variational energy of the projected Fermi
sea, $P_G |FS\rangle$, in Hubbard like models. In these and
other early papers, the projection operator, $P_G=\prod_i (1-\alpha\, \hat
n_{i\uparrow} \hat n_{i\downarrow})$, was generalised to partial
projection with the parameter $\alpha$ determined by optimising
the energy. Partially projected states were used successfully
in modelling normal liquid $^3$He
\cite{Vollhardt84,Vollhardt87} and heavy fermion systems
\cite{Rice85,Varma86}.

Here, we focus on the $t$-$J$ model (\emph{i.e.}, the
large $U$ limit of the Hubbard model). Consequently, we shall
mainly discuss the fully projected
case, {\it i.e.} $\alpha=1$. We will derive the corresponding
renormalisation factors (Gutzwiller renormalisation factors) in
this limit, and will not not discuss the generalisation to finite double
occupancy. The latter case is obtained easily following the same
reasoning. The GA,
\begin{equation}
\frac{\langle \Psi_0|\,P_G \hat{O}\,P_G\,|\Psi_0\rangle}{\langle
\Psi_0|P_GP_G|\Psi_0\rangle}\ \approx \ g_O\, \frac{ \langle \Psi_0|
\hat{O} |\Psi_0\rangle}{\langle \Psi_0|\Psi_0\rangle}~\ , \label{Gutz1}
\end{equation}
approximates the expectation value within the projected state
$P_G\,|\Psi_0\rangle$ by a corresponding statistical weight $g_O$
multiplying the matrix element within the unprojected wave
function $|\Psi_0\rangle$. To determine the Gutzwiller
renormalisation factor $g_O$ we can either invoke Hilbert space
counting arguments \cite{Zhang88b,Vollhardt84,Ogawa75}, or
consider the limit of infinite dimensions ($d=\infty$)
\cite{Metzner88,Gebhard90,Buenemann05a,Buenemann05b}. In the
following, we review both techniques and compare the respective
results.

\subsubsection[Counting arguments]{Gutzwiller renormalisation factors by counting arguments}
\label{countingGA}

Hilbert space counting arguments enable us to derive the
renormalisation factor $g_O$ through simple physical reasoning. We
may use,
\begin{equation}
g_O \ \approx \ \frac{ \langle \hat{O} \rangle_{\Psi} } {
\langle \hat{O} \rangle_{\Psi_0} } \label{def_g}
\end{equation}
with $|\Psi\rangle \equiv P_G|\Psi_0\rangle$, as  defining
the factor $g_O$; $\langle...\rangle_{\Psi}$ denotes the
expectation value with respect to the (projected) wave function $|\Psi\rangle$.
In the GA, the ratio in \eq{def_g} is determined
by neglecting correlations in the wave functions $|\Psi\rangle$
and $|\Psi_0\rangle$. The physical quantity
which determines the theory is the occupancy at any site $i$. Thus,
one calculates the probabilities for a site $i$ to be empty, singly
occupied with spin $\sigma$, and doubly occupied,
respectively. These probabilities are obtained by considering the
Hilbert space restrictions and are summarised for $|\Psi\rangle$
and $|\Psi_0\rangle$ in \tab{prob}. In this context, we should
note that the densities before projection ($n^0_i$,
$n^0_{i\downarrow}$ and $n^0_{i\uparrow}$) and after projection
($n_i$, $n_{i\downarrow}$ and $n_{i\uparrow}$) may differ. This is
due to the projection operator, $P_G=\prod_i (1-\hat
n_{i\downarrow}\hat n_{i\uparrow})$, which can, {\it e.g.} remove
more terms with an $\uparrow$-electron than a
$\downarrow$-electron on site $i$. Such effects become of
importance for Gutzwiller projection in antiferromagnetic, charge
ordered, or grand-canonical states. Keeping this caveat in mind,
the expectation values in \eq{def_g} can be calculated
approximately by considering the probability amplitudes of `bra'- and
`ket'-configurations that contribute. We obtain the Gutzwiller
renormalisation factor by calculating the ratio between these
approximate expectation values. Although, we neglect any
off-site correlations in the derivation of the Gutzwiller
renormalisation factor, the GA itself \eq{Gutz1} incorporates
additional correlations by the expectation value of $\hat O$ in
$|\Psi_0\rangle$. Extensions of the GA, which incorporate more correlation effects,
were proposed by Ogata and Himeda (\cite{Ogata03b}, see also \sect{AF_RMFT})
and Hsu \cite{Hsu90}.

\begin{table}[b]
\center
\begin{tabular}{l||l|l}
\hline \hline  & \multicolumn{2}{|l}{probabilities} \\
\cline{2-3} \raisebox{1.5ex}[-1.5ex]{occupancy on site $i$} & in $|\Psi\rangle$& in $|\Psi_0\rangle$ \\
\hline \hline $\langle (1-\hat n_{i\downarrow})(1-\hat n_{i\uparrow}) \rangle$ &
 $1-n_i$ & $(1-n^0_{i\downarrow})(1-n^0_{i\uparrow})$ \\
\hline  $\langle \hat n_{i\downarrow}(1- \hat n_{i\uparrow}) \rangle$ &
 $n _{i\downarrow}$ & $n^0_{i\downarrow}(1-n^0_{i\uparrow})$ \\
\hline  $\langle \hat n_{i\uparrow}(1- \hat n_{i\downarrow}) \rangle$ &
 $n_{i\uparrow}$ & $n^0_{i\uparrow}(1-n^0_{i\downarrow})$ \\
\hline  $\langle \hat n_{i\downarrow} \hat n_{i\uparrow} \rangle$ & $0$ &
 $n^0_{i\downarrow}n^0_{i\uparrow}$ \\
\hline
\end{tabular}
\caption{Probability for different occupancies on site $i$ in
$|\Psi\rangle$ and $|\Psi_0\rangle$. We distinguish between the densities before
projection ($n^0_i$, $n^0_{i\downarrow}$ and $n^0_{i\uparrow}$)
and after projection ($n_i$, $n_{i\downarrow}$ and $n_{i\uparrow}$).} \label{prob}
\end{table}

\begin{figure}
\centering
\includegraphics*[width=0.5\textwidth]{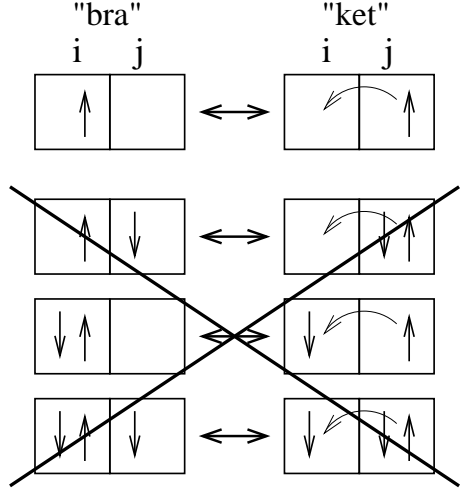}
\caption{Required bar- and ket-configurations, so that $\langle
(1-\hat n_{i\downarrow} ) c^\dagger_{i\uparrow} (1-\hat
n_{j\downarrow} ) c_{j\uparrow} \rangle$ contributes in
$|\Psi\rangle$ and $|\Psi_0\rangle$. Configurations that do not
contribute to $\langle (1-\hat n_{i\downarrow} )
c^\dagger_{i\uparrow} (1-\hat n_{j\downarrow} ) c_{j\uparrow}
\rangle$ are crossed out.} \label{hopping_Gutz}
\end{figure}

To illustrate the above scheme, we consider the expectation value of
the hopping element, $\langle c^\dagger_{i\uparrow} c_{j\uparrow}
\rangle$. For a projected state, $|\Psi\rangle=P_G|\Psi_0\rangle$,
we can write
\begin{equation}
\langle c^\dagger_{i\uparrow} c_{j\uparrow} \rangle_{\Psi}=\langle (1-\hat n_{i\downarrow} )
 c^\dagger_{i\uparrow} (1-\hat n_{j\downarrow} ) c_{j\uparrow} \rangle_{\Psi}\ .
\label{projElement}
\end{equation}
We then perform the GA for the right hand side of
\eq{projElement}, which is written in terms of projected operators
$(1- \hat n_{i\downarrow} ) c^\dagger_{i\uparrow}$ and $(1- \hat
n_{j\downarrow} ) c_{j\uparrow}$. It is convenient to rewrite the
matrix elements in this manner before performing the GA, since it
guarantees agreement with the infinite dimensions approach. Next
we consider the probability for $\langle (1-\hat n_{i\downarrow} )
c^\dagger_{i\uparrow} (1- \hat n_{j\downarrow} ) c_{j\uparrow}
\rangle$ in $|\Psi\rangle$ and $|\Psi_0\rangle$. Configurations
can only contribute if the bra-vector
has a single $\uparrow$-electron on site $i$ and a vacancy on site $j$.
For the ket-vector
the interchanged occupancies are necessary, {\it i.e.} a single
$\uparrow$-electron on site $j$, and a vacancy on site $i$. The
corresponding hopping process is illustrated in
\fig{hopping_Gutz}. With the help of \tab{prob} we find the
amplitudes of the bra- and ket-contribution, and the product gives
the probability in ${|\Psi\rangle}$,
\begin{equation}
\left[n_{i\uparrow}(1-n_{j})\right]^{1/2} \cdot
\left[n_{j\uparrow}(1-n_{i})\right]^{1/2} \ , \label{hop_proj}
\end{equation}
and in ${|\Psi_0\rangle}$,
\begin{equation}
\left[ n^0_{i\uparrow}(1-n^0_{i\downarrow})
(1-n^0_{j\downarrow})(1-n^0_{j\uparrow}) \right]^{1/2} \cdot
\left[ n^0_{j\uparrow}(1-n^0_{j\downarrow})
 (1-n^0_{i\downarrow})(1-n^0_{i\uparrow}) \right]^{1/2} \ .  \label{hop_unproj}
\end{equation}
The square roots stem from the fact that both
bra- and ket-vector only provide amplitudes; the probability
is obtained by a product of two amplitudes.

Combining \eq{hop_proj} and \eq{hop_unproj} yields
\begin{eqnarray}
&&\frac{\langle (1-\hat n_{i\downarrow} ) c^\dagger_{i\uparrow} (1-\hat n_{j\downarrow} )
c_{j\uparrow} \rangle_{|\Psi\rangle}}
{\langle (1-\hat n_{i\downarrow} ) c^\dagger_{i\uparrow} (1-\hat n_{j\downarrow} )
c_{j\uparrow} \rangle_{|\Psi_0\rangle}} \nonumber \\
 && \qquad\approx \tilde g_t=\frac 1
{(1-n^0_{i\downarrow}) (1-n^0_{j\downarrow})} \cdot \frac
{\left[n_{i\uparrow}(1-n_{j})n_{j\uparrow}(1-n_{i})\right]^{1/2}}
{\left[ n^0_{i\uparrow} (1-n^0_{j\uparrow})
n^0_{j\uparrow}(1-n^0_{i\uparrow}) \right]^{1/2}} \
\label{hopp_GutzTilde}.
\end{eqnarray}
The expectation value in $|\Psi\rangle$ is now obtained by renormalising
the unprojected value by \eq{hopp_GutzTilde},
\begin{deqarr}
\langle c^\dagger_{i\uparrow} c_{j\uparrow} \rangle_{\Psi}&=&
\arrlabel{hopping_GA} \langle (1-\hat n_{i\downarrow} )
c^\dagger_{i\uparrow} (1-\hat n_{j\downarrow} ) c_{j\uparrow}
\rangle_{\Psi} \\ &\approx& \tilde g_t \, \, \langle (1-\hat
n_{i\downarrow} ) c^\dagger_{i\uparrow} (1-\hat n_{j\downarrow} )
c_{j\uparrow} \rangle_{\Psi_0}  \\
& \approx& \underbrace{\tilde g_t \, \,(1-n^0_{i\downarrow}
)(1-n^0_{j\downarrow} )}_{=g_t} \, \, \langle
c^\dagger_{i\uparrow}  c_{j\uparrow} \rangle_{\Psi_0} \
\label{hopping_GA3}.
\end{deqarr}
In the last row of \eq{hopping_GA} we decoupled the densities in
${|\Psi_0\rangle}$. The Gutzwiller renormalisation factor is then,
\begin{equation}
 g_t=\frac
{\left[n_{i\uparrow}(1-n_{j})n_{j\uparrow}(1-n_{i})\right]^{1/2}}
{\left[ n^0_{i\uparrow} (1-n^0_{j\uparrow})  n^0_{j\uparrow}(1-n^0_{i\uparrow})
 \right]^{1/2}} \label{hopping_factor_den} \ .
\end{equation}
We emphasise that the decoupling in \eq{hopping_GA3} is controlled in
the limit of infinite dimensions, {\it viz.} all neglected
decouplings yield off-site correlations of higher
order\footnote{Strictly speaking, we violate this rule by
neglecting decouplings which include on-site pairing, $\langle
c^\dagger_{i\uparrow}c^\dagger_{i\downarrow}\rangle $. However, we
work in the fully projected limit, {\it i.e.} $|\Psi\rangle$ does
not allow for on-site pairing. It is thus reasonable to prohibit
on-site pairing in $|\Psi_0\rangle$ as well and to set $\langle
c^\dagger_{i\uparrow}c^\dagger_{i\downarrow}\rangle\equiv 0$.} and
thus vanish for $d=\infty$. Violating this rule causes deviations
from the mathematical thoroughness of the infinite dimension
scheme.

For the full determination of the Gutzwiller renormalisation
factor in \eq{hopping_factor_den}, it is necessary to evaluate the
dependence of the densities after projection relative to the
densities prior to projection. The situation is particularly
simple for a homogeneous wave functions with fixed particle number
and spin symmetry, where
$n^0_{i\uparrow}=n^0_{i\downarrow}=n^0_i/2=n/2$ on each site $i$.
Then, $n_{i\uparrow}=n_{i\downarrow}=n/2$, and the Gutzwiller
factor simplifies to the well-known result,
\begin{equation}
 g_t=\frac {1-n}{1-n/2} \label{hopping_simplest} \ ,
\end{equation}
which  incorporates the fact that the kinetic energy in $|\Psi
\rangle$ is connected to the motion of holes, vanishing in the
undoped case.

However, the relation of the $n_{i\sigma}$ with respect to the
$n_{i\sigma}^0$ become more subtle, if we consider, {\it e.g.} an
antiferromagnet with sublattice magnetisation $m$, where
$n^0_{A\sigma}=n/2 ± \,m$, and, $n^0_{A\sigma}=n^0_{B-\sigma}$
(sublattices $A$ and $B$, $\sigma=\uparrow,\downarrow$). In this
case, $n_{i\sigma} \neq n^0_{i\sigma}$, and we must invoke
counting arguments to determine $n_{i\sigma}$. We consider a
canonical ensemble, where the overall particle density is the same
before and after projection ($n_{i}=n_{i\uparrow} +
n_{i\downarrow} = n^0_{i\uparrow} + n^0_{i\downarrow}=n^0_{i}=n$).
Furthermore, the density $n_{i\sigma}$ is necessarily related to
the probability of finding a single $\sigma$-electron at site $i$ in
$|\Psi_0\rangle$. Thus, $n_{i\sigma} \propto n^0_{i\sigma}
(1-n^0_{i-\sigma})$. Due the conserved particle density,
\begin{equation}
n_{i\uparrow}+n_{i\downarrow}=n=n^0~,
\end{equation}
and so,
\begin{equation}
n_{i\sigma}=n^0_{i\sigma} (1-n^0_{i-\sigma})
 \frac{n}{n-2 n^0_{i\uparrow}n^0_{i\downarrow}} \ . \label{renorm_nsigma}
\end{equation}
Inserting this expression in the numerator of
\eq{hopping_factor_den} gives the Gutzwiller renormalisation
factor,
\begin{equation}
 g_t=\frac {1-n}{1-2 n^0_{\uparrow}n^0_{\downarrow}/n} \  , \label{hopping_AF}
\end{equation}
where $n^0_{\uparrow}$ and $n^0_{\downarrow}$ are from the same
site. We note that \eq{renorm_nsigma} is valid for sites $i$ and
$j$ on the same as well as on different sublattices as one can
show easily and reduces to, $g_t=(1-n)/(1-n/2)$, in the
non-magnetic limit, $n^0_{\sigma}=n/2$.

The situation becomes yet more complicated if we consider states
with an inhomogeneous particle density, where it is difficult to
determine $n_{i}$, $n_{i\downarrow}$, and $n_{i\uparrow}$.
Therefore, most authors assume $n_{i}=n^0_{i}$. However, this assumption is
incorrect, because the operator
$P_G=\sum_i (1-\hat n_{i\uparrow}\, \hat n_{i\downarrow})$ allows
for changes in the local particle density. An elegant solution is
to redefine the operator $P_G$, so that $n_{i}=n^0_{i}$ or even
$n_{i\sigma}=n^0_{i\sigma}$. This conservation of local particle
densities can be achieved by incorporating appropriate fugacity
factors (that describe the local chemical potential)
into a new operator $\tilde P_G$ (Gutzwiller correlator),
which is then not a projection operator any more. The redefined
operator $\tilde P_G$ still allows to present any projected wave
function as $|\Psi\rangle=\tilde P_G|\tilde \Psi_0\rangle$,
however, the unprojected wave function $|\tilde \Psi_0\rangle$
will generally differ from $|\Psi_0\rangle$ defined by
$|\Psi\rangle=P_G|\Psi_0\rangle$. The use of $\tilde P_G$ instead
of $P_G$ is often not explicitly stated in literature, although
the assumed conservation of densities is only valid for a
generalised Gutzwiller correlator $\tilde P_G$. Such a clear
distinction between $\tilde P_G$ and $P_G$ becomes
particularly important when results from the GA are compared to VMC
calculations that implement the original Gutzwiller
projector $P_G$. The non-conserving of local particle densities by
the operator $P_G$ also explains discrepancies between VMC
calculations in the canonical and the grand canonical scheme
\cite{Edegger05a}. This will be discussed in more detail in
\sect{fugacity}.

Before turning to the $d=\infty$ scheme, let us discuss the
the Gutzwiller renormalisation factor $g_S$ for the
superexchange interaction, defined by
\begin{equation}
\langle {\mathbf S}_{i} {\mathbf S}_{j}
\rangle_{\Psi}\,=\,g_S\,\langle  {\mathbf S}_{i} {\mathbf S}_{j}
\rangle_{\Psi_0}\ . \label{Gutz_SS}
\end{equation}
We first consider the GA for the contribution $\langle S^+_{i}
S^-_{j} \rangle$, {\it i.e.}
\begin{equation}
\langle S^+_{i} S^-_{j} \rangle_{\Psi}\,=\,g^{±}_S\,\langle
S^+_{i} S^-_{j} \rangle_{\Psi_0}\ . \label{Gutz_SpSm}
\end{equation}
The procedure resembles the derivation of $g_t$. We note that the
process $S^+_{i} S^-_{j}$ requires, an $\uparrow$-spin on site $i$
and a $\downarrow$-spin on site $j$ in the bra-vector, and the
reverse in the ket-vector. Therefore, the probability becomes
\begin{equation}
(\,n_{i\uparrow}n_{j\downarrow}n_{i\downarrow}n_{j\uparrow}\,)^{1/2}
\end{equation}
in the state ${|\Psi\rangle}$, while it is
\begin{equation}
[\,n^0_{i\uparrow}(1-n^0_{i\downarrow})n^0_{j\downarrow}(1-n^0_{j\uparrow})
n^0_{i\downarrow}(1-n^0_{i\uparrow})n^0_{j\uparrow}(1-n^0_{j\downarrow})\,]^{1/2}
\end{equation}
in the state ${|\Psi_0\rangle}$. Using $n_{i,\sigma}$ from
\eq{renorm_nsigma} yields,
\begin{equation}
g^±_S=\frac 1 {(1-2 n^0_{\uparrow}n^0_{\downarrow}/n)^{2}}\ .
\label{SiSj_GutzGeneral}
\end{equation}
One can show again, that above formula results also for the case
of sites belonging to the same sublattice.

Next we evaluate the GA for the diagonal contribution to the
superexchange,
\begin{equation}
\langle S^z_{i} S^z_{j} \rangle_{\Psi}\,=\,g^z_S\,\langle S^z_{i}
S^z_{j} \rangle_{\Psi_0}\ . \label{Gutz_SzSz}
\end{equation}
Here, we use, $S^z_{i}=1/2 (\hat n_{i\uparrow}-\hat
n_{i\downarrow})$, and write,
\begin{align}
4\,\langle S^z_{i} S^z_{j} \rangle &= \langle \hat n_{i\uparrow}
(1-\hat n_{i\downarrow})\hat n_{j\uparrow} (1-\hat
n_{j\downarrow}) \rangle + \langle \hat n_{i\downarrow} (1-\hat
n_{i\uparrow})\hat n_{j\downarrow} (1-\hat
n_{j\uparrow}) \rangle \nonumber \\
& - \langle \hat n_{i\uparrow} (1-\hat n_{i\downarrow})\hat
n_{j\downarrow} (1-\hat n_{j\uparrow}) \rangle - \langle \hat
n_{i\downarrow} (1-\hat n_{i\uparrow})\hat n_{j\uparrow} (1-\hat
n_{j\downarrow}) \rangle \ , \label{SzSz_densities}
\end{align}
which is valid for any wave function. The
Gutzwiller approximations of the terms in \eq{SzSz_densities} give
a common renormalisation factor,
\begin{equation}
g^z_S=\frac 1 {(1-2 n^0_{\uparrow}n^0_{\downarrow}/n)^{2}}\ .
\label{SzSz_GutzGeneral}
\end{equation}
This is seen by considering the term $\langle \hat n_{i\uparrow}
(1-\hat n_{i\downarrow})\hat n_{j\uparrow} (1-\hat
n_{j\downarrow}) \rangle$ in \eq{SzSz_densities}, as an example.
By applying the probabilities from \tab{prob}, we obtain,
\begin{equation}
\frac{ \langle \hat n_{i\uparrow} (1-\hat n_{i\downarrow})\hat
n_{j\uparrow} (1-\hat n_{j\downarrow}) \rangle_{\Psi}} { \langle
\hat n_{i\uparrow} (1-\hat n_{i\downarrow})\hat n_{j\uparrow}
(1-\hat n_{j\downarrow}) \rangle_{\Psi_0} } \approx \frac{
n_{i\uparrow}  n_{j\uparrow} } {
 n^0_{i\uparrow} (1-n^0_{i\downarrow})
n^0_{j\uparrow} (1- n^0_{j\downarrow})  } \equiv g^z_S \ ,
\end{equation}
where using \eq{renorm_nsigma} for $n_{i\uparrow}  n_{j\uparrow}$
directly confirms \eq{SzSz_GutzGeneral}. Since all density terms
of \eq{SzSz_densities} renormalise in exact the same manner,
$g^z_S$ gives the correct renormalisation factor for $\langle
S^z_{i} S^z_{j} \rangle_{\Psi}$ in \eq{Gutz_SzSz}.

From \eq{SiSj_GutzGeneral} and \eq{SzSz_GutzGeneral}, we find a
common Gutzwiller renormalisation factor, $g_S=g^±_S=g^z_S$, for
\eq{Gutz_SS}, which simplifies to,
\begin{equation}
g_S=\frac{1}{(1-n/2)^2}\ , \label{SiSj_Gutz}
\end{equation}
in the non-magnetic limit, $n^0_{\sigma}=n/2$. At half-filling,
$n=1$ and $g_S \to 4$, the magnetic correlations are four times as
pronounced in $|\Psi \rangle$ than in $|\Psi_0 \rangle$. We note
that Gutzwiller approximations for other quantities are easily
obtained by following the same reasoning as for $g_t$ and $g_S$.

\subsubsection[Infinite dimensions]{Gutzwiller renormalisation factors
in infinite dimensions}

The effects of the Gutzwiller correlator can be evaluated exactly
in the limit of infinite dimensions \cite{Metzner88,Gebhard90}.
Gebhard \cite{Gebhard90} showed that a simple diagrammatic
evaluation is possible for $d=\infty$. Using the Gutzwiller
renormalisation factors from $d=\infty$ for finite dimensions
corresponds to a mean field approximation. Thus the $d=\infty$
approach provides a systematic way to calculate Gutzwiller
factors. Typically, one is interested in the doping
dependence of such factors and here, the results from $d=\infty$ are in
qualitative agreement with Gutzwiller factors calculated using counting
arguments. Discrepancies between the two methods are merely quantitative.
Here, we summarise the calculation of Gutzwiller factors in the limit $d=\infty$,
for fully projected states. The reader is referred to recent works of Bünemann, \etal \
\cite{Buenemann05a,Buenemann05b} for a detailed account.

To simplify calculations, the Gutzwiller projector $P_G$ is
reformulated as a Gutzwiller correlator $\tilde P_G$ within the
$d=\infty$ scheme. This redefinition agrees with the one
discussed earlier and ensures that local densities are conserved,
{\it viz.} $n_{i\sigma}=n_{i\sigma}^0$. The Gutzwiller correlator,
$\tilde P_{G}=\prod_i \tilde P_{G,i}$, is written as a product of
local correlators,
\begin{equation}
\tilde P_{G,i}=\lambda^0_i \ (1-\hat n_{i\downarrow})(1- \hat n_{i\uparrow})\,+
\,\lambda^\uparrow_i\ \hat n_{i\uparrow}(1- \hat n_{i\downarrow})\,+\,
\lambda^\downarrow_i \  \hat n_{i\downarrow}(1- \hat n_{i\uparrow}) \ .
\end{equation}
Physically, the parameters $\lambda^0_i$ and $\lambda^\sigma_i$ allow to weight
locally the probabilities to find empty sites and sites occupied with a spin $\sigma$,
respectively. The $\lambda^0_i$, $\lambda^\uparrow_i$, and $\lambda^\downarrow_i$
are determined by the constraints,
\begin{eqnarray}
\langle {\tilde P_{G,i}}^2 \rangle_{\tilde \Psi_0} &\equiv& 1 \ , \label{condPnorm} \\
\langle {\tilde P_{G,i}}\, \hat n_{i\sigma} {\tilde P_{G,i}}
\rangle_{\tilde \Psi_0} &\equiv&
\langle \hat n_{i\sigma} \rangle_{\tilde \Psi_0} =  n^0_{i\sigma}
\label{condPdensity} \ .
\end{eqnarray}
Equation \eq{condPnorm} guarantees the normalisation, $\langle
\Psi | \Psi\rangle=\langle \Psi_0|\tilde P_G \tilde P_G
|\Psi_0\rangle =1$, of the projected wave function and equation
\eq{condPdensity} provides the conservation of local densities.
Evaluating these equations, we find,
\begin{eqnarray}
\lambda^0_i&=&\sqrt{\frac{1-n_i}{(1- n_{i\downarrow})(1- n_{i\uparrow})}} \ , \\
\lambda^\sigma_i&=&\sqrt{\frac 1 {(1- n_{i-\sigma})}} \ .
\end{eqnarray}
Using these parameters in the Gutzwiller correlator $\tilde P_G$
guarantees via \eq{condPnorm} a conserved norm and via
\eq{condPdensity} conserved spin densities for any projected wave
function, $|\Psi\rangle\equiv \tilde P_G |\tilde \Psi_0\rangle$.
The GA for an operator $\hat O_{ij}$ acting on the sites $i$ and
$j$ is now obtained by neglecting all correlations except those
between sites $i$ and $j$. This procedure becomes exact in
infinite dimensions and is written as,
\begin{equation}
\langle\tilde \Psi_0| \tilde P_G \hat O_{ij} \tilde P_G |\tilde \Psi_0\rangle \label{dInf_Gutz}
=\langle\tilde \Psi_0| \tilde P_{G,i} \tilde P_{G,j} \hat O_{ij} \tilde P_{G,i} \tilde
P_{G,j}  |\tilde \Psi_0\rangle \ .
\end{equation}
Decoupling the right hand site and neglecting all off-site
correlations of higher order, provides the exact solution for
$d=\infty$, which agrees with the results from counting arguments
presented in \sect{countingGA}.

As an example, we consider the hopping process,
$\langle c^\dagger_{i\uparrow} c_{j\uparrow} \rangle_{\tilde P_G |\Psi_0\rangle}$.
Using \eq{dInf_Gutz}, we find,
\begin{deqarr}
\langle\tilde \Psi_0|\tilde P_{G} c^\dagger_{i\uparrow} c_{j\uparrow} \tilde P_{G}
 |\tilde \Psi_0\rangle \arrlabel{hopping_infinite}
=\langle\tilde \Psi_0|\tilde P_{G,i} c^\dagger_{i\uparrow}
 \tilde P_{G,i}\tilde P_{G,j} c_{j\uparrow}
\tilde P_{G,j} |\tilde \Psi_0\rangle \, \, &&\\
= \lambda^\uparrow_i \lambda^0_i \lambda^0_j \lambda^\uparrow_j \
\langle\tilde \Psi_0| (1-\hat n_{i\downarrow} )
c^\dagger_{i\uparrow}
(1-\hat n_{j\downarrow} ) c_{j\uparrow} |\tilde \Psi_0\rangle \, \, && \\
= \underbrace{\lambda^\uparrow_i \lambda^0_i \lambda^0_j
\lambda^\uparrow_j (1- n_{i\downarrow} ) (1-n_{j\downarrow}
)}_{=g_t} \ \langle\tilde \Psi_0|  c^\dagger_{i\uparrow}
c_{j\uparrow} |\tilde \Psi_0\rangle  \, , &&
\end{deqarr}
where we decoupled the densities in the last row as already done
in the discussion using counting arguments. Equation
\eq{hopping_infinite} is exact in infinite dimensions, and gives
the Gutzwiller renormalisation factor,
\begin{equation}
 g_t=\sqrt{\frac
{(1-n_{j})(1-n_{i})}
{(1-n_{j\uparrow})  (1-n_{i\uparrow})}}  \ ,
\end{equation}
which agrees with \eq{hopping_factor_den} if we assume locally
conserved densities. However, we note that this result differs
from \eq{hopping_AF}, which incorporates the changed spin
densities due to the projection operator $P_G$. The scheme presented
above is applicable to any kind of operator, and gives the exact
result for $d=\infty$. It provides an useful check for results
derived from counting arguments. Nevertheless we must keep in mind
that results may differ depending on our choice of $P_G$ and
$P_G|\Psi_0\rangle$ or $\tilde
P_G$ and $\tilde P_G|\tilde \Psi_0\rangle$, in the counting
arguments leading to the derivation of the Gutzwiller renormalisation
factors.

\subsection[Canonical and grand canonical scheme]{Gutzwiller approximation in the
canonical and the grand canonical scheme}
\label{fugacity}

In this subsection we follow Edegger, \etal \ \cite{Edegger05a}
and study the effects of projection on superconducting BCS wave
functions,
\begin{equation}
| \Psi_0 \rangle=|{\rm BCS} \rangle \equiv  \prod_{\bk}\left(
u_\bk+v_\bk c_{\bk\uparrow}^\dagger c_{-\bk\downarrow}^\dagger
\right)|0\rangle~ \ .
\end{equation}
Since $|\Psi_0\rangle=|{\rm BCS} \rangle$ exhibits particle number
fluctuations, the projection operator $P_G$ can change the average
particle number $N$ of the wave function, {\it i.e.} in general,
\begin{equation}
{\langle \Psi_{{0}}| \hat{N} |\Psi_{{0}}\rangle \over\langle
\Psi_{{0}}|\Psi_{{0}}\rangle} \neq {\langle
\Psi_{{0}}|P_G~\hat{N}~P_G |\Psi_{{0}}\rangle \over\langle
\Psi_{{0}}|P_G^2 |\Psi_{{0}}\rangle}~.
\end{equation}
In the above equation, the equality between the l.h.s. and r.h.s.
could be recovered by replacing
the Gutzwiller projector $P_G$ by a Gutzwiller correlator $\tilde
P_G$ which conserves local densities as discussed in the previous
subsection. Here, we follow a different route to
compensate the effects of projection by using a fugacity factor in
the wave function. This ansatz explains differences observed
between VMC calculations in the canonical framework (fixed
particle number) and the grand canonical ensemble (fluctuating
particle number) using the corresponding GA.

\subsubsection{Incorporation of a fugacity factor}

We first examine the particle number distributions
to illustrate the effect of the projection operator $P_G$
in the projected Hilbert space. Towards this end, we write
the average numbers, $\bar{N}^{(0)}(\bar{N})$ in the
unprojected (projected) Hilbert space, as,
\begin{equation}
\bar{N}^{(0)} = \sum_N N  \, \rho^{(0)}_N~,\qquad
\bar{N} = \sum_N N \, \rho_N~, \label{mean_after1}
\end{equation}
where,
\begin{equation}
\rho^{(0)}_N = \frac{\langle \Psi_{{0}}|\, P_N
\,|\Psi_{{0}}\rangle}{\langle \Psi_{{0}}|\Psi_{{0}}\rangle} , \qquad
{\rho}_N = \frac{\langle \Psi_{{0}}|\,P_G~P_N~P_G
\,|\Psi_{{0}}\rangle}{\langle
\Psi_{{0}}|P_G~P_G~|\Psi_{{0}}\rangle}~,
\end{equation}
are the particle number distributions in the unprojected and
projected BCS wave functions respectively. Here, the operator
$P_N$ describes the projection onto terms with particle number $N$.
As discussed in \cite{Edegger05a}, we can relate
the particle number distributions before and after projection by
\begin{equation}
\underbrace{\frac{\langle \Psi_{{0}}|\,P_G~P_N~P_G
\,|\Psi_{{0}}\rangle}{\langle
\Psi_{{0}}|P_G~P_G~|\Psi_{{0}}\rangle}}_{\rho_N} = g_N \,
\underbrace{\frac{\langle \Psi_{{0}}|\, P_N
\,|\Psi_{{0}}\rangle}{\langle
\Psi_{{0}}|\Psi_{{0}}\rangle}}_{\rho^{(0)}_N} \, \label{gutzP} ,
\end{equation}
with
\begin{equation}
g_N=\underbrace{\frac{\langle
\Psi_{{0}}|\Psi_{{0}}\rangle}{\langle
\Psi_{{0}}|P_G~P_G~|\Psi_{{0}}\rangle}}_{=C(=\rm const)}
\frac{\langle \Psi_{{0}}|\, P_G~P_N~P_G
\,|\Psi_{{0}}\rangle}{\langle
\Psi_{{0}}|\,P_N\,|\Psi_{{0}}\rangle}  \ .
\end{equation}
The above equation \eq{gutzP} describes the GA for the projection operator $P_N$
with the corresponding renormalisation factor $g_N$. The parameter $C$ is an
irrelevant constant, which does not depend on $N$.
It follows that if we were to impose the condition that the average
particle numbers before and after projection be identical,
a factor $g_N^{-1}$ needs to be included in \eq{mean_after1}.
Then, from \eq{mean_after1} and \eq{gutzP}, we obtain the particle
number after projection $\bar{N}_{{\rm new}}$,
\begin{equation}
\bar{N}_{{\rm new}} \equiv\   \sum_N N\,\frac{1}{g_N}\, \rho_N \
=\ \sum_N\, N \, \frac{g_N\,\rho^{(0)}_N}{g_N} \ =\ \bar{N}^{(0)}
~, \label{mean_after2}
\end{equation}
which is the desired result.


This procedure can be implemented for the
wave function $|\Psi_{{0}}\rangle$. Since the BCS wave function is
a linear superposition of states with particle numbers $\ldots,
N-2, N, N+2, \ldots $, we consider the effect of projection on two
states whose particle numbers differ by two. Then, the ratio is
\begin{equation}
f^2 \ \equiv\  \frac{g_{N+2}}{g_N} \ \approx\
\left(\frac{L-N}{L-N/2}\right)^2\, \to \left( \frac{1-n}{1-n/2} \right)^2=g_t^2
\label{define_f} \end{equation}
where the factors $g_N$ were evaluated combinatorially in the
thermodynamic limit \cite{Edegger05a}. We note that the fugacity
factor $f$ is equal to $g_t$, the Gutzwiller renormalisation factor
for the hopping term (see also \sect{GutzPrinc}).
Equation \eq{define_f} shows that the
projection operator acts unequally on the $N$ and $N+2$ particle
states; the renormalisation of the weight of the $N+2$ particle
states ${g_{N+2}}$, is ${g_t^2}$ times the weight of the $N$
particle states, ${g_N}$. This effect can be rectified as in
\eq{mean_after2} by multiplying every Cooper pair
$c_{\bk\uparrow}^\dagger c_{-\bk\downarrow}^\dagger$ by a
amplitude $\frac{1}{g_t}$ in the BCS wave function.
Alternatively (following Anderson \cite{Anderson04c}), we
can multiply every empty state by the factor $g_t$, and write
\begin{equation}
|{\Psi}^{(f)}_{{0}}\rangle  = \prod_\bk \frac{\left(g_t\,u_\bk + v_\bk c_{\bk \uparrow}^
\dagger c_{-\bk \downarrow}^\dagger \right)} {\sqrt{g_t^2
|u_\bk|^2 +|v_\bk|^2}}\,| 0 \rangle ~.\label{pwa_BCS}
\end{equation}
Then again by construction, the fugacity factor $f=g_t$ in
\eq{pwa_BCS} ensures that the projected wave function
$P_G|\Psi^{(f)}_{{0}}\rangle$ and the unprojected wave function
$|\Psi_{{0}}\rangle$ without fugacity factor have the same
particle number. The denominator in \eq{pwa_BCS} is the new
normalisation factor. Two points deserve further attention. The
first is that a relative phase factor between $N$ and $N+2$ particle
states in the projected BCS wave function can be absorbed into the
definition of the fugacity factor. The second point is that
our ansatz for the fugacity factor assumes that the effects of
the projection operator are independent of $k$; \textit{i.e.}, the
fugacity factor we obtain is independent of $k$. We do not see
\textit{a priori} why the fugacity factor cannot depend on $k$. This
is an interesting line of investigation since a $k$- dependent
fugacity factor in the Gutzwiller - BCS wave function would lead to
(experimentally) verifiable consequence.

\subsubsection[Singular particle number renormalisation]
{Singular particle number renormalisation close to half-filling}

We saw that the inclusion of the fugacity factor is necessary
for the average particle number in a BCS wave function to remain
unchanged when projecting out all doubly occupied sites.
Alternatively, one might ask what is the effect of the projection
operator on a BCS wave function, if projection changes
the mean particle number, because no fugacity factor is introduced.

\begin{figure}[t]
  \centering
 \includegraphics*[width=0.67\textwidth]{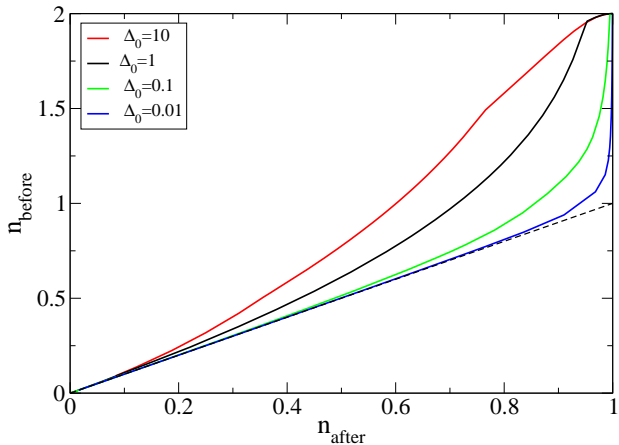}
 \caption{
Particle density before projection $n_{{\rm before}}$, equation \eq{p_before},
as a function of the particle density after projection $n_{\rm after}$,
equation \eq{p_after2}, for different $d$-wave order parameters $\Delta$. The dashed line
indicates the Fermi liquid
result $n_{\rm before}=n_{\rm after}$. From \cite{Edegger05a}. \label{fig:nf_na}}%
\end{figure}

In this situation, as shown by Edegger, \etal \ \cite{Edegger05a}, the
the particle density after projection is determined by the
self-consistent equation,
\begin{equation}
\bar N_{\rm after} \equiv  \frac{\langle
{\Psi}_{{0}}|\,P_G\,\hat{N}\,P_G
 \,|{\Psi}_{{0}}\rangle}
{\langle {\Psi}_{{0}}|\,P_G\,|{\Psi}_{{0}}\rangle} \approx
 2\,\sum_\bk \frac{g_t^2 |{v}_\bk|^2} {|{u}_\bk|^2
+g_t^2 |{v}_\bk|^2} \label{p_after2}~ ,
\end{equation}
which can be solved iteratively with
$g_t$ specified by the particle number in
$\bar N_{\rm after}$. Since the particle density in
the state $|{\Psi}_{{0}}
\rangle$ before projection is given by,
\begin{equation}
\bar n_{\rm before}
\equiv \frac {\bar N_{\rm before}} L =  \frac 2 L \,\sum_\bk |{v}_\bk|^2~,  \label{p_before}
\end{equation}
equation \eq{p_after2} provides a way to calculate the
particle number in the state $P_G|{\Psi}_{{0}} \rangle$
\emph{after} projection, whenever the particle number in the
state $|{\Psi}_{{0}}\rangle$ \textit{before} projection
is known as a function of the identical factors $u_\bk$ and $v_\bk$.

In the following, we discuss numerical solutions of
\eq{p_after2}, where we use the standard BCS
expressions for a $d$-wave superconductor,
\begin{equation}
v^2_\bk =\frac{1}{2} \left( 1 - \frac{\xi_\bk}{E_k} \right)~,
\quad u^2_\bk =\frac{1}{2} \left( 1 + \frac{\xi_\bk}{E_k}
\right)~, \label{start_dwave}
\end{equation}
with,
\begin{deqarr}
E_\bk &=&\sqrt{ \Delta_\bk^2 + \xi_\bk^2 },\\
\Delta_\bk &=&\Delta \,(\, \cos k_x - \cos k_y \,),\\
\xi_\bk &=&-2\,(\, \cos k_x + \cos k_y \,)-\mu~. \arrlabel{end_dwave}
\end{deqarr}
Therefore, the only free parameters, which must be specified within the calculations,
are the chemical potential $\mu$ and the order parameter $\Delta$.

The particle numbers (before and after projection) for fixed
values of the order parameter $\Delta$ can now be determined as a
function of the chemical potential. The results for the particle
densities are shown in \fig{fig:nf_na}. The results clearly show
that the particle density before projection attains its maximal
value ($n_{\rm before}=2$), when $n_{\rm after}=1$ (half-filling).
This result holds for any finite value of the order parameter
$\Delta$. The case of half-filling, $n_{\rm after} \to 1$, is
therefore singular in the grand canonical scheme and large
deviations with respect to the canonical framework can be
expected. In the opposite limit, {\it viz.} low densities of
electrons, $n_{\rm before}$ converges to the value of $n_{\rm
after}$ as expected. As illustrated by the results in
\fig{fig:nf_na}, the size of the intermediate region depends on
the magnitude of the order parameter $\Delta$, {\it i.e.} the
effects of projection are largest for wavefunction with a large
$\Delta$ and consequently large particle number fluctuations.

\begin{figure}
  \centering
  \includegraphics*[width=0.72\textwidth]{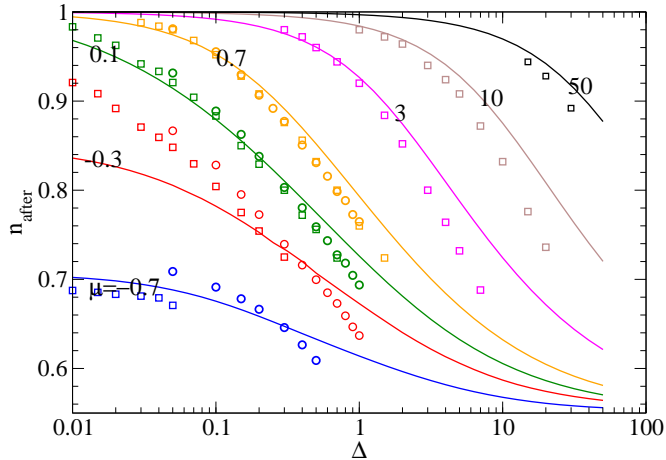}
  \caption{\label{fig:compare_Yoko}%
The particle density  after projection $n_{\rm after}$ as a
function of the parameter $\Delta$ for a $d$-wave BCS
state at various chemical potentials $\mu$. The figure shows a
comparison between results from equation \eq{p_after2} (solid lines)
and the VMC results of Yokoyama and Shiba \cite{Yokoyama88} [for
$6×6$ - (circles) and $8×8$ -lattices (squares)].
Numbers in the figure denote the chemical potentials of the
corresponding curves. From \cite{Edegger05a}.}
\end{figure}

To check the accuracy of equation \eq{p_after2} we compare it with
VMC calculations of Yokoyama and Shiba (YS) \cite{Yokoyama88}, who
numerically studied a projected BCS wave functions with
fluctuating particle number (but without a fugacity factor). YS
determined the particle density of the projected $d$-wave state
$P_G\,|\Psi_{{0}} \rangle$ as a function of the parameter $\Delta$
for various fixed chemical potentials $\mu$ within a grand
canonical scheme (see VMC data in \fig{fig:compare_Yoko}). Since
the unprojected wave function $|\Psi_{{0}} \rangle$ was specified
through \eq{start_dwave}-\eq{end_dwave} in the VMC calculation, we
can also determine the relation between $n_{\rm after}$ and
$\Delta$ by \eq{p_after2}. As shown in \fig{fig:compare_Yoko}, the
results from \eq{p_after2} are in good qualitative agreement with
the VMC data of YS. Small discrepancies are mostly explained by
finite size corrections in the VMC calculation (VMC calculation
only for $6×6$ and $8×8$-lattices). Figure \ref{fig:compare_Yoko}
clearly reveals the singular effect of the projection near the
insulating phase (half filling), where the chemical potential
diverges to infinity.

\subsubsection[Gutzwiller renormalisation factors]
{Gutzwiller renormalisation factors
in the canonical and the grand canonical ensemble}

\begin{figure}
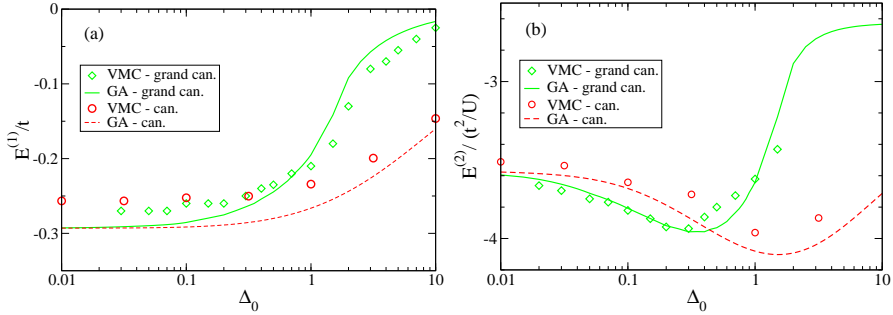

  \centering
  \includegraphics[width=0.48\textwidth]{./Figures/fig2a_YpG.eps}
  \includegraphics[width=0.48\textwidth]{./Figures/fig2b_YpG.eps}
  \caption{(a) The kinetic energy $E^{(1)}$
and (b) the energy of the remaining terms $E^{(2)}$ per site of
the $t$-$J$ model as a function of $\Delta$ for the $d$-wave state
at a filling $n=0.9$. Fixed particle (can., circles) VMC data
\cite{Gros88} and grand canonical (grand can., squares) VMC data
\cite{Yokoyama88} are compared. The dashed/solid lines represent
the corresponding Gutzwiller approximations (GA). From
\cite{Edegger05a}.} \label{compare_VMC}
\end{figure}

Next we discuss the differences between the Gutzwiller
approximation in the canonical and grand canonical scheme. The
validity of the analytical expressions derived in this section can
be confirmed \cite{Edegger05a} by a comparison with numerically
exact VMC calculations \cite{Gros88,Yokoyama88}.

We first consider the canonical case, where we are interested
in the expectation value of an operator $\hat O$ calculated within
a projected wave function $P_N P_G |\,\Psi_{{0}}\rangle$ with
fixed particle number. The GA
corresponding to the operator $\hat O$ is given by,
\begin{deqarr}
&&\frac {\langle \Psi_{{0}}|\, P_G~P_N~\hat{O}~P_N~P_G \,
|\,\Psi_{{0}}\rangle}{\langle
  \Psi_{{0}}|\,  P_G~P_N~P_G\, |\,\Psi_{{0}}\rangle} \label{fixedNa} \\
&\approx & \ g_O\, \frac {\langle \Psi_{{0}}|\, P_N~\hat{O}~P_N
\,|\,\Psi_{{0}}\rangle}{\langle
  \Psi_{{0}}| \,P_N \,|\,\Psi_{{0}}\rangle}   \label{fixedNb}\\
&= & \ g_O\, \frac {\langle \Psi_{{0}}|\, \hat{O}\,
|\,\Psi_{{0}}\rangle}{\langle
  \Psi_{{0}}|\,\Psi_{{0}}\rangle}\  ,
\arrlabel{fixedN}
\end{deqarr}
with the Gutzwiller renormalisation factor $g_O$ and the projector
$P_N$ onto terms with particle number $N$. The term \eq{fixedNa}
represents a quantity which can be calculated exactly by the VMC
scheme with fixed particle number \cite{Gros88,Paramekanti01}.
Since the particle number is fixed, the usual Gutzwiller
approximation can be invoked, leading to  \eq{fixedNb}. The
equality to the last row is guaranteed only when $N$ is equal to
the average particle number of ${|\,\Psi_{{0}}\rangle}$ ($N=\bar
N$). Under this condition a transformation from a canonical to a
grand canonical ensemble is valid in the pre-projected
Hilbert-space.

In the grand canonical scheme the expectation
value of $\hat O$ is calculated
with a particle number non-conserving wave
function. Therefore, this scheme must be modified as follows,
\begin{equation}
\frac {\langle \Psi^{(f)}_{{0}}|\, P_G~\hat{O}~P_G\,
|\,\Psi^{(f)}_{{0}}\rangle}{\langle
  \Psi^{(f)}_{{0}}|\, P_G~P_G \,|\, \Psi^{(f)}_{{0}}\rangle}
\ \approx\   \ g_O\, \frac {\langle \Psi_{{0}}|\, \hat{O}
\,|\,\Psi_{{0}}\rangle}{\langle
  \Psi_{{0}}|\,\Psi_{{0}}\rangle}\  ,
\label{grandC}
\end{equation}
where ${P_G |\, \Psi^{(f)}_{{0}}\rangle}$ is the projected
$d$-wave state including a fugacity factor \eq{pwa_BCS}. This
correction is essential to guarantee the validity of the
Gutzwiller approximation, because without it,
the left hand side (lhs) and
the right hand side (rhs) of \eq{grandC} would correspond to
states with different mean particle numbers.
Equation \eq{grandC} shows that a fugacity factor
must be included into wave functions with particle number fluctuations
when used for a Gutzwiller approximation.

\begin{figure}
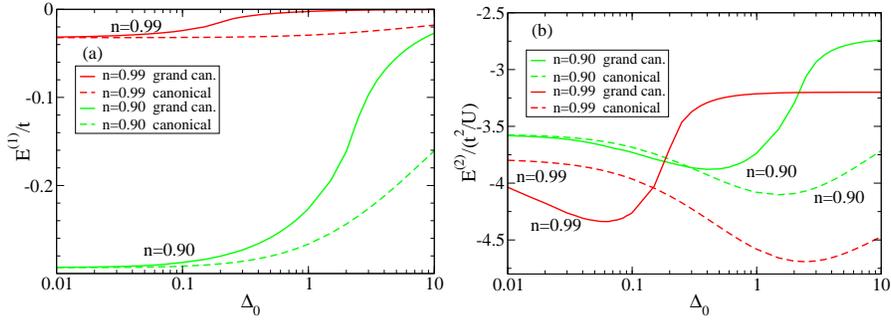

  \centering
  \includegraphics[width=0.48\textwidth]{./Figures/fig2a_YpGNEW.eps}
  \includegraphics[width=0.48\textwidth]{./Figures/fig2b_YpGNEW.eps}
  \caption{Comparison of the canonical and grand canonical scheme at
filling (dashed lines) $n=0.9$ and (solid lines) $n=0.99$. (a) The
kinetic energy $E^{(1)}$ and (b) the energy of the remaining terms
$E^{(2)}$ per site of the $t$-$J$ model are shown as a function of
$\Delta$ for the $d$-wave state. The results are obtained by
Gutzwiller approximations, {\it i.e.} via \eq{fixedN} and
\eq{grandC} for the canonical and the grand canonical scheme,
respectively. The calculations for the grand canonical scheme
follow the steps given by \cite{Edegger05a}.}
\label{compare_VMC_halffilled}
\end{figure}

A comparison of \eq{fixedN} and \eq{grandC} gives the important result,
\begin{equation}
\frac{\langle \Psi_{{0}}| P_G P_N \hat{O} P_N P_G
|\,\Psi_{{0}}\rangle}{\langle
  \Psi_{{0}}|  P_G P_N P_G|\,\Psi_{{0}}\rangle}
\ \approx \ \frac{\langle \Psi^{(f)}_{{0}}| P_G \hat{O}
P_G|\,\Psi^{(f)}_{{0}}\rangle}{\langle
  \Psi^{(f)}_{{0}}| P_G^2 |\, \Psi^{(f)}_{{0}}\rangle}\ . \label{comp_eq}
\end{equation}
This equation shows that to obtain identical results in the grand canonical
(rhs) and canonical (lhs) scheme, one has to use different wave functions.
The wave function ${|\,\Psi^{(f)}_{{0}}\rangle}$ is a $d$-wave
state including a fugacity factor, whereas
${|\,\Psi_{{0}}\rangle}$ is a pure $d$-wave state.

The arguments leading up to \eq{grandC} and \eq{comp_eq} can be
verified by a comparison with VMC studies. In \fig{compare_VMC}, we
show Gutzwiller approximations from Edegger, \etal \ \cite{Edegger05a}
together with the corresponding VMC calculations, {\it i.e.}
fixed particle number VMC \cite{Gros88} for the canonical scheme and
particle number non-conserving VMC \cite{Yokoyama88} for the grand canonical scheme.
The figure shows that canonical
and grand canonical approaches yield different energies, however the GA qualitatively
matches the corresponding VMC results. The differences in the two schemes
are due the projection operator $P_G$, which changes the particle number in a
grand canonical wave function. For these two methods to yield the same
results, a fugacity factor must be incorporated in the wave function when
working in a grand canonical ensemble. We note that the discrepancies
between the canonical and the grand canonical scheme increase significantly
towards half-filling as illustrated in
\fig{compare_VMC_halffilled}. This effect is due to the strong renormalisation
of the particle density in this limit (see \fig{fig:nf_na}).

\subsection[Partially projected states]{Gutzwiller approximation for
partially projected states}
\label{GA_partial}

Thus far, we discussed the Gutzwiller renormalisation scheme
for fully projected wave functions. It is however necessary to
consider sometimes, partially projected states of the form,
\begin{equation}
 |\Psi_l'\rangle \ =\ P_l'\ |\Psi_0\rangle , \qquad
 P_l' = \prod_{i\neq l}(1-\hat n_{i\uparrow} \hat n_{i\downarrow})~.
\label{Psi_l}
\end{equation}
The wave function $|\Psi_l'\rangle$ describes a state where double
occupancies are projected out on all sites except the site $l$,
which we call the ``reservoir'' site. The reason for the appearance of
reservoir sites can be seen as follows. Consider, for example, the
operator $P_G c_{l\uparrow}$, which can be rewritten as
$c_{l\uparrow} P_l'$. Such commutations become necessary, {\it
e.g.} for the calculation of the quasiparticle weight (discussed
in section \ref{EX_RMFT}), where partially projected states arise
inevitably.

Before discussing $|\Psi_l'\rangle$ in more detail, we remark that
the notation `partially projected' is also used for a projection
operator,
\begin{equation}
 P_\alpha = \prod_{i}(1-\alpha \, \hat n_{i\uparrow} \hat n_{i\downarrow})
 \ ,
\end{equation}
with $\alpha \in [0,1]$. The operator $P_\alpha$ is  used for
studying Hubbard-like model with `partially projected' wave
functions $P_\alpha |\Psi_0 \rangle$
 (see also \sect{GutzPrinc}). Here, the
parameter $\alpha$ controls the total number of double
occupancies, whereas $P'_l$ in \eq{Psi_l} yields a fully projected
state with only a single unprojected reservoir site $i$. We
emphasise that the respective Gutzwiller approximations for these
two projection operators are fundamentally different.

Below we follow the work of Fukushima, \etal \ \cite{Fukushima05},
who developed an analytical method to calculate expectation values
for partially projected states [as defined in \eq{Psi_l}]. The
calculations rest on counting arguments, however, similar results
can in principle be obtained within the infinite dimensions
approach. We first determine the local occupancy of the reservoir
site, which is then used to derive the Gutzwiller renormalisation
factors of specific expectation values. We also provide a
comparison to VMC calculations to test the validity of the
approximation and determine its limitations.

We are interested in expectation values such as
\begin{equation}
{\langle \Psi_l'| \hat{O} |\Psi_l'\rangle \over\langle
\Psi_l'|\Psi_l'\rangle }\ =\ g_O'\, {\langle \Psi_0| \hat{O}
|\Psi_0\rangle \over\langle \Psi_0|\Psi_0\rangle }~,
\label{renorm_l}
\end{equation}
that generalise the Gutzwiller approximation to partially projected
wave functions. Note that the reservoir site does not
have a special role in
the unprojected wave function $|\Psi_0\rangle$. This is
in contrast to the impurity problem,
where an impurity site would break the translational
invariance of both the unprojected and of the projected
wave function.

\subsubsection{Occupancy of the reservoir site}

The Gutzwiller approximation in \eq{renorm_l} can be performed by counting arguments
as in the fully projected case (see \sect{countingGA}). However, the
occupancy of the reservoir will differ from the occupancy of a fully projected site,
an effect that must be considered when deriving Gutzwiller renormalisation
factors.

Fukushima, \etal \ \cite{Fukushima05} showed that the
probabilities for the reservoir site to be empty, single occupied,
or double occupied are,
\begin{eqnarray} \label{psi_l_empty}
 \langle (1-\hat n_{l\uparrow})(1-\hat n_{l\downarrow})\rangle _{\Psi_l'} & =&
X(1-n) \approx \frac{(1-n)^2}{(1-n_\uparrow)(1-n_\downarrow)} \ , \\
\label{psi_l_single}
 \langle \hat n_{l\sigma}(1-\hat n_{l-\sigma})\rangle _{\Psi_l'} & =&
X n_\sigma \approx \frac{(1-n) n_\sigma}{(1-n_\uparrow)(1-n_\downarrow)} \ ,\\
\label{psi_l_double}
 \langle d\rangle _{\Psi_l'} \ \equiv\
 \langle \hat n_{l\uparrow} \hat n_{l\downarrow}\rangle _{\Psi_l'} & =&
1-X\approx \frac{n_\uparrow n_\downarrow} {(1-n_\uparrow)(1-n_\downarrow)} \ ,
\end{eqnarray}
respectively. Here
\begin{equation}
X\ =\ {\langle \Psi_0|P_G P_G |\Psi_0\rangle \over \langle \Psi_0|P_l' P_l'
  |\Psi_0\rangle }\   = {\langle \Psi |\Psi \rangle \over \langle \Psi_l'|\Psi_l'\rangle}~
\label{def_X}
\end{equation}
is defined as the ratio between the normalisations
$\langle \Psi|\Psi\rangle $ and $\langle \Psi_l'|\Psi_l'\rangle $.
This ratio can be estimated by a Gutzwiller approximation
\cite{Fukushima05} and becomes
\begin{equation}
X\ \approx \ {1-n\over (1-n_\uparrow)(1-n_\downarrow)}~
\label{X_Gutzwiller}
\end{equation}
in the thermodynamic limit, where $n_\sigma = N_\sigma/L$ ($\sigma=\
\uparrow,\downarrow$) and $n=n_\uparrow+n_\downarrow$ are the respective
particle densities. We note that $X$ vanishes at half-filling.
Consequently the reservoir becomes
exactly doubly occupied,
{\it i.e.} $\lim_{n \to 1 } \ \langle d\rangle _{\Psi_l'} \ = \ 1$.

\subsubsection{Renormalisation of mixed hopping terms}

\begin{figure}
\centering
\includegraphics*[width=0.73\textwidth]{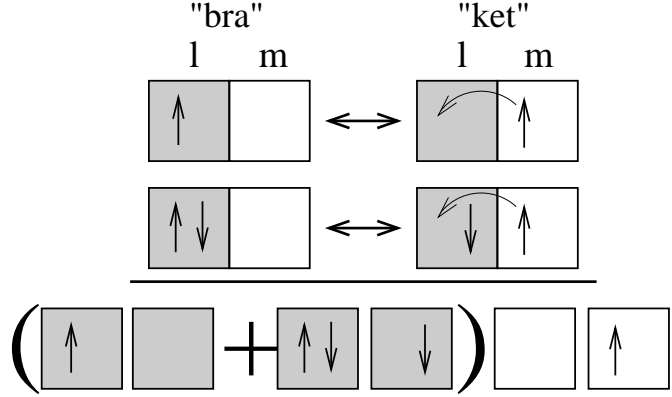}
\caption{Required bra- and ket-configurations, so that $\langle
 c^\dagger_{l\uparrow} (1-\hat
n_{m\downarrow} ) c_{m\uparrow} \rangle$ contribute in $|\Psi'_l
\rangle$ when $l$ is a reservoir site (indicated by a shaded
background). The last row presents the sum from the two possible
contributions as used in \eq{prop_exX}. Boxes with white
background indicate the fully projected site $m$.}
\label{hopping_Gutz_part}
\end{figure}

The occupancies of the reservoir site,
\eq{psi_l_empty}-\eq{psi_l_double}, directly enter the respective
Gutzwiller renormalisation factor $g_O'$. We consider here, as an example,
the mixed hopping term,
\begin{equation}
\frac{\langle \Psi_0|P_{l}' c_{l\sigma}^\dagger
c_{m\sigma}^{\phantom{\dagger}} P_{l}'|\Psi_0\rangle} {\langle
\Psi_0|P_l' P_l'|\Psi_0\rangle} \, \approx g'_t \, \frac{\langle \Psi_0|
c_{l\sigma}^\dagger c_{m\sigma}^{\phantom{\dagger}}
|\Psi_0\rangle} {\langle \Psi_0|\Psi_0\rangle}.
\label{example_part}
\end{equation}
where $l$ denotes the reservoir site and $m \neq l$
is a fully projected site. Following the arguments leading to
\eq{projElement}, we rewrite,
\begin{deqarr}
\langle c^\dagger_{l\uparrow} c_{m\uparrow}
\rangle_{\Psi'_l}&=&\langle c^\dagger_{l\uparrow} (1-\hat
n_{m\downarrow} ) c_{m\uparrow} \rangle_{\Psi'_l}\  \label{Gutz_motifieda} \\
&\approx& \tilde g'_t \ \langle c^\dagger_{l\uparrow} (1-\hat
n_{m\downarrow} ) c_{m\uparrow} \rangle_{\Psi_0}\   \label{Gutz_motifiedb} \\
&\approx& \underbrace{\tilde g'_t (1-n_{m\downarrow} )}_{= g'_t} \ \langle
c^\dagger_{l\uparrow} c_{m\uparrow} \rangle_{\Psi_0}\ \ .
\label{Gutz_motifiedc}
\end{deqarr}
As for \eq{projElement}, we perform the GA for the rhs of \eq{Gutz_motifieda}
to guarantee agreement with the infinite dimensions approach.
However, the decoupling of $(1-n_{m\downarrow} )$ in \eq{Gutz_motifiedc}
becomes exact in infinite dimensions and we
can recover the GA of \eq{example_part}.

In analogy with the calculations in \sect{GutzPrinc}, we consider
the probability for $\langle c^\dagger_{l\uparrow} (1-\hat
n_{m\downarrow} ) c_{m\uparrow} \rangle$ in $|\Psi'_l\rangle$ to
determine the corresponding Gutzwiller renormalisation factor
$\tilde g'_t$ entering in \eq{Gutz_motifiedc}. We illustrate the
two configurations that can contribute, together with the
resulting probability from combining the bra- and ket-vectors, in
\fig{hopping_Gutz_part}. Using \eq{psi_l_empty}-\eq{psi_l_double}
for the partially projected site (grey in
\fig{hopping_Gutz_part}), we find the probability,
\begin{eqnarray}
& &\left(\left[X n_{l\uparrow} X (1-n_l)\right]^{1/2}+\left[(1-X)
X n_{l\downarrow}\right]^{1/2}\right) \cdot
\left [(1-n_{m})n_{m\uparrow}\right ]^{1/2}  \nonumber \\
&=& X \left(\left[ n_{l\uparrow} (1-n_l)\right ]^{1/2}+\left[\frac
{1-X} X n_{l\downarrow}\right]^{1/2}\right)  \cdot
\left [(1-n_{m})n_{m\uparrow} \right ]^{1/2}, \label{prop_exX}
\end{eqnarray}
in $|\Psi'_l\rangle$. With $n_{l\sigma}=n_{m\sigma}=n_\sigma$ and
\eq{X_Gutzwiller}, the above expression simplifies to,
\begin{equation}
X \left[n_\sigma(1-n) + n^2_\sigma\right]=X\,{n_\sigma
(1-n_\sigma)} \ . \label{part_Gutz_illus1}
\end{equation}
For the respective probabilities in $|\Psi_0\rangle$, we use
\tab{prob} and obtain,
\begin{equation}
\left[ n^0_{l\uparrow}(1-n^0_{l\downarrow}) \right]^{1/2} \cdot
\left[ n^0_{m\uparrow}(1-n^0_{m\downarrow})
(1-n^0_{m\downarrow})(1-n^0_{m\uparrow}) \right]^{1/2} \ .
\label{part_Gutz_illus2}
\end{equation}

\begin{figure}
\center
\includegraphics[width=0.7\textwidth,keepaspectratio]{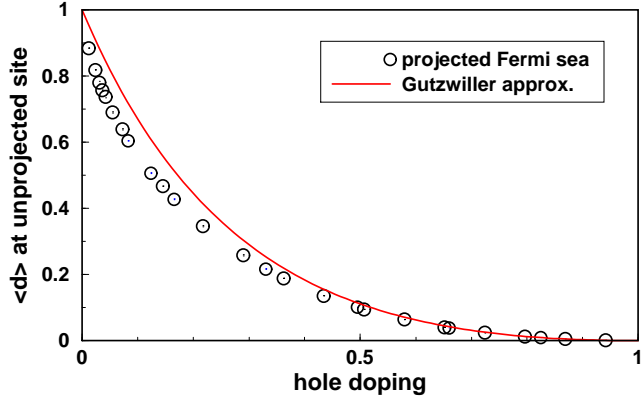}
\caption{Double occupancy of the reservoir site $\langle d\rangle _{\Psi_l'}=\langle
\hat n_{l\uparrow} \hat n_{l\downarrow} \rangle _{\Psi_l'}$ as a function of
doping, for the partially projected Fermi sea.
Note the good agreement between the Gutzwiller result (solid
line), equation \eq{psi_l_double}, and the VMC
results for the projected Fermi sea (open circles). Statistical
errors and finite-size corrections are estimated to be smaller
than the symbols. From \cite{Fukushima05}.}
\label{d_Gutz_doping}
\end{figure}

As pointed out in \sect{GutzPrinc},
$n^0_{l\sigma}=n^0_{m\sigma}=n_\sigma$, for non-magnetic wave
function. We then get the renormalisation factor $\tilde g'_t$
from the ratio of \eq{part_Gutz_illus1} and \eq{part_Gutz_illus2},
{\it i.e.}
\begin{equation}
\tilde g'_t =  \frac X {1-n_\sigma} \ .
\end{equation}
Together with \eq{Gutz_motifiedc}, we obtain the renormalisation
factor,
\begin{equation}
g'_t =  (1-n_\sigma) \tilde g'_t = X = \frac
{1-n}{(1-n_\uparrow)(1-n_\downarrow)}\ ,
\end{equation}
for the GA in \eq{example_part}. Other
expectation values in partially projected states can be calculated
similarly. In \chap{EX_RMFT}, we will use the same
scheme in calculating the quasiparticle weight.

\subsubsection[Comparison with VMC calculations]{Comparison of the
GA for partially projected states with VMC calculations}

\begin{figure}
\center
\includegraphics[width=0.7\textwidth,keepaspectratio]{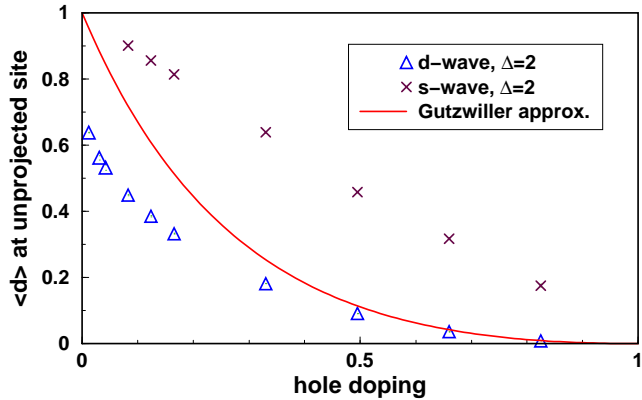}
\caption{Double occupancy of the reservoir site $\langle d\rangle _{\Psi_l'}=\langle
\hat n_{l\uparrow} \hat n_{l\downarrow}\rangle _{\Psi_l'}$ as a function of
doping, for the partially projected BCS wave function. The solid line is the
GA result from \eq{psi_l_double}. The
parameterisation follows \cite{Gros89a}. Statistical errors and
finite-size corrections for the VMC results
are estimated to be smaller than the
symbols. From \cite{Fukushima05}.} \label{d_BCS_doping}
\end{figure}
Before concluding this section, we illustrate a comparison
between \eq{psi_l_double} and VMC results for $\langle d\rangle
_{\Psi_l'}=\langle \hat n_{l\uparrow} \hat n_{l\downarrow} \rangle
_{\Psi_l'}=1-X$. Fukushima, \etal \cite{Fukushima05} found that
the results obtained by a generalised Gutzwiller approximation are
in excellent qualitative agreement with the VMC results for a
partially projected Fermi sea as shown in \fig{d_Gutz_doping}. We
also used VMC to obtain the same quantity using projected
$s$/$d$-wave BCS states\footnote{The BCS states are defined by,
$|v_\bk|^2\, =\, 1/2 \, (1 - \xi_\bk/E_\bk )$, and $u_\bk v^*_\bk
\, = \, \Delta_\bk/(2 E_\bk)$, where $\xi_\bk=-2\,(\cos k_x + \cos
k_y)-\mu$ and $E_\bk=\sqrt{|\Delta_\bk|^2+\xi_\bk^2}$ [s-wave:
$\Delta_\bk=\Delta$, d-wave: $\Delta_\bk=\Delta \, (\cos k_x -
\cos k_y)$].} as variational states in the simulation. The results
for $\langle d\rangle _{\Psi_l'}$ in BCS states are shown in
\fig{d_BCS_doping}. In contrast to the projected Fermi sea, a
clear deviation from the Gutzwiller approximation is seen. This
underscores the importance of pairing correlations in the
unprojected wave function that are not completely taken into
account by the Gutzwiller approximation scheme. It also explains
to a certain extent, the discrepancies between the VMC
calculations and the GA for the quasiparticle weight as we will
discuss in section \ref{EX_VMC}.

To clarify the limitations of the Gutzwiller approximation
for projected superconducting states in more detail, VMC can be used to
calculate the hole density in the vicinity of the reservoir site.
In the half-filled limit the reservoir site is double occupied
and therefore a single hole is distributed among the
remaining fully projected sites. The VMC calculations of Fukushima \
\etal \ \cite{Fukushima05} show
very different density oscillations
for the projected Fermi sea and the projected $d$-wave state.

\begin{figure}
\center
\includegraphics[width=6cm,keepaspectratio]{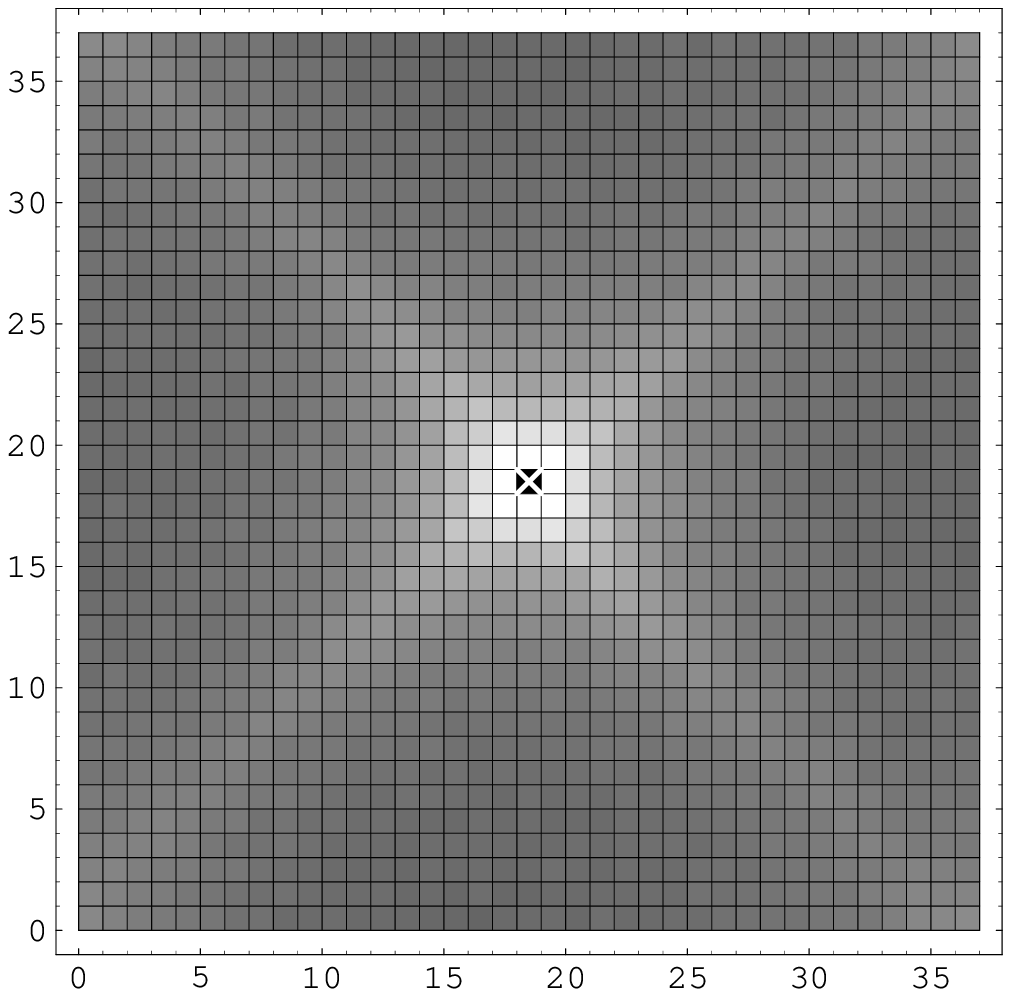}
 \includegraphics[width=6cm,keepaspectratio]{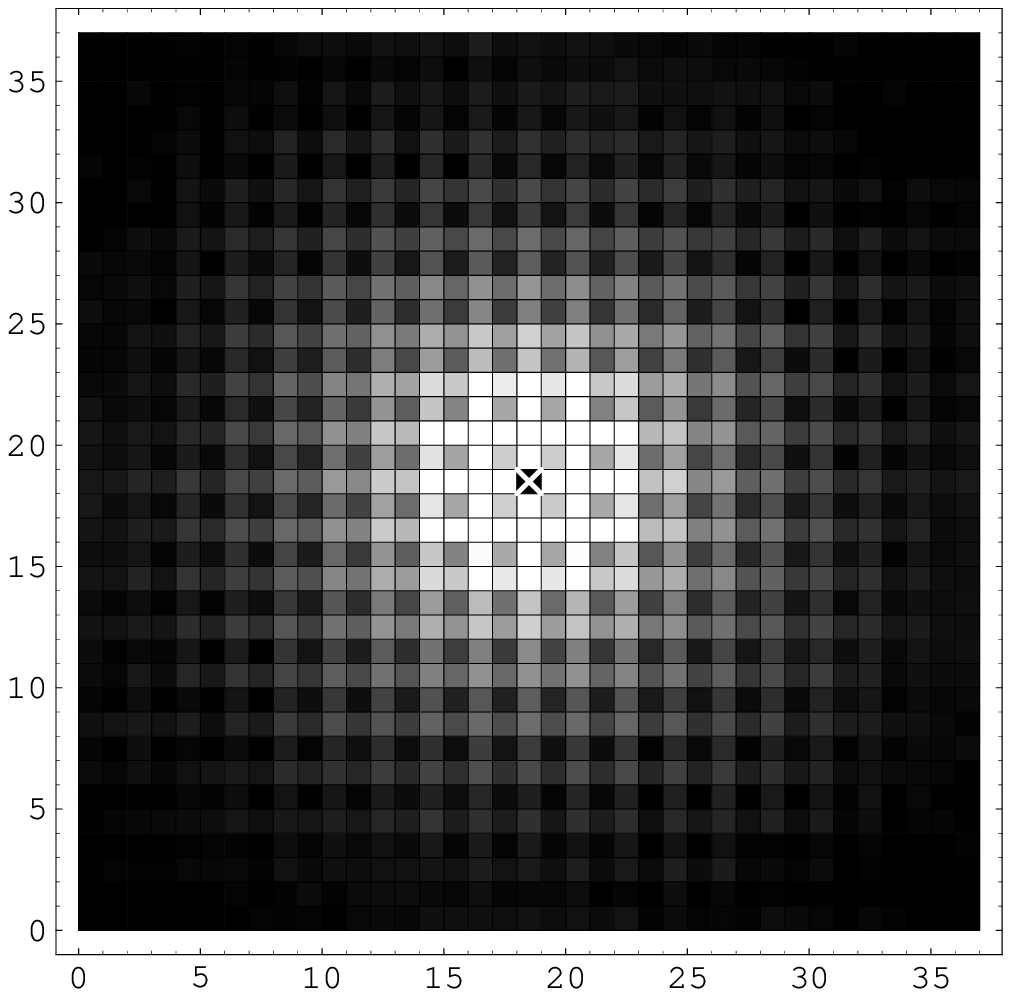}
\caption{VMC results for the hole density
$ n_h(m) = \langle (1-n_m)\rangle _{\Psi_l'}$
(colour coding: white/black correspond to
high/low values of $n_h(m)$) in the partially projected
state $|\Psi_l'\rangle $, for sites $m$ other than
the reservoir site $l$ (marked by the cross).
Left: Fermi sea. Right: $d$-wave state. From \cite{Fukushima05}
\label{density_plots}}
\end{figure}

Figure \ref{density_plots} shows
VMC results for the hole density
$$
n_h(m)\  =\ \langle 1-n_m \rangle_{\Psi_l'}~,
$$
in the partially projected state $|\Psi_l'\rangle$.
The sites $m$ are distinct from the reservoir site $l$
(marked by a cross in the figure). All results shown correspond
to half filling; \textit{viz.} $n_{\uparrow}=n_{\downarrow}=0.5$.
We choose $\Delta=1$ for the projected BCS $d$-wave state.
For the Fermi sea, we see that the hole is distributed
more uniformly than in the $d$-wave case even though the diagonal
direction has a larger probability of being occupied by a hole.
The $d$-wave has a quasi checker-board pattern
where only one of four sites is black, and the
hole tends to be near the reservoir site.
The VMC results for the projected BCS wave functions are strikingly
different in that the hole density
is \textit{not} uniform. On the other hand, the
Gutzwiller approximation would be exact, if all states in the
Hilbert space contribute equally to the wave function. That would
correspond to a uniform density of holes. Clearly, some limitations
in the Gutzwiller approximation show up when treating projected superconducting
wave functions. This is in agreement with our previous considerations, where
we found that the functional form of $X$ [\eq{X_Gutzwiller}, derived using Gutzwiller
approximation] agrees well with the VMC calculations only for the
projected Fermi sea, but not for the projected BCS state (see \fig{d_Gutz_doping}
and \fig{d_BCS_doping}).

\section[Renormalised mean field theory]
{Renormalised mean field theory: Basic ideas and recent
extensions}
\label{RMFT}

On the basis of the GA,  Zhang, \etal \ \cite{Zhang88b} derived a
renormalised mean field theory (RMFT) for the $t$-$J$ model. In
this section, we present an overview on this approach, which plays
a central role within Gutzwiller-RVB theory. We illustrate
successes and recent extensions of the RMFT for the HTSC, derive
the RMFT gap equation, and review its applications to the Hubbard
model in the strong coupling limit. Further extensions to
antiferromagnetic and inhomogeneous phases provide a
quantitative description of the Cuprate phase diagram.

\subsection{Overview on the RMFT method}
\label{sect_RMFToverview}

\begin{figure}
\centering
\includegraphics*[width=0.95\textwidth]{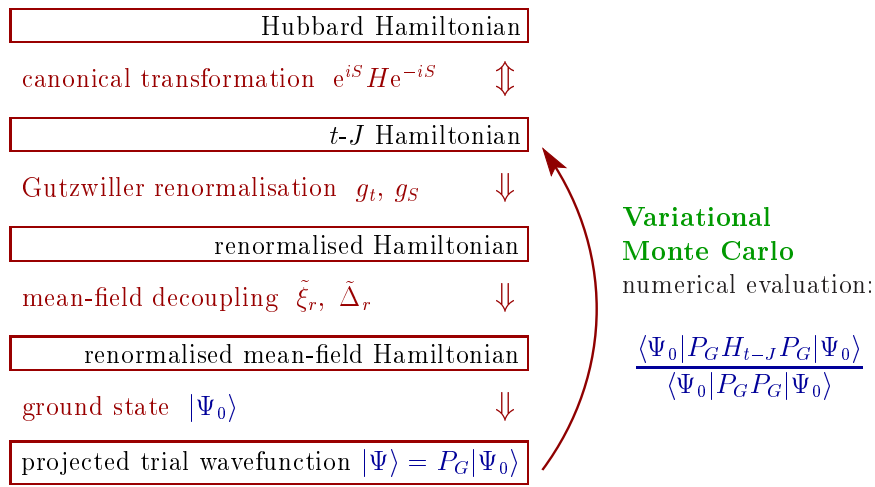}
\caption{Schematic illustration of the RMFT method; see text for a
detailed description.} \label{RMFToverview}
\end{figure}

We start with an overview and discuss how RMFT allows for
a systematic treatment of the Hubbard Hamiltonian in the
strong-coupling limit. We present the basic concepts in this subsection,
and the method itself
will be discussed in detail in the corresponding subsections.
Figure \ref{RMFToverview} summarises the main steps necessary for
the strong-coupling treatment of the Hubbard model within the
RMFT. In the following we refer to the individual steps
illustrated in \fig{RMFToverview}.

As shown in \fig{RMFToverview}, the first step is to apply a
canonical transformation $e^{-iS}$ to the Hubbard Hamiltonian
removing hopping processes that change the number of doubly
occupied sites. Doing so, we obtain the $t$-$J$ Hamiltonian, which
is defined in the subspace excluding double occupancy. The $t$-$J$
Hamiltonian provides an effective low energy Hamiltonian for the
Hubbard model in the strong coupling limit as already discussed in
detail in \sect{transtJ}.

Next we invoke the Gutzwiller approximation to remove the
restriction to projected states within the $t$-$J$ Hamiltonian. As
we will discuss in \sect{renorm_ham}, this procedure results in a
renormalised Hamiltonian with terms weighted by the corresponding
doping-dependent Gutzwiller renormalisation factors (see also
\sect{GutzPrinc}).

We then perform a mean field decoupling for the renormalised
Hamiltonian, focusing on hopping amplitudes $\tilde
\xi_{r}\equiv\sum_\sigma\, \langle c^\dagger_{i\sigma}
c_{i+r\sigma} \rangle_{\Psi_0}$ and pairing amplitudes $\tilde
\Delta_{r}\equiv\,\langle c^\dagger_{i\uparrow}
c^\dagger_{i+r\downarrow} -c^\dagger_{i\downarrow}
c^\dagger_{i+r\uparrow} \rangle_{\Psi_0} $ . Following this way,
we find self-consistent gap equations for the mean field
amplitudes (see \sect{gapeq}).

Solving the gap equations provides us with the mean field ground
state $|\Psi_0 \rangle$ of the renormalised $t$-$J$ Hamiltonian in
the `pre-projected' Hilbert space. Once $|\Psi_0 \rangle$ is
known, we can construct the Gutzwiller projected state $|\Psi
\rangle = P_G |\Psi_0\rangle$, which then provides an
approximate wave function for ground state of the projected
$t$-$J$ Hamiltonian. We can control the RMFT by using $|\Psi
\rangle$ as a projected trial wave function within the VMC
technique (see \chap{VMC}). Thus, the above scheme provides a
consistent framework to study Gutzwiller projected wave functions by
a combination of RMFT and VMC methods.

The projected wave function $|\Psi \rangle $ allows for the
calculation of relevant physical quantities as well as for the
definition of excited states within the $t$-$J$ model. To
determine observables within the Hubbard Hamiltonian, we can
employ the re-transformed wave function $e^{-iS}|\Psi \rangle$ for
the calculation of expectation values. We will discuss this
approach in \sect{RMFT_extens}.

\subsection{Derivation of the RMFT gap equations}
\label{RMFT_GapEq}

In this subsection we review the work of Zhang, \etal \
\cite{Zhang88b} and develop a renormalised mean field theory
(RMFT) for the $t$-$J$ model based on the Gutzwiller
renormalisation scheme ($\hat =$ GA, see \sect{GutzPrinc}). To
illustrate the RMFT, we start with the simplest form of the
$t$-$J$ Hamiltonian,
\begin{equation}
H_{t-J}\ =P_G \left[ \,-t \sum_{\langle i,j \rangle,\sigma} \left(
c_{i,\sigma}^\dagger c_{j,\sigma} + c_{j,\sigma}^\dagger
c_{i,\sigma} \right) \, + \, J \, \sum_{\langle i,j \rangle} \,
{\bf S}_i \, {\bf S}_j \right] P_G\ . \label{simple_tJ}
\end{equation}
We restrict ourselves to nearest neighbour hopping $t$, and a
superexchange interaction $J$. We neglect any further hopping
parameters as well as additional contributions in the Hamiltonian
like the density-density term and the correlated hopping terms,
see equation \eq{tJH}. The effects of such extensions are
discussed in \sect{RMFT_extens}, where we consider an RMFT for the
Hubbard model, including next nearest neighbour hopping matrix elements.

\subsubsection{Derivation of the renormalised $t$-$J$ Hamiltonian}
\label{renorm_ham}

Two steps are necessary to obtain explicit analytic expressions
for the ground state of the $t$-$J$ model \eq{simple_tJ} for
various doping levels $x$, where $x=1-n$. The first is the
Gutzwiller approximation, where the effects of the projection
$P_G$ are taken into account by appropriate renormalisation
factors. We search for a Gutzwiller projected state
$P_G|\Psi_0\rangle$ that minimises the energy expectation value,
\begin{eqnarray}
E_0&=&\frac {\langle\Psi_0| P_G H_{t-J}  P_G |\Psi_0\rangle}
{\langle\Psi_0| P_G P_G |\Psi_0\rangle} \nonumber \\
&=& -t \sum_{\langle i,j \rangle,\sigma}  \frac {\langle\Psi_0|P_G ( c_{i,\sigma}^\dagger c_{j,\sigma}
 + c_{j,\sigma}^\dagger
c_{i,\sigma} ) P_G |\Psi_0\rangle}{\langle\Psi_0| P_G P_G |\Psi_0\rangle}
\, \nonumber  \\
&& + \, J \, \sum_{\langle i,j \rangle} \, \frac {\langle\Psi_0|P_G{\bf S}_i \,
{\bf S}_j P_G |\Psi_0\rangle}{\langle\Psi_0| P_G P_G |\Psi_0\rangle} \label{ground_state_tJ} \ .
\end{eqnarray}
By invoking a GA for \eq{ground_state_tJ}, we get rid of the
projection operator $P_G$ and obtain,
\begin{equation}
E_0\approx -g_t t \sum_{\langle i,j \rangle,\sigma}   \frac {\langle\Psi_0|
( c_{i,\sigma}^\dagger c_{j,\sigma} + c_{j,\sigma}^\dagger
c_{i,\sigma} )  |\Psi_0\rangle}{\langle\Psi_0 |\Psi_0\rangle}
\, + \, g_S \, J \, \sum_{\langle i,j \rangle} \, \frac {\langle\Psi_0|{\bf S}_i \, {\bf S}_j
  |\Psi_0\rangle}{\langle\Psi_0 |\Psi_0\rangle}  \ . \label{approx_E0}
\end{equation}
The GA for the hopping term [first term in \eq{approx_E0}] has a
renormalisation factor, $g_t=(1-n)/(1-n/2)$, which was derived in
the previous section, equation \eq{hopping_simplest}. For the
superexchange term [second term in \eq{approx_E0}], the
renormalisation factor is $g_S=1/(1-n/2)^2$, where we assume a
homogeneous state without any sublattice magnetisation, see
\eq{SiSj_Gutz}.

We may now determine the variational ground state by searching for
the state $|\Psi_0\rangle$, that minimises the renormalised
$t$-$J$ Hamiltonian, $\tilde H_{t-J}$, defined as,
\begin{equation}
\tilde H_{t-J}= -g_t t \sum_{\langle i,j \rangle,\sigma}
 ( c_{i,\sigma}^\dagger c_{j,\sigma} + c_{j,\sigma}^\dagger c_{i,\sigma} )
\, + \, g_S \, J \, \sum_{\langle i,j \rangle} \, {\bf S}_i \, {\bf S}_j
 \ . \label{renorm_tJH}
\end{equation}
Once $|\Psi_0\rangle$ is known, we may consider the projected
state, $P_G |\Psi_0\rangle$, as a trial ground-state of $H_{t-J}$.

\subsubsection{Mean field decoupling of the renormalised Hamiltonian}
\label{gapeq}

The next step in the derivation of the RMFT, see
\sect{sect_RMFToverview}, is the realisation that $\tilde H_{t-J}$
allows for several types of molecular-fields
\cite{Edegger06a,Zhang88b}: For simplification we only concentrate
on the singlet pairing amplitude,
\begin{equation}
\tilde \Delta_{r}\equiv\,\langle c^\dagger_{i\uparrow}
c^\dagger_{i+r\downarrow} -c^\dagger_{i\downarrow}
c^\dagger_{i+r\uparrow} \rangle_{\Psi_0} \ , \label{delta_amp}
\end{equation}
and the hopping amplitude,
\begin{equation}
\tilde \xi_{r}\equiv\sum_\sigma\, \langle c^\dagger_{i\sigma}
c_{i+r\sigma} \rangle_{\Psi_0} \ , \label{hopping_amp}
\end{equation}
where $r=\hat x, \hat y \, \hat = \, (1,0),(0,1)$ connects nearest neighbour sites.
This decoupling scheme of the renormalised
Hamiltonian leads to a BCS ground state,
\begin{equation}
|\Psi_0 \rangle = \prod_\bk (u_\bk + v_\bk
c^\dagger_{\bk\uparrow}c^\dagger_{-\bk\downarrow})~|0\rangle \ ,
\end{equation}
with,
\begin{equation}
v_\bk^2=\frac 1 2 \left(1- \frac {\xi_\bk} {E_\bk} \right)
\label{MF_vk} \ ,
\end{equation}
and $u_\bk^2=1-v_\bk^2$.
The resulting gap equations are
\begin{eqnarray}
\tilde \Delta_r&=&1/L \sum_\bk \cos (\bk\,r) {\Delta_\bk}/{E_\bk} \ , \label{equivBCS}\\
\tilde \xi_r&=&-{1}/L \sum_\bk \cos (\bk\,r) {\xi_\bk}/{E_\bk} \ , \label{equivhopping}
\end{eqnarray}
together with the condition, $x= {1}/L \sum_\bk {\xi_\bk}/{E_\bk}$,
for the hole-doping concentration.
The dispersion of the mean field excitations
is given by, $E_\bk = \sqrt{\xi_\bk^2 + \Delta^2_\bk}$, where
\begin{eqnarray}
\Delta_\bk &=& \frac {3 g_S J} 4\,
\left(\, \tilde \Delta_x \cos k_x +  \tilde \Delta_y \cos k_y \, \right)~ \,  \\
\xi_\bk &=& -\left(2g_t t+ \frac {3 g_S J} 4 \tilde \xi_x \right) \cos k_x
-\left(2g_t t+ \frac {3 g_S J} 4 \tilde \xi_y \right) \cos k_y  - \mu \ .
\label{zetak_simple}
\end{eqnarray}
Equation \eq{equivBCS} resembles the usual BCS gap equation,
except that we consider independent
pairing along the $x$- and the $y$-direction. Together with
\eq{equivhopping}, we have four coupled gap equations (for
$\tilde \Delta_{x}$, $\tilde \Delta_{y}$,
$\tilde \xi_{x}$, $\tilde \xi_{y}$), which must be
solved self-consistently. The $\Delta_\bk$ is obviously
related to pairing in the state $|\Psi_0\rangle$, however, it is
not identical to the superconducting order parameter in
$P_G|\Psi_0\rangle$ as will be shown below.
The $\xi_\bk$ becomes the
renormalised dispersion in the absence of pairing and includes a
chemical potential $\mu$ to regulate the particle density.

\subsubsection{Solutions of the RMFT gap equations}
\label{sol_simple}
\begin{figure}
\centering
\includegraphics*[width=0.65\textwidth]{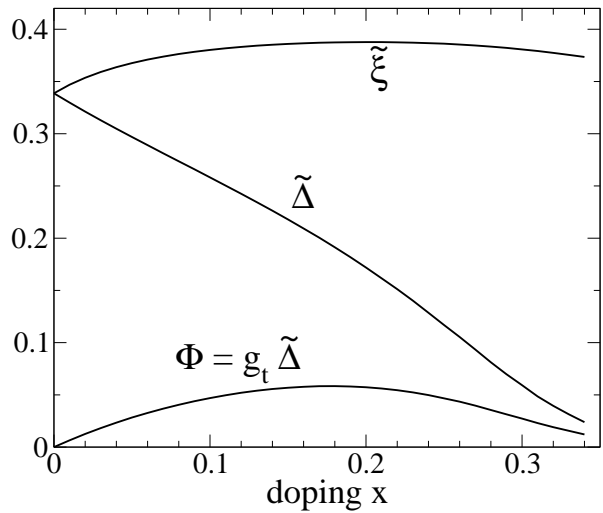}
\caption{Doping dependence of the $d$-wave pairing amplitude
$\tilde \Delta$, the hopping amplitude $\tilde \xi$, and the
superconducting order parameter $\Phi$, see \eq{order_para}, in
the $d$-wave ground state for the $t$-$J$ model \eq{simple_tJ}
with $J=t/3$. The results were obtained self consistently
by solving the RMFT gap equations.}
\label{simpleOrderPara}
\end{figure}

The gap equation can be solved numerically. We present results
obtained for $J/t=1/3$, which is a
reasonable choice for HTSC. However, we emphasise that the
results presented below are not sensitive to this particular choice
of $J$, {\it i.e.} the results stay quite similar for $J/t=0.2-0.5$.
We find that a $d$-wave pairing state is stable for
$x\leq0.35$. In this case, $\tilde \Delta\equiv|\tilde
\Delta_x|=|\tilde \Delta_y|$ with $\tilde \Delta_x=-\tilde
\Delta_y$ and $\tilde \xi\equiv \tilde \xi_x= \tilde \xi_y \,$. We
illustrate the doping dependence of these quantities in
\fig{simpleOrderPara}. The superconducting order parameter,
\begin{equation}
\Phi\equiv|\langle c^\dagger_{i\uparrow}c^\dagger_{i+\tau
\downarrow}- c^\dagger_{i\downarrow}c^\dagger_{i+\tau
\uparrow}\rangle_\Psi| \ , \label{order_para}
\end{equation}
is an expectation value in the projected ground states,
$|\Psi\rangle\equiv P_G|\Psi_0\rangle$, where $\tau$ is a
neighbouring site. Evaluating $\Phi$ by the GA (\sect{GutzPrinc})
one finds that $\Phi$ is renormalised as the hopping amplitude
by $g_t$, namely $\Phi \approx g_t \,\tilde  \Delta$. As illustrated in
\fig{simpleOrderPara}, $\Phi$ vanishes linearly near $x=0$, while
$\tilde  \Delta$ continuously increases towards half-filling. These
results are in good agreement with VMC results
\cite{Paramekanti04,Gros89a,Paramekanti01}
and the experimentally observed $T_c$ for
the $d$-wave pairing in the HTSC.

For the renormalised Hamiltonian, the $E_\bk$ corresponds to the
dispersion of the Bogoliubov quasiparticles, $|\Psi_{\bk,0}^\sigma
\rangle\,\equiv\, \gamma^\dagger_{\bk \sigma}|\Psi_0\rangle$, with
$\sigma=\uparrow \downarrow\,$, where the corresponding Bogoliubov
operators are defined by, $\gamma^\dagger_{-\bk\downarrow}\equiv
u_\bk c^\dagger_{-\bk\downarrow}+v_\bk  c_{\bk\uparrow}$, and,
$\gamma^\dagger_{\bk\uparrow}\equiv u_\bk
c^\dagger_{\bk\uparrow}-v_\bk c_{-\bk\downarrow}$, respectively.
However, $E_\bk$ also describes the excitation energy of
the corresponding projected Gutzwiller-Bogoliubov quasiparticles,
\begin{equation}
|\Psi_{\bk \sigma} \rangle \equiv
P_G|\Psi_{\bk,0}^\sigma\rangle\,= \,P_G
\gamma^\dagger_{\bk\sigma}|\Psi_0\rangle \ . \label{GB_QP}
\end{equation}
To see why, one evaluates the expectation value of the $t$-$J$
Hamiltonian with respect to $|\Psi_{\bk \sigma} \rangle$. Because,
$|\Psi_{\bk \sigma} \rangle=P_G|\Psi_{\bk,0}^\sigma \rangle$, is
renormalised exactly as, $|\Psi\rangle=P_G|\Psi_0\rangle$, we
recover the renormalised Hamiltonian $\tilde H_{t-J}$,
\eq{renorm_tJH},  by invoking a GA. The state
$|\Psi_{\bk,0}^\sigma \rangle$ is now acting onto $\tilde
H_{t-J}$, yielding in mean field decoupling a Bogoliubov
quasiparticle with excitation energy $E_\bk$. Therefore, the gap
$\Delta_\bk$ in $E_\bk = \sqrt{\xi_\bk^2 + \Delta^2_\bk}$
corresponds to the quasiparticle gap in the projected
superconducting state and is directly proportional to the mean
field amplitude $\tilde \Delta$ in \fig{simpleOrderPara}. We note
that RMFT \cite{Zhang88b} correctly reported the doping dependence
of the $d$-wave gap, {\it i.e.} an increasing gap with decreasing
doping, well before this behaviour was experimentally
established.

The above calculations follow the original work of Zhang \emph{et al.} \cite{Zhang88b}
Note that the results are restricted to a homogeneous and non magnetic phase.
Therefore, the results cannot adequately describe the
antiferromagnetic region of the phase diagram near half-filling
as well as inhomogeneous phases observed in HTSC. However, we emphasise that
the Gutzwiller - RVB being a variational mean field theory can be
extended to study such phases as well. In the following subsections,
we describe some attempts made in this direction, that provide
a more detailed
description of the phase diagram of the HTSC.


\subsubsection{Local $SU(2)$ symmetry in the half-filled limit}

In the limit of half filling, the
kinetic energy renormalises to zero, since $g_t \to 0$ as $x \to
0$. The $t$-$J$ model reduces to the antiferromagnetic Heisenberg
model, which is conserved under local $SU(2)$ gauge
transformations \cite{Zhang88b,Affleck88},
\begin{equation}
c^\dagger_{i\uparrow} \to
\alpha_i c^\dagger_{i\uparrow} + \beta_i c_{i\downarrow} ~,
\quad c_{i\downarrow} \to - \beta^*_i
c^\dagger_{i\uparrow} + \alpha^*_i c_{i\downarrow} ~,
\label{SU2eq}
\end{equation}
where $\alpha_i \alpha_i^* + \beta_i \beta_i^* = 1$. The
invariance of the Hamiltonian is due the spin operator ${\mathbf
S_i}$, which is invariant under $SU(2)$ transformations, as can
be proved by applying \eq{SU2eq} to the operators $S^±_i$ and
$S^z_i$. For $S^+_i$ we find,
\begin{eqnarray}
S^+_i=c^\dagger_{i\uparrow}c_{i\downarrow} &\to& (
\alpha_i c^\dagger_{i\uparrow} + \beta_i c_{i\downarrow})
( - \beta^*_i c^\dagger_{i\uparrow} + \alpha^*_i c_{i\downarrow})
\nonumber
\\
&\to& \alpha_i \alpha^*_i \, S^+_i + \beta_i \beta^*_i \,
S^+_i = S^+_i ~.
\end{eqnarray}
The invariance of $S_i^-$ and $S^z_i$ under \eq{SU2eq} can be
shown analogously.

\begin{table}[b]
\center
\begin{tabular}{l||l}
\hline \hline $d$-wave pairing \phan & $\tilde \Delta_x=-\tilde
\Delta_y = \tilde \xi_x= \tilde \xi_y = C/\sqrt{2} $ \\
\hline  $d$-wave density matrix$^*$ \phan & $\tilde \Delta_x= \tilde
\Delta_y
= \tilde \xi_x=- \tilde \xi_y = C/\sqrt{2} $ \\
\hline chiral state \phan & $\tilde \Delta_x=-\,i \,\tilde
\Delta_y = C
\, , \quad \tilde \xi_x=\tilde \xi_y = 0 $ \\
\hline anisotropic state \phan & $\tilde \Delta_x= \tilde \xi_y =
C \, ,
\quad  \tilde \Delta_y=\tilde \xi_x = 0$\\
\hline \hline
\end{tabular}
\caption{Examples of degenerate states of the renormalised mean
field Hamiltonian at half-filling, see also \cite{Zhang88b}. The
general constant $C=0.479$ is determined by the RMFT gap
equations, \eq{equivBCS} and \eq{equivhopping}. $^*$We note that
the $d$-wave density matrix is not the $d$-density wave (DDW) order discussed in
\cite{Chakravarty02}.} \label{SU2states}
\end{table}

Owing to the local $SU(2)$ gauge symmetry, the renormalised mean
field Hamiltonian has a large degeneracy in the representation of
ground states at half-filling, as may be seen by transforming the
mean field amplitudes $\tilde \Delta_r$ and $\tilde \xi_r$,
\eq{delta_amp} and \eq{hopping_amp}, under \eq{SU2eq}. Some of the
resulting (degenerate) states, that are related to each other by
$SU(2)$ transformations, are summarised in
\tab{SU2states}. Another example among the degenerate states in
the $SU(2)$ manifold is
the staggered $\pi$-flux state \cite{Affleck88},
\begin{equation}
\xi_{\mathbf {ij}}=|\xi_0|^2 \exp\left(i (-1)^{i_x+i_y} \frac \pi
4 \right), \label{piflux}
\end{equation}
with a complex hopping amplitude $\xi_{\mathbf {ij}}\equiv \langle
c^\dagger_{{\mathbf i}\sigma} c_{{\mathbf i}\sigma}\rangle_0$, but
a vanishing pairing amplitude $\Delta_{\mathbf {ij}}\equiv \langle
c^\dagger_{{\mathbf i}\uparrow} c_{{\mathbf
i}\downarrow}\rangle_0=0$.

It is important to note that above degeneracy is not true in terms
of the projected wave function, since it only results from using
an under-determined representation. In other words, the states that
are degenerate are the unprojected states $|\Psi_0\rangle $, but
not the physical states $P_G|\Psi_0\rangle$ \cite{Zhang88b}.
Therefore, the entire set of degenerate grounds states in the
renormalised mean field Hamiltonian correspond (modulo a trivial phase
factor) to a single projected state; for a proof see
\cite{Zhang88b}.

All states listed in the \tab{SU2states} and \eq{piflux} have the
same superexchange energy, even at finite doping, due to the $SU(2)$
invariance of this term. However, the kinetic energy $T$, \eq{kin_tJ}, and the
3-site term $H_3$, \eq{H2_tJ}, which only vanish  at half-filling,
are not invariant under the $SU(2)$ transformation \eq{SU2eq}.
Therefore the degeneracy in $|\Psi_0\rangle $ is lifted at finite
doping, where the $d$-wave pairing state is selected due to its
lower kinetic energy.

The $SU(2)$ gauge symmetry of the superexchange term led to the
speculation that at finite doping, when the degeneracy is lifted,
some among the `degenerate' states may compete with the $d$-wave
state (also one of the degenerated states at half-filling). In
particular, it was argued that staggered flux states could serve
as a `competing' and/or as a `normal' state in the underdoped
regime of the HTSC \cite{Lee06,Ivanov03}. Besides this competing
order scenario, a spin-charge locking mechanism \cite{Anderson05}
resting on the presence of the $SU(2)$ gauge symmetry at
half-filling was suggested. The consequence of the degeneracy in the
unprojected mean field ground states at half-filling
and its possible relationship with competing or coexisting order are not yet fully understood.

\subsection{RMFT for the Hubbard model and application to HTSC}
\label{RMFT_extens}

The RMFT presented so far can be improved by considering all terms
contributing to the $t$-$J$ Hamiltonian \eq{tJH}, {\it i.e.}
including the density-density term \eq{Jterm} as well as the
correlated hopping terms \eq{H2_tJ}. The inclusion of these terms
is also necessary if one were to use RMFT for the Hubbard Hamiltonian,
since the unitary transformation $H_{t-J}=P_G
e^{iS} H e^{-iS} P_G $ (discussed in \sect{transtJ})
between the Hubbard and $t-J$ Hamiltonians lead to these terms.
Physical quantities for the Hubbard model can be evaluated by considering expectation
values in the re-transformed wave function $e^{-iS} P_G |\Psi_0
\rangle$ \cite{Edegger06a}. Below we will use this ansatz to study the
superconducting order parameter. In the following, we will include
next nearest neighbour hopping ($t'$) into the Hubbard Hamiltonian
in order to allow for quantitative comparison with experimental
data for the HTSC.

\subsubsection{Generalised gap equations for the strong coupling limit}

We obtain the RMFT gap equations for the Hubbard model in the
strong coupling limit by considering the corresponding effective
Hamiltonian, {\it i.e.} the full $t$-$J$ Hamiltonian. The gap
equations for this $t$-$J$ Hamiltonian, which includes all terms
from equation \eq{tJH}, can then be derived in the same way
as described in the previous subsection.

First we invoke the GA to obtain the renormalised Hamiltonian for
\eq{tJH}. We note that all (nearest as well as further neighbours)
hopping terms are renormalised by $g_t=(1-n)/(1-n/2)$ and all
superexchange terms by $g_S=1/(1-n/2)^2$. Since the
density-density term commutes with the projection operator $P_G$,
it does not pick up any Gutzwiller renormalisation factor. The new
correlated hopping terms, equation \eq{H2_tJ}, are of the following
form,
\begin{eqnarray}
\langle  c_{i+\tau_1,\uparrow}^\dagger c_{i,{\downarrow}}^\dagger
c_{i,{\downarrow}} c_{i+\tau_2,\uparrow}\rangle_{P_G \Psi_0}
&\approx& g_{3} \, \langle c_{i+\tau_1,\uparrow}^\dagger
n_{i,\downarrow}
(1-n_{i,{\uparrow}})c_{i+\tau_2,\uparrow}\rangle_{\Psi_0}
\nonumber \ ,\\
\langle  c_{i+\tau_1,\downarrow}^\dagger c_{i,\uparrow}^\dagger
 c_{i,\downarrow} c_{i+\tau_2,\uparrow} \rangle_{P_G \Psi_0}
&\approx& g_{3} \, \langle c_{i+\tau_1,\downarrow}^\dagger
c^\dagger_{i,\uparrow} c_{i,\downarrow}
c_{i+\tau_2,\uparrow}\rangle_{\Psi_0} \arrlabel{Gutz_3} \ ,
\end{eqnarray}
involve three sites ($i$, $i+\tau_1$, and $i+\tau_2$), and are
renormalised by a factor $g_3=(1-n)/(1-n/2)^2$. For a derivation
of the GA for the correlated hopping terms we refer to the
appendix of \cite{Edegger05a}.

Next we decouple the resulting renormalised Hamiltonian by the
same scheme discussed in the previous subsection, obtaining
therefore the same gap equations, \eq{equivBCS} and
\eq{equivhopping} as before. However, the dispersion relation,
$E_\bk = \sqrt{\xi_\bk^2 + \Delta^2_\bk}$, with \cite{Edegger06a},
\begin{eqnarray}
\xi_\bk &=& -\left(2g_t t+ J\,\frac {\tilde \xi} 4
\,x_1\,+J_3\,\frac{\tilde \xi'} 4\,x_2\right)
(\cos k_x+\cos k_y)\nonumber \\
& &-\left(2g_t t'+ J'\,\frac {\tilde \xi'} 4 \,x_1\,+J_3\,\frac{\tilde \xi}
4\,x_2\right)\,2\,\cos
k_x \cos k_y \nonumber\\
& &-x_D \sum_{\tau_1\neq\tau_2} \frac{t_{\tau_1} t_{\tau_2}}{4U}
\cos\left[\bk (\tau_1-\tau_2) \right]\,-\mu \ , \label{epsk}\\
\Delta_\bk &=& J\,\frac {\tilde \Delta} 4\,[3g_s+1-(3+x)g_3]\,(\cos k_x-\cos
k_y)~ , \label{Delta_tUK}
\end{eqnarray}
incorporates the effects of further neighbour hopping and
correlated hopping terms. These expressions for $\xi_\bk$ and
$\Delta_\bk$ are valid for $\tilde \Delta\equiv|\tilde
\Delta_x|=|\tilde \Delta_y|$, $\tilde \Delta_x=-\tilde \Delta_y$,
$\tilde \xi\equiv \tilde \xi_x= \tilde \xi_y$, {\it i.e.} for the
$d$-wave pairing state, which is the most stable solution of the
gap equations \eq{equivBCS} and \eq{equivhopping}. Note that we
defined new hopping amplitudes for next nearest neighbours, $\tilde
\xi'\equiv \tilde \xi_{x+y}= \tilde \xi_{x-y}$. The last sum in
\eq{epsk} is a sum over all pairs of non-identical neighbouring
sites $\tau_1$ and $\tau_2$, where $t_{\tau_1}$ and $t_{\tau_2}$
are nearest and next nearest neighbour hopping terms. We defined,
$J=4t^2/U$, $J_3=4t't/U$, and $J'=4t'^2/U$ and abbreviated,
$x_1=3g_s-1+3(3-x)g_3$, $x_2=4(3-x)g_3$, and $x_D=(1-x^2)g_3$ in
\eq{epsk}.

As in \sect{RMFT_GapEq}, the ground state $|\Psi_0 \rangle$ of the
renormalised $t$-$J$ Hamiltonian results from above equations. By
including a projection operator $P_G$ into the wave function we
obtain $P_G |\Psi_0 \rangle$, which corresponds to a variational
wave function for the ground state of the $t$-$J$ Hamiltonian in
the fully projected Hilbert space. Invoking the canonical
transformation $e^{-iS}$ then provides an approximate ground
state $e^{-iS} P_G |\Psi_0 \rangle$ for the Hubbard model.

\subsubsection{Results from the generalised gap equations}
\label{result_ex}

For a comparison with experiments, we follow \cite{Edegger06a}
and consider a ratio
$t^\prime/t=-1/4$ between next nearest and nearest neighbour
hopping amplitudes, a value used widely in the modelling of the
band structure of various HTSC \cite{Dagotto94}. Furthermore, we
choose an on-site repulsion $U=12\,t$, {\it i.e.} we work in the
strong coupling regime $U\gg t,t'$, where the transformation from
the Hubbard to the {$t$-$J$} model is valid approximately. The above
choice of the model parameters reduces the number of free
parameters to one energy scale, $t\approx
300-500$ meV, for the HTSC.

\begin{figure}
\centering
\includegraphics*[width=0.65\textwidth]{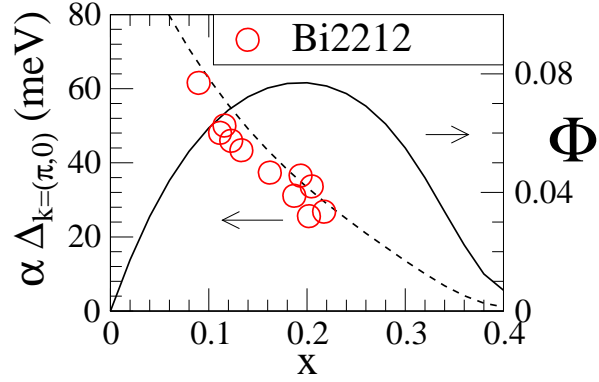}
\caption{Doping dependence of (solid line, right scale) the
superconducting order parameter, $\Phi$, and (dashed line, left
scale) the superconducting gap, $|\Delta_\bk|$, at $\bk=(\pi,0)$
for $t=300$ meV. The RMFT superconducting gap is scaled by a
factor $\alpha=1/2$ for comparison with experimental data (red
circles, Bi2122 \cite{Campuzano04}). From \cite{Edegger06a}.}
\label{RMFTphasediagram}
\end{figure}

In \fig{RMFTphasediagram} we show the doping dependence of the
superconducting gap, $|\Delta_\bk|$ at $\bk=(\pi,0)$, within RMFT,
which resembles experimental observations quite well. However, the
magnitude of the gap is overestimated by a factor of about 2 (see
scaling factor $\alpha = 1/2$ in \fig{RMFTphasediagram}) within
mean field theory. This overall mismatch is attributed to the fact
that dynamical \cite{Anderson02a} and long-range correlations are
neglected within RMFT, which is based on a local and static molecular-field
approximation.

As mentioned in \sect{sol_simple}, the superconducting gap is not
identical to the true superconducting order parameter,
$\Phi\equiv|\langle c^\dagger_{i\uparrow}
c^\dagger_{i+\tau\downarrow}-c^\dagger_{i\downarrow}
c^\dagger_{i+\tau\uparrow}\rangle|$ \cite{Zhang88b,Paramekanti01}.
Here, we determine the expectation value of $\Phi$ within the
re-transformed wave function, $e^{-iS}P_G |\Psi_0\rangle$.
Following \sect{transtJ}, we evaluate the canonical transformation
$e^{-iS}$ in order ${\mathcal{O}(t/U)}$. Doing so, provides
systematic $t/U$-corrections to the result from
\sect{sol_simple}, where we used the wave function $P_G |\Psi_0\rangle$ in
calculating the expectation value of the superconducting order
parameter.

The calculations for the expectation value of a general
observable $\hat O$ within the Hubbard model are summarised by,
\begin{deqarr}
 \langle \, \hat O \,  \rangle_{e^{-iS} P_G \Psi_0 } \,
&=& \, \langle \, e^{iS} \hat O e^{-iS} \,  \rangle_{P_G \Psi_0}\, \\
& \approx &
\, \langle \, \hat O + \, i [S, \hat O ] \,  \rangle_{P_G
\Psi_0}\,~, \label{order_para_this}
\end{deqarr}
where the last step corresponds to the evaluation of $e^{-iS}$ in
order ${\mathcal{O}(t/U)}$; compare with \sect{transtJ}. Note that
\eq{order_para_this} corresponds to an expectation value of the
operator $O + \, i [S, \hat O ]$ in the projected state $P_G
|\Psi_0\rangle$. We can therefore use a generalised Gutzwiller
approximation by invoking the counting arguments given in
\sect{GutzPrinc}.

\begin{figure}
\centering
\includegraphics*[width=0.55\textwidth]{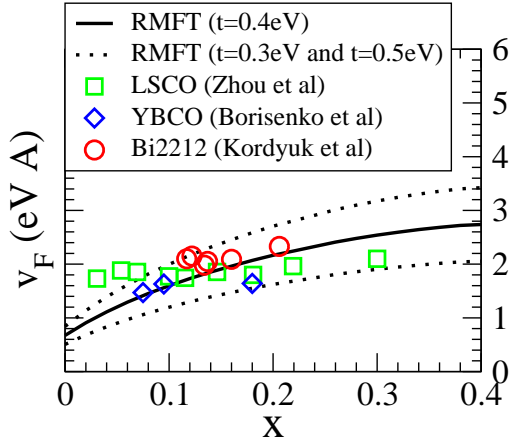}
\caption{Doping dependence of  Fermi velocity, $v_F$. The RMFT
results are compared with  experimental data from Zhou, \etal \
\cite{Zhou03}, Borisenko, \etal \ \cite{Borisenko06}, and Kordyuk,
\etal \ \cite{Kordyuk05}. From \cite{Edegger06a}.}
\label{fig_vFRMFT}
%
%
\end{figure}

Setting, $\hat O=c^\dagger_{i\uparrow}
c^\dagger_{i+\tau\downarrow}-c^\dagger_{i\downarrow}
c^\dagger_{i+\tau\uparrow}$, the superconducting
order parameter $\Phi$ for the Hubbard model
can be calculated using by above scheme. One
finds,
\begin{equation}
\Phi \approx \,g_t \tilde \Delta + \frac t U
\,g_3\,{(6-x)}\,\tilde \Delta\, \tilde \xi~. \label{ordertUeq}
\end{equation}
In deriving \eq{ordertUeq} we considered $t' \approx 0$ within
$S$, for simplicity, since $|t'| \ll |t|$. As shown in
\fig{RMFTphasediagram}, $\Phi$ vanishes as $x \to 0$, and the
$t/U$-corrections do not qualitatively change the result of Zhang,
\etal \ \cite{Zhang88b} near half-filling. We emphasise that the above
procedure can be used to calculate the expectation value of
any observable and provides a
systematic way to study the Hubbard model in the strong coupling
limit.

Next we consider the nature of the low lying excitations, {\it
i.e.} the quasiparticles created at the nodal point, $k_F \equiv
k_{F,x}=k_{F,y}$. The nodal dispersion around $k_F$ is
characterised by the velocity, $v_F$, which directly influences a
number of experimentally accessible quantities. The Fermi velocity
$v_F$ can be directly calculated by the gradient of $\xi_\bk$
along the direction, $(0,0) \to (\pi,\pi)$ within RMFT. The
results obtained by using  $\xi_\bk$ from equation \eq{epsk} are
presented in \fig{fig_vFRMFT} (for $t=0.3,0.4,0.5\ \text{eV}$ and
$a_0=4 Å$). Edegger \etal \ \cite{Edegger06a} also  found that the
theoretically $v_F$ is well approximated by the formula,
\begin{equation}
v_F/a_0 \approx \sqrt 2 \sin k_F \left[2g_t(t+2t^\prime \cos k_F)
+ x_1\frac{J}{4}\tilde \xi \right]~. \label{vf_tU}
\end{equation}
where $J'$ and $J'$ are set to zero for
simplicity. As we will discuss later,
the effective values of $J'$ and $J''$ within the GA become zero
at half filling. So, ignoring their effect on the dispersion
modifies the result only weakly. As seen in
\fig{fig_vFRMFT}, $v_F$ increases with $x$, but remains finite as
$x \to 0$. In addition, we can infer from \eq{vf_tU} that
the energy scale of the nodal velocity at $x=0$
is determined by $J$, {\it i.e.}
${v_F}/({a_0 J})\approx\sqrt{2}\sin(k_F) \frac {11} 4 \tilde \xi
\approx 1.5$
(with $\tilde \xi\approx0.38$ and $k_F\approx \frac
\pi 2$). The observed doping dependence stems from the effects of
the Gutzwiller projection $P_G$. As $x$ increases, holes gain
kinetic energy by direct hopping, {\it viz.} $g_t$ increases with
doping; but $g_s$ decreases, leading to the doping dependence of
$v_F$ seen in \fig{fig_vFRMFT}. Note, that the RMFT results
presented in \fig{fig_vFRMFT} are absolute in value. No rescaling
has been made for comparison with the experiments, contrary to the
results for the gap $|\Delta_{\bk=(\pi,\pi)}|$ presented in
\fig{RMFTphasediagram}.

\begin{figure}
\centering
\includegraphics*[width=0.7\textwidth]{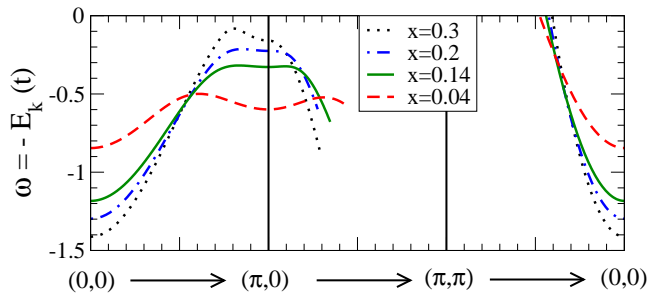}
\caption{Energy dispersion,
$\omega=-E_\bk=\sqrt{\xi_\bk+\Delta_\bk}$, of the
Gutzwiller-Bogoliubov quasiparticle for different doping $x$. From
\cite{Edegger06a}.} \label{fig_Ekdispersion}
\end{figure}


The above results are in agreement with the VMC results of
Paramekanti, \etal , who extracted $v_F$ from the discontinuity of
the first moment of the spectral function in the repulsive $U$
Hubbard model (\cite{Paramekanti01,Paramekanti04}, also
\sect{SPgroundstateVMC}) and of Yunoki, \etal, who obtain $v_F$
from the quasiparticle dispersion in the $t$-$J$ model
(\cite{Yunoki05a}, also \sect{QPenergy}). These results also yield
a good fit to the ARPES data
\cite{Zhou03,Borisenko06,Kordyuk05,Bogdanov00,Johnson01}, as
illustrated in \fig{fig_vFRMFT}. We note that the doping
dependence of $v_F$ in the severely underdoped regime remains to
be settled experimentally. While some groups report a nearly
constant Fermi velocity (see data for LSCO in \fig{fig_vFRMFT}),
others observe a slight increase with doping (see data for YBCO
and Bi2212 in \fig{fig_vFRMFT}). We further emphasise that the
energy scales $t$ and $J$ might be extracted from the ARPES data
in $v_F$, whenever data with high accuracy becomes available. By
using $\tilde \xi\approx0.38$ and setting $k_F$ and the ratio
$t'/t$ to the experimentally observed values, $t$ and $J$ can be
fitted by \eq{vf_tU}. In addition, the RMFT calculations find that
the nodal properties remain essentially unchanged when $\tilde
\Delta$ is set to $0$, {\it i.e.} the doping dependence of $v_F$
results from the vicinity of the RVB state to a Mott insulator,
rather than the occurrence of superconductivity itself.

\begin{figure}
\centering
\includegraphics*[width=0.55\textwidth]{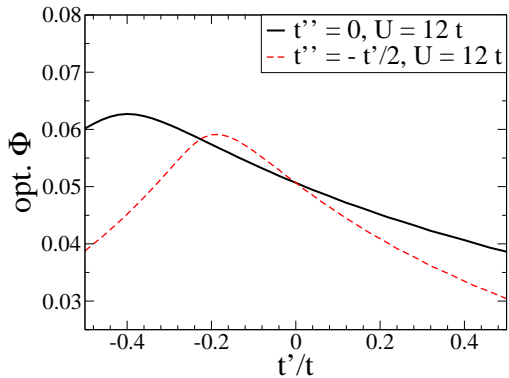}
\caption{The magnitude of the optimal superconducting order parameter
(opt. $\Phi$) as a function of the ratio $t'/t$ as obtained
in the RMFT calculations for the Hubbard model. Opt. $\Phi$ provides the
value of $\Phi$ at optimal doping, {\it i.e.} the maximal value
of $\Phi$ for this particular set of model parameters ($t'/t$, $U=12t$).
The dashed line shows calculations, where third nearest
neighbour hopping has been taken into account into dispersion
by setting $t''=-t'/2$.
}

\label{optSC}
\end{figure}

In \fig{fig_Ekdispersion}, we present the energy dispersion,
$\omega=-E_\bk$, of the Gutzwiller-Bogoliubov QP along the
directions, $(0,0)\to(\pi,0)$, $(\pi,0)\to(\pi,\pi)$, and
$(\pi,\pi)\to(0,0)$ for different doping levels $x$. The
dispersion is flattened when approaching half-filling and the gap
around $(\pi,0)$ becomes large. We emphasise that these RMFT
calculations  adequately describe only the low energy sector of
the HTSC, and do not seek to explain the `kink' at higher energies
\cite{Zhou03,Borisenko06,Kordyuk05,Bogdanov00,Johnson01}.

Equations  \eq{epsk} and \eq{Delta_tUK} can also be used to study
the effects of $t'$ onto the magnitude of the superconducting
order parameter $\Phi$. Figure \ref{optSC} shows the value
of $\Phi$ at optimal doping
as a function of the ratio $t'/t$ in the
bare non-interacting dispersion.
We observe a maximum at about $t'/t=-0.35$
depending slightly on whether we set\footnote{We note that the inclusion of $t''=-t'/2$ into the
bare dispersion is sometimes
used to get a better fit with band-structure calculations
and ARPES data.} $t''=0$  or $t''= - t'/2$.
This observation is in agreement with
band-structure calculations \cite{Pavarini01}
where it is found empirically that
compounds with ratio $t'/t \approx -0.1$ in the
dispersion (determined from band-structure
calculations) have smaller $T_c$ (corresponds to smaller $\Phi$)
than compounds with a ratio $t'/t \approx -0.35$. The RMFT
calculations in \fig{optSC} also match VMC results
of Shih \etal \ \cite{Shih04b}, which we will discuss in more detail
in \sect{increase_var}.

To summarise, the calculational scheme we described above, presents
a systematic way to study the Hubbard
model in the strong coupling limit. It is based on the
validity of the $t$-$J$ model as an effective Hamiltonian
(in the large $U$ limit) and on
determining expectation values within the re-transformed trial
wave function, $e^{-iS}P_G |\Psi_0\rangle$; a scheme that can be
extended to excited states as will be described in
\chap{EX_RMFT} (see also \cite{Edegger06a}).

\subsection{Possible extensions and further applications}

\subsubsection{Incorporation of antiferromagnetism}
\label{AF_RMFT}
\begin{figure}
\centering
\includegraphics*[width=0.6\textwidth]{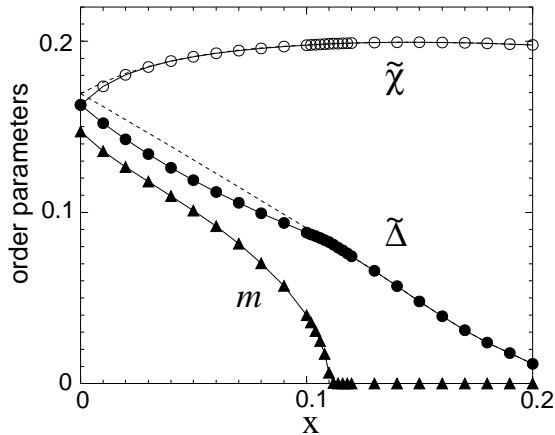}
\caption{RMFT calculations including antiferromagnetism from Ogata
and Himeda \cite{Ogata03b}. The self-consistent parameters $\tilde
\Delta$ (pairing amplitude), $\tilde \xi$ (hopping amplitude), and
$m$ (staggered magnetisation) are shown as a function of the
doping rate $x=1-n$ for $J/t=0.3$ and $t'=J'=0$. The dashed lines
represent the results when the antiferromagnetic order is
suppressed, i.e, $m$ is fixed to zero. From \cite{Ogata03b}.}
\label{OgataAF}
\end{figure}

The incorporation of antiferromagnetism is an example of a
possible extension of the RMFT. In order to describe an
antiferromagnet with finite sublattice magnetisation $m$, we have
to allow for an additional degree of freedom in the wave function.
When deriving the corresponding gap equation we must keep in mind
that the antiferromagnetic correlation affects the GA as discussed
in \sect{GutzPrinc}. However, Himeda and Ogata \cite{Himeda99}
showed by VMC calculations that even the formulas from
\sect{GutzPrinc} do not adequately describe all aspects in a
magnetic ordered state. They determined effective Gutzwiller
renormalisation factors by comparing the numerically obtained
expectation values in the projected state with the respective mean
field values before projection. It was found that the
$z$-component of the Gutzwiller renormalisation factor $g^z_S$ is
enhanced compared with those of the $xy$-component $g^±_S$.

Ogata and Himeda \cite{Ogata03b} argued that the discrepancies
stem from spatial correlations neglected by the GA. They derived
extended Gutzwiller renormalisation factors by considering a
cluster around the sites $i$ and $j$ to incorporate further
inter-site correlations. Applying these renormalisation factors
and solving the gap equations including antiferromagnetism yields
the results of \fig{OgataAF}. We see that for doping,
$\delta<0.1$, long range antiferromagnetic order coexists with
superconductivity. For higher doping the magnetisation $m$
vanishes and solely the superconducting order remains. This result
is obtained neglecting the next nearest neighbour hopping ($t'=0$)
and agrees with previous VMC results
\cite{Himeda99,Chen90,Giamarchi91,Ivanov04}. We note that the
extended Gutzwiller renormalisation factors of Ogata and Himeda
are essential for reproducing the VMC calculations. However,
\fig{OgataAF} does not quantitatively agree with the
experimentally phase diagram of hole-doped Cuprates, where
antiferromagnetism disappears at about $3-5 \%$ doping. A
better match may be obtained by considering the effects of $t'$
as done in VMC \cite{Shih04a} and quantum cluster studies
\cite{Senechal05,Aichhorn05}.

\subsubsection{Applications to inhomogeneous systems}

The RMFT described has also been used to study inhomogeneous phases such as
stripes and checkerboard charge order
\cite{Anderson04b,Huang05,Poilblanc05,Li06}, vortex cores
\cite{Himeda97,Tsuchiura03}, magnetic and non-magnetic
impurities \cite{Tsuchiura99,Tsuchiura00,Tsuchiura01}. These
investigations throw light on the interplay between
antiferromagnetic correlations, $d$-wave superconductivity, and
charge order and can be compared with STM data.

However, such studies require an unrestricted Hartree-Fock
treatment of the renormalised $t$-$J$ Hamiltonian \eq{renorm_tJH},
{\it i.e.} local expectation values such as, $\tilde
\Delta_{ij}\equiv\,\langle c^\dagger_{i\uparrow}
c^\dagger_{j\downarrow}\rangle_{\Psi_0}$, and, $\tilde
\xi_{ij\sigma}\equiv\, \langle c^\dagger_{i\sigma} c_{j\sigma}
\rangle_{\Psi_0}$, must be considered
independently for each bond \cite{Wang06}.
Furthermore, the local charge densities $n_{i\sigma}$ generally
differ from site to site, and thus the Gutzwiller renormalisation
factors of the renormalised Hamiltonian depend on the site indices
$i$ and $j$ ($g_t^{ij}$, $g_S^{ij}$). Special attention must be
paid when deriving these Gutzwiller renormalisation factors,
because the local charge densities can differ between the
projected and unprojected state (see discussion in
\sect{GutzPrinc}). For inhomogeneous systems the RMFT gap
equations generalise to the so-called Bogoliubov-de Gennes
equations, which must then be solved self-consistently.

The investigation of charge modulations within above framework
\cite{Anderson04b,Huang05,Poilblanc05,Li06} provides an
understanding of the $4×4$ checkerboard patterns
seen in the STM data of the HTSC. These studies neglect long
range antiferromagnetism and assume $\tilde \xi_{ij}=\tilde
\xi_{ij\uparrow}=\tilde \xi_{ij\downarrow}$ and
$n_{i\uparrow}=n_{i\downarrow}$. This is a reasonable assumption
since the authors concentrated on doping levels where
antiferromagnetism is not observed experimentally. The
renormalised mean field Hamiltonian can then be written as
\cite{Poilblanc05},
  \begin{eqnarray}
   H_{\rm MF}= &-& t \sum_{\langle ij\rangle\sigma} g^{ij}_t
      (c^{\dagger}_{i,\sigma}c_{j,\sigma}+ {\rm h.c.})-\mu\sum_{i\sigma}n_{i,\sigma}\nonumber \\
   &-&\frac{3}{4} J \sum_{\langle ij\rangle\sigma}g^{ij}_S
(\tilde \xi_{ji}c^{\dagger}_{i,\sigma}c_{j,\sigma} + {\rm  h.c.} -|\tilde \xi_{ij}|^2)\nonumber \\
   &-&\frac{3}{4} J \sum_{\langle ij\rangle\sigma}g^{ij}_S
(\tilde \Delta_{ji}c^{\dagger}_{i,\sigma}c^\dagger_{j,-\sigma}
   + {\rm h.c.} -|\tilde \Delta_{ij}|^2) \ .
  \end{eqnarray}

However, we must abandon above constraints,  $\tilde \xi_{ij}=\tilde
\xi_{ij\uparrow}=\tilde \xi_{ij\downarrow}$  and
$n_{i\uparrow}=n_{i\downarrow}$, for investigations
around vortex cores or impurities, where antiferromagnetic
correlations are essential. Doing so and solving the Bogoliubov-de
Gennes equations under an uniformly applied magnetic field shows
that significant antiferromagnetic correlations develop inside
vortex cores \cite{Tsuchiura03} in agreement with experimental
observations
\cite{Lake01,Lake02,Khaykovich02,Miller02,Mitrovic01,Kakuyanagi02}.
Tsuchiura, \etal \ \cite{Tsuchiura99,Tsuchiura00,Tsuchiura01} also
studied the effects of magnetic and  non-magnetic impurities onto
the local density of states in HTSC within above approach. The
results obtained resemble the STM data
\cite{Hudson99,Yazdani99,Pan03} quite well\footnote{Nevertheless
VMC calculations \cite{Liang02} for a non-magnetic impurity report
some minor discrepancies to the corresponding RMFT study
\cite{Tsuchiura01}.}. The self-consistent determination of order
parameters within the renormalised Bogoliubov-de Gennes theory was
also applied to study surface effects in 2D superconducting states
\cite{Tanuma98,Tanuma99}.

To analyse the above problems within an unrestricted Hartree-Fock
theory, most authors consider a large (but finite) unit cell, which
exhibits a certain charge ordering pattern or which has a vortex
core or an impurity site in the middle. The corresponding
renormalised Bogoliubov-de Gennes equations can then be solved by
assuming a lattice of unit cells ({\it e.g.} $N_c=20×20$) and
making use of Fourier transformations. While most studies
use the Gutzwiller factors derived in \chap{GutzApp} some recent
works \cite{Tsuchiura03,Tsuchiura01} use the extensions
proposed by Ogata and Himeda \cite{Ogata03b} (see \sect{AF_RMFT}).
Finally, we note that all these studies concentrate on the ground state properties ($T=0$).
It would be very interesting to consider finite
temperature effects within a renormalised unrestricted
Hartree-Fock theory, and to our knowledge,
such studies have not yet been carried out.

\subsubsection{Gossamer superconductivity}

Another class of renormalised mean field theories considers
a modified version of the Hubbard model, which includes a
superexchange interaction $J$ like in the $t$-$J$ Hamiltonian.
This $t$-$J$-$U$ model was proposed by Zhang \cite{Zhang03} to
study the so-called gossamer superconductivity \cite{Laughlin02}.
Here, the form of the GA, which includes finite double occupancy,
must be used for the renormalised Hamiltonian \cite{Zhang03}. The
RMFT gap equations are obtained  in a straightforward manner,
and the number of double occupancies is determined by
optimising the ground state energy. Within this approach, at
half-filling, there is a first order phase transition from a Mott
insulating phase at large Coulomb repulsion U to a gossamer
superconducting phase at small U. Away from half-filling the Mott
insulator evolves into an RVB state, which is adiabatically
connected to the gossamer superconductor \cite{Gan05a}. Some
authors follow this approach to study HTSC
\cite{Gan05a,Kopec04,Normand04,Yuan05,Wang06}
while others have used it in the phenomenology of
organic superconductors \cite{Powell05,Gan05b}.

\subsubsection{Time-dependent Gutzwiller approximation}

The studies discussed so far mainly focused on the superconducting
state. Seibold and Lorenzana \cite{Seibold01} considered the Hubbard model
without superconducting pairing and developed a time-dependent
Gutzwiller approximation analogous to the time-dependent
Hartree-Fock theory. This formalism incorporates ground state
correlations beyond the GA within the random phase approximation
and allows for a computation of the dynamical density-density
response function. The scheme successfully describes several
interesting features of HTSC, such as the dynamics of stripes
\cite{Lorenzana03} or the dispersion of magnetic excitations
\cite{Seibold04,Seibold05,Seibold06a} and was recently applied to
investigate checkerboard inhomogeneities \cite{Seibold06b}. It
would be very useful if this scheme could be adapted to study
the dynamics of a homogeneous superconducting phase.

\section[Variational Monte Carlo calculations]
{Variational Monte Carlo calculations for HTSC - an overview}
\label{VMC}

The VMC technique allows for an accurate evaluation of expectation
values in Gutzwiller projected wave functions. In this section we
present technical details of the VMC method and review the
variational search for the optimal ground state energy in the
Hubbard and $t$-$J$ model. In this context, we also discuss the
coexistence of superconductivity with antiferromagnetism  and flux
states as well as the improvement of the trial wave function by
further variational parameters and Jastrow factors. Further we
consider doping dependent features of projected wave functions and
compare them to experimental findings in HTSC. Finally,
we discuss a recent numerical study dealing with the
tendency towards a spontaneous breaking of the Fermi surface
symmetry.

\subsection{Details of the VMC method}

The VMC method was first applied to the study the projected Fermi
sea \cite{Gros87,Horsch83}, which has a fixed particle number.
However, superconducting BCS wave functions $|\Psi_0\rangle$ are
generally defined in a grand canonical ensemble, where the wave
function shows particle number fluctuations. These particle number
fluctuations are also present in the projected BCS wave function,
$|\tilde \Psi\rangle \equiv P_G\,|\Psi_0\rangle$. Within the VMC
scheme we have now two possibilities, we can either directly work
with $|\tilde \Psi\rangle$ (grand canonical ensemble)
\cite{Yokoyama88}, or we can project out the particle number
fluctuations by a projector $P_N$ that fixes the particle number
and work with $|\Psi\rangle \equiv P_N |\tilde \Psi\rangle$
(canonical ensemble) \cite{Gros88}. In this review, we only
present the method of Gros \cite{Gros88,Gros89a}, used in
most recent VMC calculations, since it avoids complications due to
the fluctuating particle number. The possible discrepancies between
the grand canonical and the canonical VMC scheme
have been discussed in detail in \sect{fugacity}.

\subsubsection{Real space representation of the trial wave function}
\label{realspaceRVB}

Before performing a VMC calculation we have to rewrite the wave
function in an appropriate way. By inserting a complete set of
states $\{\, |\alpha\rangle \, \}$ for the subspace that excludes
double occupancy (and with a fixed particle number $N$), we can remove
the projection operator $P_N$ and $P_G$ from the wave function
$|\Psi \rangle$,
\begin{equation}
|\Psi\rangle \equiv P_N P_G |\Psi_0\rangle = \sum_\alpha \langle
\alpha |\Psi_0\rangle \,|\alpha\rangle \ . \label{eq51}
\end{equation}
The most suitable choice of $|\alpha\rangle$ is given by a
straightforward real space representation in terms of fermion
creation operators,
\begin{equation}
|\alpha\rangle \, = \, c^\dagger_{\bR_1,\uparrow} \dots
c^\dagger_{\bR_{N_\uparrow},\uparrow} c^\dagger_{\bR_1,\downarrow}
\dots c^\dagger_{\bR_{N_\downarrow},\downarrow} |0\rangle \ .
\label{alpha}
\end{equation}
The state \eq{alpha} is specified by two disjoint sets $\{\bR_1
\dots \bR_{N_\uparrow}\}$ and $\{\bR_1 \dots
\bR_{N_\downarrow}\}$, which determine the positions of the up-
and down-spin electrons on a finite lattice.

Next we have to calculate the overlap $\langle \alpha
|\Psi_0\rangle$. To determine this quantity by a Monte Carlo
calculation, we write the BCS wave function $P_N|\Psi_0\rangle$ as
\cite{Anderson87},
\begin{deqarr}
 P_N|\Psi_0\rangle \, &\equiv& \,P_N
\prod_{\bk}
\left(u_\bk+v_\bk \ c_{\bk\uparrow}^\dagger
 c_{-\bk\downarrow}^\dagger \right)|0\rangle~
  \arrlabel{fixedN_sect5}\\
&\propto&\,P_N \prod_{\bk}
\left(1+a_\bk \ c_{\bk\uparrow}^\dagger
 c_{-\bk\downarrow}^\dagger \right)|0\rangle~ \label{wave_b} \\
&\propto&\, \left(\sum_{\bk} a_\bk \
c_{\bk\uparrow}^\dagger c_{-\bk\downarrow}^\dagger
 \right)^{N/2}|0\rangle~  \label{wave_c}\\
&=&\,\left(\sum_{\bR_{j,\downarrow},\bR_{j,\uparrow}}\,
a(\bR_{j,\downarrow}-\bR_{j,\uparrow}) \
c_{\bR_{j,\uparrow},\uparrow}^\dagger
 c_{\bR_{j,\downarrow},\downarrow}^\dagger \right)^{N/2}|0\rangle~ \
 .\label{wave_d}
\end{deqarr}
In \eq{wave_b} we defined the quantity $a_\bk\equiv
v_\bk/u_\bk$, which can be written as
\begin{equation}
a_\bk=\frac{\Delta_\bk}{\xi_\bk+\sqrt{\xi_\bk^2+\Delta_\bk^2}}~, \
\label{ak}
\end{equation}
using the mean field result from \eq{MF_vk}. Due to the
projection operator $P_N$ we can then represent the wave function by
a product of $N/2$ pairs, where we use
$N_\sigma=N_\uparrow=N_\downarrow=N/2$, valid for a BCS
wave function. In \eq{wave_d} we assumed $a_\bk=a_{-\bk}$, applied
a Fourier transformation, and defined
\begin{equation}
a(\br) \equiv \sum_\bk \, a_\bk\, \cos(\bk \cdot \br) \ .
\label{a_realspace}
\end{equation}
Finally, we arrive at the real space representation of
$P_N|\Psi_0\rangle$ as in \eq{wave_d}.

Since all configurations $\alpha$ in \eq{eq51} have the same to
particle number $N$, $\langle \alpha |\Psi_0\rangle=\langle \alpha
|P_N|\Psi_0\rangle$. Making use of \eq{wave_d} one finds that the
overlap, $\langle \alpha |\Psi_0\rangle$, is given
\cite{Gros88,Gros89a} by the determinant of the matrix $A_\alpha$
which has the form,
\[  \left( \begin{array}{cccc}
a(\bR_{1,\downarrow}-\bR_{1,\uparrow}) & a(\bR_{1,\downarrow}-\bR_{2,\uparrow}) &
 \dots & a(\bR_{1,\downarrow}-\bR_{N_\sigma,\uparrow}) \\
a(\bR_{2,\downarrow}-\bR_{1,\uparrow}) & a(\bR_{2,\downarrow}-\bR_{2,\uparrow}) &
 & a(\bR_{2,\downarrow}-\bR_{N_\sigma,\uparrow}) \\
 \vdots    &         &   \ddots   &   \vdots    \\
a(\bR_{N_\sigma,\downarrow}-\bR_{1,\uparrow}) &
 a(\bR_{N_\sigma,\downarrow}-\bR_{2,\uparrow})  &
\dots & a(\bR_{N_\sigma,\downarrow}-\bR_{N_\sigma,\uparrow})
\end{array} \right)\ . \] To see this we must expand \eq{wave_d} and
gather all terms contributing to the configuration $\alpha$, which
has down-electrons on
$\{\bR_{1,\downarrow},\bR_{2,\downarrow}\dots
\bR_{N_\sigma,\downarrow}\}$ and up-electrons on
$\{\bR_{1,\uparrow},\bR_{2,\uparrow}\dots
\bR_{N_\sigma,\uparrow}\}$. The number and functional form of
these terms are obviously the same as those for $|A_\alpha|$. Next
we must order up- and  down-electrons in the same way for all
terms. By doing so we pick up relative signs, which are exactly
reproduced by the determinant of $A_\alpha$.

We note that the above real space representation can be extended
\cite{Weber06a} to wave functions, which allow for a staggered
magnetisation and an unequal number of up- and down-electrons,
$N_\uparrow \neq N_\downarrow$. Then, the $a(\br)$ in \eq{wave_d}
and \eq{a_realspace} becomes spin and site dependent, {\it i.e.}
$a(\br) \to
a(\bR_{i,\sigma_i},\bR_{j,\sigma_j},\sigma_i,\sigma_j)$. The
values of $a(\bR_{i,\sigma_i},\bR_{j,\sigma_j},\sigma_i,\sigma_j)$
depend on the particular choice of the mean field wave function
and can be evaluated numerically. The overlap $\langle \alpha |
\Psi_0 \rangle$ is then determined by \cite{Weber06a},
\begin{equation}
\langle \alpha | \Psi_0 \rangle = P_f({\mathbf Q}) \ ,
\label{Pfaffian}
\end{equation}
where $P_f({\mathbf Q})$ is the Pfaffian\footnote{The Pfaffian is
the analogue of a determinant which is defined only for
antisymmetric matrices. For an antisymmetric matrix $A$, the
square of the Pfaffian is equivalent to its determinant, {\it
viz.} ${P_f(A)}^2=|A|$.} of the matrix
\begin{equation}
{\mathbf Q}=a(\bR_{i,\sigma_i},\bR_{j,\sigma_j},\sigma_i,\sigma_j)
-a(\bR_{j,\sigma_j},\bR_{i,\sigma_i},\sigma_j,\sigma_i) \ .
\end{equation}
The positions of the electrons, $\bR_{i,\sigma_i}$ and
$\bR_{j,\sigma_j}$ determine the real space configuration
$\alpha$. For a simple BCS wave function with
${a(\bR_{i,\uparrow},\bR_{j,\uparrow},\uparrow,\uparrow)}
= a(\bR_{i,\downarrow},\bR_{j,\downarrow},\downarrow,\downarrow)=0$
and
$a(\bR_{i,\uparrow}-\bR_{j,\downarrow})
=a(\bR_{i,\uparrow},\bR_{j,\downarrow},\uparrow,\downarrow)$, the
overlap $\langle \alpha | \Psi_0 \rangle$ in equation
\eq{Pfaffian} reduces to the previously discussed determinant
$|A_\alpha|$.

\subsubsection{Implementation of the Monte Carlo simulation}

Using \eq{eq51}, we may write the expectation value of an
operator $\hat O$ in $|\Psi \rangle$ as,
\begin{deqarr}
\langle \hat O \rangle_\Psi &=& \frac{\langle \Psi_0 |P_G P_N \hat
O P_N P_G |\Psi_0 \rangle}{\langle \Psi_0| P_G P_N P_G |\Psi_0
\rangle}  \\
 &=&\sum_{\alpha,\beta} \langle \alpha|\hat O |\beta \rangle
\frac{\langle \Psi_0 | \alpha \rangle \langle \beta | \Psi_0
\rangle} {\langle \Psi_0|P_N P_G |\Psi_0
\rangle}  \\
&=&\sum_\alpha \left( \sum_\beta \frac{\langle  \alpha|\hat
O|\beta \rangle \langle \beta | \Psi_0 \rangle} {\langle \Psi_0 |
\alpha \rangle} \right) \frac {|\langle \Psi_0 | \alpha
\rangle|^2} {\langle \Psi_0| P_N P_G |\Psi_0
\rangle}  \\
&=& \sum_\alpha f(\alpha)\ p(\alpha) \ , \arrlabel{VMCeq}
\end{deqarr}
with,
\begin{deqarr}
f(\alpha) & = & \sum_\beta \frac{\langle  \alpha|\hat O|\beta
\rangle \langle \beta | \Psi_0 \rangle} {\langle \Psi_0 | \alpha
\rangle}  \ ,\\
p(\alpha) & = & \frac {|\langle \Psi_0 | \alpha
\rangle|^2}{\langle \Psi_0| P_N P_G |\Psi_0 \rangle} \
\label{palpha}.
\end{deqarr}
Here, $\alpha$ and $\beta$ are real space configurations \eq{alpha}. Since,
\begin{equation}
p(\alpha) \geq 0\, , \qquad \sum_\alpha p(\alpha)=1\ ,
\end{equation}
are the features of a probability distribution, we can evaluate
$\langle \hat O \rangle_\Psi$ by a random walk through the
configuration space with weight $p(\alpha)$. Therefore, we can
analyse \eq{VMCeq} by a standard Metropolis Monte Carlo
calculation. We note that the norm ${\langle \Psi_0| P_N P_G
|\Psi_0 \rangle}$ in \eq{palpha} is not of relevance within the
Monte Carlo calculation, since only relative probabilities
$p(\alpha)$ enter the transition probability.

Next, we make comments on the updating procedure and the
calculation of the determinant $|A_\alpha|$. Most VMC calculations
generate a new configuration $\alpha'$ by randomly interchanging
two electrons with opposite spin or moving an electron to an empty
site. The random walks thus constructed are ergodic. In general, to
optimise the numerical performance, the rules for generating the
random walk through the configuration space should be chosen so as to
maximise the acceptance rate, $T(\alpha \to \alpha')$.

The calculation of $|A_\alpha|$ is numerically expensive and is
required at each Monte Carlo step for the computation of
$p(\alpha)$. Therefore it is advantageous to determine the ratio
$|A_{\alpha'}|/|A_{\alpha }|$ between new and old determinant (new
and old configuration $\alpha'$ and $\alpha$) instead of directly
evaluating $|A_{\alpha'}|$ for every configuration. According to
Ceperley, \etal \ \cite{Ceperley77}, this ratio can be efficiently
computed within $\mathcal{O}(N_\sigma^2)$ computation steps, while
a direct evaluation of $|A_{\alpha'}|$ requires
$\mathcal{O}(N_\sigma^3)$ steps. The trick is to store not only
the matrix $A_\alpha$, but also its inverse $A_\alpha^{-1}$. For
the commonly used updating procedures mentioned above, $\alpha'$
differs from $\alpha$ only by the interchange of two electrons
with opposite spins or the interchange of an electron and an empty
site. Thus, the matrices $A_{\alpha'}$ and $A_{\alpha}$ differ
only by one row and one column, and
$|A_{\alpha'}|/|A_{\alpha}|=|A_{\alpha'} A_{\alpha}^{-1}|$ , which
enters the transition rate $T(\alpha \to \alpha')$, can be easily
computed.

\begin{figure}
\centering
\includegraphics*[width=0.7\textwidth]{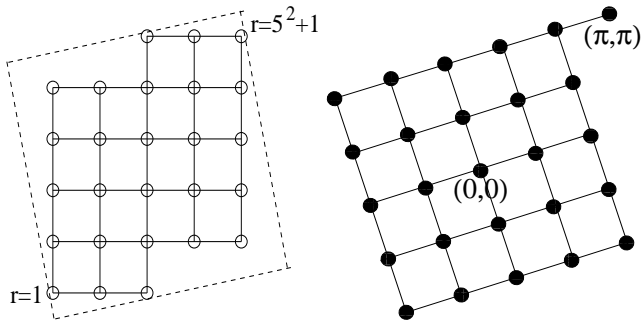}
\caption{(left) Real space picture of the $L^2+1$ lattice for
$L=5$, with periodic boundary conditions, $(5,1)$ and $(-1,5)$,
applied along the opposite edges of the tilted square indicated by
dashed lines. (right) The $\bk$-space Brillouin zone of the
`tilted lattice' for $L=5$. From \cite{Paramekanti04}.}
\label{boundary}
\end{figure}

A general advantage of the Monte Carlo method is the possibility
to estimate the numerical accuracy systematically with the error
being proportional to the inverse square root of the number of
Monte Carlo steps $N_r$. Present computer capacities allow us to
consider sufficiently large clusters, where finite size effects
play a minor role. However, $a_\bk=v_\bk/u_\bk$,
defined in equation \eq{ak} becomes singular
whenever $\Delta_\bk=0$ and $\xi_\bk\leq0$. In
particular, this is problematic for a $d$-wave order parameter,
for which $\Delta_\bk=0$ for all $k$-points along the Brillouin zone
diagonals, {\it i.e.} $|k_x|=|k_y|$. It is thus convenient to
avoid these $\bk$-points by an appropriate choice of boundary
conditions. There are three different approaches discussed in
literature. The first possibility is a tilted lattice with
periodic boundary conditions (PBCs). Such a lattice has $L^2+1$
sites with odd $L$, preserves the fourfold rotational symmetry of
the lattice, and does not introduce any twist in the boundary
conditions. An example of these (widely used) boundary conditions
(see, {\it e.g.}
\cite{Gros88,Paramekanti04,Paramekanti01,Gros89a}) is illustrated
in \fig{boundary}. Another choice is an $L× L$ lattice with even
$L$ and periodic and antiperiodic boundary conditions in the $x$-
and the $y$-direction, respectively. Finally it is possible to use
a rectangular $L_x × L_y$ lattice  with PBCs and mutually coprime
dimensions $L_x$ and $L_y$, {\it i.e.} the greatest common divisor
of $L_x$ and $L_y$ being $1$.

\subsection{Improvements of the trial wave function}
\label{VMC_groundstate}

The early VMC calculations for projected BCS states of Gros
\cite{Gros88} and Yokoyama and Shiba \cite{Yokoyama88} were
carried out to check whether a Gutzwiller projected
superconducting wave function is indeed a variationally good
trial state for the $t$-$J$ model. To limit the number of
variational parameters, these authors used a dimensionless dispersion,
$\xi_\bk=-2\,(\cos k_x + \cos k_y) - \,\mu$, and various
superconducting gap functions $\Delta_\bk$ to calculate
$a_\bk={\Delta_\bk}/(\xi_\bk+\sqrt{\xi_\bk^2+\Delta_\bk^2}) $. In
his original work, Gros \cite{Gros88} compared variational
energies of $s$-wave, $\Delta_\bk=\Delta$, $d$-wave,
$\Delta_\bk=\Delta \,(\cos k_x - \cos k_y)$, and extended
$s$-wave, $\Delta_\bk=\Delta (\cos k_x + \cos k_y)-\mu$,
functions. By optimising solely\footnote{Due to the fixed particle
number, the chemical potential $\mu$ becomes an additional free
parameter. However, this parameter was fixed in \cite{Gros88} by
setting the chemical potential $\mu$ to those of the unprojected
wave function.} the variational parameter $\Delta$, he found that
a $d$-wave gap can substantially lower the energy compared to
projected Fermi sea ($\Delta_\bk=0$) at half-filling as well as at
finite doping. The result is consistent with other early works
such as the VMC calculations of Yokoyama and Shiba
\cite{Yokoyama88} or the mean field theories of Zhang, \etal \
\cite{Zhang88b} and Kotliar and Liu \cite{Kotliar88}.

More detailed studies showed that the optimal superconducting
state remains a pure $d$-wave even when mixed states of $s$- and
$d$-wave pairing are considered \cite{Yokoyama96}. The optimal
variational parameter $\Delta$ decreases when going away from
half-filling and vanishes at about $30 \%$ doping. The exact
dimension of the superconducting region in the phase diagram
depends on the choice of $J/t$ as well as on the inclusion of the
correlated hopping term \cite{Yokoyama96}.

\subsubsection{Antiferromagnetism and flux states}

\begin{figure}
\centering
\includegraphics*[width=0.7\textwidth]{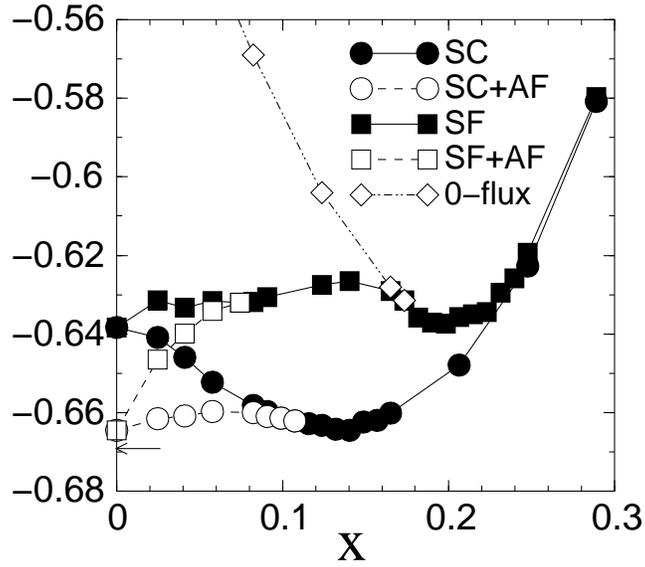}
\caption{Optimal energies per site (in units of $J$) for five
different Gutzwiller projected wave functions with a linear part
subtracted ($E-\mu_\text{sep}$) as a function of doping $x$. Wave
functions: superconducting without antiferromagnetism (SC, solid
circles), superconducting with antiferromagnetism (SC+AF, empty
circles), staggered-flux without antiferromagnetism (SF, solid
squares), staggered-flux with antiferromagnetism (SF+AF, empty
squares), and zero-flux (projected Fermi sea, empty diamonds). The
arrow in the panel shows the best variational estimate for the
half-filled system ($E=-0.669J$ per site)
\cite{Trivedi90,Buonaura98}. Only nearest neighbour hopping is
considered and $J/t=0.3$. From \cite{Ivanov04}.} \label{Ivanov}
\end{figure}

Further extensions
\cite{Himeda99,Chen90,Giamarchi91,Ivanov04,Shih04a} considered the
incorporation of antiferromagnetism for a more accurate
description of the $t$-$J$ model near half-filling. These studies
show a coexistence between superconductivity and antiferromagnetic
long-range order (AFLRO) for doping $x \leq 0.1$. At half-filling,
the optimal so-constructed wave function has a staggered
magnetisation of $0.75$ and a variational energy of $-0.664J$ per
site; impressively close to the best numerical estimate of
$-0.669J$ per site by Green's function Monte Carlo techniques
\cite{Trivedi90,Buonaura98}. A comparison of the variational
energies of the different wave functions is given in \fig{Ivanov}.
The figure also reveals an upward convexity of the ground state
energy (SC+AF state) as a function of doping. This indicates a
phase separation at $x_\text{sep}=0.13$ (see \fig{Ivanov}).
However, Ivanov \cite{Ivanov04} demonstrated that a sufficiently
strong nearest neighbour Coulomb repulsion can suppress the
formation of separated phases. Further VMC calculations showed
that the coexistence of superconductivity and AFLRO is nearly
absence if next and second nearest neighbour hopping are
included \cite{Shih04a}. For these more realistic model parameters,
the AFLRO disappears at about $6 \%$ doping in better agreement
with experimental observations \cite{Shih04a}.

Apart from the superconducting states, the projected
staggered-flux state has also been studied
as a competitive variational
state; however, its energy lies above those of the $d$-wave for
all dopings (\fig{Ivanov}). As discussed in \chap{RMFT}, the flux
state becomes identical to the superconducting state at
half-filling explaining the collapse of the energies in
\fig{Ivanov} (see also \cite{Ivanov03}). This behaviour is due to
$SU(2)$-symmetry, which is also responsible for the
occurrence of staggered-vorticity correlations of current in the
$d$-wave state at small dopings \cite{Ivanov00}.

\subsubsection{Increasing the number of variational parameters}
\label{increase_var}

\begin{figure}
\centering
\caption{Momentum distribution $n(\bk)$ in the first Brillouin
zone for doping $x=$ (a) 0.19, (b) 0.31 (c) 0.42 (d) 0.49 for $12×
12$ lattice with $J/t=0.3$, $t'=-0.3t$, and $t''=0$. (e) optimal
parameters $t'_v$ (squares) and $\Delta$ (circles). From
\cite{Shih04b}.} \label{Shih}
\end{figure}

In recent VMC calculations, the chemical potential $\mu$ as
well as the next nearest neighbour hopping $t'_\text{var}$ are
chosen as additional variational parameters, which are optimised
numerically. While the chemical potential has minor influence on
the optimal state \cite{Yokoyama96}, a variational $t'_\text{var}$
can significantly effect the shape of the Fermi surface. Himeda
and Ogata \cite{Himeda00} reported that for a bare dispersion
$t'=0$ and a doping level of $x=0.15$ the lowest energy is
provided by a variational $t'_\text{var}=-0.1t$, causing a
spontaneous deformation of the Fermi surface. More detailed VMC
studies \cite{Shih04b} include next nearest ($t'$) and next next
nearest ($t''$) neighbour hopping in the bare dispersion and also
use variational parameters $t'_{\rm var}$ and $t''_{\rm var}$. The
obtained momentum distribution $n(\bk)$ (related to the Fermi
surface, see \sect{FSHTSC}) together with the optimal variational
$\Delta$ and $t'_\text{var}$ from these calculations are
illustrated in \fig{Shih}. This work of Shih, \etal \
\cite{Shih04b} also revealed that the more negative the bare ratio
$t'/t$, the higher the superconducting pairing in the optimal
variational state of the $t$-$J$ model. This is in agreement with
band-structure calculations \cite{Pavarini01} that suggest that
the ratio $t'/t$ is essential to raise $T_c$. Similar
trends can be inferred from the RMFT
calculations for the Hubbard model discussed in \sect{result_ex}.

VMC studies of inhomogeneous phases
\cite{Himeda02} find that around $x=1/8$, stripe states with
fluctuating $d$-wave superconductivity can lower the
variational ground state energy in the two-dimensional $t$-$J$
model. More recent studies report that at $x=1/8$, a bond-order
modulated staggered flux state can also overcome the RVB
superconductor for sufficiently large short range Coulomb
repulsion \cite{Weber06b}. However, the energy gains within these
studies are often quite small and sensitively depend on model
parameters. Nevertheless, these VMC calculations show that the
slightly doped $t$-$J$ model exhibits tendencies towards various
inhomogeneities, which could be relevant for explaining
some of the experimental observations in the underdoped HTSC.

The energy of the projected $d$-wave state can be improved further
by the incorporation of Jastrow factors (see \sect{other_trial}).
Sorella, \etal \ \cite{Sorella02a} showed numerically that
such wave functions lower the variational energy and still
exhibit long range superconducting order. Nevertheless there is
still debate (see
\cite{Shih98,Lee02,Sorella02b}) whether the superconductivity
within the VMC scheme results only from a biased choice of the
wave function or is indeed a ground state property of the $t$-$J$
model. In our opinion, this debate does not pose an obstacle towards
our understanding of the HTSC, since we are primarily
interested in physical properties of projected wave functions
rather than in proving them to be exact ground states of a particular Hamiltonian.
In this, we follow the point of view espoused by
Anderson, \etal \
\cite{Anderson04a}, \emph{`The philosophy of this
method is analogous to that used by BCS for superconductivity, and
by Laughlin for the fractional quantum Hall effect: simply guess a
wave function. Is there any better way to solve a non-perturbative
many-body problem?'}

\subsubsection{Investigation of the Pomeranchuk instability}

\begin{figure}
\centering
 \includegraphics[width=0.8\textwidth]{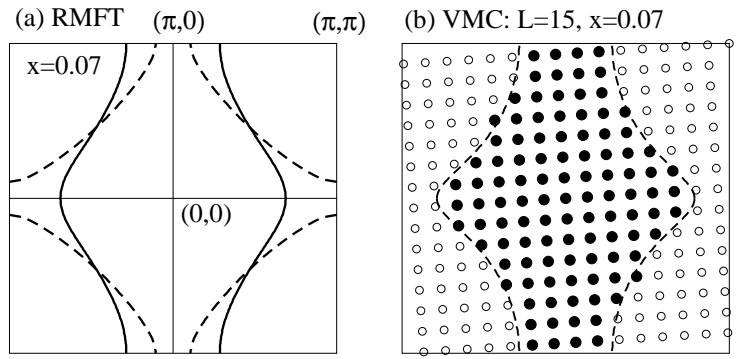}
 \caption{Fermi surface of the isotropic $t$-$J$ model
 with $J=0.3t$ and $t'=-0.3t$
 at $x=0.07$ (a) RMFT results for the Fermi surface
 of the normal state with $\Delta_\bk\equiv0$
 (quasi 1D state, solid line) and the
 optimal $d$-wave state (isotropic, dashed line).
 (b) Best quasi 1D state on a $(15^2+1)$-sites lattice by VMC;
 filled circles indicate
 the Fermi surface. From
\cite{Edegger06b}.}
 \label{FS}
\end{figure}

The possibility that strong correlations may break the
symmetry of the underlying Fermi surface was studied
recently in \cite{Edegger06b}. As illustrated in \fig{FS}(a),
this instability results in a deformation of the Fermi surface,
which becomes quasi one dimensional, although the underlying two
dimensional (2D) lattice is still isotropic. Motivated by the
Fermi surface depicted in \fig{FS}(a), the state resulting from
the tetragonal symmetry breaking can be called a `quasi 1D state'.
This phenomenon is also called a Pomeranchuk instability
of the Fermi surface.

\begin{figure}
  \centering
 \includegraphics*[width=0.8\textwidth]{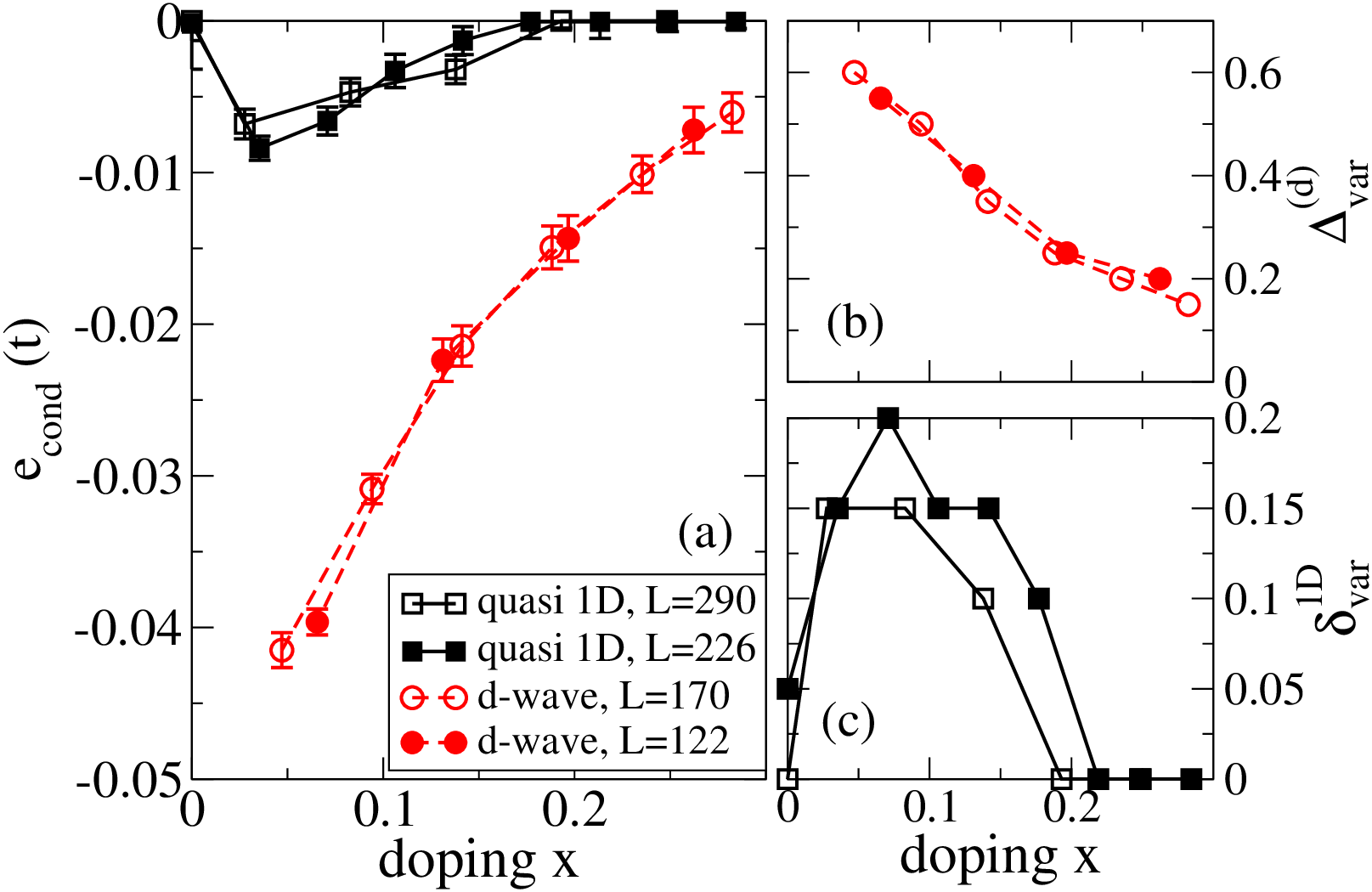}
 \caption{
 (a) VMC results for condensation energies per site $e_{\rm cond}$
 of the quasi 1D state ($\delta^{\rm 1D}_{\rm var}\neq 0$, $\Delta_\bk\equiv 0$) and
 the $d$-wave state ($\Delta_\bk \neq 0$, $\delta^{\rm 1D}_{\rm var} \equiv 0$) with
 $t'=-0.3\,t$; see \eq{XiKpom} and \eq{DKpom} for the definition of these states.
 The optimal variational $\Delta^{(d)}_{\rm var}$ of the $d$-wave is shown in (b),
 the optimal variational asymmetry $\delta^{\rm 1D}_{\rm var}$ of the quasi 1D state
 is given in (c). The errors in (b) and (c) are $\Delta \Delta^{(d)}_{\rm var} =0.05$ and
$\Delta \delta^{\rm 1D}_{\rm var} = 0.05$, respectively. System
sizes:
 $L=11^2+1=122$, $L=13^2+1=170$, $L=15^2+1=226$, and
 $L=17^2+1=290$. From \cite{Edegger06b}.
 }
 \label{compareDwave}
\end{figure}

To investigate instabilities towards quasi 1D states, we have to
extend the variational space by an additional order parameter,
which allows for a finite asymmetry in the wave function. A
possible choice was proposed by Edegger \etal \ \cite{Edegger06b},
who determined
$a_\bk={\Delta_\bk}/(\xi_\bk+\sqrt{\xi_\bk^2+\Delta_\bk^2})$ by
using
\begin{eqnarray}
\xi_\bk=&-&2\,[(1+\delta_{\rm var}^{\rm 1D}) \cos k_x +
(1-\delta_{\rm
var}^{\rm 1D})\cos k_y] \nonumber \\
&-& 4 t'_{\rm var} \cos k_x \cos k_y\,-\, \mu_{\rm var}
\label{XiKpom}
\end{eqnarray}
and
\begin{equation}
\Delta_\bk=\Delta_{\rm var}^{(d)}(\cos {k_x}-\cos{k_y})+
\Delta_{\rm var}^{(s)}(\cos {k_x}+\cos {k_y})\ . \label{DKpom}
\end{equation}
The five variational parameters, {\it viz.} the asymmetry
$\delta^{\rm 1D}_{\rm var}$ between the $x-$ and the
$y-$direction, the variational next nearest neighbour hopping term
$t'_{\rm var}$, a variational chemical potential $\mu_{\rm var}$,
and variational parameters for $d-$ and $s-$wave pairing,
$\Delta_{\rm var}^{(d)}$ and $\Delta_{\rm var}^{(s)}$, can be
optimised by determining the energy expectation values for
different choices of variational parameters within a standard VMC
technique.

VMC calculations for the isotropic $t$-$J$ model \cite{Edegger06b}
show in agreement with previous studies
\cite{Yamase00a,Yamase00b,Halboth00}, that the optimal variational
state remains a pure $d$-wave without any anisotropy (see
\fig{compareDwave}). However, when restricting solely to
non-superconducting solutions, {\it i.e.} setting $\Delta_\bk
\equiv 0 $ by using $\Delta_{\rm var}^{(d)} = \Delta_{\rm
var}^{(s)} = 0$, a projected anisotropic Fermi sea provides a
better energy than the isotropic one. In \fig{compareDwave}, this
effect is shown for a VMC calculation \cite{Edegger06b} in an
isotropic $t$-$J$ model.  The figure also illustrates that the
 optimal $d$-wave state has much better
energy than the quasi 1D state, which is the best state in the
variational subspace  $\Delta_\bk \equiv 0$.

\begin{figure}
  \centering
 \includegraphics*[width=0.8\textwidth]{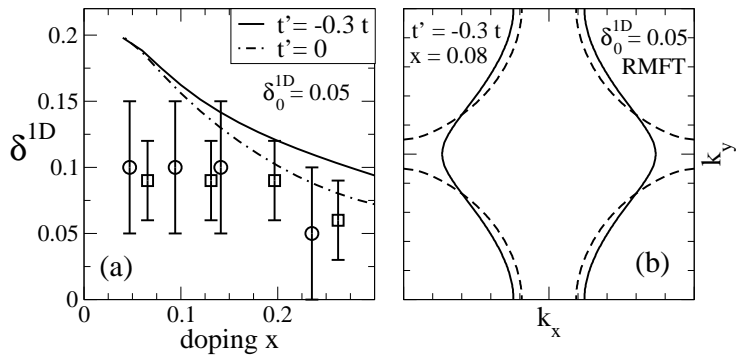}
 \caption{RMFT and VMC results for the $d+s$-wave ground state
 of the anisotropic $t$-$J$ model with $J=0.3 t$ and $\delta_0^{\rm 1D}\equiv (t_x-t_y)/(t_x+t_y)=0.05$.
 (a) Effective asymmetry $\delta^{\rm 1D}\equiv(\tilde t_x-\tilde t_y)/(\tilde
 t_x+\tilde t_y)$ from RMFT
 as a function of hole doping $x$ for (dashed) $t'=0$ and (solid)
 $t'=-0.3t$. VMC results for $t'=0$ are given by squares
 and circles for $L=122$ and
 $L=170$, respectively.
 (b) RMFT Fermi surface (solid lines) of the $d+s$-wave ground
 state and the tight binding dispersion (dashed) at $x=0.08$ with $t'=-0.3t$ and $\delta^0_{\rm
 1D}=0.025$. From
\cite{Edegger06b}.
 }\label{anisotrop}
\end{figure}

The situation can be quite different when the underlying lattice
structure is anisotropic. In this case, the tendency towards a
quasi 1D state is present even in the superconducting state. RMFT,
VMC \cite{Edegger06b}, and SBMFT \cite{Yamase00a,Yamase00b}
calculations predict an optimal state in which the bare
anisotropy $\delta^{\rm 1D}_0$ of the lattice can be significantly
enhanced due to the electron correlations. As seen in
\fig{anisotrop}(a), the bare asymmetry of $\delta^{\rm 1D}_0=0.05$
increases within the RMFT calculations up to about $\delta^{\rm
1D}_{\rm opt}=0.2$ in the underdoped regime. These results are
confirmed to some extent by VMC calculations [\fig{anisotrop}(a),
circles and squares], that show an increase of the asymmetry up to
about $\delta^{\rm 1D}_{\rm var}\approx 0.1$. Furthermore, the
enhancement of anisotropy may even lead to a change in the
topology of the underlying Fermi surface as can be inferred from
\fig{anisotrop}(b).

The PI is one out of several possible instabilities in the $t$-$J$
model that arise from the effects of superexchange and that can be
revealed by VMC and RMFT techniques. Since $J\propto 4t^2/U$, a
small asymmetry in the bare hopping integral $t$ becomes twice as
large in the superexchange energy. Hence, it is natural that the
effects discussed in this subsection are largest in the underdoped
regime, where the dispersion is mainly determined by $J$.

\subsection{Ground state properties - VMC results}
\label{VMC_Para}

Within this subsection, we discuss ground state properties of the
HTSC by considering observables in a Gutzwiller projected
superconducting state. We follow, in part, Paramekanti, \etal \
\cite{Paramekanti04,Paramekanti01}, who studied the Hubbard model
in the strong coupling limit using the re-transformed trial wave
function, $e^{-iS}P_G |\Psi_0\rangle$ (see \chap{RMFT}). By
evaluating the canonical transformation $e^{-iS}$ to
$\mathcal{O}(t/U)$, this ansatz can be treated within the VMC
scheme. The $t/U$-corrections due to $e^{-iS}$ provide a more
accurate description of the HTSC, however, the qualitative nature
of the results is not changed compared to the $t$-$J$ model. In
the following, we ignore the possibility of the superconducting
state coexisting with a flux state,
antiferromagnetism, or a charge ordered state.

\subsubsection{Superconducting gap and order parameter}

\begin{figure}
\centering
\includegraphics*[width=0.5\textwidth]{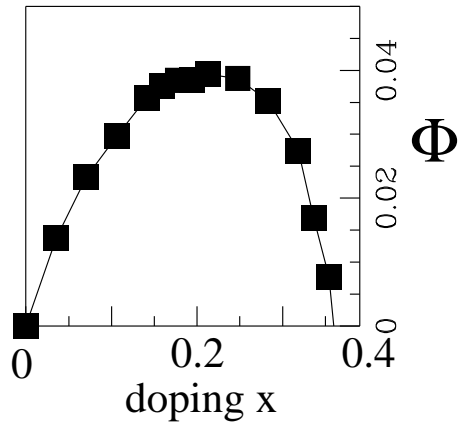}
\caption{Doping dependence of the superconducting order parameter
$\Phi$ from VMC calculations of Paramekanti, \etal \
\cite{Paramekanti01}. Model parameters: $U=12t$, $t'=-t/4$.}
\label{order_Para}
\end{figure}

In the previous subsection, we saw that the variational
parameter $\Delta$, which is proportional to the superconducting
gap $\Delta_\bk$, increases in the limit of half-filling. The
doping dependence of $\Delta$ is illustrated in \fig{Shih} and
resembles the RMFT result (\fig{RMFTphasediagram}) as well as the
experimental observed gap at $\bk=(\pi,0)$. However, we cannot
deduce the relevant energy scale of the gap from $\Delta$, since
it is a dimensionless parameter within the VMC calculations. For
detailed statements about the gap we have to consider the energy
of excited states as we will do in sections \ref{EX_RMFT} and
\ref{EX_VMC}.

When considering the variational parameter $\Delta$, we must
realise as discussed in \chap{RMFT}, that it does not correspond to the true superconducting
order $\Phi\equiv|\langle c^\dagger_{i\uparrow}
c^\dagger_{i+\tau\downarrow}-c^\dagger_{i\downarrow}
c^\dagger_{i+\tau\uparrow}\rangle|$.
The relevant physical quantity here is the off-diagonal long range order
(ODLRO) \cite{Gros88,Gros89a,Paramekanti01} defined by,
$$F_{\alpha,\beta}  (\br - \br') = \langle c_{\uparrow}^{\dagger}(\br)
c_{\downarrow}^{\dagger}(\br
+ \hat \alpha) c_{\downarrow}(\br) c_{\uparrow}(\br + \hat \beta)
\rangle\ ,$$
where $\hat \alpha,\hat \beta = {\hat x}, {\hat y}$. In the limit
of large $| \br - \br' |$, $F_{\alpha,\beta}$ is related to
$\Phi^2$ via $F_{\alpha,\beta} \to ± \Phi^2$ with $+$ ($-$) sign
obtained for $\hat a \parallel$ ($\perp$) to $\hat b$, indicating
$d$-wave superconductivity \cite{Paramekanti01}. The doping
dependence of the superconducting order parameter $\Phi$ is
depicted in \fig{order_Para} (VMC calculations of Paramekanti,
\etal \ \cite{Paramekanti01}). It is not identical to $\Delta$ as
first noted by Gros \cite{Gros88,Gros89a}. The VMC calculation
match the RMFT result (\fig{RMFTphasediagram}), where $\Phi$
vanishes linearly as $x \to 0$. The vanishing order parameter
$\Phi$ indicates a Mott insulating phase at $x=0$, where
superconductivity is destroyed by the suppression of particle
number fluctuations. At finite doping $x$ a superconducting state
is realised in the range $0<x<0.35$.

\subsubsection[Derivation of spectral features]
{Derivation of spectral features from ground state properties}
\label{SPgroundstateVMC}

\begin{figure}
\centering
\includegraphics*[width=0.9\textwidth]{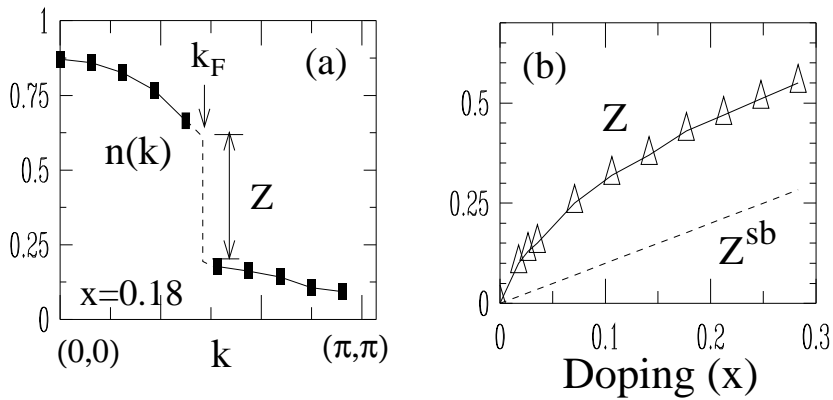}
\caption{(a) The momentum distribution function $n(\bk)$ plotted
along the diagonal $\bk=(k,k)$ showing the jump at $k_F$ which
implies a gapless nodal quasiparticle with spectral weight $Z$.
(b) Nodal quasiparticle weight $Z(x)$ as a function of doping $x$
compared with the simple SBMFT result $Z^\text{sb}(x)=x$. Model
parameters: $U=12t$, $t'=-t/4$. From \cite{Paramekanti01}. }
\label{moment_para}
\end{figure}

Next we follow \cite{Paramekanti01} and analyse the one-particle
spectral function $A(\bk,\omega)$ by calculating the moments,
\begin{equation}
M_l(\bk) = \int_{-\infty}^0 d\omega \, \omega^l \, A(\bk,\omega) \
,
\end{equation}
in the projected $d$-wave ground state at $T=0$. This ansatz
allows to obtain information about $A(\bk,\omega)$ from ground
state expectation values without the need for explicit
representations of the excited states. We first concentrate on the
zeroth moment $M_0(\bk) \equiv n(\bk)$, which is equivalent to the
moment distribution $n(\bk)$. Figure \ref{moment_para}(a) shows
that $n(\bk)$ has a jump along $(0,0)$ to $(\pi,\pi)$. This
implies the existence of gapless quasiparticles and allows us to
write the low energy part of the spectral function along the
diagonal as,
\begin{equation}
A(\bk,\omega) = Z \delta(\omega - {\xi}_k) + A^{\rm inc} \ ,
\end{equation}
where, ${\xi}_k = v_F(k - k_F)$, is the quasiparticle dispersion and
$A^{\rm inc}$, a smooth incoherent part. The location of the
discontinuity determines the Fermi point $k_F$ and its magnitude,
the quasiparticle weight, $Z$. Figure \ref{moment_para}(b)
reveals a significant doping dependence and shows that $Z$
vanishes when approaching the Mott-Hubbard insulator $x=0$. This
behaviour is in agreement with more direct calculations, which
explicitly include quasiparticle states (sections \ref{EX_RMFT}
and \ref{EX_VMC}) as well as experiments.

\begin{figure}
\centering
\includegraphics*[width=0.7\textwidth]{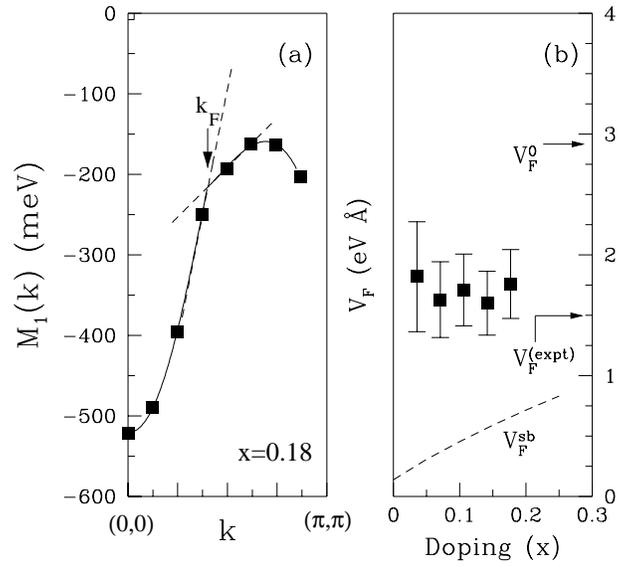}
\caption{(a) The first moment $M_1(\bk)$ of the spectral function
along the zone diagonal, with smooth fits for $k < k_F$ and $k >
k_F$, showing a discontinuity of $Z v_F$ in its slope at $k_F$.
(b) Doping dependence of the nodal quasiparticle velocity obtained
from the slope discontinuity of $M_1(\bk)$. Error bars come from
fits to $M_1(\bk)$ and errors in $Z$. Also shown are the bare
nodal velocity $v^0_F$, the slave boson mean field $v^{\rm
sb}_F(x)$ (dashed line), and the ARPES estimate $v^{\rm (expt)}_F$
\protect\cite{Damascelli03,Campuzano04}. Model parameters:
$U=12t$, $t'=-t/4$. From \cite{Paramekanti01}.} \label{vF_para}
\end{figure}

To determine the nodal Fermi velocity $v_F$, we have to evaluate
the first moment $M_1(\bk) = \langle c_{\bk\sigma}^{\dag} [{H},
c_{\bk\sigma}]\rangle$ along the nodal direction. Due to the
singular behaviour of $A(\bk,\omega)$ at $k_F$, it can be written
as,
\begin{equation}
M_1(\bk) = Z {\xi}_k \Theta(-{\xi}_k) + {\rm smooth\ part} \ .
\end{equation}
Since the slope $dM_1(\bk)/dk$ has a discontinuity of $Z v_F$ at
$k_F$, Paramekanti, \etal \ extracted the nodal Fermi velocity
$v_F$ as shown in \fig{vF_para}(a). The doping
dependence of $v_F$ together with its
bare value $v^0_F$ are shown in \fig{vF_para}(b).
We see that Fermi velocity is only weakly
doping dependent, a result which is consistent with the ARPES data. However, this
estimate of $v_F$ is rather inaccurate compared to the direct
evaluation \cite{Yunoki05a} from the quasiparticle excitation
energies, which will be discussed in section \ref{EX_VMC}.

\begin{figure}
\centering
\includegraphics*[width=0.7\textwidth]{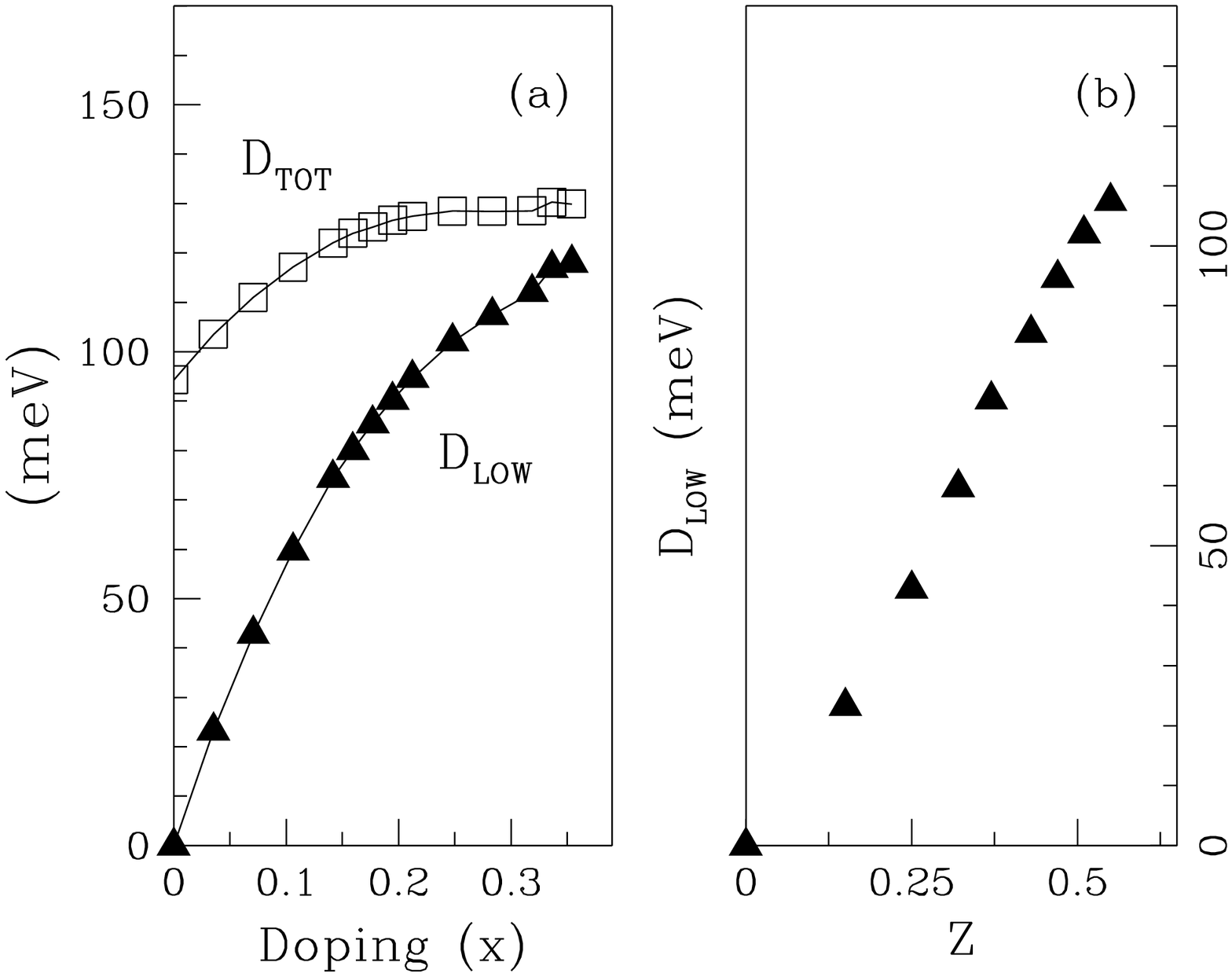}
\caption{(a) Doping dependence of the total ($D_{\rm tot}$) and
low energy ($D_{\rm low}$) optical spectral weights (b) The
optical spectral weight $D_{\rm low}$ versus the nodal
quasiparticle weight $Z$. Model parameters: $U=12t$, $t'=-t/4$.
From \cite{Paramekanti01}.} \label{Dopt_para}
\end{figure}

Ground state expectation values also provide important information
about the optical conductivity in the Hubbard and the $t$-$J$
model. The total optical spectral weight $D_\text{tot}(x)$ can be
calculated by \cite{Paramekanti04}
\begin{equation}
\int_{0}^\infty d\omega \ \Re \, \sigma(\omega) = \pi \sum_{\bk}
m^{-1}(\bk) n(\bk) \equiv \pi D_{\rm tot}/2 \ , \label{Dtot}
\end{equation}
where $m^{-1}(\bk) = $ $\left(\partial^2 \epsilon(\bk)/
\partial\bk_x
\partial\bk_x \right)$
is the non-interacting mass tensor. $\epsilon(\bk)$ is the
non-interacting dispersion (we set $\hbar=c=e=1$). Since the
integral in \eq{Dtot} goes from $0$ to $+\infty$, it includes
contributions from the upper Hubbard band and is thus finite even
at $x=0$ as shown in \fig{Dopt_para}(a).

Paramekanti, \etal \ \cite{Paramekanti04,Paramekanti01} emphasised
that the low frequency optical weight, or Drude weight
\cite{Scalapino93},
\begin{equation}
D_\text{low}=\partial^2\langle {H}_A\rangle/
\partial A^2 \ \label{DrudeWeight},
\end{equation}
is more interesting, because the upper cutoff is chosen smaller
than $U$ and thus excludes the upper Hubbard band. In
\eq{DrudeWeight}, $A$ is the electron-magnetic vector potential,
which is introduced into the Hamiltonian \eq{HubbardH} in terms of
a Peierls substitution \cite{Scalapino93},
\begin{equation} c^\dagger_{i\sigma} c_{j\sigma} \to
c^\dagger_{i\sigma} c_{j\sigma} \, \exp \left( i \, {\rm e}
{\mathbf A}\, (\bR_i-\bR_j) \right) \ , \label{Peierls}
\end{equation}
where we used ${\mathbf A}=(A,0)$ and set $\hbar=c=1$ for
simplicity. As shown in \fig{Dopt_para}(a), the Drude weight
$D_\text{low}$ vanishes linearly for $x \to 0$. This demonstrates
that the Gutzwiller projected superconductor indeed describes an
insulator in the half-filled limit, which can be argued to be a
general property of projected states \cite{Paramekanti04}. The VMC
results for the Drude weight $D_\text{low}$ resemble experimental
data in magnitude as well as in the doping dependence quite well
\cite{Paramekanti04,Yang06}.
By plotting $D_\text{low}$ versus $Z$ (from
the nodal point) Paramekanti, \etal \ also illustrated that
$D_\text{low} \propto Z$, see \fig{Dopt_para}(b).

The Drude weight $D_\text{low}$ also provides an upper bound to
the superfluid stiffness $D_s$, {\it i.e.} $D_s \leq D_{\rm low}$
\cite{Paramekanti98}. It follows that $D_s \to 0$ as $x\to 0$ in
agreement with experiments \cite{Uemura89}. Since the penetration
depth $\lambda_\text{L}$ is related to $D_s$ by
 $\lambda^{-2}_{\rm L} = 4 \pi e^2 D_s/\hbar^2
c^2 d_c$ , where $d_c$ is the mean-interlayer spacing along the
$c$ axis in a layered compound, Paramekanti, \etal \
\cite{Paramekanti04} could also estimate a lower bound for
$\lambda_\text{L}$ which is again consistent with experimental
data.

These VMC calculations based on a Gutzwiller projected
superconducting ground state describe several key features of HTSC
remarkable well. The results are in general agreement with RMFT
and confirm the usefulness of projected wave functions in the
context of HTSC. Although restricted to $T=0$, the above ansatz
gives us some hints about the finite temperature regime. The
superconducting order parameter $\Phi$ resembles the doping
dependence of $T_c$ and vanishes at half-filling, while the
superconducting gap (expected to scale with $\Delta$) remains
finite. This suggests that the underdoped regime exhibits strong
pairing and that the superconducting transition may be determined
by the onset of phase coherence (rather than the vanishing of
pairing amplitude).

\section[Quasiparticle states within RMFT]
{Quasiparticle states within renormalised mean field theory}
\label{EX_RMFT}

Extending the RMFT to excited states requires the consideration of
Gutz­willer­-Bogoliubov quasiparticles within the $t$-$J$ and the
Hubbard model. These Gutz­willer­-Bogoliubov excitations then
allow for a systematic analysis of the single particle spectral
function and explain momentum- and doping-dependent features in
ARPES and STM experiments.  Apart from these key results,
in this section, we also
discuss the renormalisation of the current carried by
Gutzwiller-Bogoliubov quasiparticles and the consequences for the
suppression of the superfluid density. We also discuss
the discrepancies between different approaches in determining the underlying Fermi surface
of a Gutzwiller projected superconductor.

\subsection{Coherent and incoherent spectral weight}
\label{coherent_weight}

To model the spectral features of HTSC, we need to study
the excited states of a Gutzwiller projected superconductor.
In this subsection, we consider
the transfer of spectral weight from coherent quasiparticles
(QPs) to an incoherent background. Stimulated by STM, which
reveals a striking particle-hole asymmetry in the spectra of
underdoped HTSC \cite{Renner98,Hanaguri04,McElroy05}, this problem
has received much attention recently and
investigated using both RMFT
\cite{Fukushima05,Anderson04c,Randeria05} and VMC methods
\cite{Fukushima05,Paramekanti04,Paramekanti01,Yunoki05b,Nave06,Yang07,Chou06,Bieri06,Yunoki06}.
As nicely seen in the experiments, {\it e.g.} in figure 1(c) and
3(e) of \cite{Hanaguri04} or \fig{STM_asymmetry}, the spectral
weights for hole and particle addition show a distinct asymmetry.

\subsubsection{Sum rules for the spectral weight}

The asymmetry in the STM spectra may be explained qualitatively by
considering sum rules
\cite{Randeria05,Harris67,Meinders93,Eskes94a,Eskes94b} for the
one-particle spectral function,
\begin{deqarr}
A(\bk,\omega)&=&\sum_m \langle 0 | c_{\bk\sigma}^\dagger |m
\rangle \langle m | c_{\bk\sigma} |0 \rangle \, \delta\left(
\omega +
(E_m - E_0) \right)  \label{spectralAkwa}\\
&&  + \sum_m \langle 0 | c_{\bk\sigma} |m \rangle \langle m |
c_{\bk\sigma}^\dagger |0 \rangle \, \delta\left( \omega - (E_m
-E_0) \right) \label{spectralAkwb} \ ,
\end{deqarr}
with,
\begin{equation}
\int_{-\infty}^\infty d\omega \, A(\bk,\omega) \, = \, 1 ~ .
\end{equation}
In \eq{spectralAkwa} and \eq{spectralAkwb}, we use the $T=0$
spectral representation of $A(\bk,\omega)$, where $|m\rangle$ are
the exact many-body eigenstates with energies $E_m$. The ground
state is given by $m=0$, and $\omega$ is measured with respect to
the chemical potential. We are now interested in the low energy
spectral weight of a doped Mott insulator described by a
Gutzwiller projected ground state, {\it i.e.} $|0\rangle \sim
|\Psi \rangle \equiv P_{G}|\Psi_0\rangle$.

When removing a hole from the ground state [as in
\eq{spectralAkwa}] it is clear that
no doubly occupied sites are created.
Thus, the resulting state is situated in the so-called `lower
Hubbard band' (LHB) and involves only low energy excitations, {\it
i.e.} $0<E_m-E_0 \ll U $ (excitation energies much smaller than
the Hubbard $U$). Thus, on the hole side, the low energy spectral
weight corresponding to momentum $\bk$ and spin $\sigma$ is given
by,
\begin{equation}
\int^0_{-\infty} d\omega \, A(\bk,\omega) = \langle
0|c^\dagger_{\bk\sigma} c_{\bk\sigma}|0 \rangle = \langle
n_{\bk\sigma} \rangle_{\Psi} \ .
\end{equation}
By summing over all spin and momenta, we obtain the total low
energy spectral weight for the hole side,
\begin{equation}
{1 \over L} \sum_{\bk,\sigma} \int^0_{-\infty} d\omega \,
A(\bk,\omega) = {1 \over L} \sum_{\bk,\sigma} \langle
n_{\bk\sigma} \rangle_\Psi = n \ . \label{tot_hole}
\end{equation}
We note that similar sum rules can be derived for the dynamical
conductivity, {\it viz.} the $f$-sum rule \cite{Baeriswyl87}.

The situation is different when adding an electron to the ground
state [as in \eq{spectralAkwb}]. In such a process a part of the
resulting state is located in the `upper Hubbard band' (UHB), {\it
i.e.} a doubly occupied site may be created. Therefore, we have to
choose an upper cutoff $\Omega_L$ (located between LHB and UHB) to
extract the low energy spectral weight. By integrating
$A(\bk,\omega)$ from $0$ to $\Omega_L$, we restrict ourselves
solely to the Gutzwiller projected eigenstates out of all $|m
\rangle$, and we obtain\footnote{For a more detailed reasoning
leading to this step, we refer to \cite{Randeria05}.},
\begin{equation}
\int_0^{\Omega_L} d\omega \, A(\bk,\omega) = \langle
0|c_{\bk\sigma}P_G c^\dagger_{\bk\sigma}|0 \rangle = \langle P_G
c_{\bk\sigma}P_G c^\dagger_{\bk\sigma} P_G \rangle_{\Psi_0} \ .
\end{equation}
Summing again over all spin and momenta and making use of Fourier
transformation, we find the total low energy spectral weight for
the electron side to be,
\begin{deqarr}
{1 \over L} \sum_{\bk,\sigma} \int^{\Omega_L}_{0} d\omega \,
A(\bk,\omega) &=& {1 \over L} \sum_{\bk,\sigma} \langle P_G
c_{\bk\sigma} \, P_G \, c^\dagger_{\bk\sigma} \, P_G \, \rangle_{\Psi_0}  \\
&=& {1 \over L} \sum_{l,\sigma} \langle P_G \, c_{l \sigma}
(1-n_{l-\sigma}) \, c^\dagger_{l \sigma} \, P_G \rangle_{\Psi_0} \label{tot_elb} \\
&=& {1 \over L} \sum_{l,\sigma} \langle  (1-n_{l-\sigma}) (1-n_{l
\sigma})  \rangle_{\Psi}  \\
&=& 2 \cdot (1-n) \ , \arrlabel{tot_electron}
\end{deqarr}
where we used, $P_G \, c^\dagger_{l \sigma} \, P_G (1-n_{l-\sigma}) \, c^\dagger_{l \sigma} \, P_G$
(for a site $l$), to get \eq{tot_elb}.

From \eq{tot_hole} and \eq{tot_electron}, we find that it is more
difficult to add an electron to the LHB than to extract one in a
doped Mott insulator. This asymmetry increases as one approaches
half-filling. For a hole density, $x=1-n$, the total spectral
weight one the particle side is reduced to $2x=2(1-n)$, while the
hole side of the spectral weight is not much affected. Note that
these sum rules while explaining the particle-hole asymmetry of
the total spectral weight, tell us very little about the energy
distribution of spectral weight within the LHB.

We further note that the total spin-integrated spectral weight is $2$, and
the integrated spectral weight of the upper Hubbard band is
consequently, $2 - n-2 \cdot (1-n)=n$, which agrees with the
Hubbard-I approximation for the paramagnetic case
\cite{Hubbard63}.

\subsubsection{Definition of coherent quasiparticle excitations}
\label{Def_coh_Ex}

To explain the distribution of spectral weight at low energies, we
approximate the eigenstates $|m\rangle$ by the
Gutzwiller­-Bogoliubov quasiparticles, equation \eq{GB_QP},
derived from RMFT \cite{Edegger06a,Randeria05}. We formulate
particle-like Gutz­willer-Bogo­liubov QPs by,
\begin{equation}\label{qp2}
    |\Psi_{\bk \sigma}^\mathrm{N+1}\rangle
    =P_\mathrm{N+1} P_G \gamma_{\bk \sigma}^{\dagger}|\Psi_0
    \rangle\ ,
\end{equation}
as well as hole-like Gutzwiller-Bogoliubov QPs with the same
momentum and spin by,
\begin{equation}\label{qp3}
    |\Psi_{\bk \sigma}^\mathrm{N-1}\rangle
    =P_\mathrm{N-1}P_{G}\gamma_{\bk \sigma}^{\dagger}|\Psi_0
    \rangle\ .
\end{equation}
In the following, we fix the particle number $\mathrm{N}$ by the operator
$P_\mathrm{N}$ and thus the ground state is
$|\Psi^\mathrm{N}\rangle =P_\mathrm{N} P_{G}|\Psi_0 \rangle$. To
avoid confusion, we include an index $\mathrm{N}$ for the particle
number in the wave function. At the level of mean field theory, the energies
corresponding to the states \eq{qp2} and \eq{qp3} are given by the
RMFT excitations $E_\bk$, as discussed in \sect{RMFT_GapEq}.

Using \eq{qp2} and \eq{qp3} in \eq{spectralAkwa} and
\eq{spectralAkwb} yields,
\begin{equation}
A(\bk,\omega) = Z^+_\bk u^2_\bk \; \delta(\omega-E_\bk)+ Z^-_\bk
v^2_\bk \; \delta(\omega+E_\bk) + A^\text{inc}(\bk,\omega)\ ,
\label{Akw_RMFT}
\end{equation}
with the QP weights $\tilde Z^±_{\bk\sigma}$ given by,
\begin{equation}\label{qp4}
     \tilde Z^+_{\bk\sigma} \equiv Z_{\bk}^{+} u^2_\bk    = \frac{|\langle\Psi_{\bk\sigma}^{\mathrm{N+1}}|c_{\bk\sigma}^{\dagger}
    |\Psi_{0}^{\mathrm{N}}\rangle|^{2}}
    {\langle\Psi_{\bk\sigma}^{\mathrm{N+1}}
    |\Psi_{\bk\sigma}^{\mathrm{N+1}}\rangle\langle\Psi_{0}^{\mathrm{N}}|\Psi_{0}^{\mathrm{N}}\rangle}
    \ ,
\end{equation}
and
\begin{equation}\label{qp5}
    \tilde Z^-_{\bk\sigma} \equiv Z_{\bk}^{-} v^2_\bk    =  \frac{|\langle\Psi_{-\bk-\sigma}^{\mathrm{N-1}}|c_{\bk\sigma}|\Psi_{0}^{\mathrm{N}}\rangle|^{2}}
    {\langle\Psi_{-\bk-\sigma}^{\mathrm{N-1}}|\Psi_{-\bk -\sigma}^{\mathrm{N-1}}\rangle
    \langle\Psi_{0}^{\mathrm{N}}|\Psi_{0}^{\mathrm{N}}\rangle} \ .
\end{equation}
Here, we distinguish between the QP weight $\tilde
Z^±_{\bk\sigma}$ mostly used in VMC calculations and the QP weight
renormalisation $Z^±_{\bk\sigma}$ often given within RMFT studies.
In \eq{Akw_RMFT}, the Gutzwiller-Bogoliubov QPs lead to
$\delta$-peaked excitations, that are associated with the
coherent peaks, {\it e.g.} seen in ARPES. For projected wave
functions, the weight of these coherent excitations is
renormalised (due to Gutzwiller projection) by a factor $Z_\bk^±$.
Thus, by construction, the full spectral weight is not exhausted by the
Gutzwiller-Bogoliubov excitations, \eq{qp2} and \eq{qp3},
demanding the presence of an incoherent background
$A^\text{inc}(\bk,\omega)$.

It is not settled yet if the asymmetry in the HTSC comes
from the incoherent part as dictated by the spectral sum rules, or if
such an asymmetry is present in the coherent QP spectrum
\cite{Fukushima05,Anderson04c,Randeria05,Yang07,Chou06}. As we
show below, recent works based on the GA support the former proposal
\cite{Edegger06a,Randeria05}, {\it i.e.} a particle-hole symmetric
quasiparticle weight renormalisation,
\begin{equation}
Z_\bk = Z_{\bk}^{+} = Z_{\bk}^{-} \ . \label{symmetry}
\end{equation}
However, recent VMC calculations \cite{Yang07,Bieri06} (discussed in
\sect{VMC_QPW}) claim that this symmetry is exactly fulfilled
only for $\bk$ at the (underlying) Fermi surface. Therefore, zero
(or very low) energy excitations would still exhibit particle-hole
symmetry, whereas coherent excitations at higher energies could
lead to an asymmetry in spectral weight.
This asymmetry in VMC results is most pronounced in the
underdoped region and disappears in the limit of zero pairing
(projected Fermi sea) \cite{Chou06,Bieri06}.

\subsubsection{Incoherent background of the spectral weight}

Next, we discuss the incoherent background of the hole
spectrum. By using the spectral representation, \eq{spectralAkwa} and
\eq{spectralAkwb}, together with $A(\bk,\omega)$ from
\eq{Akw_RMFT}, we find the relation,
\begin{equation}
\langle n_{\bk\sigma} \rangle_\Psi = Z^-_\bk u^2_\bk \ + \
n^\text{inc}_{\bk\sigma} \ ,
\end{equation}
with
\begin{equation}
n^\text{inc}_{\bk\sigma}=\int_{-\infty}^0 d\omega \,
A^\text{inc}(\bk,\omega) \ .
\end{equation}
Thus, the momentum distribution, $\langle n_{\bk\sigma}\rangle$,
provides the total spectral weight with momentum $\bk$ and spin
$\sigma$ at the hole side, {\it i.e.} the coherent weight $Z^-_\bk
u^2_\bk$ overlaid by the incoherent background
$n^\text{inc}_{\bk\sigma}$. We will calculate these quantities in
\sect{EX_RMFTtU_QP} by the GA and show their behaviour
in the first Brillouin zone.

\subsubsection{Divergent $\bk$-dependent self-energy}
\label{QP_halffilled}

In \sect{VMC_Para}, we discussed VMC results for the
QP weight renormalisation at the nodal point $\bk_F$. These
calculations show that $Z \to 0$ for $x \to 0$, where
$Z=Z^+_{\bk_F}=Z^-_{\bk_F}$. Before extending our considerations
to all $\bk$-points, let us discuss some consequences for the self-energy
in the half-filled limit. Due to the vanishing gap along the nodal
direction, $(0,0)\to (\pi,\pi)$, we can approximate the Green's
function in the vicinity of $\bk_F$ by
$G^{-1}(\bk,\omega)=\omega-\epsilon(\bk)-\mu-\Sigma(\bk,\omega)$,
where $\Sigma \equiv \Sigma^\prime+i \Sigma^{\prime\prime}$.
Standard arguments then lead to the results \cite{Randeria04},
\begin{equation}
Z=\left( 1 - \frac{\partial \Sigma^\prime}{\partial\omega}
\right)^{-1}~, \quad v_{_F}= Z \left( v_{_F}^0
+\frac{\partial\Sigma^\prime} {\partial \bk}\right) \ ,
\end{equation}
where the right hand side is evaluated at the node
$(\bk_{F},\omega=0)$. Since $Z \to 0$ for $x \to 0$,
$|\partial\Sigma^\prime/\partial\omega|$ diverges like $1/x$ in
this limit. Due to the finite Fermi velocity $v_F$ (see
\sect{RMFT_extens} and \sect{VMC_Para}), a compensating divergence
in the $\bk$-dependence of the self-energy with
\begin{equation}
\frac{\partial\Sigma^\prime}{\partial \bk} \sim \frac 1 x ~.
\label{divSelf}
\end{equation}
automatically shows up. This limiting behaviour of $v_F$ and $Z$ is
also experimentally observed and transcends conventional Landau-Fermi
liquid behaviour, where the $\bk$-dependence of the self-energy is
usually small.

Equation \eq{divSelf} constitutes a key experimental result for
the HTSC, since ARPES shows unambiguously that $v_F \to {\rm
const}$ and $Z \to 0$ for $x \to 0$
\cite{Damascelli03,Campuzano04,Feng00,Zhou03}. The fact that
\eq{divSelf} arises naturally within the Gutzwiller-RVB framework
provides a strong argument for the basic premise of the theory. It is a
consequence of the vanishing of the number of free charge carriers
$\sim 1 - n$ due to the projection close to half-filling. The
number of charge carriers is, in contrast, $\sim n$ and not
singular within normal Fermi liquid theory. These considerations
lead to further consequences for higher-energy features of the
one-particle self-energy, which have been explored by Randeria,
\etal \ \cite{Randeria04}.

\subsection{Calculation of the quasiparticle weight within RMFT}
\label{detQP}

To evaluate the QP weight in \eq{qp4} and \eq{qp5} within RMFT,
one can follow \cite{Fukushima05} and use the GA for partially
projected states as presented in \sect{GA_partial}. Here, we
briefly discuss the mean steps required for this calculation, and
refer to Fukushima, \etal \ \cite{Fukushima05} for more details.
For simplicity, one may work with a particle excitation,
\begin{equation}\label{qp10}
    |\Psi_{\bk \sigma}^\mathrm{N+1}\rangle
    =P_\mathrm{N+1} P_G c_{\bk \sigma}^{\dagger}|\Psi_0
    \rangle\ ,
\end{equation}
and a hole excitation,
\begin{equation}\label{qp11}
    |\Psi_{\bk \sigma}^\mathrm{N-1}\rangle
    =P_\mathrm{N-1}P_{G}c_{-\bk -\sigma}|\Psi_0
    \rangle\ .
\end{equation}
Note that this redefinition does not effect the final results
since all calculations include norms and, $\gamma_{\bk
\sigma}^{\dagger}|\Psi_0 \rangle \sim c_{\bk
\sigma}^{\dagger}|\Psi_0 \rangle \sim c_{-\bk -\sigma}
|\Psi_0\rangle$, for a BCS wave function $|\Psi_0 \rangle$.

The first step in the calculation of the QP weight is the determination
of the  norms $$N_{\bk\sigma}^\mathrm{N
± 1}=\langle \Psi_{\bk\sigma}^\mathrm{N ± 1}
|\Psi_{\bk\sigma}^\mathrm{N ± 1}\rangle \ $$
of the excitations $|\Psi_{\bk\sigma}^\mathrm{N ±
1}\rangle $. Invoking GA for partially projected states
(see \sect{GA_partial}), one finds \cite{Fukushima05},
\begin{equation}
{ N_{\bk\sigma}^\mathrm{N + 1} \over N^\mathrm{N}_G}  \approx g_{t}\left( 1-
n_{\bk\sigma}^0 \right)~, \qquad
{N_{-\bk-\sigma}^\mathrm{N-1} \over N^\mathrm{N}_G}
 \ =\  { n^0_{\bk\sigma}\over g_{t}}~, \label{norm_psi}
\end{equation}
where $g_{t}=(1-n)/(1-n_\sigma)$, $N^\mathrm{N}_G=\langle
\Psi^\mathrm{N}|\Psi^\mathrm{N}\rangle$, and $n_{\bk\sigma}^0 \langle  c_{\bk\sigma}^\dagger
c_{\bk\sigma}^{\phantom{\dagger}}\rangle _{\Psi_0}$
is the
momentum distribution function in the unprojected wave function.
It should be noted that the above result was derived for the
non-magnetic case $n_\sigma=n_\uparrow=n_\downarrow=n/2$.

After calculating the normalisation, one can use the same techniques
to get \cite{Fukushima05},
\begin{eqnarray}
{\langle \Psi_0| c_{\bk\sigma}^{\phantom{\dagger}} P_G
P_\mathrm{N+1} c_{\bk\sigma}^\dagger P_\mathrm{N} P_G
|\Psi_0\rangle \over N^\mathrm{N}_G}   &\approx& g_{t}(1-n_{\bk\sigma}^0)~, \label{denom1}\\
{\langle \Psi_0| c_{\bk\sigma}^\dagger P_\mathrm{N-1} P_G
c_{\bk\sigma}^{\phantom{\dagger}} P_G P_\mathrm{N}|\Psi^\mathrm{N}_0\rangle
\over N^\mathrm{N}_G}
 &\approx& n_{\bk\sigma}^0~, \label{denom2}
\end{eqnarray}
for the numerators in the equations for the QP weights, \eq{qp4}
and \eq{qp5}. Using \eq{denom1}, \eq{denom2}, and the
normalisations in \eq{qp4} and \eq{qp5}, we find the the QP
weights of particle- and the hole-like excitations,
\begin{eqnarray}
Z_{\bk}^{+} u^2_\bk
&=&\frac{|\langle\Psi_{\bk\sigma}^{\mathrm{N+1}}|c_{\bk\sigma}^{\dagger}
    |\Psi_{0}^{\mathrm{N}}\rangle|^{2}}
    {\langle\Psi_{\bk\sigma}^{\mathrm{N+1}}
    |\Psi_{\bk\sigma}^{\mathrm{N+1}}\rangle\langle\Psi_{0}^{\mathrm{N}}|\Psi_{0}^{\mathrm{N}}\rangle}
    \  \approx \ g_{t}(1-n_{\bk\sigma}^0)~, \label{M++} \\
Z_{\bk}^{-} v^2_{\bk}
&=&\frac{|\langle\Psi_{-\bk-\sigma}^{\mathrm{N-1}}|c_{\bk
\sigma}|\Psi_{0}^{\mathrm{N}}\rangle|^{2}}
    {\langle\Psi_{-\bk-\sigma}^{\mathrm{N-1}}|\Psi_{-\bk-\sigma}^{\mathrm{N-1}}\rangle
    \langle\Psi_{0}^{\mathrm{N}}|\Psi_{0}^{\mathrm{N}}\rangle} \ \approx \ g_{t} \, n_{\bk\sigma}^0~,
\label{M--}
\end{eqnarray}
respectively. Since $n_{\bk\sigma}^0=v^2_\bk=1-u^2_\bk$, it follows
$Z_{\bk}^{+} \approx Z_{\bk}^{-} \approx g_t$, which
vanishes at half filling $n\to 1$.

The above results show that within the Gutzwiller
approximation, the coherent QP weight does not cause a
particle-hole asymmetry, {\it i.e.} $Z_{\bk}^{+} \approx
Z_{\bk}^{-}$. It seems
therefore that the asymmetric DOS observed in STM can only be
explained by the incoherent spectrum of Gutzwiller projected
superconductors. A symmetric spectral weight for coherent QP
excitations is also obtained in calculations for the Hubbard model
(include transformation $e^{-iS}$) \cite{Edegger06a,Randeria05}.
However, the RMFT results for $ \tilde Z^+_\bk = Z^+_\bk u_\bk^2$
and $\tilde Z^-_\bk = Z^-_\bk v_\bk^2$ do not exactly match recent
VMC calculations \cite{Yang07,Chou06,Bieri06,Yunoki06}, which  directly evaluate $ \tilde
Z^+_\bk$ and $ \tilde Z^-_\bk$ (see \sect{VMC_QPW}). Nevertheless,
the general doping dependence of above QP weight qualitatively
agrees with VMC results and with the coherent weight seen in ARPES
measurements \cite{Edegger06a}.

\subsection[Quasiparticle weight for the Hubbard model]
{Quasiparticle weight for the Hubbard model
in the strong coupling limit}
\label{EX_RMFTtU_QP}
\begin{figure}
  \centering
 \includegraphics*[width=0.55\textwidth]{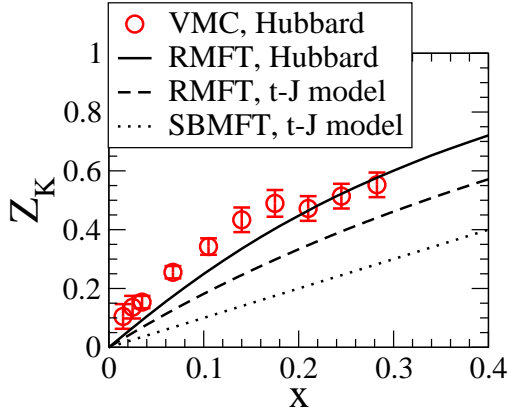}
 \caption{
Renormalisation $Z_\bk$ of the Gutzwiller-Bogoliubov nodal
quasiparticle as a function of doping $x$. The model parameters
are $t=-t'/4$ and $U=12t$. RMFT results for the Hubbard and the
$t$-$J$ model are compared with VMC data for the Hubbard model
(from \cite{Paramekanti01}) and with the SBMFT result in the
$t$-$J$ model. From \cite{Edegger06a}.
 } \label{ZKtU}
\end{figure}

In the previous subsection, we illustrated how one can determine
the QP within the GA. Here, we follow \cite{Edegger06a,Randeria05}
and extend this calculation to the Hubbard Hamiltonian, in
analogy to the extensions of the RMFT discussed in
\sect{RMFT_extens}. By using a re-transformed ground state,
$|\Psi\rangle \,\equiv\,e^{-iS} P_G P_\mathrm{N}|\Psi_0\rangle$ as
well as re-transformed excited states,
\begin{equation}
|\Psi_{\bk\sigma}^\mathrm{N ± 1}\rangle \, \equiv \,e^{-iS} P_G
P_\mathrm{N ± 1} \gamma^\dagger_{\bk\sigma}|\Psi_0\rangle \ ,
\end{equation}
we can systematically study the QP weight renormalisation within
the Hubbard model in the strong coupling limit. Evaluating the
canonical transformation $e^{-iS}$ in order $\mathcal{O}(t/U)$
gives the following particle-hole symmetric QP weight
renormalisation \cite{Edegger06a}, $Z_\bk=Z^+_\bk=Z^-_\bk$,
\begin{equation}
Z_\bk\approx g_t+ \frac{g_3} U \left(\frac{1-x^2}2
\epsilon^0_{\bk} + \frac{3-x} L \sum_{\bk'} v^2_{\bk'}
\epsilon^0_{\bk'}\right) \ , \label{RMweight}
\end{equation}
with $\epsilon^0_\bk=2 t (\cos k_x+\cos k_y) + 4\,t' \cos k_x \cos
k_y$. Equation \eq{RMweight} also includes corrections from the
next nearest neighbour hopping term $t'$. The renormalisation
$Z_\bk$ of the nodal QP weight is plotted as a solid line in
\fig{ZKtU}, and agrees well with VMC results for the Hubbard model
\cite{Paramekanti01}. The dashed line corresponds to the RMFT
result for the $t$-$J$ model, $Z_\bk = g_t$, which is compared to
the dotted line, $Z_\bk = x$, from slave boson mean field theory
(SBMFT).

\begin{figure}
\centering
\includegraphics[width=0.75\textwidth]{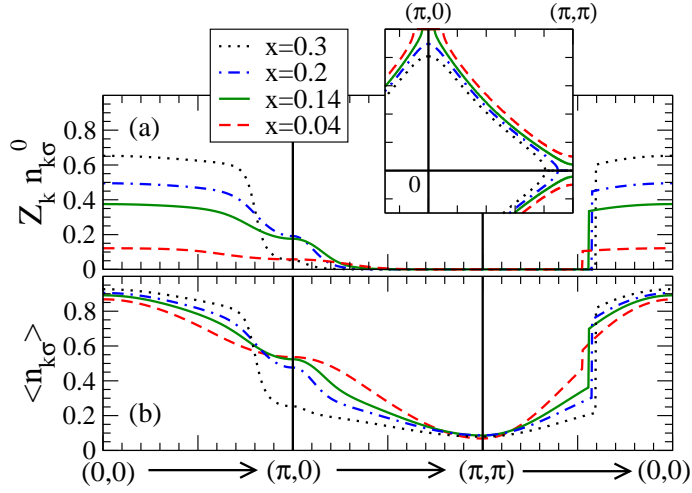}
\caption{(a) Quasiparticle weight $Z_\bk n_{\bk\sigma}^0$ and (b)
momentum distribution $\langle n_{\bk\sigma} \rangle$ of the
Gutzwiller-Bogoliubov quasiparticle for different doping $x$; The
corresponding Fermi surface, $\xi_\bk=0$, is shown in the inset of
(a).  The model parameters are $t=-t'/4$ and $U=12t$. From
\cite{Edegger06a}. \label{ZKNKFS_RMFT} }
\end{figure}

The spectral weight of the coherent peak, measured in ARPES, is
related to the QP weight $\tilde Z^-_\bk=Z^-_\bk n^0_{\bk
\sigma}$; it is shown in \fig{ZKNKFS_RMFT}(a) along the
directions, $(0,0)\to(\pi,0)$, $(\pi,0)\to(\pi,\pi)$, and
$(\pi,\pi)\to(0,0)$ for different $x$. As seen in the figure, the
QP spectral weight is severely modified by Gutzwiller projection.
It decreases with doping, and vanishes at half filling. This
causes a shift of spectral weight to an incoherent background as
seen in the momentum distribution function, $\langle
n_{\bk\sigma}\rangle\approx Z_k\,v^2_k+n^{\rm
inc}_{\bk\sigma}+\mathcal{O}(t/U)^2$. While the first term
corresponds to the coherent QP weight, the second gives the
distribution of the incoherent part. One obtains \cite{Edegger06a},
\begin{align}
&n^{\rm
inc}_{\bk\sigma}\approx\frac{(1-x)^2}{2(1+x)}\,+\sum_{\tau} \frac
{t_{\tau}} {2U} \cos(\bk \tau) \left[\frac{(1-x)^3}{1+x} +
\left(\frac {3g_s+1} {2} \right. \right. \nonumber\\
 &\quad- \left.g_3 \frac{3+x}2 \right) |\tilde \Delta_\tau|^2
\left.+\left(\frac {3g_s-1}{2}-g_3 \frac{3-x}2 \right) \tilde \xi^2_\tau
\right]~,
\end{align}
which is a smooth function of $\bk$, where $\tilde \Delta_\tau$
and $\tilde \xi_\tau$ are the pairing and hopping amplitudes between
nearest and next nearest neighbour sites,
$\tau=(± 1,0)$, $(0,± 1)$, $(± 1,± 1)$,
as defined in \sect{RMFT}. Results are
shown in \fig{ZKNKFS_RMFT}(b) . The incoherent weight is spread
over the entire Brillouin zone, and overlies the coherent part
from the Gutzwiller-Bogoliubov quasiparticles. At half-filling,
all weight becomes incoherent. These results are in qualitative
agreement with calculations for the $t$-$J$ model (recovered by neglecting the
$t/U$-corrections in above equations).

\subsubsection{Non monotonic behaviour of the QP weight at $(\pi,0)$}

\begin{figure}
  \centering
 \includegraphics*[width=0.75\textwidth]{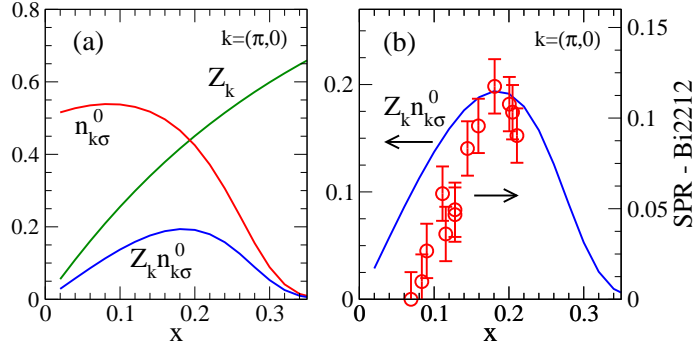}
 \caption{Doping dependence at the antinodal point, ${\bk}=(\pi,0)$:
(a) QP renormalisation $Z_\bk$ the unrenormalised QP weight,
$n^0_{\bk\sigma}=v^2_\bk$, and
 the renormalised coherent QP weight $Z_\bk v^2_\bk$; (b) coherent weight
$Z_\bk\,v^2_\bk$ compared with the experimentally determined
Superconducting Peak Ratio (SPR, ratio of coherent spectral weight
relative to the total spectral weight) for Bi2212 \cite{Feng00};
see also \fig{SPR_Science}. The model parameters are $t=-t'/4$ and
$U=12t$. From \cite{Edegger06a}. }\label{pi0}
\end{figure}

Here, we consider the coherent QP weight $Z_\bk v^2_\bk$ at the
antinodal point ${\bk}=(\pi,0)$ within the Hubbard model in the
strong coupling limit ($U=12t$). The RMFT theory predicts
a non monotonic behaviour as a function of doping, shown in \fig{ZKNKFS_RMFT}(a)
and \fig{pi0}. This effect arises from a
combination of the effects due to the Gutzwiller projection and to
the topology change [see insert of \fig{ZKNKFS_RMFT}(a)]  of the
underlying Fermi surface (FS); \fig{pi0}(a) illustrates this
clearly. While the QP weight renormalisation, $Z_\bk$, increases
with increasing doping, $n^0_\bk=v^2_\bk$, decreases due to the
topology change, which occurs at $x\approx0.15-0.20$ for our
choice of hopping parameters ($t'=-t/4$). The change of
the FS seems to be a generic feature of hole doped Cuprates
\cite{Kaminski06,Yoshida05}, although the exact doping
concentration $x$, for which this occurs, is sensitive to the
ratio between various hopping parameters. The combined effect of
strong correlations and topology change leads to a maximum of the
QP weight for the doping level, $x$, at which the underlying FS
changes topology.
Indications for such a behaviour have been seen in ARPES
\cite{Feng00,Ding01}. Feng, \etal \ \cite{Feng00} extracted the
superconducting peak ratio [SPR, illustrated in \fig{pi0}(b)]
which is proportional to the coherent QP spectral weight,
$Z_\bk\,v^2_\bk$. They found that the SPR increases with small
$x$, attains a maximum value around $x \approx 0.2$ where it
begins to decrease. Ding, \etal \ \cite{Ding01}, reported similar
results from ARPES.  In \fig{pi0}(b),
the SPR experimentally drops below the
theoretical prediction for underdoped samples. This is likely
to be the effect of inhomogeneities and
of the resulting gap variations \cite{McElroy05},
which cause a strong scattering of quasiparticles near the antinodes.

Although the topology change does not
influence the stability of the superconducting state within RMFT,
the superconducting pairing parameter $\Phi$ (related to $T_c$)
and the QP weight $Z_\bk v^2_\bk$ show some similarity as a
function of doping. However, we emphasise that this similarity
does not result from any direct relation between these two
quantities.

\subsection{Quasiparticle current renormalisation}
\label{QPcurrent}

\begin{figure}
  \centering
 \includegraphics*[width=0.76\textwidth]{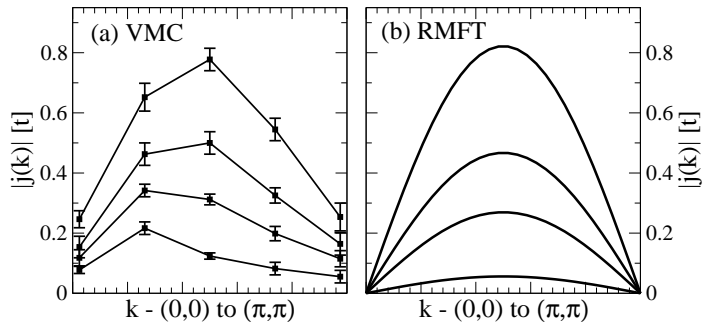}
 \caption{The magnitude of the current $|\mathbf{j}(\mathbf{k})|$ along
 the nodal direction, $(0,0)$ to $(\pi,\pi)$. (a) VMC calculations on a $10 × 10$-lattice
 (data taken from~\cite{Nave06}) and (b) RMFT results are compared for the
 doping levels $x=0.01, 0.05, 0.09, 0.17$ (increasing magnitude)\label{nodal_VMCvsRMFT}.
 From  \cite{Edegger06c}. }
\end{figure}

An important issue in the theory of the high temperature
superconductors are the properties of the nodal quasiparticle (NQP)
excitations, in particular the renormalisation of
the respective quasiparticle current \cite{Paramekanti02} and their role
in suppressing the superfluid density $\rho_s$.
As pointed out by several authors \cite{Lee97,Wen98,Ioffe02}, the
proliferation of NQPs at finite temperatures decreases
$\rho_s(T)$ \cite{Lee97},
\begin{equation}
\frac{\rho_s(T)} m \ =\ \frac{\rho^{(0)}_s} m \ -\ \frac{2 \ln
2}\pi \alpha^2 \left(\frac{v_F}{v_2}\right)\, T \ , \label{eq_sf}
\end{equation}
where $v_F$ and $v_2$ are the NQP velocities in the longitudinal
and transverse directions respectively, and $\rho^{(0)}_s$, the
zero temperature superfluid density. The renormalisation factor
$\alpha$ (also called effective charge \cite{Ioffe02}) relates the
current carried by the quasiparticle to its velocity,
\begin{equation}
\mathbf{j}(\mathbf{k}) = - e \alpha \mathbf{v(k)} \ .\nonumber
\label{cur1}
\end{equation}
Assuming that superconductivity is destroyed by thermal NQPs ,
$T_c$ is determined by simply setting \eq{eq_sf} to zero, {\it
i.e.} determining the temperature at which the superfluid density
vanishes \cite{Lee97,Wen98}. The behaviour of $T_c$ as a function
of doping is then governed by the doping dependencies of the
various quantities in \eq{eq_sf}. The latter can be calculated
within the framework of the RVB theory.
Numerical \cite{Paramekanti01} calculations show that
$\rho^{(0)}_s \to 0$ as $x \to 0$. The nodal velocity $v_F$ is
approximately constant \cite{Edegger06a}, whereas the transverse
velocity $v_2$ increases as the insulator ($x=0$) is approached.
The situation is rather unclear for the renormalisation factor
$\alpha$. While some theories argue for a constant $\alpha$
\cite{Ioffe02}, recent experimental (measurement of the superfluid
density \cite{Broun06})  as well as theoretical results
\cite{Edegger06c,Nave06} seem to support the conclusion that
$\alpha$ decreases as $x \to 0$.

\begin{figure}
  \centering
 \includegraphics*[width=0.75\textwidth]{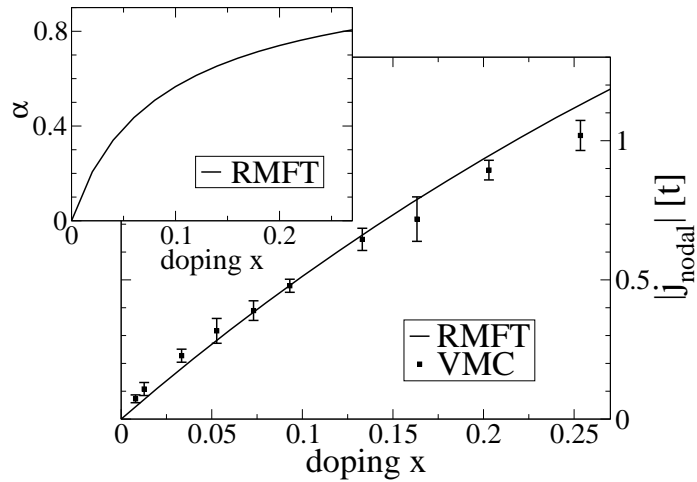}
 \caption{Magnitude of the nodal current as a function of doping.
RMFT results are compared with the VMC calculations on a $20
× 20$-lattice (data taken from~\cite{Nave06}). Insert: RMFT
result for the current renormalisation factor, $\alpha \equiv
|{\rm {\bf j}_{nodal}}|/{v_F}$, as a function of doping $x$.
 From  \cite{Edegger06c}. } \label{current_renorm}
\end{figure}

To clarify this issue, Edegger, \etal \
\cite{Edegger06c} used RMFT to calculate the current
renormalisation for the $t$-$J$ model with $J=t/3$. For the
superfluid density at zero temperature, RMFT yields a doping
dependence of,
\begin{equation}
\rho^{(0)}_s \sim g_t \equiv \frac{2x}{1+x}~,
\end{equation}
where we used \cite{Scalapino93},
\begin{equation}
\rho^{(0)}_s \sim \left \langle \sum_\sigma t_\tau
(c^\dagger_{i+\tau,\sigma} c_{i,\sigma} + c^\dagger_{i,\sigma}
c_{i+\tau,\sigma}) \right \rangle_\Psi ~ , \label{ScalopinoDs}
\end{equation}
and evaluated \eq{ScalopinoDs}, invoking the Gutzwiller approximation.
Here, we used $\tau=\hat x, \hat y$ and neglected corrections due
to the re-transformation $e^{-iS}$ of the wave function to the
Hubbard model, {\it i.e.} we set $e^{-iS}=1$. Using linear
response theory for the superfluid density \cite{Scalapino93} and
restricting ourselves to low temperatures, we recover \eq{eq_sf}
within RMFT \cite{Edegger06c}. The renormalisation factor $\alpha$
can be derived by considering the current carried by the
Gutzwiller projected Bogoliubov quasiparticle states $|\Psi_{\bk
\sigma}\rangle$,
\begin{equation}
\mathbf{j}(\mathbf{k}) \equiv i\,e\,\langle \sum_{\langle ij
\rangle,\sigma}\,t_{ij}\,\left(c^\dagger_{i,\sigma}\,c_{j,\sigma}-
c^\dagger_{j,\sigma}\,c_{i,\sigma}\right)\rangle_{\Psi_{\bk
\sigma}} \ .\label{cur2}
\end{equation}
By invoking the Gutzwiller renormalisation scheme, we find,
\begin{equation}
\mathbf{j}(\mathbf{k}) \,=\,-\,e\,g_t\frac{d}{d k} \epsilon^0
({\mathbf k}) , \label{cur3}
\end{equation}
where $\epsilon^0({\mathbf k})$ is the unrenormalised tight
binding dispersion relation; again we set $e^{-iS}=1$ for
simplicity, {\it i.e.} we neglect any $t/U$-corrections in
\eq{cur2} and \eq{cur3}. Combining \eq{cur1} and \eq{cur3} allows
us to extract $\alpha$. At the nodal point, one finds $\alpha = g_t
v^{0}_F/v_F$, where $v^{0}_F$ is the unrenormalised Fermi
velocity. The results are shown in \fig{nodal_VMCvsRMFT} and
\fig{current_renorm}, along with VMC data taken
from \cite{Nave06}. As can be see, both methods are in excellent agreement and
show that the renormalisation factor $\alpha \to 0$, as $x\to 0$.
\begin{figure}[t]
  \centering
 \includegraphics*[width=0.75\textwidth]{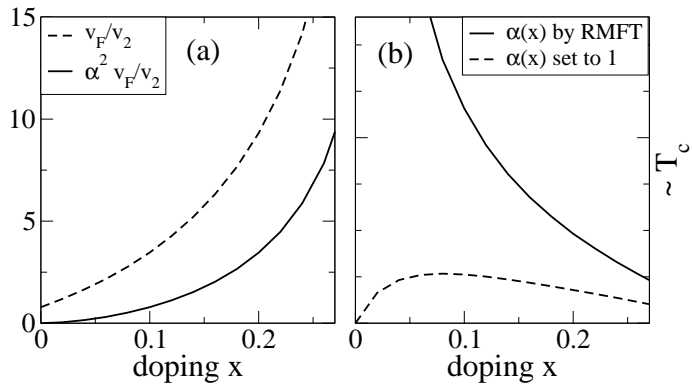}
 \caption{(a) Doping dependence of $v_F/v_2$ and  $\alpha^2
 v_F/v_2$ from RMFT with $t'=-0.2t$ and $U=12t$ in the large U Hubbard model.
 (b) Doping dependence of $T_c$ from setting \eq{eq_sf} to zero:
 (i) $\alpha(x)$ from RMFT (ii) $\alpha(x)$ is set equal to 1 by hand.
 From  \cite{Edegger06c}. \label{vFv2_Tc} }
\end{figure}
Since the $x$ dependence of the superfluid stiffness can be
obtained experimentally, it is important to study
$d\rho_s(T)/dT \propto \alpha^2 v_F/v_2$.
We show the results for this quantity in
\fig{vFv2_Tc}(a).
Note that $v_F/v_2\propto v_F/\Delta_{SC}$
already shows a significant $x$-dependence and may explain the
experimentally observed doping dependence of $d\rho_s(T)/dT$
\cite{Broun06}. However, multiplication by $\alpha^2$ leads to a
slope $d\rho_s(T)/dT$ that vanishes as $x \to 0$, {\it i.e.} as $x
\to 0$, the effective NQP charge vanishes faster than the
superfluid density does. Therefore, we get a meaningless estimate for
$T_c$ by setting \eq{eq_sf} to zero as shown in \fig{vFv2_Tc}.

This problem was noted by Lee and Wen \cite{Lee97,Wen98} in the
context of the $U(1)$ gauge theory of the $t$-$J$ model. They
argued that an $SU(2)$ formulation may resolve the problem,
yielding a constant $\alpha$. However, a constant $\alpha$ does
not completely agree with the experimentally observed
$x$-dependence of the superconducting dome either [maximal $T_c$
at $x\approx0.08$, see \fig{vFv2_Tc}(b)]. There are several
possible reasons for the discrepancy. It may be that the RMFT
result for $\alpha$ is indeed correct, in which case, the issue
can be resolved by more experiments explicitly extracting $\alpha$
in the underdoped regime. This would automatically mean that $T_c$
is not determined by NQPs, {\it i.e.} \eq{eq_sf}, and one needs to
look for other possibilities such as vortex proliferation as
mechanisms that set the scale for $T_c$.

Another possibility is that the theoretical framework behind the
Gutzwiller RVB theory misses a crucial ingredient in the
derivation of \eq{eq_sf} and the calculation of the effective
current renormalisation $\alpha$. Indeed, the applicability of the
standard Kubo formula for $\rho_s$ \cite{Scalapino93} in a
projected Hilbert space may be questioned and one needs to
reexamine this calculation carefully to check whether \eq{eq_sf}
is indeed correct.

A recent more phenomenological approach argues that the overall
temperature dependence of the superfluid density at low dopings is
well described by a three-dimensional strongly anisotropic weakly
interacting Bose gas \cite{Herbut05}. However, more work is
necessary to connect such phenomenological models to the
RVB theory we outlined so far.

\subsection[Fermi surface features in HTSC]{Determining the underlying Fermi
surface of strongly correlated superconductors}
\label{FSHTSC}

The underlying Fermi surface (FS) in the
HTSC was studied recently by us \cite{Gros06}
and Sensarma, \etal \ \cite{Sensarma06}.
These results clarify the notion of a FS in a superconducting
state and what does it mean when we say that
ARPES measures the FS of a superconductor.

In the case of the HTSC, due to the large superconducting gap
(pseudogap) below (above) the superconducting transition
temperature, an FS can be defined only along the nodal directions
(the so-called Fermi arcs
\cite{Damascelli03,Campuzano04,Loeser96,Marshall96,Norman98}). The
full `underlying FS' emerges only when the pairing interactions
are turned off, either by a Gedanken experiment, or by raising the
temperature. Its experimental determination presents a great
challenge since ARPES is more accurate at lower temperatures.
Therefore, it is of importance to know what is exactly measured by
ARPES in a superconducting or in a pseudogap state of the HTSC.

\subsubsection{Fermi vs.\ Luttinger surface}

We follow \cite{Gros06} and
begin by highlighting the differences between a Fermi and a
Luttinger surface. The FS is determined by the poles of the one
electron Green's function $G(\bk,\omega)$, {\it viz.} by ${\rm
Re}\, G(\bk,\omega=0)\equiv± \infty$ \cite{Pines_Nozieres}. The
Luttinger surface is defined as the locus of points in reciprocal
space, where the real part of the one particle Green's function
changes sign \cite{Dzyaloshinskii03}. In the Fermi liquid state of
normal metals, the Luttinger surface coincides with the FS. In a
Mott-Hubbard insulator the Green's function changes sign due to a
characteristic $1/\omega$-divergence of the single particle self
energy \cite{Gros94,Stanescu07,YangRice06} at momenta $\bk$ of the
non-interacting Fermi surface. In the HTSC the gapped states
destroy the FS but only mask the Luttinger surface. Hence, it
seems natural to relate the Luttinger surface of the
superconducting and of the pseudogap states with the concept of an
`underlying FS', and ask if such a surface can be determined by
ARPES.

The single particle Green's function is given by,
\begin{eqnarray}
G(\bk,\omega)\equiv
\sum_n \frac{|\langle n|c_{\bk\sigma}^\dagger|0\rangle|^2}
{\omega-(E_n-E_0)+i0^+}+
\sum_n \frac{|\langle n | c_{\bk\sigma} |0 \rangle|^2}{\omega+(E_n-E_0)+i0^+} \ ,
\label{GreensF}
\end{eqnarray}
where $E_n$ are the eigenvalues corresponding the eigenstates
$|n\rangle$ of the Hamiltonian; the ground state and its energy
are given by $|0\rangle$ and $E_0$, respectively. In order to
perform explicit analytic calculations one can approximate
the coherent part of \eq{GreensF} by the RMFT results for the Hubbard model
(see \sect{RMFT_extens} and \sect{EX_RMFTtU_QP}). In analogy to
\sect{Def_coh_Ex} for the spectral function $A(\bk,\omega)$, we
can use, $Z_\bk u^2_\bk=|\langle n|c_{k\sigma}^\dagger|0\rangle|^2$,
$Z_\bk v^2_\bk=|\langle n|c_{k\sigma}|0\rangle|^2$, and,
$E_\bk=E_n-E_0$. Thus, the RMFT result for the coherent
part of the Green's function becomes
\begin{equation}
G(\bk,\omega)\approx
\frac{Z_\bk u^2_k}{\omega-E_\bk+i0^+}+\frac{Z_\bk v^2_\bk}{\omega+E_\bk+i0^+} ~ .
\label{RMFT_Green}
\end{equation}
Within RMFT the elementary excitations
in the superconducting $d$-wave
ground state are given by the dispersion relation,
\begin{equation}
E_\bk\ =\ \sqrt{\xi_\bk^2+\Delta_\bk^2} \ ,
\label{E_k}
\end{equation}
where $\xi_\bk$ and $\Delta_\bk$ are determined
by \eq{epsk} and \eq{Delta_tUK}, respectively.
Evaluating ${\rm Re}\, G(\bk,\omega=0)$ by \eq{RMFT_Green} one finds
\begin{equation}
{\rm Re}\, G(\bk,\omega=0) = \frac {Z_\bk} {E_\bk} (v^2_\bk-u^2_\bk) = -
 \frac {Z_\bk} {E^2_\bk}\, \xi_\bk \ ,  \label{realGreen}
\end{equation}
where the right hand side follows from the mean field
relation, $v^2_\bk{=1-u^2_\bk}=(1-\xi_\bk/E_\bk)/2$ [see \eq{MF_vk}].
The poles of ${\rm Re}\, G(\bk,\omega=0)$, which
determine the FS, are therefore given by
\begin{equation}
E_{\bk} \equiv 0 \label{eqFS} \ .
\end{equation}
However, for a $d$-wave superconductor, equation \eq{eqFS} is
fulfilled only at the nodal points; consequently a FS is well
defined solely at these points.  Alternatively, one can consider
the Luttinger surface, defined by sign changes in the Green's
functions at $\omega=0$. From \eq{realGreen}, sign changes are
found whenever
\begin{equation}
\xi_{\bk} \equiv 0 \ . \label{LSeq}
\end{equation}
From above equations, we conclude that the Luttinger surface is
determined by the condition $\xi_\bk\equiv0$, which is also the
definition of the normal state FS when $\Delta_\bk\equiv0$.

\subsubsection{Fermi surface determination}

\begin{figure}[t]
\center
\caption{The zero frequency spectral intensity (deduced from the
inverse of $E_\bk$, which was determined by RMFT with model
parameters $t'=-t/4$ and $U=12t$) in the first Brillouin zone for
hole dopings $x=0.05$ (left) and $x=0.25$ (right). The colour
coding blue/red corresponds to the low/high zero frequency
spectral intensity. The ridges of maximal intensity are indicated
by the (dashed) red and (dashed-doted) orange lines respectively,
the Luttinger surface by the black line. From \cite{Gros06}.}
\label{density_plotsEK}
\end{figure}
There are several ways to determine the FS in practice. However,
these methods do not coincide with the underlying FS, {\it viz.}
the Luttinger surface, in the HTSC due the large superconducting
gap (or large pseudogap) in the underdoped regime.

To demonstrate this fact, we follow \cite{Gros06,Sensarma06} and
discuss the so-called `maximal intensity method' in more detail.
In this approach the intensity of ARPES spectra at zero frequency
is used to map out the underlying FS. This quantity is determined
by $A(\bk,\omega=0)= -\frac 1 \pi {\rm Im}\, G(\bk,\omega=0)$,
which becomes
\begin{equation}
\sim \frac{\Gamma_\bk}{E_\bk^2+\Gamma_\bk^2}~, \label{spInt}
\end{equation}
if one replaces $0^+$ by a finite broadening $\Gamma_\bk$ in \eq{RMFT_Green}.
The $\Gamma_\bk$ is determined both by the experimental
resolution and the width of the quasiparticle peak. When the
momentum dependence of $\Gamma_\bk$ is small compared to that of
$E_\bk$ (as is usually the case), the maximal intensity is given by
the set of momenta $\hbar \bk$ for which $E_\bk$ is minimal.

This method in determining the underlying FS was examined in
\cite{Gros06} by calculating \eq{spInt} within RMFT for a strongly
correlated $d$-wave superconducting state. All calculations in
\cite{Gros06} were done with model parameters for HTSC using RMFT
\cite{Edegger06a,Zhang88b}, for which the quasiparticle dispersion
$E_\bk$ retains the form of \eq{E_k}. Figure \ref{density_plotsEK}
shows results for the spectral intensity at zero frequency as well
as the locus of the Luttinger surface, where the former is deduced
from the inverse of $E_\bk$.

For large hole doping, $x=0.25$, the superconducting gap is small
and the Luttinger surface is close to the points in momentum space
for which the zero frequency intensity is maximal. But for smaller
doping, $x=0.05$, the gap is substantial and the Luttinger surface
deviates qualitatively from the maximal intensity surface due to
the momentum dependence of $\Delta_\bk$ (see ridges in
\fig{density_plotsEK}). It follows that when the gap or the
pseudogap is large, the criterion of maximal spectral intensity
alone does not suffice to identify the correct FS and it is
necessary to supplement the analysis of the zero frequency ARPES
intensity, \eq{spInt}, with a dispersion relation such as
\eq{E_k}. These considerations explain why the (outer) maximal
intensity ridges seen in ARPES (at low temperatures in the
underdoped regime) may yield an underlying FS whose volume is too
large. In particular, this effect is seen in
Ca$_{2-x}$Na$_x$CuO$_2$Cl$_2$ \cite{Shen05}, which also exhibits
quite a large pseudogap \cite{Kohsaka04}.

As discussed by Gros \etal \ \cite{Gros06}, even larger deviations
from the underlying FS are present in the `maximal gradient
method'. This method is based on the fact that the FS is given by
the set of $\bk$-values for which the momentum distribution
function $n_\bk$ shows a jump discontinuity. When this
discontinuity is smeared out, say, by thermal broadening or a
small gap, the gradient of $n_\bk$, $|\nabla_\bk n_\bk|$, is
assumed to be maximal at the locus of the underlying FS. However,
this method is very sensitive to the presence of even small gaps
\cite{Gros06}  and cannot be used to determine the underlying FS
unambiguously from numerical \cite{Putikka98,Maier02} or ARPES
data \cite{Campuzano96,Ronning98}. Furthermore, we note that even
the Luttinger surface in the HTSC can slightly violate the
Luttinger count. This surprising result is discussed in
\cite{Gros06} and \cite{Sensarma06} in more detail.

\begin{figure}[t]
\centering
\includegraphics*[width=0.95\textwidth]{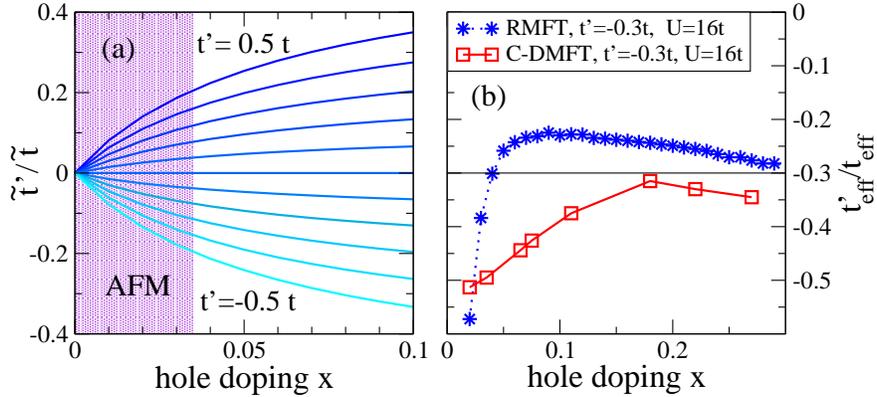}
\caption{(a) Renormalisation of the next nearest neighbour
 hopping amplitude, $t'\to\tilde t'$, as a function
 of hole doping concentration $x$ for various values of bare $t'$.
 All effective $\tilde t'$ are renormalised to zero at half
 filling by the large Coulomb repulsion. We highlight the region
 for which we expect the superconducting $d$-wave state to become
 unstable against antiferromagnetism (AFM) due to the nearly
 perfect nesting of the Luttinger surface. RMFT calculations with
 $U=12t$. From \cite{Gros06}.
 (b) Determination of $t'_{\rm eff}$ by a fit to the maximal intensity surface,
 see \fig{density_plotsEK}. We compare RMFT calculations with
 Cluster-DMFT (C-DMFT) calculations from Civelli, \etal \
 \cite{Civelli05} for $t'=-0.3 t$
 (indicated by the black line) and $U=16t$.} \label{nesting}
\end{figure}

Bieri and Ivanov \cite{Bieri06} recently proposed
an alternative definition of the
underlying Fermi surface $\bk_F$ by the
condition $\tilde Z_{\bk_F}^-=\tilde Z_{\bk_F}^+$,
{\it viz.} that the quasi-particle and the quasi-hole weight coincide at the
FS, as they do for a Fermi liquid state (see \sect{coherent_weight}
for the definition of $\tilde Z_\bk^±$).
This definition also agrees with the Luttinger surface,
$\xi_\bk \equiv 0$, within RMFT. However, when considering a
Gutzwiller-projected superconducting state within the VMC technique
deviations from the Luttinger surface are observed \cite{Bieri06}.
This deviations stem from the asymmetry
between $\tilde Z_\bk^+$ and $\tilde Z_\bk^-$, which shows up
in the VMC calculations only.

\subsubsection[Renormalisation of the Fermi surface]
{Renormalisation of the Fermi surface towards perfect nesting}
\label{FSrenorm}

The presence of strong electron-electron
interactions also changes the geometry of the Luttinger surface close to
half filling. The Cu-O planes of the HTSC are characterised by a
nearest neighbour (NN) hopping parameter $t\approx 300~\mbox{meV}$
and a next nearest neighbour (NNN) hopping parameter $t'\approx
-t/4$. These parameters are the bare parameters, and determine the
dispersion relation,
\begin{equation}
\epsilon_\bk\ =\ -2t (\, \cos k_x +\cos k_y \, )
-2t'\Big(\,\cos(k_x+k_y)+\cos(k_x-k_y)\,\Big)\ , \label{epsilon_k}
\end{equation}
in the absence of any electron-electron interaction. On the other
hand, true hopping processes are influenced by the Coulomb
interaction (here $U\ = 12\,t$) leading to a renormalisation of
the effective hopping matrix elements,
\begin{equation}
 t\ \ \to\ \  \tilde{t}=\tilde{t}(U),\qquad
t'\ \ \to\ \  \tilde{t'}=\tilde{t}'(U).
\end{equation}
One can extract $\tilde{t}$ and $\tilde{t'}$ from the RMFT
dispersion $\xi_\bk$ in \eq{epsk}, see \sect{RMFT_extens}, and
finds close to half-filling $\tilde{t}\propto J=4t^2/U$ and
$\tilde{t}'\to 0$, {\it i.e.} the NNN hopping is renormalised to
zero. This behaviour is illustrated in \fig{nesting}(a). The
resulting Luttinger surface renormalises to perfect nesting. A
similar behaviour has been observed in recent variational studies
of organic charge transfer-salt superconductors \cite{Liu05}.

At half filling the Hubbard model reduces to a spin-model with NN
$J=4t^2/U$ and a frustrating NNN $J'=4(t')^2/U$. The ground state
wave function obeys the so-called Marshall sign
rule\footnote{Marshall \cite{Marshall55} showed that the ground
state of the spin-$\frac 1 2$ Heisenberg Hamiltonian on any
bipartite lattice will be a singlet. Furthermore, the ground state
wave function picks up a sign whenever two antiparallel spins from
different sublattices are interchanged. This is the Marshall sign
rule.}
in the absence of frustration, $J'=0$, {\it viz.} when the
underlying Fermi surface is perfectly nested by the reciprocal
magnetic ordering vector $Q=(\pi,\pi)$ (in units of the inverse
lattice constant). Hence, any deviation from the Marshall sign
rule as a function of the frustrating $J'$ can be used to
determine the degree of effective frustration present in the
ground state. We emphasise this is a qualitative statement of the
ground state wave function. A numerical study has found, that the
Marshall sign rule remains valid even for small but finite $J'$,
{\it viz.} the effective frustration renormalises to zero
\cite{Richter94}. Such a behaviour is in agreement with the
results presented in \fig{nesting}(a).

However, we note that the renormalisation to perfect nesting was
not seen in Cluster-DMFT (C-DMFT) studies, {\it e.g.} by Civelli,
\etal \ \cite{Civelli05}. We believe that these discrepancies stem
from the way of fitting the effective NNN hopping $t'_{\rm eff}$.
Within the C-DMFT study of Civelli, \etal \ the FS is determined
very similarly than within the maximal intensity method.
Therefore, deviations from the effective $\tilde t'$ of the
Luttinger surface [shown in \fig{nesting}(a)] are unsurprising. In
\fig{nesting}(b), we compare the effective NNN hopping $t'_{\rm
eff}$ determined by a fit to the maximal intensity surface (see
\fig{density_plotsEK}, outer ridges) with the C-DMFT results
\cite{Civelli05}. Both methods show a qualitatively very similar
doping dependence. When approaching half-filling $|t'_{\rm eff}|$
first decreases, but then starts to grow rapidly in the underdoped
regime. We associate this effect with the increasing influence of
the $d$-wave gap in the maximal intensity surface at small doping.
Above considerations show that the determination of the underlying
FS in the HTSC is a tricky task, where special care is required
when comparing data from different approaches.

\section[Quasiparticle states in the VMC scheme]
{Quasiparticle states within the variational Monte Carlo scheme}
\label{EX_VMC}

VMC calculations for the QP weight in the $t$-$J$ model
only agree qualitatively with
the approximative RMFT results. Minor deviations from the RMFT
studies may explain a contribution of the coherent excitations to
the distinct particle-hole asymmetry seen in the STM spectra.
Apart from the QP weight, we also discuss excitation energies
determined by VMC calculations, which match well with previous
RMFT results.

\subsection{Direct calculation of the quasiparticle weight}
\label{VMC_QPW}

RMFT together with GA is an useful tool to analyse QP features in
strongly correlated superconducting states. However, the RMFT and
GA are approximate methods and it is desirable to check their
predictions numerically by VMC calculations. This consideration
motivated several authors
\cite{Yunoki05b,Nave06,Yang07,Chou06,Bieri06,Yunoki06} to
calculate the QP weight, \eq{qp4} and \eq{qp5} directly by
evaluating appropriate expectation values within the projected
wave function $|\Psi\rangle$. These VMC studies confirm the RMFT
prediction, that the QP weight decreases towards half-filling,
where it finally vanishes. However, as we will show below, the VMC
results reveal some limitations of the RMFT concerning the
determination of the detailed doping- and $\bk$-dependence of the
QP weight. We note that most of the VMC calculations presented
below do not include a re-transformed trial wave function and
describe observables in the $t$-$J$ model. These calculations can
be directly compared to the RMFT results from \sect{detQP}.

\begin{figure}
\centering
\caption{VMC result for Gutzwiller projected $d$-wave BCS state on
a $18×18$ lattice with 42 holes $(x\simeq0.13)$ and
$\Delta/t=0.1$. (a) Momentum distribution function $\langle
n_{\bk} \rangle$. (b) Total QP weight $\tilde Z^{\rm tot}_\bk$.
(c) QP weight in the $(0,0)-(\pi,\pi)$ direction. (d) QP weight in
the $(0,0)-(0,\pi)$ direction (total $\tilde Z^{\rm tot}_\bk$, add
$\tilde Z^{+}_\bk$, remove $\tilde Z^{-}_\bk$). Results correspond to
the $t$-$J$ model since the re-transformation of the wave
function was neglected. From \cite{Yang07}.} \label{Yang_VMCZk}
\end{figure}

To calculate the QP weight within the VMC scheme, most authors use
two helpful exact relations for Gutzwiller projected
wave functions. First, one finds for the QP weight
$\tilde Z_{\bk\sigma}$ of
electron-like excitations, that
\cite{Yunoki05b,Nave06,Yang07,Chou06,Bieri06},
\begin{equation}
\tilde Z_{\bk\sigma}^{+}=\frac{1+x}{2}-\langle n_{\bk\sigma}
\rangle_{\Psi^\mathrm{N}}, \label{e:relation}
\end{equation}
can be derived without any approximation and assumption. Thus
$\tilde Z_{\bk\sigma}^{+}$ can be calculated from the
momentum distribution of the ground state
$|\Psi^\mathrm{N}\rangle$ \cite{Yunoki05b,Nave06}.

For the QP weight $\tilde Z_{\bk\sigma}^{-}$, there is no exact relation
corresponding to \eq{e:relation}. However, several authors showed
\cite{Chou06,Bieri06,Yunoki06} that $\tilde Z_{\bk\sigma}^{+}$ and
$\tilde Z_{\bk\sigma}^{-}$ combined satisfy the exact relation,
\begin{equation}
\tilde Z_{-\bk-\sigma}^{+} \cdot \tilde Z_{\bk\sigma}^{-} = \frac{
\mid\langle \Psi^\mathrm{N} |
c^{\dagger}_{\bk\sigma}c^{\dagger}_{-\bk-\sigma}\mid
\Psi^\mathrm{N-2}\rangle\mid^2}{\langle \Psi^\mathrm{N}\mid
\Psi^\mathrm{N}\rangle \langle \Psi^\mathrm{N-2}\mid
\Psi^\mathrm{N-2}\rangle} \equiv P_{\bk} \label{e:pairing} \ .
\end{equation}
This relation is very useful, because the matrix elements
contributing to $P_{\bk}$ only involve ground states with
different particle numbers. The quantity $P_{\bk}$ is closely
related to the off-diagonal long-range order in the pairing
correlation and can be calculated in a straightforward way by VMC
techniques (see, {\it e.g.} \cite{Chou06}). Equation
\eq{e:pairing} was also  confirmed numerically \cite{Yunoki06}.
However, we note that \eq{e:relation} and \eq{e:pairing} are only
valid for the projected wave functions $P_G|\Psi \rangle$, and
cannot be used for the re-transformed wave function $e^{-iS}
P_G|\Psi \rangle$, since the canonical transformation  $e^{-iS}$
does not commute with the electron and the projection operators.

\begin{figure}
\centering
\includegraphics[width=0.45\textwidth]{./Figures/nk_RMFT.eps}
\includegraphics[width=0.45\textwidth]{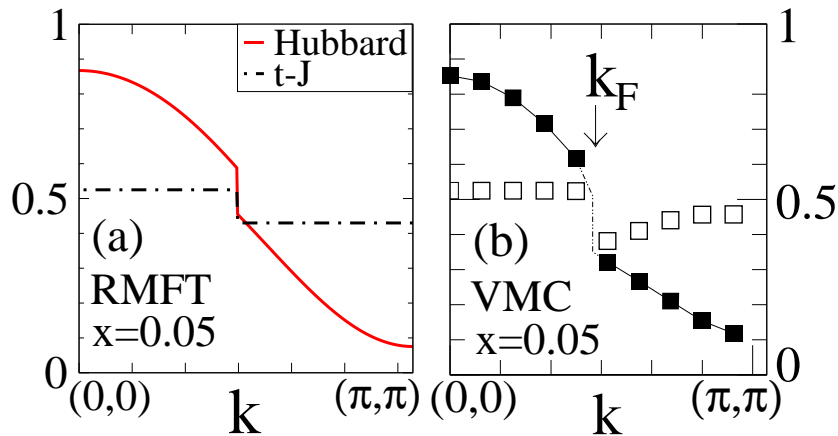}
\caption{The momentum distribution $\langle n_\bk \rangle$ along
the nodal direction $\bk=(k,k)$ for the Hubbard [black squares in
(b)] and the $t$-$J$ model [white squares in (b)] from (a) RMFT
and (b) VMC calculations, respectively. The calculations are based
on the full $t$-$J$ Hamiltonian \eq{tJH} with $t'=-t/4$ and $U=12t$ at
a doping level $x=0.05$. Expectation values for the Hubbard model
are evaluated within a re-transformed wave function, see
\eq{order_para_this}, whereas these corrections are neglected in
the $t$-$J$ model. The RMFT calculations are based on the results
from \sect{detQP} and \sect{EX_RMFTtU_QP}; the VMC data are taken
from \cite{Paramekanti04}.} \label{nk_all}
\end{figure}

\subsubsection{Momentum dependence of the quasiparticle weight}

VMC results for the QP weights $\tilde Z_\bk^+$ (adding an
electron) and $\tilde Z_\bk^-$ (removing an electron), together
with the total weight, $\tilde Z^\text{tot}_\bk=\tilde Z_\bk^+ +
\tilde Z_\bk^-$, are summarised in \fig{Yang_VMCZk}. These
calculations show that $\tilde Z^\text{tot}_\bk$ is continuous
over the whole Brillouin zone, thus supporting the idea that
$Z_\bk^+ = Z_\bk^-$ at the (underlying) Fermi surface
\cite{Yang07}. However, away from the Fermi surface,
\fig{Yang_VMCZk} also exhibits some
deviations from the simple RMFT calculations [$\tilde Z_\bk^+=g_t
u_\bk^2$ and $\tilde Z_\bk^-=g_t v_\bk^2$ with $g_t=2x/(1+x)$].
For instance, inside the Brillouin zone
and along the nodal direction,
RMFT gives a constant QP weight $\tilde Z_\bk^-$ (since
$\langle n_{\bk\sigma}\rangle=v_\bk^2$ is constant along
the nodal direction in the $t$-$J$ model, see \fig{nk_all})
whereas the VMC calculations (green triangles in \fig{Yang_VMCZk}(c), see
also \cite{Bieri06})
clearly show a non-constant behaviour.

In the absence of a superconducting gap the quasiparticle weight
at the Fermi surface is determined by the jump in the moment
distribution $\langle n_{\bk\sigma} \rangle$, as discussed in
\sect{SPgroundstateVMC}. Furthermore, $\tilde Z^+_\bk$ is
generally related via \eq{e:relation} to $\langle n_{\bk\sigma}
\rangle$ for the $t$-$J$ model.  Due to this relation
between $\langle n_{\bk\sigma}\rangle$ and $\tilde Z^+_\bk$ we
re-consider the moment dependence of $\langle n_{\bk\sigma}
\rangle$ in the VMC and the RMFT calculations. In
\fig{nk_all} we show RMFT as well as VMC results for the moment
dependence of $\langle n_{\bk\sigma} \rangle$ along the nodal
direction determined within the Hubbard and the $t$-$J$ model,
respectively. We note that expectation values for the Hubbard
model are obtained by applying a re-transformed wave function
$e^{-iS}|\Psi \rangle$, which can be evaluated in order ${\cal
O}(t/U)$; see \eq{order_para_this} in \sect{RMFT_extens}. On the
contrary, the re-transformation $e^{-iS}$ is neglected for the
calculation of observables in the $t$-$J$ model. Figure
\ref{nk_all}(a) shows that, except for the jump at the Fermi point
$\bk_F$, the RMFT gives a constant $\langle n_{\bk\sigma} \rangle$
along the nodal direction for the $t$-$J$ model. However, VMC
calculations at the same doping level ($x=0.05$) and for the same
model parameters exhibit a non-monotonic behaviour near the Fermi
point, see white squares in \fig{nk_all}(b). This effect comes
from the correlated hopping nature of the electron in the
projected Hilbert space and is not obtained within RMFT.
This result also explains the origin of the discrepancies between the RMFT
and the VMC methods in
determining the quasiparticle weight and reveals some limitations
of the RMFT in calculating momentum dependent quantities. However,
including the re-transformation of the wave function for the
Hubbard model, removes the non-monotonic behaviour of $\langle
n_{\bk\sigma} \rangle$ in the VMC data, \fig{nk_all}(b).
Thus, RMFT and VMC are in better qualitative
agreement, when $\langle n_{\bk\sigma} \rangle$ is calculated
within the Hubbard model, compare solid line in \fig{nk_all}(a)
with black squares in (b).

\subsubsection{Doping dependence of the mean quasiparticle weight}

\begin{figure}
\centering
\includegraphics[width=0.75\textwidth]{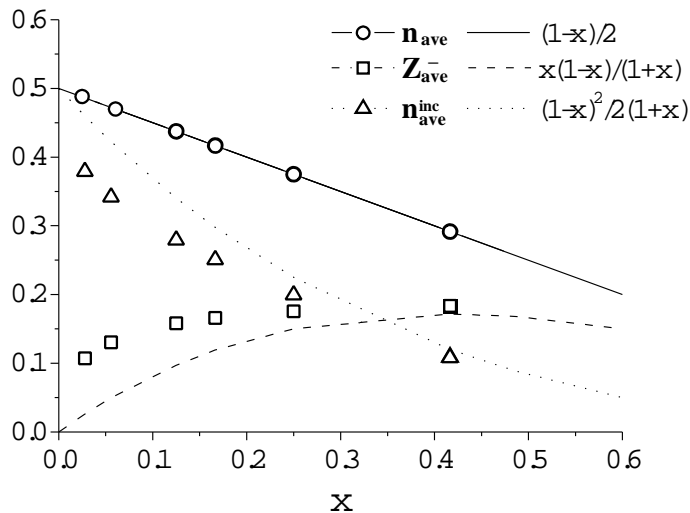}
\caption{The doping dependence of average QP weights $Z^{-}_{\rm
ave}$ for removing an electron in a $d$-wave state obtained by VMC
calculations ($12×12$ lattice, $t'=0$) and by RMFT,
respectively. The results are for the $t$-$J$ model (no re-transformation
of the wave function). The squares (triangles) are the VMC results for
$Z^{-}_{\rm ave}$ ($n_{\rm ave}^{\rm inc}=n_{\rm ave}-Z^{-}_{\rm
ave}$) with $n_{\rm ave}=1/L \sum_\bk \langle n_{\bk\sigma}
\rangle_\Psi =(1-x)/2$. The dashed and dotted lines without data
points represent results by RMFT. From \cite{Chou06}.}
\label{Chou_VMCinc}
\end{figure}

Some discrepancies between VMC and RMFT in the doping dependence
of the coherent QP weight have been discussed by Chou, \etal \
\cite{Chou06}. The authors calculate the average coherent QP
weight for removing an electron,
\begin{equation}
\tilde Z^{-}_{\rm ave}\equiv \frac 1 L \sum_{\bk} \tilde
Z_{\bk\sigma}^{-}, \label{e:incoherent}
\end{equation}
by the VMC scheme and compare it with the RMFT results. As shown
in \fig{Chou_VMCinc}, VMC calculations give a significantly larger
coherent QP weight than RMFT at the hole side, which is directly
related to a reduction of the (average) incoherent background
$n^{\rm inc}_{\rm ave}$ by the same amount.

On the other hand the average QP weight for adding an
electron,
\begin{equation}
\tilde Z_{\rm ave}^{+}\equiv \frac 1 L \sum_{\bk} \tilde
Z_{\bk\sigma}^{+}=\frac {1+x} 2 - \frac 1 L \sum_\bk \langle
n_{\bk\sigma} \rangle_\Psi = \frac {1+x} 2 - \frac {1-x} 2 = x ~,
\label{Z+}
\end{equation}
is exactly the same in the RMFT and the VMC scheme, where we used
\eq{e:relation} in \eq{Z+}. Thus, it was argued \cite{Chou06},
that the increased coherent weight at the hole side seen in the
VMC calculations, can explain the particle-hole asymmetry in the
tunnelling experiments. However, considering the large asymmetry
in the experiments and the predictions from sum rules, it is
likely that at least at higher energies a considerable part of
the asymmetry is caused by the incoherent background.

\subsection{VMC calculations for the quasiparticle energy}
\label{QPenergy}

\begin{figure}
\begin{center}
\includegraphics*[width=0.52\textwidth,angle=-90]{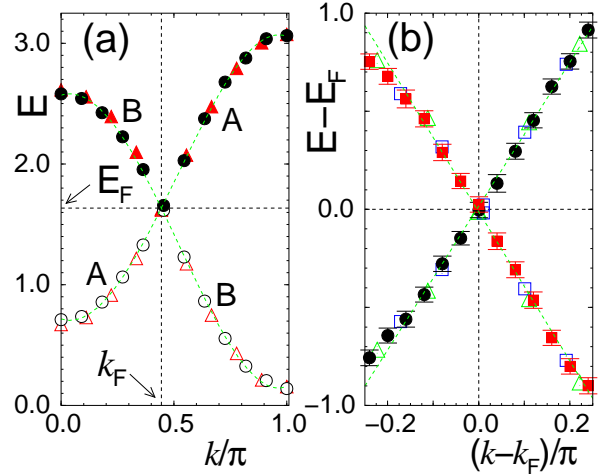}
\caption{ Dispersion $E$ in the nodal direction for the 2D $t$-$J$
model with $J/t$=$0.3$ and $t'/t$=$-0.2$ at $x$=$0.099$. (a) Full
dispersion for $L$=$162$ (triangles) and $242$ (circles). The
electron removal (addition) spectrum is denoted by open (solid)
symbols. The dashed lines are tight binding fits. (b) Same as (a)
but focusing on the excitations near $E_{\rm F}$. In addition to
the data for $L$=$162$ (open triangles) and $242$ (open squares),
results for $L=1250$ (solid squares and circles) are also plotted.
From \cite{Yunoki05a}.} \label{Yunoki_disp}
\end{center}
\end{figure}

In the previous section, we discussed how the
spectral weight of Gutzwiller-Bogoliubov QP excitations can
be determined directly using VMC and how such
results compare with RMFT.
Now we turn to the excitation energies
$E_\bk$, of the QP. Here again, RMFT
results can be checked by directly calculating the energy
corresponding to the excited state $|\Psi^\mathrm{N±
1}_{\bk,\sigma} \rangle$, equations \eq{qp2} and \eq{qp3}, within
the $t$-$J$ model. Subtracting the ground state energy, we obtain
the excitation energy,
\begin{equation}
E=\langle H_{t-J}\rangle_{\Psi^\mathrm{N±
1}_{\bk,\sigma}}-\langle H_{t-J}\rangle_{\Psi^\mathrm{N}} \,
 \label{exVMC_Yunoki}
\end{equation}

We discuss now the VMC calculations of Yunoki, \etal \
\cite{Yunoki05a}, who also included a Jastrow factors into the
wave functions to improve the ground state energy. Figure
\ref{Yunoki_disp} illustrates a typical dispersion along the nodal
direction obtained by determining $E_\bk=|E|$ for every
$\bk$-point separately. As shown in the figure, a tight-binding
dispersion fits well to the numerical data, and it is possible to
extract interesting quantities like the nodal Fermi velocity $v_F$
or the nodal Fermi point $|\bk_F|$.

\begin{figure}
\begin{center}
\includegraphics*[width=0.55\textwidth,angle=-90]{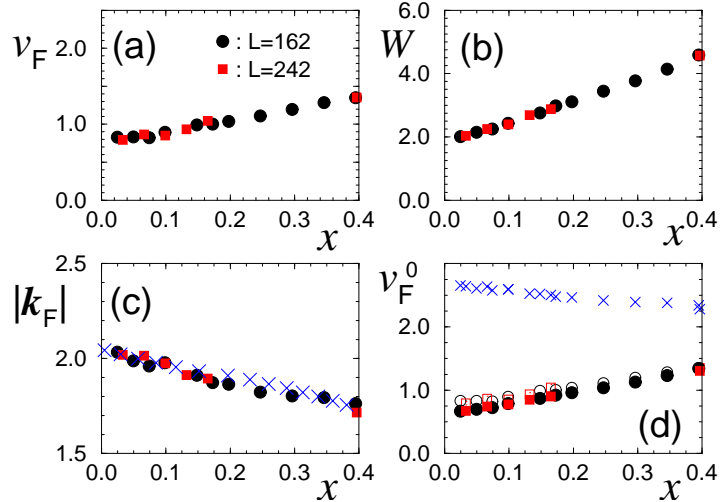}
\caption{ (a) Nodal Fermi velocity $v_{\rm F}$, (b) bandwidth $W$,
(c) nodal Fermi momentum $|{\bf k}_{\rm F}|$, and (d)
unrenormalised Fermi velocity $v_{\rm F}^0$ (crosses) compared to
$v_F$ for the 2D $t$-$J$ model with $J/t$=$0.3$ and $t'/t$=$-0.2$
at different $x$. From \cite{Yunoki05a}.} \label{Yunoki_vf}
\end{center}
\end{figure}

By repeating the calculation from \fig{Yunoki_disp} for various
electron densities, one can determine the doping dependence of
$v_F$ and $|\bk_F|$. Figure \ref{Yunoki_vf}(a) illustrates that
the Fermi velocity only slightly decreases when approaching
half-filling as already seen from RMFT (\cite{Edegger06a},
\sect{RMFT_extens}). The results of Yunoki, \etal \ also agree
with previous VMC calculations utilising the moments of the spectral
function (\cite{Paramekanti01}, \sect{VMC_Para}), as well as with
ARPES experiments
\cite{Zhou03,Borisenko06,Kordyuk05,Bogdanov00,Johnson01}. In
\fig{Yunoki_vf}(c), we see the doping dependence of the nodal
Fermi point $|\bk_F|$, which matches experimental and RMFT
predictions. The renormalised band width $W$ is given in
\fig{Yunoki_vf}(b); it is tightly related to $v_F$. Figure
\ref{Yunoki_vf}(d) illustrates a comparison between the Fermi
velocity $v_F$ and the unrenormalised velocity $v_F^0$, revealing
the strong renormalisation effects due to the Gutzwiller
projection. However, it is important to note that, in contrast to
the QP  weight $\tilde Z_\bk$, the Fermi velocity does not vanish
in the half-filled limit

While Yunoki \etal \ \cite{Yunoki05a} only considers the nodal
dispersion, Yunoki recently extended these VMC calculations to the
whole Brillouin zone \cite{Yunoki06}. His results agree quite well
with the RMFT dispersion, giving further support to the
Gutzwiller-Bogoliubov QP picture. To conclude this section,
VMC calculations for the spectral weight and the QP excitations
are in good qualitative agreement with RMFT. This is important
because it shows that the simple analytical approach
of RMFT together with the GA can be used reliably. Two key features emerge
consistently from these two approaches: a finite and constant
Fermi velocity contrasting with a vanishing QP weight in the
half-filled limit.

\section{Summary and outlook}

In this review, we attempted to summarise the basic idea of using
Gutzwiller projected wave functions in the description of
high temperature superconductivity. Projected wave functions provide a straightforward
implementation of the RVB picture wherein superexchange leads to pair
correlations and doping the Mott insulator leads to a superconducting
ground state.

Projected wave functions can be studied both analytically and
numerically. A superconducting state with $d$- wave symmetry
arises as the best variational wave function within the Gutzwiller -RVB
theory. Incorporation of antiferromagnetic order and
next-nearest neighbour hopping then allows for a quantitative
description of the Cuprate phase diagram within the $t$-$J$ and
the Hubbard model. Sophisticated variational Monte Carlo
calculations (VMC) give detailed information about the size of the
antiferromagnetic region and the stability against phase
separation. These ground state properties seen in the VMC
technique were recently confirmed by various quantum cluster
methods, lending further support to the Gutzwiller-RVB picture.

Besides the VMC technique, the effect of projection can be treated
by Gutz­willer approximation, which then allows for a formulation
of a renormalised mean field theory (RMFT). The RMFT results agree
in general with VMC calculations and provide systematic analytic
expressions for doping-dependent features.

Within the Gutzwiller-RVB picture, high temperature
superconductors are viewed as doped Mott insulators, {\it i.e.}
restriction to single occupied orbitals due to strong correlation
effects. That causes a significant decrease in the mobility of
electrons (holes) near half-filling as correctly described within
above microscopic calculations. The resulting renormalisation of
the kinetic energy explains the decrease of the superconducting
order parameter, of the superfluid density, and of the Drude
weight when approaching half-filling. RMFT and VMC calculations
also explain the large superconducting gap and the small
quasiparticle weight in the underdoped Cuprates. Further, the
modelling of charge ordered states, impurity sites, or vortex
cores qualitative agrees with experiments.

While the quasiparticle weight $Z$ vanishes in the half-filled
limit, the nodal Fermi velocity $v_F$ stays finite. RMFT and VMC
calculations explain this interesting experimental observation by
the effect of the superexchange interaction on the dispersion.
However, in the half-filled limit, such a behaviour ($Z \to 0$ and
$v_F={\rm const}$) immediately results in a divergence of the
$\omega$- as well as of the $\bk$-dependence of the self-energy.
The consequences of these divergences for any Fermi liquid
description at finite doping are not fully understood. In a recent
paper, Anderson \cite{Anderson06a} suggested that projected wave
functions contain the essential physics to explain the non-Fermi
liquid behaviour of the normal state in the cuprate
superconductors, \emph{i.e.} the region in the cuprate phase
diagram above the pseudogap temperature scale.
One reason why there
has not been much progress on this issue is that we need a scheme
to calculate the single particle Green's function directly in a
Gutzwiller projected state. The standard technique of introducing
a complete set of orthogonal excitations works as long as we only
consider the contribution of the ``projected quasiparticle
(hole)''. However, as we discussed earlier, the total spectral
weight of a photohole (say) is not exhausted by the projected
quasihole excitation. The effect can be understood most
transparently as the non commutativity of a photohole state
$c_{i\sigma} P_G |\Psi_0 \rangle$ and projected excitations of the
form $P_G c_{i\sigma} |\Psi_0 \rangle$. It follows that a
photohole is a mixture of a projected single hole excitation and a
multiparticle excitation which signifies the backflow of
say,$\downarrow$ spins accompanying a propagating $\uparrow$ spin
hole. A consistent scheme to treat this effect has not been
devised yet.

Another important open question is the role played by phase fluctuations
in Gutzwiller projected BCS wave functions.
It was noted in the early papers of Anderson and collaborators
that phase fluctuations are expected to play an important role as one
approaches the Mott insulator in the phase diagram. The recent
experiments of Ong and collaborator point to the existence of a
vortex liquid phase above $T_c$ in the underdoped superconductors.
A description of this phase within the Gutzwiller RVB
theory has not yet been formulated. It should be pointed out that
a large corpus of literature exists on fluctuating
$d$ - wave superconductors, but to our knowledge, no one
has attempted to derive an effective ``phase - only'' model from
a microscopic Hamiltonian for Mott Hubbard superconductors.
Müller-Hartmann and collaborators addressed this issue
many years ago, using a high temperature expansion \cite{drzazga_88}.
Using standard functional integral techniques,
these authors performed a Ginzburg Landau expansion for the
free energy functional of an RVB state \cite{Baskaran87b}.
They find that singlet pairing sets in at temperature
scales higher than the (mean field) transition temperature
$T_c$. As hole concentration goes to zero,
the local $U(1)$ gauge symmetry in the theory leads to
phase fluctuations that destroy off diagonal long range
order. However, their calculations show that an extended $s$ - wave
state is favoured, which result is inconsistent with
the Gutzwiller RVB theory described in this article. It would
be very interesting to revisit this problem and attempt to
derive a Ginzburg Landau expansion of the $d$ - wave
Gutzwiller RVB state. Such a step is necessary to extend
the Gutzwiller - RVB framework to the
description of phase degrees of freedom, and the effect
of the latter in destroying superconducting correlations, both as a function
of temperature and doping.

A related issue is the understanding of the pseudogap state within
the RVB theory. The view we advocated was that local singlet
pairing exists at temperature scales $T < T^*$. Much support for
this idea comes from the experimental observation that the BCS
ratio, $\Delta/(2 k_B T^*)$, is constant and in agreement with
mean field theory for all doping levels, when we use the onset
temperature of the pseudogap $T^*$ instead of $T_c$. While this is
certainly suggestive, there is no direct way of proving this
within the theory, again because we do not yet have a method to
describe finite temperature effects within the Gutzwiller
framework. Extending the Gutzwiller-RVB theory to the description
of finite temperature phases is an important step that needs to be
taken to complete our understanding of the pseudogap state. In
this context, we note that a a finite temperature scheme for
Gutzwiller projected Fermi liquids was developed by Seiler
\emph{et al.} to study $^3$He \cite{Seiler86}. Whether a similar
scheme can be developed to study projected $d$ - wave
superconductors at finite temperatures remains to be seen. A
related issue is the investigation of finite frequency excitations
in the Gutzwiller - RVB scheme.
In particular, it will be very useful to study the collective
excitations of the Gutzwiller superconductor along the lines of
Anderson's original work on equations of motion for collective
modes in a BCS superconductor \cite{pwa_58}.

\section*{Acknowledgements}
The authors thank P.W. Anderson for several discussions and
comments on this manuscript. VNM thanks P. W. Anderson, G.
Baskaran, G. Levine, N. P. Ong, T. V. Ramakrishnan, D. Schmeltzer
and Z.-Y. Weng for several discussions over the years.




\addcontentsline{toc}{section}{Bibliography}
\begin{thebibliography}{99}

\bibitem{Bednorz86} J.G. Bednorz and K.A. Müller, Z. Phys. \textbf{64}, 189 (1986).

\bibitem{Norman03} M.R. Norman and C. Pepin, Rep. Prog. Phys. \textbf{66}, 1547 (2003).

\bibitem{Tsuei00} C.C. Tsuei and J.R. Kirtley,
                  Rev. Mod. Phys. \textbf{72}, 969 (2000).

\bibitem{Scalapino07} D.J. Scalapino,  in
         `Handbook of High Temperature Superconductivity',
         J.R. Schrieffer (Ed.), Springer (2006).

\bibitem{Edegger05a} B. Edegger, N. Fukushima,
C. Gros, and V.N. Muthukumar, Phys. Rev. B \textbf{72}, 134504
(2005).

\bibitem{Fukushima05} N. Fukushima, B. Edegger, V.N. Muthukumar,
and C. Gros, Phys. Rev. B \textbf{72}, 144505 (2005).

\bibitem{Edegger06a} B. Edegger, V.N. Muthukumar, C. Gros, and P.W. Anderson,
Phys. Rev. Lett. \textbf{96}, 207002 (2006).

\bibitem{Edegger06b} B. Edegger, V.N. Muthukumar, C. Gros, Phys. Rev. B {\bf 74}, 165109
(2006).

\bibitem{Gros06}
C. Gros, B. Edegger, V.N. Muthukumar, and P.W. Anderson, PNAS
{\bf 103}, 14298 (2006).

\bibitem{Edegger06c} B. Edegger, V.N. Muthukumar, and C. Gros, Physica C, to be published.

\bibitem{Pauling38} L. Pauling, Phys. Rev. {\bf 54}, 899 (1938).

\bibitem{Pauling48} L. Pauling, Nature, {\bf 161}, 1019 (1948).

\bibitem{Anderson73} P.W. Anderson, Mat. Res. Bull \textbf{8}, 153 (1973).

\bibitem{Fazekas74} P. Fazekas and P.W. Anderson, Philos. Mag. \textbf{30},
432 (1974).

\bibitem{Anderson87} P.W. Anderson, Science \textbf{235}, 1196 (1987).

\bibitem{GrosJoynt87} C. Gros, R. Joynt, and T.M. Rice,
         Z. Phys. {\bf 68}, 425 (1987).

\bibitem{Gros88} C. Gros, Phys. Rev. B {\bf 38}, 931 (1988).

\bibitem{Kotliar88} G. Kotliar and J. Liu,
         Phys. Rev. B \textbf{38}, 5142 (1988).

\bibitem{Lee06} P.A. Lee, N. Nagaosa, and X.-G. Wen,
          Rev. Mod. Phys. {\bf 78}, 17 (2006).

\bibitem{Zhang88b} F.C. Zhang, C. Gros, T.M. Rice, and H. Shiba,
Supercond. Sci. Tech. {\bf 1}, 36 (1988).

\bibitem{Paramekanti01} A. Paramekanti, M. Randeria, and N. Trivedi,
         Phys. Rev. Lett. \textbf{87}, 217002 (2001).

\bibitem{Norman05} M.R. Norman, D.P. Pines, and C. Kallin, Adv. Phys. \textbf{54}, 715
(2005).

\bibitem{Damascelli03} A. Damascelli, Z. Hussain, and Z.-X. Shen,
Rev. Mod. Phys. {\bf 75}, 473 (2003).

\bibitem{Campuzano04} J.C. Campuzano, M.R. Norman, and M. Randeria,
in \emph{Physics of Conventional and Unconventional
Superconductors}, (Springer Verlag 2004);
cond-mat/0209476.

\bibitem{Zhang88a} F.C. Zhang and T.M. Rice, Phys. Rev. B \textbf{37}, 3759 (1988).

\bibitem{Ong04} N.P. Ong, Y. Wang, S. Ono, Y. Ando, and S. Uchida, Ann. Phys.
\textbf{13}, 9 (2004).

\bibitem{Ong03} Y. Wang, S. Ono, Y. Onose, G. Gu, Y. Ando, Y. Tokura, S.
Uchida, and N.P. Ong, Science \textbf{299}, 86 (2003).

\bibitem{Kivelson03} S.A. Kivelson, E. Fradkin, V. Oganesyan, I.P. Bindloss, J.M.
Tranquada, A. Kapitulnik, and C. Howald, Rev. Mod. Phys.
\textbf{75}, 1201 (2003).

\bibitem{Tranquada05} J.M. Tranquada, to appear as a chapter in
`Treatise of High Temperature Superconductivity' by J. Robert
Schrieffer; cond-mat/0512115.

\bibitem{Varma89} C.M. Varma , P.B. Littlewood ,
S. Schmitt-Rink, E. Abrahams, and A.E. Ruckentstein,
Phys. Rev. Lett. \textbf{63} 1996 (1989).

\bibitem{Dagotto05} E. Dagotto, Science \textbf{309}, 257 (2005).


\bibitem{Doniach82} S. Doniach and E. Sondheimer,
Green's Functions for Solid State Physicists (Benjamin, Reading, Mass., 1982).


\bibitem{Mesot99} J. Mesot {\it al.},
         Phys. Rev. Lett. \textbf{83}, 840 (1999).

\bibitem{Campuzano99} J. C. Campuzano {\it al.},
         Phys. Rev. Lett. \textbf{83}, 3709 (1999).


\bibitem{Won94} H. Won and K. Maki, Phys. Rev. B \textbf{49}, 1397 (1994).

\bibitem{Kugler01} M. Kugler, O. Fischer, Ch. Renner, S. Ono, and Y. Ando,
 Phys. Rev. Lett. \textbf{86}, 4911 (2001).

\bibitem{Feng00} D.L. Feng, D.H. Lu, K.M. Shen, C. Kim, H. Eisaki,
A. Damascelli, R. Yoshizaki, J.-i. Shimoyama, K. Kishio,
G.D. Gu, S. Oh, A. Andrus, J. O'Donnell,
J.N. Eckstein, and Z.-X. Shen, Science {\bf 289} 277 (2000).

\bibitem{Ding01} H. Ding, J.R. Engelbrecht, Z. Wang, J.C. Campuzano,
S.-C. Wang, H.-B. Yang, R. Rogan, T. Takahashi, K. Kadowaki, and
D.G. Hinks, Phys. Rev. Lett. {\bf 87}, 227001 (2001).

\bibitem{Zhou03} X.J. Zhou {\it et al.},
Nature \textbf{423}, 398 (2003).

\bibitem{Chiao00} M. Chiao, R.W. Hill, C. Lupien, L. Taillefer, P. Lambert, R. Gagnon,
  and P. Fournier, Phys. Rev. B \textbf{62}, 3554 (2000).

\bibitem{Norman98} M.R. Norman, H. Ding, M. Randeria, J. C. Campuzano,
T. Yokoya, T. Takeuchi, T. Takahashi,T. Mochiku, K. Kadowaki, P.
Guptasarma, and D.G. Hinks, Nature {\bf 392}, 157 (1998).

\bibitem{Loeser96} A.G. Loeser, Z.-X. Shen,
D.S. Dessau, D.S. Marshall, C.H. Park, P. Fournier, and A.
Kapitulnik, Science {\bf 273}, 325 (1996).

\bibitem{Marshall96} D.S. Marshall, D.S. Dessau, A.G. Loeser, C-H. Park,
A.Y. Matsuura, J.N. Eckstein, I. Bozovic, P. Fournier, A.
Kapitulnik, W.E. Spicer, and Z.-X. Shen, Phys. Rev. Lett. {\bf
76}, 4841 (1996).

\bibitem{Ding96} H. Ding, T. Yokoya, J.C. Campuzano, T. Takahashi, M. Randeria,
         M.R. Norman, T. Mochiku, K. Kadowaki, and  J. Giapintzakis,
         Nature {\bf 382}, 51 (1996).

\bibitem{Kanigel06} A. Kanigel {\it et al.},
Nature Physics {\bf 2}, 447 (2006).

\bibitem{Hudson99}  E.W. Hudson, S.H. Pan,  A.K. Gupta, K.-W. Ng, J.C. Davis,
Science \textbf{285}, 88 (1999).

\bibitem{Yazdani99} A. Yazdani, C.M. Howald, C.P. Lutz, A. Kapitulnik, and D. M. Eigler,
Phys. Rev. Lett. \textbf{83}, 176 (1999).

\bibitem{Pan03} S.H. Pan, E.W. Hudson, K.M. Lang, H. Eisaki, S. Uchida, and J.C. Davis,
Nature \textbf{403}, 746 (2003).

\bibitem{Maggio95} I. Maggio-Aprile, Ch. Renner, A. Erb, E. Walker, and Ø. Fischer,
Phys. Rev. Lett. \textbf{75}, 2754 (1995).

\bibitem{Renner98} Ch. Renner, B. Revaz, K. Kadowaki,
I. Maggio-Aprile, and Ø. Fischer, Phys. Rev. Lett. \textbf{80}, 3606 (1998).

\bibitem{Pan00} S.H. Pan, E.W. Hudson, A.K. Gupta, K.-W. Ng,
H. Eisaki, S. Uchida, and J.C. Davis, Phys. Rev. Lett. \textbf{85}, 1536 (2000).

\bibitem{Vershinin04} M. Vershinin,
 S. Misra, S. Ono, Y. Abe, Y. Ando, and A. Yazdani,
 Science \textbf{303}, 1995 (2004).

\bibitem{Hanaguri04} T. Hanaguri, C. Lupien, Y. Kohsaka, D.-H. Lee, M. Azuma,
M. Takano, H. Takagi, and J.C. Davis, Nature \textbf{430}, 1001 (2004).

\bibitem{McElroy05} K. McElroy, D.-H. Lee, J.E. Hoffman, K.M. Lang,
J. Lee, E.W. Hudson, H. Eisaki, S. Uchida, and J.C. Davis,
Phys. Rev. Lett. \textbf{94}, 197005 (2005).

\bibitem{Emery87} V.J. Emery, Phys. Rev. Lett. \textbf{58}, 3759 (1987).

\bibitem{Varma87} C.M. Varma, S. Schmitt-Rink, and E. Abrahams,
 Solid State Commun. \textbf{62}, 681 (1987).

\bibitem{Dagotto94} E. Dagotto, Rev. Mod. Phys. \textbf{66}, 763 (1994).

\bibitem{Moriya00} T. Moriya and K. Ueda, Adv. Phys. \textbf{49}, 555 (2000).

\bibitem{Yanase03} Y. Yanase, T. Jujo, T. Nomura, H. Ikeda, T. Hotta, and
K. Yamada, Phys. Rep. \textbf{387}, 1 (2003).

\bibitem{Chubukov02} A.V. Chubukov, D. Pines, and J. Schmalian,
in `The Physics of Conventional and Unconventional
Superconductors', K.H. Bennemann and J.B. Ketterson (Eds.),
Springer (2002).

\bibitem{Kivelson05} S.A. Kivelson and E. Fradkin, to appear
as a chapter in `Treatise of High Temperature Superconductivity'
by J. Robert Schrieffer; cond-mat/0507459.

\bibitem{Carlson02} E.W. Carlson, V.J. Emery, S.A. Kivelson, and D. Orgad,
in `The Physics of Conventional and Unconventional
Superconductors', K.H. Bennemann and J.B. Ketterson (Eds.),
Springer (2002).

\bibitem{Demler04}  E. Demler, W. Hanke, and S.-C. Zhang,
Rev. Mod. Phys. \textbf{76}, 909 (2004).

\bibitem{Maier05} T. Maier, M. Jarrell, T. Pruschke, and M.H. Hettler,
Rev. Mod. Phys. \textbf{77}, 1027 (2005).

\bibitem{Varma06} C.M. Varma, Phys. Rev. B \textbf{73}, 155113 (2006).

\bibitem{Chakravarty02} S. Chakravarty, R.B. Laughlin, D.K. Morr, and C. Nayak,
Phys. Rev. B \textbf{63}, 094503 (2001).

\bibitem{Randeria98} Mohit Randeria,
in `Proceedings of the International School of Physics Enrico
Fermi Course CXXXVI on High Temperature Superconductors',
G. Iadonisi, J.R.  Schrieffer, and M.L. Chiafalo (Eds.),
 IOS Press (1998), p. 53.

\bibitem{Chen05} Q. Chen, J. Stajic, S. Tan, and K. Levin, Phys. Rep. \textbf{412}, 1
(2005).

\bibitem{Endoh88} Y. Endoh {\it et al.},
Phys. Rev. B {\bf 37}, 7443 (1988).

\bibitem{Chakravarty89} S. Chakravarty, B.I. Halperin, and D.R. Nelson,
Phys. Rev. B {\bf 39}, 2344 (1989).

\bibitem{Liang88} S. Liang, B. Doucot, and P. W. Anderson, Phys. Rev. Lett.
{\bf 61}, 365 (1988).

\bibitem{Kane89} C.L. Kane, P.A. Lee, and N. Read,
Phys. Rev. B {\bf 39}, 6880 (1989).

\bibitem{Gros89b} C. Gros and M.D. Johnson
Phys. Rev. B {\bf 40}, 9423 (1989).

\bibitem{Anderson02b} P.W. Anderson,  Physica B {\bf 318}, 28 (2002).

\bibitem{lee_88} T. K. Lee and S. Feng, Phys. Rev. B \textbf{38}, 11809 (1988).

\bibitem{hsu_90} T. C. Hsu, Phys. Rev. B \textbf{41}, 11379 (1990).

\bibitem{ho_01} C.-M. Ho, V. N. Muthukumar, M. Ogata, and P.W. Anderson,
Phys. Rev. Lett. \textbf{86}, 1626 (2001).

\bibitem{shimizu_03} Y. Shimizu, K. Miyagawa, K. Kanoda, M. Maesato, and
G. Saito, Phys. Rev. Lett. \textbf{91}, 107001 (2003).

\bibitem{motrunich_05} O. Motrunich, Phys. Rev. B \textbf{72}, 045105 (2005).

\bibitem{sslee_05} Sung-Sik Lee and Patrick A. Lee, Phys. Rev. Lett. \textbf{95}, 036403 (2005).

\bibitem{Zhitomirsky96} M.E. Zhitomirsky and K. Ueda, Phys. Rev. B {\bf 54}, 9007 (1996).

\bibitem{Capriotti00} L. Capriotti and S. Sorella, Phys. Rev. Lett. {\bf 84}, 3173
(2000).

\bibitem{Takano03} K. Takano, Y. Kito, Y. Ono, and K. Sano, Phys. Rev. Lett. {\bf 91},
197202 (2003).

\bibitem{Capriotti01} L. Capriotti, F. Becca, A. Parola, and S. Sorella, Phys. Rev.
Lett. {\bf 87}, 097201 (2001).

\bibitem{Mambrini06} M. Mambrini, A. Laeuchli, D. Poilblanc, and F. Mila,
Phys. Rev. B {\bf 74}, 144422 (2006).

\bibitem{Weber06a} C. Weber, A. Laeuchli, F. Mila, T.
Giamarchi, Phys. Rev. B {\bf 73}, 014519 (2006).

\bibitem{Valla06}
T. Valla, A.V. Fedorov, J. Lee, J.C. Davis, and G.D. Gu, Science
{\bf 314}, 1914 (2006).

\bibitem{Baskaran03} G. Baskaran, Phys. Rev. Lett. {\bf 91}, 097003 (2003).

\bibitem{Kumar03} B. Kumar and B.S. Shastry, Phys. Rev. B
{\bf 68}, 104508 (2003).

\bibitem{Ogata03a} M. Ogata, J. Phys. Soc. Jpn. {\bf 72}, 1839
(2003).

\bibitem{Imada98} M. Imada, A. Fujimori, and Y. Tokura,
Rev. Mod. Phys. {\bf 70}, 1039 (1998).

\bibitem{Powell05} B.J. Powell, R.H. McKenzie,
Phys. Rev. Lett. \textbf{94}, 047004 (2005).

\bibitem{Gan05b} J.Y. Gan, Y. Chen, Z.B. Su,
and F.C. Zhang, Phys. Rev. Lett. \textbf{94}, 067005 (2005).

\bibitem{Liu05} J. Liu, J. Schmalian, and N. Trivedi,
                Phys. Rev. Lett. {\bf 94}, 127003 (2005).

\bibitem{Baskaran87} G. Baskaran, Z. Zou, and P. W. Anderson, Solid State Commun.
\textbf{63}, 973 (1987).

\bibitem{Suzumura88} Y. Suzumura, Y. Hasegawa, and H. Fukuyama, J.
Phys. Soc. Jpn. \textbf{57}, 2768 (1988).

\bibitem{Yokoyama88} H. Yokoyama and H. Shiba,
J. Phys. Soc. Jpn {\bf 57}, 2482 (1988).

\bibitem{Anderson04b} P.W. Anderson, cond-mat/0406038 (unpublished).

\bibitem{Huang05} H.-X. Huang, Y.-Q. Li, F.-C. Zhang,
Phys. Rev. B \textbf{71}, 184514 (2005).

\bibitem{Poilblanc05} D. Poilblanc, Phys. Rev. B \textbf{72}, 060508 (R) (2005).

\bibitem{Li06} C. Li, S. Zhou, Z. Wang, Phys. Rev. B \textbf{73}, 060501(R) (2006).

\bibitem{Himeda99} A. Himeda and M. Ogata, Phys. Rev. B \textbf{60}, R9935 (1999).

\bibitem{Chen90} G.J. Chen, R. Joynt, F.C. Zhang, and G. Gros,
Phys. Rev. B \textbf{42}, 2662 (1990).

\bibitem{Giamarchi91} T. Giamarchi and C. Lhuillier, \textbf{43}, 12943 (1991).

\bibitem{Ivanov04} D.A. Ivanov, Phys. Rev. B \textbf{70}, 104503 (2004).

\bibitem{Shih04a} C.T. Shih, Y.C. Chen, C.P. Chou,
and T.K.~Lee, Phys. Rev. B \textbf{70}, 220502(R) (2004).

\bibitem{Ogata03b} M. Ogata and A. Himeda, J. Phys. Soc. Jpn. \textbf{72}, 374 (2003).

\bibitem{Tsuchiura99} H. Tsuchiura, Y. Tanaka, M. Ogata, and S. Kashiwaya,
J. Phys. Soc. Jpn. \textbf{68}, 2510 (1999).

\bibitem{Tsuchiura00} H. Tsuchiura, Y. Tanaka, M. Ogata, and S. Kashiwaya,
Phys. Rev. Lett. \textbf{84}, 3165 (2000).

\bibitem{Tsuchiura01} H. Tsuchiura, Y. Tanaka, M. Ogata, and S. Kashiwaya
Phys. Rev. B \textbf{64}, 140501 (2001).

\bibitem{Himeda97} A. Himeda, M. Ogata, Y. Tanaka, and S. Kashiwaya,
 J. Phys. Soc. Jpn. \textbf{66}, 3367 (1997).

\bibitem{Tsuchiura03} H. Tsuchiura, M. Ogata, Y. Tanaka, and S. Kashiwaya,
 Phys. Rev. B \textbf{68}, 012509 (2003).

\bibitem{Anderson05} P.W. Anderson, Phys. Rev. Lett. \textbf{96}, 017001 (2005).

\bibitem{Emery95} V.J. Emery and S.A. Kivelson, Nature {\bf 374},
434 (1995).

\bibitem{Lee00} D.-H. Lee, Phys. Rev. Lett. {\bf 84}, 2694 (2000)

\bibitem{Gros87} C. Gros, R. Joynt, and T.M. Rice, Phys. Rev. B {\bf 36}, 381 (1987).

\bibitem{Paramekanti04} A. Paramekanti, M. Randeria, and N. Trivedi,
         Phys. Rev. B \textbf{70}, 054504 (2004).

\bibitem{Gutzwiller63} M.C. Gutzwiller, Phys. Rev. Lett. \textbf{10}, 159 (1963).

\bibitem{Brinkman70} W.F. Brinkman and T.M. Rice, Phys. Rev. B \textbf{2}, 4302 (1970).

\bibitem{Vollhardt84} D. Vollhardt,  Rev. Mod. Phys. \textbf{56}, 99 (1984).

\bibitem{Seiler86} K. Seiler, C. Gros, T.M. Rice, K. Ueda, and D. Vollhardt,
 J. Low. Temp. Phys. {\bf 64}, 195 (1986).

\bibitem{Ogawa75} T. Ogawa, K. Kanda, and T. Matsubara, Prog. Theor. Phys. \textbf{53},
614 (1975).

\bibitem{Metzner88} W. Metzner and D. Vollhardt, Phys. Rev. B \textbf{37}, 7382 (1988).

\bibitem{Gebhard90} F. Gebhard, Phys. Rev. B \textbf{41}, 9452 (1990).

\bibitem{Buenemann05a} J. Bünemann, F. Gebhard,
K. Radnoczi, and P. Fazekas, J. Phys. Condens. Matter \textbf{17}
3807 (2005).

\bibitem{Buenemann05b} J. Bünemann, F. Gebhard, T. Ohm, S. Weiser,
and W. Weber, in `Frontiers in Magnetic Materials', A. Narlikar (Ed.),
Springer (2005).

\bibitem{Barnes76} S.E. Barnes, J. Phys. F \textbf{6}, 1375 (1976).

\bibitem{Coleman84} P. Coleman, Phys. Rev. B \textbf{29}, 3035 (1984).

\bibitem{Kotliar86} G. Kotliar and A.E. Ruckenstein,
Phys. Rev. Lett. \textbf{57}, 1362 (1986).

\bibitem{Baskaran87b}  G. Baskaran and P. W. Anderson,
                       Phys. Rev. B {\bf 37}, 580 (1988).

\bibitem{Wen96} X.-G. Wen and P.A. Lee,
                Phys. Rev. Lett. {\bf 76}, 503 (1996).

\bibitem{weng_99} Z. Y. Weng, D. N. Sheng, and C. S. Ting,
Phys. Rev. B \textbf{59}, 8943 (1999).

\bibitem{vnm_02a} V. N. Muthukumar and Z. Y. Weng,
Phys. Rev. B \textbf{65}, 174511 (2002).

\bibitem{vnm_02b} Z. Y. Weng and V. N. Muthukumar,
Phys. Rev. B \textbf{66}, 094509 (2002).

\bibitem{zhou_03} Yi Zhou, V. N. Muthukumar, and Zheng-Yu Weng,
Phys. Rev. B \textbf{67}, 064512 (2003).

\bibitem{Ivanov06} D.A. Ivanov, Phys. Rev. B {\bf 74}, 024525 (2006).

\bibitem{Capello05} M. Capello, F. Becca, M. Fabrizio, S. Sorella,
and E. Tosatti, Phys. Rev. Lett. {\bf 94}, 026406 (2005).

\bibitem{Watanabe06} T. Watanabe, H. Yokoyama, Y.Tanaka, and J. Inoue,
J. Phys. Soc. Jpn {\bf 75}, 074707, (2006).

\bibitem{Hellberg91} C.S. Hellberg, and E.J. Mele, Phys. Rev. Lett.
{\bf 67}, 2080 (1991).

\bibitem{Valenti92} R. Valenti, and C. Gros,
Phys. Rev. Lett. {\bf 68}, 2402 (1992).

\bibitem{lee_90} T.K. Lee, and L. N. Chang,
                 Phys. Rev. B \textbf{42}, 8720 (1990).

\bibitem{Ivanov03} D.A. Ivanov, and P.A. Lee,
                   Phys. Rev. B \textbf{68}, 132501 (2003).

\bibitem{Ran07} Y. Ran, M. Hermele, P.A. Lee, and X.-G. Wen,
                Phys. Rev. Lett. {\bf 98}, 117205 (2007).

\bibitem{Jastrow55} R. Jastrow, Phys. Rev. {\bf 98}, 1479 (1955).

\bibitem{Sorella02a} S. Sorella, G.B. Martins, F. Becca, C. Gazza,
L. Capriotti, A. Parola, and E. Dagotto, Phys. Rev. Lett. {\bf
88}, 117002 (2002).

\bibitem{Yunoki05a} S. Yunoki, E. Dagotto, and S. Sorella,
Phys. Rev. Lett. \textbf{94}, 037001 (2005).

\bibitem{Gros89a} C. Gros, Ann. Phys. \textbf{189}, 53 (1989).

\bibitem{Lugas06} M. Lugas, L. Spanu, F. Becca, and S. Sorella,
Phys. Rev. B {\bf 74}, 165122 (2006).

\bibitem{Vollhardt87} D. Vollhardt, P. Wölfle, and P.W. Anderson, Phys. Rev. B \textbf{35},
6703 (1987).

\bibitem{Rice85} T.M. Rice and K. Ueda, Phys. Rev. Lett. \textbf{55}, 995
(1985).

\bibitem{Varma86} C.M. Varma, W. Weber, and L.J. Randall, Phys. Rev. B \textbf{33}, 1015
(1986).

\bibitem{Hsu90} T.C. Hsu, Phys. Rev. B {\bf 41}, 11379 (1990).

\bibitem{Anderson04c} P.W. Anderson and N.P. Ong,
                      J. Phys. Chem. Solids {\bf 67}, 1 (2006).

\bibitem{Affleck88} I. Affleck, Z. Zou, T. Hsu,
and P.W. Anderson, Phys. Rev. B. \textbf{38}, 745 (1988).

\bibitem{Anderson02a} P.W. Anderson, J. Phys. Chem. Solids \textbf{63}, 2145 (2002).

\bibitem{Borisenko06} S.V. Borisenko {\it et al.},
Phys. Rev. Lett. \textbf{96}, 117004 (2006).

\bibitem{Kordyuk05} A.A. Kordyuk, S.V. Borisenko, A. Koitzsch,
 J. Fink, M. Knupfer, and H. Berger, Phys. Rev. B {\bf 71}, 214513 (2005); A.A. Kordyuk,
 private communication.

\bibitem{Bogdanov00} P.V. Bogdanov, A. Lanzara, S.A. Kellar, X.J. Zhou,
E.D. Lu, W.J. Zheng, G. Gu, J.-I. Shimoyama, K. Kishio, H. Ikeda,
R. Yoshizaki, Z. Hussain, and Z. X. Shen, Phys. Rev. Lett. {\bf 85}, 2581 (2000).

\bibitem{Johnson01}  P.D. Johnson, T. Valla, A.V. Fedorov, Z. Yusof,
B.O. Wells, Q. Li, A.R. Moodenbaugh, G.D. Gu, N. Koshizuka, C.
Kendziora, Sha Jian, and D.G. Hinks, Phys. Rev. Lett. {\bf 87},
177007 (2001).

\bibitem{Pavarini01} E. Pavarini, I. Dasgupta, T. Saha-Dasgupta, O. Jepsen, and O.K.
Andersen, Phys. Rev. Lett. {\bf 87} 047003 (2001).

\bibitem{Shih04b} C.T. Shih, T.K. Lee, R. Eder, C.-Y. Mou, and Y.C. Chen, Phys. Rev.
Lett. \textbf{92}, 227002 (2004).

\bibitem{Senechal05} D. Senechal, P.L. Lavertu, M.A. Marois, and A.M.S.
Tremblay, Phys. Rev. Lett. \textbf{94}, 156404 (2005).

\bibitem{Aichhorn05} M. Aichhorn and E. Arrigoni, Europhys. Lett. \textbf{71}, 117 (2005).

\bibitem{Wang06} Q.-H. Wang, Z.D. Wang, Y. Chen, and F.-C. Zhang,
Phys. Rev. B \textbf{73}, 092507 (2006).

\bibitem{Lake01} B. Lake {\it et al.},
Science \textbf{291}, 1759 (2001).

\bibitem{Lake02} B. Lake {\it et al.},
Nature \textbf{415}, 299 (2002).

\bibitem{Khaykovich02} B. Khaykovich, Y.S. Lee, R.W. Erwin,
                       S.-H. Lee, S. Wakimoto, K.J. Thomas,
                       M.A. Kastner, and R.J. Birgeneau,
Phys. Rev. B \textbf{66}, 014528 (2002).

\bibitem{Miller02} R.I. Miller, R.F. Kiefl, J.H. Brewer, J.E. Sonier,
J. Chakhalian, S. Dunsiger, G.D. Morris, A.N. Price,
D.A. Bonn, W.H. Hardy, and R. Liang, Phys. Rev. Lett. \textbf{88}, 137002 (2002).

\bibitem{Mitrovic01} V.F. Mitrovic, E.E. Sigmund, M. Eschrig, H.N. Bachman, W.P. Halperin,
A.P. Reyes, P. Kuhns,  and W.G. Moulton, Nature  \textbf{413}, 501 (2001).

\bibitem{Kakuyanagi02} K. Kakuyanagi, K.I. Kumagai,
and Y. Matsuda, Phys. Rev. B \textbf{65}, 060503(R) (2002).


\bibitem{Liang02} S.-D. Liang and T.K. Lee,  Phys. Rev. B \textbf{65}, 214529 (2002).


\bibitem{Tanuma98} Y. Tanuma, Y. Tanaka, M. Ogata, and S. Kashiwaya,
J. Phys. Soc. Jpn. \textbf{67}, 1118 (1998).

\bibitem{Tanuma99} Y. Tanuma, Y. Tanaka, M. Ogata, and S. Kashiwaya,
 Phys. Rev. B \textbf{60}, 9817 (1999).

\bibitem{Zhang03} F.-C. Zhang,
Phys. Rev. Lett. \textbf{90}, 207002 (2003).

\bibitem{Laughlin02} R.B. Laughlin, cond-mat/0209269 (unpublished).

\bibitem{Gan05a} J.Y. Gan, F.C. Zhang, Z.B. Su, Phys. Rev. B \textbf{71}, 014508 (2005).

\bibitem{Kopec04} T.K. Kopec, Phys. Rev. B \textbf{70}, 054518 (2004).

\bibitem{Normand04} B. Normand, A.M. Oles, Phys. Rev. B \textbf{70}, 134407 (2004).

\bibitem{Yuan05} F. Yuan, Q. Yuan, and C.S. Ting, Phys. Rev. B \textbf{71}, 104505 (2005).

\bibitem{Seibold01}  G. Seibold and J. Lorenzana,
Phys. Rev. Lett. \textbf{86}, 2605 (2001).

\bibitem{Lorenzana03} J. Lorenzana and G. Seibold,
 Phys. Rev. Lett. \textbf{90}, 066404 (2003).

\bibitem{Seibold04} G. Seibold, F. Becca, P. Rubin, and J. Lorenzana,
Phys. Rev. B \textbf{69}, 155113 (2004).

\bibitem{Seibold05}  G. Seibold, J. Lorenzana, Phys. Rev. Lett. \textbf{94}, 107006 (2005).

\bibitem{Seibold06a} G. Seibold, J. Lorenzana,  Phys. Rev. B \textbf{73}, 144515 (2006).

\bibitem{Seibold06b} G. Seibold, J. Lorenzana, and M. Grilli,
cond-mat/0606010 (unpublished).

\bibitem{Horsch83} P. Horsch and T.A. Kaplan, J. Phys. C \textbf{16}, L1203 (1983).

\bibitem{Ceperley77} D. Ceperley, G. V. Chester, and M.H. Kalos,
Phys. Rev. B \textbf{16}, 3081 (1977).

\bibitem{Yokoyama96} H. Yokoyama and M. Ogata, J. Phys. Soc. Jpn. {\bf 65}, 3615
(1996).

\bibitem{Trivedi90} N. Trivedi and D.M. Ceperly, Phys. Rev. B {\bf 41}, 4552 (1990).

\bibitem{Buonaura98} M.C. Buonaura and S. Sorella, Phys. Rev. B {\bf 57}, 11446 (1998).

\bibitem{Ivanov00} D.A. Ivanov, P.A. Lee, and X.-G. Wen,
Phys. Rev. Lett. \textbf{84}, 3958 (2000).

\bibitem{Himeda00} A. Himeda and H. Ogata, Phys. Rev.
Lett. \textbf{85}, 4345 (2000).

\bibitem{Himeda02} A. Himeda, T. Kato, and M. Ogata,
Phys. Rev. Lett. \textbf{88}, 117001 (2002).

\bibitem{Weber06b} C. Weber, D. Poilblanc, S. Capponi, F. Mila, and C. Jaudet,
Phys. Rev. B {\bf 74}, 104506 (2006).

\bibitem{Shih98} C.T. Shih, Y.C. Chen, H.Q. Lin, and T.K. Lee,
Phys. Rev. Lett. \textbf{81}, 1294 (1998).

\bibitem{Lee02} T.K. Lee, C.T. Shih, Y.C. Chen, and H.Q. Lin,
Phys. Rev. Lett. \textbf{89}, 279702 (2002).

\bibitem{Sorella02b} S. Sorella, A. Parola, F. Becca, L. Capriotti,
C. Gazza, E. Dagotto, and G. Martins, Phys. Rev. Lett.
\textbf{89}, 279703 (2002).

\bibitem{Anderson04a} P.W. Anderson, P.A. Lee, M. Randeria, T.M. Rice, N. Trivedi,
and F.C. Zhang, J. Phys. Condens. Matter \textbf{16}, R755
(2004).

\bibitem{Yamase00a} H. Yamase, and H. Kohno,
J. Phys. Soc. Jpn. {\bf 69}, 2151 (2000).

\bibitem{Yamase00b} H. Yamase, and H. Kohno,
J. Phys. Soc. Jpn. {\bf 69}, 332 (2000).

\bibitem{Halboth00} C. J. Halboth, and W. Metzner,
Phys. Rev. Lett. {\bf 85}, 5162 (2000).

\bibitem{Scalapino93} D.J. Scalapino, and S.R. White,
S.-C. Zhang, Phys. Rev. B \textbf{47}, 7995  (1993).

\bibitem{Yang06} K.-Y. Yang, {\it et al.},
                  Phys. Rev. B {\bf 73}, 224513 (2006)

\bibitem{Paramekanti98}
A. Paramekanti, N. Trivedi and M. Randeria, Phys. Rev. B {\bf 57},
11639 (1998).

\bibitem{Uemura89} Y.J. Uemura {\it et al.},
Phys. Rev. Lett. {\bf 62}, 2317 (1989).














\bibitem{Randeria05} M. Randeria, R. Sensarma, N. Trivedi, and F.-C. Zhang,
Phys. Rev. Lett. \textbf{95}, 137001 (2005).

\bibitem{Yunoki05b} S. Yunoki, Phys. Rev. B {\bf 72}, 092505 (2005).

\bibitem{Nave06} C.P. Nave, D.A.Ivanov, and P.A. Lee,
                 Phys. Rev. B {\bf 73}, 104502 (2006).

\bibitem{Yang07} H.-Y. Yang, F. Yang, Y.-J. Jiang, and T. Li,
                 J. Phys.: Condens. Matter {\bf 19}, 016217 (2007).

\bibitem{Chou06} C.-P. Chou, T.K. Lee, and C.-M. Ho,
                 Phys. Rev. B {\bf 74}, 092503 (2006).

\bibitem{Bieri06} S. Bieri and D. Ivanov, Phys. Rev. B {\bf 75}, 035104 (2007).

\bibitem{Yunoki06} S. Yunoki, Phys. Rev. B {\bf 74}, 180504(R) (2006).

\bibitem{Harris67} A.B. Harris and R.V. Lange, Phys. Rev. \textbf{157}, 295 (1967).

\bibitem{Meinders93} M.B.J. Meinders, H. Eskes, and G.A. Sawatzky,
Phys. Rev. B \textbf{48}, 3916 (1993).

\bibitem{Eskes94a} H. Eskes and A. M. Oles, Phys. Rev. Lett. \textbf{73}, 1279 (1994).

\bibitem{Eskes94b} H. Eskes, A.M. Oles,
M.B.J. Meinders, and W. Stephan, Phys. Rev. B \textbf{50}, 17980
(1994).

\bibitem{Baeriswyl87} D. Baeriswyl, C. Gros, and T.M. Rice,
Phys. Rev. B {\bf 35}, 8391 (1987).

\bibitem{Hubbard63} J. Hubbard,
Proc. Roy. Soc. London A {\bf 276}, 238 (1963).

\bibitem{Randeria04} M. Randeria, A. Paramekanti, and N. Trivedi,
Phys. Rev. B {\bf 69}, 144509 (2004).

\bibitem{Kaminski06} A. Kaminski, S. Rosenkranz, H.M. Fretwell, M.R. Norman,
M. Randeria, J.C. Campuzano, J-M. Park, Z.Z. Li, and H. Raffy,
Phys. Rev. B \textbf{73}, 174511 (2006).

\bibitem{Yoshida05} T. Yoshida {\it et al.},
                   Phys. Rev. B {\bf 74}, 224510 (2006).

\bibitem{Paramekanti02} A. Paramekanti, and M. Randeria,
        Phys. Rev. B {\bf 66}, 214517 (2002).

\bibitem{Lee97} P.A. Lee, and X.-G. Wen,
Phys. Rev. Lett. \textbf{78}, 4111 (1997).

\bibitem{Wen98} X.-G. Wen, and P.A. Lee, Phys. Rev. Lett.
\textbf{80}, 2193 (1998).

\bibitem{Ioffe02} L.B. Ioffe, and A.J. Millis,
J. Phys. Chem. Solids, {\bf 63}, 2259 (2002).

\bibitem{Broun06} D.M. Broun, P.J. Turner, W.A. Huttema, S. Ozcan,
B. Morgan, R. Liang, W.N. Hardy, and D.A. Bonn,
 24th International Conference on Low Temperature Physics - LT24.
AIP Conference Proceedings {\bf  850}, 441 (2006).

\bibitem{Herbut05} I.F. Herbut, Phys. Rev. Lett. \textbf{94}, 237001
(2005).

\bibitem{Sensarma06} R. Sensarma, M. Randeria, N. Trivedi,
Phys. Rev. Lett. {\bf 98}, 027004 (2007).

\bibitem{Pines_Nozieres} D. Pines and P. Noziéres,
 {\it The Theory of Quantum Liquids} (Addison-Wesley) (1966).

\bibitem{Dzyaloshinskii03} I. Dzyaloshinskii,
Phys. Rev. B {\bf 68}, 085113 (2003).

\bibitem{Gros94} C. Gros, W. Wenzel, R. Valent\'i,
         G. Hülsenbeck, and J. Stolze, Europhys. Lett. {\bf 27}, 299 (1994).

\bibitem{Stanescu07} T.D. Stanescu, P.W. Phillips, and T.-P. Choy,
         Phys. Rev. B {\bf 75}, 104503 (2007).

\bibitem{YangRice06} K.-Y. Yang, T.M. Rice, and F.-C. Zhang,
                 Phys. Rev. B {\bf 73}, 174501 (2006).


\bibitem{Shen05} K.M. Shen, F. Ronning, D.H. Lu, F. Baumberger,
N.J.C. Ingle, W.S. Lee, W. Meevasana, Y. Kohsaka, M. Azuma, M.
Takano, H. Takagi, and Z.-X. Shen, Science {\bf 307}, 901 (2005).

\bibitem{Kohsaka04} Y. Kohsaka, K. Iwaya, S. Satow, T. Hanaguri,
 M. Azuma, M. Takano, and H. Takagi,
 Phys. Rev. Lett.  {\bf 93}, 097004 (2004).

\bibitem{Putikka98} W.O. Putikka, M.U. Luchini, and R.R.P. Singh,
  Phys. Rev. Lett. {\bf 81}, 2966 (1998).

\bibitem{Maier02} T.A. Maier, T. Pruschke, and M. Jarrell,
 Phys. Rev. B {\bf 66}, 075102 (2002).

\bibitem{Campuzano96} J.C. Campuzano, H. Ding,
 M.R. Norman, M. Randeira, A.F. Bellman, T. Yokoya,
 T. Takahashi, H. Katayama-Yoshida, T. Mochiku, and K. Kadowaki,
 Phys. Rev. B {\bf 53}, R14737 (1996).

\bibitem{Ronning98} F. Ronning, C. Kim, D.L. Feng, D.S. Marshall,
A.G. Loeser, L.L. Miller, J.N. Eckstein, I. Bozovic, and Z.-X.
Shen, Science {\bf 282}, 2067 (1998).


\bibitem{Civelli05} M. Civelli , M. Capone, S.S. Kancharla, O.
Parcollet, G. Kotliar, Phys. Rev. Lett. {\bf 95}, 106402 (2005).

\bibitem{Marshall55} W. Marshall,
Proc. Roy. Soc. (London) A {\bf 232}, 48 (1955).

\bibitem{Richter94} J. Richter, N.B. Ivanov, and K. Retzlaff,
  Europhys.  Lett. {\bf 25}, 545 (1994).


\bibitem{Anderson06a} P.W. Anderson,
Nature Physics {\bf 2}, 626 (2006).


\bibitem{drzazga_88} M. Drzazga, A. Kampf, E. Müller-Hartmann, and H. A. Wischmann,
Z. Phys. B Condensed Matter \textbf{74}, 67 (1989).

\bibitem{pwa_58} P. W. Anderson, Phys. Rev. \textbf{112}, 1900 (1958).

\end{thebibliography}
\end{document}